\documentclass[twoside,12pt,a4paper,here]{book}
\usepackage[authoryear]{natbib}
\bibpunct{(}{)}{;}{y}{}{,}
\usepackage[center,footnotesize,it,md,tt,oneline]{caption2}
\usepackage[english]{babel}

\usepackage{doc} 
\usepackage{makeidx}
\usepackage{latexsym}
\usepackage[dvips]{graphicx}
\usepackage{rotating}
\usepackage{fancyheadings}
\makeatletter
\newcommand\tabcaption{\def\@captype{table}\caption}
\newcommand\figcaption{\def\@captype{table}\caption}
\makeatother

\usepackage{amssymb}
\usepackage{amsmath}

\usepackage{graphicx}

\usepackage{epsf}

\usepackage{lscape}
\usepackage{rotate}

\usepackage{euscript}
\usepackage{latexsym}
\usepackage{textcomp}
\usepackage[T1]{fontenc}

\begin{document}
\pagestyle{empty}
\begin{center}
\begin{minipage}[h]{2cm}
\vspace{-3.0cm}
\includegraphics[width=3cm]{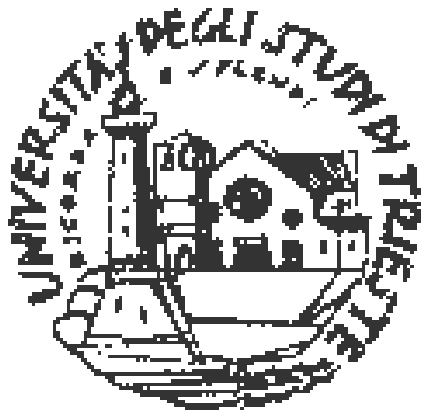}
\end{minipage}
\end{center}
\begin{center}
\smallskip
\Large {\bf UNIVERSIT\`A DEGLI STUDI DI TRIESTE}\\
\normalsize {\bf Dipartimento di Astronomia}\\
\vspace{-2.0cm}
 ~\\~\\~\\
\normalsize{XVI CICLO DEL DOTTORATO DI RICERCA IN FISICA}\\
\normalsize{Anno Accademico 2002/2003}\\
\end{center}
\vspace{1.0cm}
\begin{center}
\LARGE{\bf 
Dynamical Evolution and Galaxy Populations \\in the Cluster 
ABCG\,209 at z=0.2} 
~\\~\\~\\
\end{center}
\begin{minipage}[t]{3.6cm}
\begin{center}
DOTTORANDA \\ 
Amata Mercurio\\
\end{center}
\end{minipage}
\hfill 
\begin{minipage}[t]{13cm}
\begin{center}
COORDINATORE DEL COLLEGIO DEI DOCENTI \\
Chiar.mo Prof. Gaetano Senatore\\
Universit\`a degli Studi di Trieste\\
\smallskip
TUTORE\\
Dott.ssa Marisa Girardi\\
Universit\`a degli Studi di Trieste\\
\smallskip
CORRELATORE\\
Dott.ssa Paola Merluzzi\\
INAF--Osservatorio Astronomico di Capodimonte\\
\end{center}
\end{minipage} 


\newpage

 \ \ \

\newpage
\vspace{3.0cm}

\begin{minipage}[t]{13cm}
\begin{flushright}

Chiedete e vi sar\`a dato\\
cercate e troverete\\
bussate e vi sar\`a aperto.\\
\smallskip
Infatti chi chiede riceve\\
chi cerca trova\\
a chi bussa sar\`a aperto\\
\ \ \ \ {\it San Matteo 7,7-9}\\
\vspace{+4.0cm}
{\Large \it a zia Edda} \\

\end{flushright}
\end{minipage}
%
\begin{minipage}[t]{5cm}
\begin{flushright}
\end{flushright}
\end{minipage} 
\vspace{1cm}



\newpage

 \ \ \

\newpage

\begin{center}

{\it Delicato e faticante \`e parlare con efficacia a uno (alla sua anima,
cio\`e al fondo libero del suo essere) della ricchezza e della
bellezza, quelle interiori, per\`o, cos\`{\i} dissimili da quelle
esteriori. 

Lo si pu\`o fare soltanto e appropriatamente se l'anima di chi parla
\`e convinta e persuasa tutta e soltanto non per deduzione logica, ma
per la forza dell'aver esperito l'immensa tenuta della sua libert\`a
nella scelta della bellezza e ricchezza interiore, e nel letificante
aver allontanato ogni adesione alla legge del pi\`u forte.

Il parlare entra, allora, nel mondo vitale della sinergia liberante,
dove viene confermata la necessit\`a del direttamente, le necessit\`a
cio\`e che nel rapporto sia stretta quasi all'infinito la presenza di
mediazioni oggettive.}

\end{center}
\begin{flushright}
da ``Aprire su Paidea'', a cura di Edda Ducci
\end{flushright}

\bigskip

\vspace{2.0cm}

Dedico un ringraziamento speciale a tutti coloro che hanno saputo
parlare {\it da anima ad anima}, aiutandomi a portare in superficie la
bellezza dell'interiore. Le loro parole, ma soprattutto i loro silenzi
carichi di significato mi hanno permesso di superare tutte le
difficolt\`a con la forza del sentire.

Grazie ai miei educatori che hanno saputo {\it distinguere tra sapere
e sapere} impedendomi di diventare una semplice vite della {\it gabbia
della mera comunit\`a accademica che offende e lede la felice
agilit\`a retaggio della libert\`a del singolo.}

\tableofcontents
\pagestyle{headings}
\large
\chapter*{ \huge Introduction}
\markboth{Introduction}{Introduction}
\addcontentsline{toc}{chapter}{\numberline{}Introduction}
\vspace{2.0cm}
\normalsize
\bibliographystyle{natbib}
\newcommand{\ct}{\citeyear}
\newcommand{\cit}{\citealt*}

\bigskip

Understanding the formation and the evolution of galaxies and galaxy
clusters, has always been of great interest for astronomers.

Although the study of galaxy clusters has a long history, there are
many unresolved and still debated questions.

\begin{description}

\item[-] When  did galaxies and galaxy clusters form? And in which order?

\item[-] How do the global properties of clusters evolve?

\item[-] Which are the steps in cluster evolution that cause the
luminosity and morphological segregation of galaxies?

\item[-] How do the gas and  the galaxy stellar populations evolve in clusters?

\end{description}

The main current challenges in understanding the properties of galaxy
clusters is the evidence that many of them are still in an early stage
of formation. Cluster of galaxies are no longer believed to be simple,
relaxed structures, but are interpreted within the hierarchical growth
via mergers of smaller groups/clusters. Although many observational
evidences exist that there is a large fraction of clusters which are
really still forming, the fundamental question raised from these
observations remain: are clusters generally young or old?

To address this issue one needs to have measurements of subclustering
properties of a large sample of clusters, but at the same time it is
fundamental to precisely characterize galaxies belonging to different
structures and environments inside a single cluster.

Due to the complexity in the distribution of different cluster
constituents (galaxies, gas, dark matter), combined multi-wavelength
observations and detailed analysis are necessary to understand the
complex history of clusters formation and evolution.

Up to now complete and detailed analysis have been performed on many
nearby clusters. Moreover, the use of large ground-based telescopes
(4-8m diameter) and of the Space Telescope allowed studies on clusters
at high redshifts: up to z $\sim$ 1. This increasing amount of data
induced astronomers to perform statistical studies, which are crucial
to trace the large scale structure and to provide a measure of the
mean density of the universe.

Clusters at intermediate--redshifts (z = 0.1-0.3) seem to represent an
optimal compromise for these studies, because they allow to achieve
the accuracy needed to study the connection between the cluster
dynamics and the properties of galaxy populations and, at the same
time, to span look--back time of some Gyr. 

This thesis work is focused on the analysis of the galaxy cluster
ABCG\,209, at z$\sim$ 0.2, which is characterized by a strong
dynamical evolution. The data sample used here consists in new optical
data (EMMI-NTT B, V and R band images and MOS spectra) and archive
data (CFHT12k B and R images), plus archive X-ray (Chandra) and radio
(VLA) observations.

After a short historical overview and a brief description of cluster
properties (Chapter 1), the optical analysis of the cluster is carried
out by studying the internal structure and the dynamics of the
cluster, through the spectroscopic investigation of member galaxies
(Chapter 2). The luminosity function, which is crucial in order to
trace the photometric properties of the cluster, such as luminosity
segregation, and for understanding the evolving state of this cluster
is studied in Chapter 3. The investigation of the connection between
the internal cluster dynamics and the star formation history of member
galaxies was performed through two independent approaches : i) the
study of the colour--magnitude relation and of the luminosity function
in different environments (Chapter 4) and ii) the spectral
classification of cluster galaxies, and the comparison with stellar
evolution models (Chapter 5).

\large
\chapter{\Large Cluster properties}
\markboth{Chapter 1}{Chapter 1}
\normalsize
\bibliographystyle{natbib}

\def\lesssim{\mathrel{\hbox{\rlap{\hbox{\lower4pt\hbox{$\sim$}}}\hbox{$<$}}}}
\def\gtrsim{\mathrel{\hbox{\rlap{\hbox{\lower4pt\hbox{$\sim$}}}\hbox{$>$}}}}
\newcommand{\mincir}{\raise -2.truept\hbox{\rlap{\hbox{$\sim$}}\raise5.truept
\hbox{$<$}\ }}
\newcommand{\magcir}{\raise -2.truept\hbox{\rlap{\hbox{$\sim$}}\raise5.truept
\hbox{$>$}\ }}
\newcommand{\siml}{\raise -2.truept\hbox{\rlap{\hbox{$\sim$}}\raise5.truept
\hbox{$<$}\ }}
\newcommand{\simg}{\raise -2.truept\hbox{\rlap{\hbox{$\sim$}}\raise5.truept
\hbox{$>$}\ }}
\newcommand{\be}{\begin{equation}}
\newcommand{\ee}{\end{equation}}
\newcommand{\ba}{\begin{eqnarray}}
\newcommand{\ea}{\end{eqnarray}}
\newcommand {\h} {$\mathrm{h^{-1}}$ Mpc $\;$}
\newcommand {\kpc} {$\mathrm{h^{-1}}$ kpc}
\newcommand {\hh} {$\mathrm{h^{-1}}$ Mpc}
\newcommand {\ks} {km~s$^{-1} \;$} 
\newcommand {\kss} {km~s$^{-1}$}
\newcommand {\msun} {$\mathrm{h^{-1} \  M_{\odot} \;}$}
\newcommand {\m} {$\mathrm{M_{\odot} \;}$}
\newcommand {\ml} {$\mathrm{h \, M_{\odot}/L_{\odot} \;}$}
\newcommand {\mll} {$\mathrm{h \, M_{\odot}/L_{\odot}}$}
\newcommand{\vel}{\,{\rm km\,s^{-1}}}

Galaxy clusters are the largest gravitationally bound structures in
the universe, which have collapsed very recently or are still
collapsing. They typically contain 10$^2$-10$^3$ galaxies in a region
of few Mpc, with a total mass that can be larger than 10$^{15}$ solar
masses\footnote{\footnotesize 1 Mpc = 3.1 $\cdot$ 10$^{24}$ cm; one
solar mass corresponds to 1.989 10$^{33}$ g.}. X-ray observations show
that clusters are embedded in an ``intra--cluster medium'' of hot
plasma, with typical temperatures ranging from 2 to 14 keV, and with
central densities of $\sim$ 10$^{-3}$ atoms cm$^{-3}$. The hot plasma
is detected through X-ray emission, produced by thermal bremsstrahlung
radiation, with typical luminosity of L$_X$ $\sim$ 10$^{43-45}$ erg
s$^{-1}$.  Radio observations prove the existence of radio emission
associated with galaxy clusters. This is synchrotron emission due to
the interaction between non--thermal population of relativistic
electrons (with a power-law energy distribution) and a magnetic field.

Clusters are interesting objects both as individual units and as
tracers of large scale structures. They offer an optimal laboratory to
study many different astrophysical problems such as the form of the
initial fluctuation spectrum, the evolution and the formation of
galaxies, the environmental effects on galaxies, and the nature and
quantity of dark matter in the universe.

An increasing amount of data revealed that clusters are very complex
systems and pointed out the necessity of studying these structures
using various complementary approaches. 

In this chapter, I will give first a short historical overview of the
study of galaxy clusters, and then I will outline their basic
properties at different wavelengths.

\section{Historical Overview}

The study of galaxy clusters has a long history: early observations of
{\it ``nebulae''} , described at the beginning of the XX century
(\cit{wol02}, \ct{wol06}) already recognized the clustering of
galaxies. Later, the work of Shapley and Ames (\ct{sha32}) drew out,
for the first time, many of the features of the large scale galactic
distribution. From their studies of the distribution of galaxies
brighter than 13$^{th}$ magnitude, they were able to delineate the
Virgo cluster, several concentrations of clusters at greater
distances, and to point out the asymmetry in the distribution of
galaxies between the northern and the southern galactic hemispheres.

In 1933 Fritz Zwicky (\ct{zwi33}) first noticed that the dynamical
determinations of cluster masses require a very large amount of unseen
material (dark matter). Since then, galaxy clusters have been studied
in detail and at different wavelengths, but his original conclusion
remained unchanged. Few years later, he discussed the {\it clustering
of nebulae} from a visual inspection of photographic plates of the
survey from the 18 inch (0.5 m) Schmidt telescope at Palomar, by
noting that elliptical galaxies are much more strongly clustered than
late-type galaxies (\cit{zwi38}, \cit{zwi42}).

The first systematic optical study of clusters properties is due to
Abell (\ct{abe58}), who compiled an extensive, statistically complete,
catalogue\footnote{\footnotesize Abell's catalogue covers the northern
areas of sky with a declination greater than 20$^{o}$.} of rich
clusters.  He defined as cluster the region of the sky where the
surface numerical density of galaxies is higher than that of the
adjacent field, and obtained an estimate of the redshift
\footnote{\footnotesize The redshift z of an extragalactic object is
the shift in the observed wavelength $\lambda_{\mathrm{obs}}$ of any
feature in the spectrum of the object relative to the laboratory
wavelength $\lambda_{\mathrm{lab}}$: z = $(\lambda_{\mathrm{obs}} -
\lambda_{\mathrm{lab}})/\lambda_{\mathrm{lab}}$.} from the
magnitude of the tenth brightest galaxy in the cluster.  In the
compilation of the catalogue, he selected galaxy clusters on the basis
of the following criteria: {\bf i)} the cluster had to contain at
least 50 galaxies in the magnitude range m$_3$--m$_3$+2, where m$_3$
is the magnitude of the third brightest galaxy; {\bf ii)} these
galaxies had to be contained within a circle of radius R$_A$= 1.5
h$_{100}^{-1}$ Mpc (Abell radius)\footnote{\footnotesize h$_{100}$ =
H$_0$/100 km s$^{-1}$ Mpc$^{-1}$.}; {\bf iii)} the estimated cluster
redshift had to be in the range 0.02 $\le$ z $\le$ 0.20. Abell gave
also an estimate of position of the centre and of the distance of
clusters. The resulting catalogue, consisting of 1682 clusters,
remained the fundamental reference for studies on galaxy clusters
until Abell, Corwin, \& Olowin (\ct{abe89}) published an improved and
expanded catalogue, including the southern sky. These catalogues
constituted the basis for many cosmological studies and still now
represent important resources in the study of galaxy clusters.

Zwicky and his collaborators (Zwicky et al. \ct{zwi61}) constructed
another extensive catalogue of rich galaxy clusters
\footnote{\footnotesize Zwicky's catalogue is confined to the northern
areas of sky (declination greater than -3$^{o}$).}, by identifying
clusters as enhancements in the surface number density of galaxies on
the National Geographic Society-Palomar Observatory Sky Survey
(Minkowsky \& Abell, \ct{min63}), but using selection criteria less
strict than Abell's catalogue.

He determined the boundary (scale size) of clusters by the contour (or
``isopleth'') where the surface density of galaxies falls to twice the
local background density. This isopleth had to contain at least 50
galaxies in the magnitude range m$_1$--m$_1$+3, where m$_1$ is the
magnitude of the brightest galaxy. No distance limits were specified,
although very nearby clusters, such as Virgo, were not included
because they extended over several plates. Zwicky's catalogue gave
also an estimate of the centre coordinates, the diameter, and the
redshift.

Abell and Zwicky reported in their catalogues the richness of galaxy
clusters, that is a measure of the number of galaxies associated with
that cluster. Abell (\ct{abe58}) defined the richness as the number of
galaxies contained in one Abell radius and within a magnitude limit of
m$_3$+2, whereas Zwicky et al. (\ct{zwi61}) defined richness as the
total number of galaxies visible on the red Sky Survey plates within
the cluster isopleth.

Abell (\ct{abe58}) divided his clusters into richness groups (see
Table \ref{ricA}), using criteria that are independent from the
distance\footnote{\footnotesize Just (\ct{jus59}) has found only a
slight richness-distance correlation in Abell's catalogue, and,
moreover, this effect could be explained by an incompleteness of
$\sim$ 10\% in the catalogue for distant clusters.}. Abell richnesses,
generally, correlate well with the number of galaxies (except for some
individual cases) and are very useful for statistical studies, but
must be used with caution in studies of individual clusters. Because
of the presence of background galaxies, stating with absolute
confidence that any given galaxy belongs to a given cluster is not
possible. This is the reason why richness is only a statistical
measure of the population of a cluster, based on some operational
definition.

\begin{table}[h]
     \caption[]{Abell richness classes.}
     \label{ricA}
      $$
           \begin{array}{|c|c|c|c|c|c|c|}
            \hline
\mathrm{Class} & 0 & 1 & 2 & 3 & 4 & 5\\
            \hline                                          
\mathrm{Richness} & 30-49 & 50-79 & 80-129 & 130-199 & 200-299 & >300\\
            \hline                                          
         \end{array}
      $$
         \end{table}

In 1966, X-ray emission was detected for the first time in the centre
of the Virgo cluster (Byram et al. \ct{byr66}; Bradt et
al. \ct{bra67}). Five years later, X-ray sources were also detected in
the direction of Coma and Perseus clusters (Fritz et al. \ct{fri71};
Gursky et al. \ct{gur71a}, \ct{gur71b}; Meekins et al. \ct{mee71}) and
\citet{cav71} suggested that extragalactic X-ray
sources are generally associated with groups or clusters of galaxies.

The launch of X-ray astronomy satellites allowed surveys of the entire
sky in order to search for X-ray emissions. These observations
established a number of properties of the X-Ray sources associated
with clusters (see Sarazin \ct{sar86}, \ct{sar88} for a
review). Particularly, they showed that clusters are extremely
luminous, (L$_\mathrm{X} \sim 10^{43-45}$ erg s$^{-1}$), that X-ray
sources associated to galaxy clusters are spatially extended and do
not vary temporally in their brightness (Kellogs et al. \ct{kel72};
Forman et al. \ct{for72}; Elvis \ct{elv76}).

The analysis of X-ray spectra of clusters showed that the X-ray
emission is due to a diffuse plasma of either thermal or non-thermal
electrons. Although several emission mechanisms were proposed, this
emission is mainly produced by thermal bremsstrahlung radiation, with
L$_X$ $\sim$ 10$^{43-45}$ erg s$^{-1}$. This interpretation of
cluster X-ray spectra requires that the space between galaxies in
clusters be filled with very hot ( $\sim$ 10$^{8}$ K) gas, which could
have simply fallen into clusters and stored since the formation of the
universe (Gunn \& Gott \ct{gun72}). Remarkably, the total mass of
intra--cluster medium is comparable to the total mass of the stars of
all the galaxies in the cluster.

In 1976, X-ray line emission from iron was detected in the Perseus
cluster (Mitchell et al. \ct{mit76}) and consequently also in Coma and
Virgo (Serlemitsos et al. \ct{ser77}). The emission mechanism for this
line is thermal, and its detection confirmed the thermal
interpretation of clusters X-ray emission. However, the only known
sources of significant quantities of iron are nuclear reactions in
stars. Since no significant populations of stars have been observed
outside galaxies and the abundance of iron in the intra-cluster medium
(ICM) is similar to that in stars, a substantial portion of this gas
must have been ejected from stars in galaxies (Bahcall \& Sarazin
\ct{bah77}).

The association between radio sources and galaxy clusters was first
established by Mills (\ct{mil60}) and van den Berg (\ct{van61}).
Radio emission from Abell's clusters was also studied at 1.4 GHz by
Owen (\ct{owe75}), Jaffe \& Perola (\ct{jaf75}) and Owen et
al. (\ct{owe82}).

These observations suggested that radio emission is synchrotron
emission due to the interaction of non-thermal population of
relativistic electrons with a magnetic field (Robertson
\ct{rob85}). Even though radio emission from clusters is mainly due to
sources associated with individual radio
galaxies\footnote{\footnotesize Strong radio emission is associated
mainly with giant elliptical galaxies, which occur preferentially in
clusters.}, diffuse radio emission is detected with no apparent parent
galaxy and there seems to be a correlation between cluster radio
emission and cluster morphology.

\section{Cluster components}

Clusters of galaxies are constituted by i) galaxies, which represent
only a few percentage of the total mass, by ii) the intercluster
medium (ICM), a hot, very rarefied gas (10$^{-3}$ atoms cm$^{-3}$), which
is between 10\% and 30\% of the total mass, and finally by iii) the
dark matter, whose existence is inferred from its gravitational
effects on galaxies, perhaps representing 70--80\% of the total
mass. 

The wavelength-dependence of the cluster properties is related to the
physical phenomena that dominate in these systems. From the optical
and near IR passbands ($\lambda$=0.6--2.2$\mu$) it is possible to
derive the spatial distribution and the properties of galaxies and the
distortion of the images of background galaxies due to the
gravitational lensing of the cluster. Dynamical studies are also
needed in order to quantify the amount of dark matter and to study its
distribution. In the X--ray passbands (0.1--10 keV), the ICM is
detected and, under some assumptions, X-ray observations can lead to
estimate the total cluster binding mass. The millimeter wavebands
(200-300 Ghz) are ideal for the detection of
Sunyaev-Zeldovich\footnote{\footnotesize The Sunyaev--Zeldovich effect
\ct{sun72} is a perturbation of the microwave background
as it passes through the hot gas of galaxy clusters.} (SZ) effect,
which is independent from the redshift. The radio observations show
the interaction between radio galaxies and the ICM, revealing the
presence of large--scale magnetic fields and also a population of
relativistic particles whose origins and evolution are still under
discussion.


\subsection{Galaxies}

Clusters are characterized by the presence of a wide range of galaxy
morphologies, with the fraction of different morphological types
varying between clusters and also within individual clusters.

The first study that provided a quantitative evidence that the mixture
of morphological types varies within individual clusters was due to
Oemler (\ct{oem74}). Oemler discovered that the fraction of spirals
($f(SP)$) in centrally-concentrated clusters increases with
radius, providing the first example of a morphology-radius relation in
clusters.

The first large scale study of morphological segregation was made by
Dressler (\ct{dre80}), who obtained the morphological types of $\sim$
6000 galaxies in 55 clusters. Dressler's analysis underlined that
there is a relationship between galaxy types and the local density of
galaxies (morphology--density relation).


The necessity of classifying clusters on the basis of their properties
is connected with the need of simplifying this complex reality. A
number of different clusters' properties have been used to construct
morphological classification systems.

Clusters have been classified on the basis of the spatial distribution
of galaxies as Zwicky et al. (\ct{zwi61}), on the degree to which the
cluster is dominated by its brightest galaxies (Bautz \& Morgan
\ct{bau70}), on the basis of the nature and distribution of the ten
brightest cluster galaxies (Rood \& Sastry \ct{roo71}), and on the
content of the mix of different morphological types (Morgan \ct{mor61}
and Oemler \ct{oem74}).

These different systems are highly correlated, and it seems that
clusters can be represented as a one-dimensional sequence, running
from regular to irregular clusters (Abell \ct{abe65}, \ct{abe75}).
Regular clusters are highly symmetric, have a core with a high
concentration of galaxies toward the center, tend to be compact and
spiral-poor or cD clusters.  Subclustering is weak or absent in {\it
regular} clusters, whereas {\it irregular} clusters have little
symmetry or central concentration, are spiral rich, and often show
significant subclustering. This suggests that regular clusters are, in
some sense, dynamically relaxed systems, while the irregular clusters
are dynamically less-evolved and have preserved roughly their
structure since the epoch of formation.

\subsection{Diffuse gas}

Hot diffuse gas fills the volume between the galaxies in a cluster.
This gas is heated by the energy released during the initial
gravitational collapse. This heating can be a violent process as gas
clouds enveloping groups of galaxies collide and merge to become a
cluster. Then the gas cools slowly and forms a quasi-hydrostatic
``atmosphere''.

The virial theorem implies that the square of the ICM thermal velocity
(sound speed) is comparable to the gravitational potential. This means
that gas can support itself close to the hydrostatic equilibrium in
the gravitational field of a cluster only if its sound speed is
similar to the velocity dispersion of the cluster. Taking into account
that the cluster velocity dispersion is generally in the range
10$^2$-10$^3$ km s$^{-1}$, the gas temperature has to be
10$^7$-10$^8$ K. At these high temperatures the gas loses energy
through thermal bremsstrahlung process, which produces the diffuse
X-ray radiation.

Gas cools to low temperature due to its X-ray radiation with a time
scale t$_{\mathrm{cool}} \propto \frac{T^{\alpha}}{n}$, where $n$ is
the gas density and $\alpha$ is in the range [-1/2,1/2]. When the
cooling time is shorter than the Hubble time, a cooling flow is formed
(see \cit{fab94} for a review). The gas is believed to remain close to
the hydrostatic equilibrium because the cooling time, although shorter
than the Hubble time, is generally longer than the gravitational
free-fall time.

The observational evidence of a cooling flow is a sharp peak in the
X-ray surface brightness, because the gas density is rising steeply in
this region, and therefore the radiative cooling time is shortest. The
observed surface brightness depends upon the square of the gas
density, and only weakly on the temperature, so this result is not
model--dependent.

However, X-ray spectra of cooling flow clusters have shown that the
gas temperature drops by only a factor 3 in the central region, where
the radiative cooling time is of the order of 10$^8$ yr, so that it
does not appear to cool further (Peterson et al. \ct{pet01},
\ct{pet03}; Tamura et al. \ct{tam01}; Sakelliou et
al. \ct{sak02}). Chandra imaging and spectroscopy confirm this result
and show that the temperature generally drops smoothly to the
centre. This lack of cooler gas is known as the 'cooling flow
problem'. Some physical processes are proposed to solve this problem,
whose crucial point is to understand the source of heating of the gas.

Two major sources of heat are jets from a central massive black hole
(Tucker \& Rosner \ct{tuc83}; Binney \& Tabor \ct{bin95}; Soker et
al.\ct{sok02}; Br$\ddot{\mathrm{u}}$ggen \& Kaiser \ct{brug01}; David
et al. \ct{dav01}; Churazov et al. \ct{chu01}, \ct{chu02};
B$\ddot{\mathrm{o}}$hringer et al. \ct{boh02}; Reynolds, Heinz \& Begelman
\ct{rey02}; Nulsen et al. \ct{nul02}) and conduction from the outer hot cluster
atmosphere (Narayan \& Medvedev \ct{nar01}; Voigt et al. \ct{voi02}; Fabian et
al. \ct{fab02}; Ruszkowski \& Begelman \ct{rus02}).

Most cooling flows have a central radio source which demonstrates that
the central black hole is active, but not all those radio sources are
sufficiently powerful. In fact radio source heating can work
energetically for a cool, low-luminosity cluster, but it seems
implausible for hot, high-luminosity clusters because sources are
required to reach very high, and unobserved, levels of power. A
further issue is that the coolest gas in the cluster cores is found to
be directly next to the radio lobes (Fabian et al. \ct{fab00}; Blanton
et al. \ct{blan01}; McNamara et al. \ct{mcn01}). Conduction, on the
other hand, is another possible solution for many clusters. However,
the presence of tangled magnetic fields in cluster cores (see
e.g. Carilli \& Taylor \ct{car02}) makes the level of conductivity
uncertain.

Another heat source is explored by Fabian(\ct{fab03}): the
gravitational potential of the cluster core.  In a standard cooling
flow the gas cools to the local virial temperature (Fabian, Nulsen \&
Canizares \ct{fab84}), which declines toward the centre. As the gas
flows inward, gravitational (accretion) energy is released which is
assumed to be radiated locally. The virial theorem implies that the
gravitational energy is comparable to the initial thermal energy of
the hot gas, so is a well-matched heat source. If the gravitational
energy is not radiated locally but instead can be displaced to smaller
radii, then it can offset the radiative cooling of the inner, cooler
parts of the flow.

In this process, the ICM is assumed to be magnetized and therefore
thermally unstable and, crucially, blobs, which are denser and cooler
than the surroundings, are assumed to cool more quickly, separate from
the bulk flow and to fall inward. The gravitational energy released by
infalling blobs can offset radiative cooling in the inner parts of the
flow.

However partially resolving the blobs is currently impossible if they
are smaller than about 100 pc, so the physical process at the basis of
the gas heating is not yet understand. The cooling flow problem is
very important because of its relevance to galaxy formation (Fabian et
al. \ct{fab02}), especially in massive galaxies, and further studies
are needed.

The thermal state of ICM may be affected by the presence of magnetic
fields, and relativistic electrons and protons. Important processes
involving magnetic fields and particles are radio synchrotron
emission, and X-ray and $\gamma$-ray emission from Compton scattering
of electrons by the cosmic microwave background (CMB)
radiation. Recent observations of extended radio emissions which does
not originate in galaxies provide the main evidence for the presence
of relativistic electrons and magnetic fields. Radio measurements
yield a mean field value of few $\mu$G, under the assumption of global
energy equipartition.

Compton scattering of the radio emitting (relativistic) electrons by
the CMB yields non-thermal X-ray and $\gamma$-ray
emissions. Measurement of this radiation provides additional
information that enables direct determination of the electron density
and the mean magnetic field, without any assumption of energy
equipartition (Rephaeli et al. \ct{rep94}).

\subsection{Dark Matter}

It is generally believed that most of the matter in the universe is
dark, i.e. it cannot be observed directly from the light that it
emits, but its existence is inferred only from its gravitational
effect.

The first evidence for the presence of dark matter in the universe
came from the virial analysis of the Coma cluster performed by Zwicky
in \ct{zwi33}.  Analysing the velocities of cluster galaxies, he
pointed out that the velocity dispersions were too high if compared
with those expected considering a gravitationally bound system where
the mass is inferred only by the luminous matter. For this reason he
concluded that the existence of the Coma cluster would be impossible
unless its dynamics were dominated by dark matter.

The basic principle is that any isolated self--gravitating system
reaches the equilibrium if the gravity is balanced by the kinetic
pressure.  According to the Virial Theorem, the potential energy $U$
is related to the kinetic energy $K$ by:

\begin{equation}
2K+U=0 \ ,
\end{equation}

\noindent
where, for a spherical system, $U$ is related to the total mass $M$ of
the cluster inside a radius $r$, and $K$ depends directly on the
mean-square velocity $<v^2>$ as:

\begin{equation}
U=- \frac{\alpha G M^2}{r} \ \ \ , \ \ \ K=\frac{1}{2} M <v^2> \ ,
\end{equation}

\noindent
where $\alpha$ is a constant which reflects the radial density
profile.

These relations imply that if we measure velocities in some region of
the clusters we can derive the gravitational mass necessary in order
to stop all the objects flying apart. When such velocity measurements
are done, it turns out that the amount of inferred mass is much more
than one that can be explained by the luminous galaxies.

Gravitational lensing offers another powerful tool in order to
estimate the amount of dark matter. In fact, a gravitational field can
deflect a beam of light when it pass close to a mass clump.  Clusters
act as gravitational lenses, which magnify, distort and multiply the
images of background galaxies. A study of the properties of these
images provides information about the mass distribution of the lens
and therefore about the dark matter.

When the distances between background galaxy and cluster, and between
cluster and observer are much larger than the size of the cluster
itself, clusters behave as ``thin lenses'', which means that lensing
effects do not depend on the three--dimensional mass distribution, but
rather on the surface density ($\Sigma(r)$) on the sky. In this
situation, the deflection of light rays is given by the gradient of
the gravitational potential generated by $\Sigma(r)$, which is the
deflecting mass density integrated along the line of sight.

Rays that encounter a mass with a critical impact parameter
r$_{\mathrm{E}}$ (Einstein radius) are deflected. The angle
($\theta_{\mathrm{E}}$) through which the light is deflected depend on
the mass of the deflectors, so that the measure of this angle allow to
measure the mass of the lens (see the following section for the mass
derivation) and by comparing this mass with the luminous ones, to
derive an estimate of the amount of dark matter.

Cosmology provides additional evidence for the existence of a large
amount of dark matter in the universe and offers crucial information
about its physical nature. Information about the physical nature of
dark matter can be derived from : i) the primordial nucleosynthesis,
ii) the Hubble diagram\footnote{\footnotesize The Hubble diagram is
the plot of the apparent magnitude of objects with known absolute
luminosity (standard candles) versus redshift. A curve in this plot
depends on the pair
($\Omega_{\mathrm{\Lambda}},\Omega_{\mathrm{M}}$).}, iii) the cosmic
microwave background (CMB) (see Roncadelli \ct{ron03} for a review).

The analysis of CMB power spectrum, in particular the position of the
first peak, yields the specific value of the cosmic density parameter
$\Omega \simeq 1$ (e.g. Spergel et al. \ct{spe03}), that implies that
the universe is spatially flat. Moreover, the ratio of the height of
the first to the second peak gives a value of the cosmic baryonic
fraction $\Omega_{\mathrm{B}} \simeq 0.045$. This value is confirmed
from independent estimates of $\Omega_{\mathrm{B}}$ arising from i)
the features of the absorption lines of neutral hydrogen observed in
the Lyman-$\alpha$ forest of spectra of high--redshift quasars, and
ii) a comparison between predicted and observed abundances of light
elements like deuterium $D$, helium $He^3$, $He^4$ and lithium $Li^7$,
which have been formed during the first few minutes of the life of the
universe. Therefore cosmology provides a solid prediction of the total
amount of baryons in the universe.

A best-fit to the data of distant type-Ia supernovae, which are
believed to be good standard candles (Riess et al. \ct{rie98};
Perlmutter et al. \ct{per99}), yields:

\begin{equation}
\label{omegal}
\Omega_{\mathrm{\Lambda}} \simeq 1.33 \Omega_{\mathrm{M}} + 0.33 \ .
\end{equation}

Considering that:

\begin{equation}
\label{omega}
\Omega_{\mathrm{\Lambda}} + \Omega_{\mathrm{M}} = \Omega \simeq 1
\end{equation}

\noindent
it follows that:

\begin{equation}
\label{omegaml}
\Omega_{\mathrm{\Lambda}} \simeq 0.71 \ \ {\mathrm{and}} \ \ \Omega_{\mathrm{M}} \simeq 0.29 \ .
\end{equation}

This analysis implies the presence of dark matter and address its
nature.  In fact, the contribution from photons emitting in the
optical and in the X-ray band to $\Omega_{\mathrm{B}}$ represents only
$\sim$ 50\%. Thus, half of the existing baryons are invisible and
constitute the baryonic dark matter. However $\Omega_{\mathrm{M}}$
largely exceeds $\Omega_{\mathrm{B}}$, so dark matter is dominated by
elementary particles carrying no baryonic number. It has been
suggested that these particles can be weakly interacting massive
particles (WIMPS), but the nature of non--baryonic dark matter is
still unclear.

\subsection{Cluster masses}

The determination of the cluster mass using optical data is based on
the time-independent Jeans equation or its derivations, such as the
virial theorem (e.g., Binney \& Tremaine \ct{bin87}). Assuming
dynamical equilibrium and spherical symmetry, the total mass $M(r)$
within a radius $r$, is given by:

\begin{equation}
\rho(r) \frac{d\phi}{dr}=\rho(r)\frac{GM(r)}{r^2}=-\frac{\rho(r) \sigma_r^2}{dr}-\frac{2\rho(r)}{r}(\sigma_r^2-\sigma_t^2) \ ,
\end{equation}

\noindent
where $\rho(r)$ is the density of the cluster, $\phi(r)$ is the total
potential, and $\sigma_r$ and $\sigma_t$ are the radial and tangential
components of galaxy velocity dispersions. If we set $\beta(r) =
1-(\sigma_t^2/\sigma_r^2)$, we need to know the three functions
$\rho(r)$, $\sigma_r(r)$ and $\beta(r)$ in order to estimate the mass
distribution.

Observations of galaxies allow to estimate the projected galaxy
density, which may be inverted, under some assumptions, to give
$\rho(r)$.  By measuring radial velocities of a large sample of
galaxies we may determine the dependence of the line--of--sight
velocity dispersion from the radius. Even in this idealised situation,
both functions $\sigma_r(r)$ and $\sigma_t(r)$ cannot be determined
from the observations. Therefore it is impossible to derive $M(r)$
without further assumptions, considering, for example an isotropic
velocity distribution ($\sigma_r(r)$ = $\sigma_t(r)$ at all radii) and
assuming that galaxies trace the cluster mass ($\rho
\propto \rho_{gal}$).

Dynamical masses based on optical data have the additional drawback
that the mass distribution or the velocity anisotropy of galaxy orbits
should be known {\it a priori}.  These two quantities can be
disentangled only in the analysis of the whole velocity distribution,
which, however, requires a large number of galaxy spectra (e.g.,
Dejonghe \ct{dej87}; Merritt \ct{mer88}; Merritt \& Gebhardt
\ct{mer94}).  Moreover, without information on the relative
distribution of dark and galaxy components, the total mass is
constrained only at order-of-magnitudes (e.g., Merritt \ct{mer87}).
However, the assumption that the mass is distributed as the observed
cluster galaxies is supported by several evidences from both optical
(e.g., Carlberg, Yee, \& Ellingson \ct{car97a}) and X-ray data (e.g.,
Watt et al. \ct{wat92}; Durret et al. \ct{dur94}; Cirimele, Nesci, \&
Trevese
\ct{cir97}), as well as from gravitational lensing data (e.g., Narayan
\& Bartelmann \ct{nar99}).

In order to solve these problems, cluster masses can be derived by
using the virial theorem. The great advantage of the virial theorem is
that the global line of sight velocity dispersion $\sigma_{\mathrm{los}}$,
and consequently the total mass, are independent from any possible
anisotropy in the velocity distribution.

For spherical systems ($\sigma^2$ = 3 $\sigma_{\mathrm{los}}^2$), where
the mass is distributed as in the observed galaxies, the total virial
mass (M$_\mathrm{V}$) of the cluster is given by (Merrit \ct{mer88}):

\begin{equation}
\mathrm{M_V} = \frac{6 \pi \sigma_{\mathrm{los}}^2 \mathrm{R_{H}}}
{2 \mathrm{G}} \ ,
\end{equation}

\noindent
where $\mathrm{R_{H}}$ is the harmonic radius, which depends
on the distance between any pair of galaxies (r$_{ij}$), and $\sigma_{los}$
is the line of sight velocity dispersion.

Another assumption is that clusters are systems in dynamical
equilibrium because they are not yet relaxed. This assumption is
generally not strictly valid. However, some studies suggest that the
estimate of optical virial mass is robust against the presence of
small substructures (Escalera et al. \ct{esc94}; Girardi et
al. \ct{gir97}; see also Bird \ct{bir95} for a partially different
result), although it is affected by strong substructures (e.g.,
Pinkney et al. \ct{pin96}).

The hot ICM provides an alternative means of measuring cluster
mass. For a locally homogeneous gas in hydrostatic equilibrium within
a spherical potential well, the equation of hydrostatic equilibrium
give the total mass within a radius $r$:

\begin{equation}
\label{eq_xm}
M(r)=-\frac{krT_{gas}(r)}{G\mu m_H}\left[\frac{dlog(\rho_{gas}(r))}{dlog(r)}+\frac{dlog(T_{gas}(r)}{dlog(r))}\right] \ ,
\end{equation}

\noindent
where $T_{gas}(r)$ and $\rho_{gas}(r)$ are the temperature and the
density of the gas respectively, $m_H$ is the mass of proton, $\mu$ is
the mean molecular weight, and $k$ is Boltzmann constant.

X-ray emission from ICM allow to determine both density and
temperature of gas, but measurement of X-ray surface brightness drops
below the background level at large radii, so it becomes difficult to
measure brightness, and particularly gas temperature.

In general, X-ray masses are a result of an extrapolation to regions
beyond those for which data are reliable. This causes a mistake when
both gas emissivity and temperature show substantial large scale
asymmetries with clear evidence of substructures.

Giant luminous arcs (strong lensing) in clusters, due to gravitational
lensing, give important constraints on the mass and the mass
distribution in clusters. With this method the cluster mass is derived
by measuring the deformation in the shape of lensed background
galaxies.

Considering two planes orthogonal to the line of sight (the ``lens
plane'' and the ``source plane''), the lens equation for a spherically
symmetric system is: 

\begin{equation}
\label{lens_eq}
S \sim \left(1 - \frac{M(r)}{\pi r^2 \Sigma_{cr}}\right)I  \ ,
\end{equation}

\noindent
where S denotes a generic point in the source plane, I is the image of
S in the lens plane, $M(r)$ represents the mass inside the radius $r$,
and $\Sigma_{cr}$ is a reference value for the lens surface density,
completely fixed by the distance between the lens and the source.

The image of the point S=0 (caustic), where the line of sight
intersects the source plane, is a circle (the Einstein ring).  In the
ideal situation of perfect Einstein ring, the total mass $M(r_E)$
contained inside $r_E$ (Einstein radius) is given by:

\begin{equation}
\label{eq_lm}
\frac{M(r_E)}{\pi r_E^2}=\frac{c^2}{4\pi G}\left(\frac{D_{OS}}{D_{OS}D_{LS}}\right)
 \ ,
\end{equation}

\noindent
where the $D_{ij}$ are angular diameter distances between observer
(O), lens (L) and source (L).

For an extended source close to the caustic, the magnified image
consists of two elongated arcs, located on opposite sides. However a
tiny perturbation of the spherical symmetry leads to a strong
demagnification of one of these arcs, and a single arc is observed.

Therefore, determining observationally the radius of giant arcs and
the angular distances between observer, lens and source it is possible
to derive the total gravitational mass. With the strong lensing, only
the mass inside the Einstein radius can be derived, while a more
general technique is based on weak lensing.

Cluster produces weakly distorted images of all background galaxies
that lie sufficiently close to its position on the sky. Because
lensing compresses the image in one direction stretching it in the
orthogonal direction, the observed lensed images are called
arclets. The ellipticities of these arclets allow to derive the
strength of the gravitational field at every arclet position, from
which the total cluster mass can be derived.

The mass estimated from gravitational lensing phenomena are completely
independent from the cluster dynamical status, but a good knowledge of
cluster geometry is required in order to derive the cluster mass from
the projected mass (e.g., Fort \& Mellier \ct{for94}). Strong lensing
observations give values for the mass contained within very small
cluster regions ($\lesssim$ 100 kpc), and weak lensing observations
are generally more reliable in providing the shape of the internal
mass distribution rather than the amount of mass (e.g., Squires \&
Kaiser \ct{squ96}).

\section{Cluster mergers}

Until the 80's clusters have been modeled as virialized spherically
symmetric systems (e.g. Kent \& Gunn \ct{ken82}), resulting from a
phase of violent relaxation during the collapse of a protocluster
(Lynden-Bell \ct{lyn67}). Violent relaxation produces a galaxy density
distribution which has the form of a self-gravitating isothermal
sphere and a thermal velocity distribution. 

Subsequently, observational and theoretical evidences suggested that
many galaxy clusters have still to reach dynamical equilibrium.  In
particular, optical and X--ray studies showed that a large fraction of
clusters ($30$--$40\%$) have substructures (Girardi et
al. \ct{gir97b}; Jones \& Forman \ct{jon99}), indicating frequent
occurrence of merging events between subclusters.

The existence of substructures suggests that clusters are young
objects, since bound subgroups can survive loosely only for a few
cluster crossing times\footnote{\footnotesize A galaxy traveling
through a cluster with a velocity $v$ will cross a radius $R$ in a
crossing time $ t_{\mathrm{CR}} = R/v \sim 10^9
\mathrm{yr}\left(\frac{R}{Mpc}\right)\left(\frac{10^3
\mathrm{km/s}}{\sigma_{\mathrm{v}}}\right) \ , $ where
$\sigma_{\mathrm{v}}$ is the observed dispersion of the velocities
along the line of sight about the mean:
$\sigma_{\mathrm{v}}=<(v_{\mathrm{r}} - <v_{\mathrm{r}}>)^2>^{1/2}$.},
i.e. for a time shorter than the Hubble time.

In the framework of the scenario of hierarchical structure formation,
objects are formed from the collapse of initial density enhancements
and subsequently grow by gravitational merging and by accretion of
smaller clusters and groups. Starting from these ideas, the abundance
of substructure was suggested as an effective measure of the mean
matter density in the universe (Richstone, Loeb \& Turner \ct{ric92},
Bartelmann, Ehlers \& Schneider
\ct{bar93}, Kauffman \& White \ct{kau93}, and Lacey \& Cole \ct{lac93}). 
Indeed, the frequency of subclustering at the present epoch is set by
the mean density at the recombination epoch. The linear theory
predicts that clustering in critical density universes continues to
grow until present-day, whereas in low-density
ones\footnote{\footnotesize Assuming $\Omega_\Lambda$ = 0}, it begins
to decline after a redshift z $\sim \Omega_0^{-1}-1$, where $\Omega_0$
is the present value of the normalized cosmic matter density. This
means that clusters in a low-density universe are expected to be
dynamically more relaxed, less substructured, and less elongated. For
these reasons the analysis of substructures are fundamental in order
to discriminate between different cosmological models.

\subsection{Optical substructures}

An increasing amount of data has shown that several clusters contain
subsystems usually called substructures, which i) suggest that
clusters are still in the process of relaxation (e.g., West
\ct{wes94}); ii) affect the estimate of global cluster properties 
(e.g., Bird \ct {bir95}; Schindler \ct{sch96}; Roettiger Burns, \&
Loken \ct{roe96}; Allen \ct{all98}); iii) indicate that
cluster merger phenomena are ongoing. Moreover, the analysis of the
gravitational lensing (\cit {mir91}) has shown that the distribution
of the gravitational mass in clusters, that is the mass of all the
constituents including dark matter, is not spherically symmetric, but
has multiple peaks.

Despite of an increased improvement in the analysis of substructures,
subclustering remains difficult to investigate in a detailed and
unambiguous way. Substructure in galaxy clusters can be quantified
with the robust $\Delta$ statistic (Dressler \& Shectman \ct{dre88}),
which uses velocity fields and sky--projected positions. However, in
order to characterize subclusters better, it is necessary to study
galaxy luminosities, colours and star formation rates (Girardi \&
Biviano \ct{gir02}), because there is observational evidence for a
strong connection between the properties of galaxies and the presence
of substructure.

This kind of analyses has been applied only to a few clusters because
of the lack of deep imaging and spectroscopy in large fields of
view. Therefore, further studies of individual clusters are necessary
in order to provide new and useful information on the physics of the
cluster merger processes.

\subsection{X-ray Substructures}

X-ray observations have shown that cluster morphologies could be very
complex (e.g. multiple peaks in the X-ray luminosity, isophote
twisting with centroid shifts, elongations) and characterized by the
presence of substructures.

Cluster mergers produce supersonic shocks, compressing and heating the
intra--cluster gas, and increasing the pressure and entropy. This can be
measured as local distortions of the spatial distribution of X-ray
temperatures and surface brightness (e.g., Schindler \&
M$\ddot{\mathrm{u}}$ller \ct{sch93}; Roettinger, Loken, \& Burns
\ct{roe97}).

A quantitative analysis of the X-ray morphologies can be performed
through different approaches. A detailed structural analysis can be
made by examining the residuals obtained subtracting a smooth model
representing a relaxed cluster (usually a King $\beta$ model) from the
X-ray image (e.g., Davis \ct{dav94}; Neumann \&
B$\ddot{\mathrm{o}}$hringer \ct{boh97}). A more general method is to
perform a wavelet decomposition of the X-ray image (e. g., Slezak et
al. \ct{sle94}; Grebenev et al. \ct{gre95}; Pierre \& Stark
\ct{pie98}; Arnaud et al. \ct{arn00}).

Empirically, there is a strong statistical anti-correlation between
cooling flows and irregular morphologies, as derived by statistical
analysis of X-ray images (Buote \& Tsai \ct{buo96}). This shows
evidence that mergers could disrupt cooling flows. However it is not
clear exactly how and under which circumstance mergers disrupt
cooling flows (see Sarazin \ct{sar02} for review). More detailed studies
are necessary, using deep X-ray observations, to achieve high
signal-to-noise ratio, and performing comprehensive study of
substructures characteristic.

\subsection{Radio halos and relics}

Diffuse radio sources with no obvious connection to individual galaxy
are found in few galaxy clusters (Giovannini et al. \ct{gio93}). These
sources are separated generally into two classes: halos and relics.

Radio halos are regular in shape, with typical size of $\sim$ 1 Mpc,
low surface brightness and steep radio spectra (S$_{\nu}
\propto \nu^{\alpha}$, where $\alpha \sim$ -1), and are located at the
cluster centre. Relics are located at the cluster peripheries, show
irregular and elongated shapes, and exhibit stronger polarization than
halos. In a few clusters, both a central halo and a peripheral relic
are present. These sources demonstrate the existence of relativistic
electrons and large scale magnetic fields in the intra--cluster medium,
which probe the presence of non--thermal processes in clusters.

The knowledge of physical properties, origin and evolution of radio
sources is limited by the low number of well studied halos and relics
up to now. One of the main problems is that they are rare phenomena.
However, thanks to the sensitivity of the radio telescopes and to the
existence of deep radio surveys, the number of known halos and relics
has recently increased.

The first and the best studied example was the Coma C (e.g.,
Giovannini et al. \ct{gio93}, Deiss et al. \ct{dei97}), who was
discovered over 30 years ago (Willson \ct{wil70}). The spectra
suggested that radio halo emission arises predominantly by the
synchrotron process. However, it is not clear the reason why halos are
not present in all clusters.

A number of models have been proposed to explain the formation of
radio halos (e.g., Jaffe \ct{jaf77}; Dennison \ct{den80}; Roland
\ct{rol81}). Most of these early models suggested that ultra-relativistic
electrons originate either as relativistic electrons from cluster
radio sources, re--accelerated {\it in situ} by Fermi processes or by
turbulent galactic wakes, or as secondary electrons produced by the
interaction between relativistic protons (again from cluster radio
galaxies) and thermal protons. However, the energetics involved are
problematic and the models could not always fit the observations
(e.g., see B$\ddot{\mathrm{o}}$hringer \ct{bor95} for review).

Harris, Kapahi, \& Ekers (\ct{har80}) first suggested that radio halos
are formed in cluster mergers where the merging process creates the
shocks and turbulence necessary for the magnetic field amplification
and high-energy particle acceleration. Tribble (\ct{tri93}) showed
that the energetics involved in a merger are more than enough to power
a radio halo. The halos thus produced are expected to be transient
since the relativistic electrons lose energy on time scales of 10$^8$
yr and the time interval between mergers is of the order of 10$^9$
yr. This argument was used to explain why radio halos are rare.

The details of this process, however, remain controversial because of
the difficulty in directly accelerating the thermal electrons to
relativistic energies (e.g., Tribble \ct{tri93}; Sarazin \ct{sar01};
Brunetti et al. \ct{brun01}; Blasi \ct{blas01}). In fact, owing to
this difficulty, it is often assumed that a ``reservoir'' of
relativistic particles is established at some time in the past
evolution of the cluster, with the current merger merely serving to
re--accelerate relativistic particles from this reservoir. In this
case it is unclear whether the current or the past dynamical state of
the cluster is the primary factor in the creation of a radio halo.

X-ray observations provide circumstantial evidence for a connection
between cluster merging and radio halos (see Feretti \ct{fer02} and
references therein) because, in particular, radio halos are found only
in clusters with X-ray substructure and weak (or absent) cooling
flows. However, it has been argued (e.g., Giovannini \& Feretti
\ct{gio00}; Liang et al. \ct{lia00}; Feretti \ct{fer02}) that merging
cannot be solely responsible for the formation of radio halos because
at least 50\% of clusters show evidence for X-ray substructure (Jones
\& Forman \ct{jon99}) whereas only $\sim10$\% possess radio halos.

Results of Bacchi et al. (\ct{bac03}) confirm the correlation of radio
halos with cluster merger processes. Moreover, they found a relation
between radio halos and X-ray luminosities, arguing that clusters with
a low X-ray luminosity ($<$ 10$^{45}$ erg s$^{-1}$) would host halos
that are too faint to be detected with the present generation of radio
telescopes. Only the future generation of radio telescopes (LOFAR,
SKA) will allow the investigation of this point.

\section{The cluster ABCG\,209}

The analysis of substructured clusters plays an important role in the
understanding the large--scale structure formation, in constraining
cosmological models, and in the study of galaxy evolution.

Results from simulations of isolated clusters showed strong variations
of cluster morphologies with the cosmological parameter $\Omega_0$
(e.g. \cit{moh95}). However, Thomas et al. (\ct{tho98}) found that the
properties of relaxed clusters are very similar in all the examined
cosmologies. For this reason the study of highly structured clusters
is a fundamental issue in order to discriminate between different
cosmologies.

Recent studies at z$\lesssim$ 0.2 indicate that the fraction of blue
galaxies could increase with the degree of substructure (Caldwell \&
Rose \ct{cal97}; Metevier, Romer \& Ulmer \ct{met00}), providing
support for a connection between the Butcher--Oemler (\ct{but84})
effect, in which the fraction of blue galaxies in clusters is observed
to increase with redshift, and the hierarchical model of cluster
formation (Kauffmann \ct{kau95}).  Numerical simulations have shown
that cluster mergers can influence the evolution of the galaxy
population. Indeed mergers induce a time-dependent gravitational field
that stimulates perturbations in disk galaxies, leading to starbursts
(SBs) in the central parts of these galaxies.

Detailed cosmological simulations, which follow the formation and the
evolution of galaxies, are able to successfully predict the observed
cluster-centric star formation and colour gradient, and the
morphology--density relation (Balogh, Navarro \& Morris \ct{bal00};
Diaferio et al. \ct{dia01}; Springel et al. \ct{spr01}; Okamoto \&
Nagashima \ct{oka03}), but it is not clear whether they predict trends
with density of colour or star formation for galaxies of a fixed
morphology and luminosity.

In this context, we analyse the rich galaxy cluster ABCG\,209 at z
$\sim$ 0.2. For this purpose we collected new photometric (B--, V--
and R--band images) and spectroscopic data (multi--object
spectroscopy) at the ESO New Technology Telescope (NTT) with the ESO
Multi Mode Instrument (EMMI) in October 2001.

In order to examine the effect of the environment on the galaxy
properties, in particular the galaxy luminosity function and the
cluster red sequence, we used archive wide--field B-- and R--band
images, which allows the photometric properties of the cluster
galaxies to be followed out to radii of 3--4 Mpc h$^{-1}_{70}$.

In order to compare the dynamical status of different cluster
components, the results of the optical analysis are compared with
those obtained from the analysis of archive X-ray (Chandra) and radio
(VLA) observations.

ABCG\,209 is a rich, very X--ray luminous and hot cluster (richness
class $\mathrm{R=3}$, Abell et al. \ct{abe89}; $\mathrm{L_X(0.1-2.4\
keV) \sim 14\;h_{50}^{-2}\;10^{44}}$ erg $\mathrm{s^{-1}}$, Ebeling et
al. \ct{ebe96}; $\mathrm{T_X\sim10}$ keV, Rizza et al. \ct{riz98}
). The first evidence for its complex dynamical status came from the
significant irregularity in the X--ray emission with a trimodal peak
(Rizza et al. \ct{riz98}). Moreover, Giovannini et al. (\ct{gio99})
have found evidence for the presence of a radio halo, which is
indicative of recent cluster merger process (Feretti \ct{fer02}).

This cluster was chosen for its richness, allowing its internal
velocity field and dynamical properties to be studied in great detail,
for its substructures, allowing the effect of cluster dynamics and
evolution on the properties of its member to be examined, and for the
presence of X-ray and radio observations, allowing a multiwavelengths
analysis of the different cluster component.

Finally the redshift of this cluster represents an optimal compromise
between the necessity to achieve high accuracy, in order to study the
connection between the cluster dynamics and the properties of galaxy
populations, and the needed look--back time of some Gyr, in order to
investigate the cluster evolution from the comparison with local
clusters.

\large
\chapter{\Large Internal dynamics}
\footnotetext[1]{\footnotesize
The content of this chapter is published in Mercurio, A., Girardi, M.,
Boschin, W., Merluzzi, P., \& Busarello, G. 2003, A\&A, 397, 431.}
\setcounter{footnote}{1}
\label{cap:2}
\markboth{Chapter 2}{Chapter 2}
\normalsize

In this chapter we study the internal dynamics of ABGC\,209 on the
basis of new spectroscopic and photometric data.  The distribution in
redshift shows that ABCG\,209 is a well isolated peak of 112 detected
member galaxies at $\mathrm{z=0.209}$, characterised by a high value
of the line--of--sight velocity dispersion,
$\mathrm{\sigma_v=1250}$--$1400$
\kss, on the whole observed area (1 \h from the cluster center), that
leads to a virial mass of $\mathrm{M=1.6}$--$2.2\times 10^{15}$\msun
within the virial radius, assuming the dynamical equilibrium.  The
presence of a velocity gradient in the velocity field, the elongation
in the spatial distribution of the colour--selected likely cluster
members, the elongation of the X--ray contour levels in the Chandra
image, and the elongation of cD galaxy show that ABCG\,209 is
characterised by a preferential NW--SE direction.  We also find a
significant deviation of the velocity distribution from a Gaussian,
and relevant evidence of substructure and dynamical segregation.  All
these facts show that ABCG\,209 is a strongly evolving cluster,
possibly in an advanced phase of merging.

\section{Introduction}


In hierarchical clustering cosmological scenarios galaxy clusters form
from the accretions of subunits. Numerical simulations show that
clusters form preferentially through anisotropic accretion of
subclusters along filaments (West et al. \ct{wes91}; Katz \& White
\ct{kat93}; Cen \& Ostriker \ct{cen94}; Colberg et al \ct{col98},
\ct{col99}). The signature of this anisotropic cluster formation is
the cluster elongation along the main accretion filaments (e.g.,
Roettiger et al. \ct{roe97}).  Therefore the knowledge of the
properties of galaxy clusters, plays an important role in the study of
large--scale structure (LSS) formation and in constraining
cosmological models.
 

On small scales, clusters appear as complex systems involving a
variety of interacting components (galaxies, X--ray emitting gas, dark
matter). A large fraction of clusters (30\%-40\%) contain
substructures, as shown by optical and X--ray studies (e.g., Baier \&
Ziener
\ct{bai77}; Geller \& Beers \ct{gel82}; Girardi et
al. \ct{gir97}; Jones \& Forman \ct{jon99}) and by recent results
coming from the gravitational lensing effect (e.g., Athreya et
al. \ct{ath02}; Dahle et al. \ct{dah02}), suggesting that they are
still in the dynamical relaxation phase. Indeed, there is growing
evidence that these subsystems arise from merging of groups and/or
clusters (cf. Buote \ct{buo02}; and Girardi \& Biviano \ct{gir02} for
reviews).

Very recently, it was also suggested that the presence of radio halos
and relics in clusters is indicative of a cluster merger. Merger
shocks, with velocities larger than 10$^3$ km s$^{-1}$, convert a
fraction of the shock energy into acceleration of pre--existing
relativistic particles and provide the large amount of energy
necessary for magnetic field amplification (Feretti \ct{fer02}). This
mechanism has been proposed to explain the radio halos and relics in
clusters (Brunetti et al. \ct{brun01}).

The properties of the brightest cluster members (BCMs) are related to
the cluster merger.  Most BCMs are located very close to the center of
the parent cluster. In many cases the major axis of the BCM is
aligned along the major axis of the cluster and of the surrounding LSS
(e.g., Binggeli \ct{bin82}; Dantas et al. \ct{dan97}; Durret et
al. \ct{dur98}). These properties can be explained if BCMs form by
coalescence of the central brightest galaxies of the merging
subclusters (Johnstone et al. \ct{joh91}).

The optical spectroscopy of member galaxies is the most powerful tool
to investigate the dynamics of cluster mergers, since it provides
direct information on the cluster velocity field.  However this is
often an ardue investigation due to the limited number of galaxies
usually available to trace the internal cluster velocity.  To date, at
medium and high redshifts (z $\gtrsim$ 0.2), only few clusters are really
well sampled in the velocity space (with $>$ 100 members; e.g., 
Carlberg et al \ct{car96}; Czoske et al. \ct{czo02}).

In order to gain insight into the physics of the cluster formation, we
carried out a spectroscopic and photometric study of the cluster
ABCG\,209, at z$\sim$0.2.  In Sect.~\ref{sec:22} we present the new
spectroscopic data and the data reduction. The derivation of the
redshifts is presented in Sect.~\ref{sec:23}. In Sect.~\ref{sec:24} we
analyse the dynamics of the cluster, and in Sect.~\ref{sec:25} we
complete the dynamical analysis with the information coming either
from optical imaging or from X--ray data.  In Sect.~\ref{sec:26} we
discuss the results in terms of two pictures of the dynamical status
of ABCG\,209. Finally, a summary of the main results is given in
Sect.~\ref{sec:27}.


 We assume a flat cosmology with
$\mathrm{\Omega_M=0.3}$ and $\mathrm{\Omega_{\Lambda}=0.7}$.  For the
sake of simplicity in rescaling, we adopt a Hubble constant of 100 h
\ks Mpc$^{-1}$. In this assumption, 1 arcmin corresponds to $\sim$
0.144 Mpc.  Unless otherwise stated, we give errors at the 68\%
confidence level (hereafter c.l.).

\section{Observations and data reduction}
\label{sec:22}

The data were collected at the ESO New Technology Telescope (NTT) with 
the ESO Multi Mode Instrument (EMMI) in October 2001.

\subsection{Spectroscopy}

Spectroscopic data have been obtained with the multi--object
spectroscopy (MOS) mode of EMMI. Targets were randomly selected by
using preliminary R--band images (T$\mathrm{_{exp}}$=180 s) to
construct the multislit plates. We acquired five masks in four fields
(field of view $5^\prime \times 8.6^\prime$), with different position
angles on the sky, allocating a total of 166 slitlets (alligment stars
included).  In order to better sample the denser cluster region, we
covered this region with two masks and with the overlap of the other
three masks.  We exposed the masks with integration times from 0.75 to
3 hr with the EMMI--Grism\#2, yielding a dispersion of $\sim2.8$
\AA/pix and a resolution of $\sim 8$ \AA FWHM, in the spectral range
385 -- 900 nm.

Each scientific exposure (as well as flat fields and calibration
lamps) was bias subtracted. The individual spectra were extracted and
flat field corrected.  Cosmic rays were rejected in two steps. First,
we removed the cosmic rays lying close to the objects by interpolation
between adjacent pixels, then we combined the different exposures by
using the IRAF\footnote{IRAF is distributed by the National Optical
Astronomy Observatories, which are operated by the Association of
Universities for Research in Astronomy, Inc., under cooperative
agreement with the National Science Foundation.} task IMCOMBINE with
the algorithm CRREJECT (the positions of the objects in different
exposures were checked before).  Wavelength calibration was obtained
using He--Ar lamp spectra. The typical r.m.s. scatter around the
dispersion relation was $\sim$ 15 \ks.  The positions of the objects
in the slits were defined interactively using the IRAF package
APEXTRACT.  The exact object position within the slit was traced in
the dispersion direction and fitted with a low order polynomial to
allow for atmospheric refraction. The spectra were then sky subtracted
and the rows containing the object were averaged to produce the
one--dimensional spectra.

The signal--to--noise ratio per pixel of the one--dimensional galaxy
spectra ranges from about 5 to 20 in the region 380--500 nm.

\subsection{Photometry}

A field of $9.2^\prime \times 8.6^\prime$ (1.2 $\times$ 1.1
$\mathrm{h^{-2}}$ Mpc$^2$ at z=0.209) was observed in B--, V-- and
R--bands pointed to the center of the cluster. Two additional adjacent
fields were observed in V--band in order to sample the cluster at
large distance from the center (out to $\sim$ 1.5 \hh).  The reduction
of the photometric data will be described in detail in the next
chapter.  In order to derive magnitudes and colours we used the
software SExtractor (Bertin \& Arnouts
\ct{ber96}) to measure the Kron magnitude (Kron \ct{kro80}) in an
adaptive aperture equal to $\mathrm{a \cdot r_K}$, where
$\mathrm{r_K}$ is the Kron radius and a is a constant. Following
Bertin \& Arnouts (\ct{ber96}), we chose $\mathrm{a=2.5}$, for which
it is expected that the Kron magnitude encloses $\sim 94 \%$ of the
total flux of the source.  We use the photometric data to investigate
the colour segregation in Sect.~\ref{sec:25}.

\section{Redshifts measurements}
\label{sec:23}
 
Redshifts were derived using the cross--correlation technique (Tonry
\& Davis \ct{ton81}), as implemented in the RVSAO package.  We
adopted galaxy spectral templates from Kennicutt (\ct{ken92}),
corresponding to morphological types EL, S0, Sa, Sb, Sc, Ir.  The
correlation was computed in the Fourier domain.

Out of the 166 spectra, 112 turned out to be cluster members (seven of
which observed twice), 22 are stars, 8 are nearby galaxies, 1 is
foreground and 6 are background galaxies. In 10 cases we could not
determine the redshift.

In order to estimate the uncertainties in the redshift measurements,
we considered the error calculated with the cross--correlation
technique, which is based on the width of the peak and on the
amplitude of the antisymmetric noise in the cross correlation.  The
wavelength calibration errors ($\sim$ 15 \ks) turned out to be
negligible in this respect.  The errors derived from the
cross--correlation are however known to be smaller than the true
errors (e.g., Malumuth et al.  \ct{mal92}; Bardelli et
al. \ct{bar94}; Quintana et al. \ct{qui00}). We checked the error
estimates by comparing the redshifts computed for the seven galaxies
for which we had duplicate measurements.  The two data sets agree with
one--to--one relation, but a reasonable value of $\chi^2$ for the fit
was obtained when the errors derived from the cross--correlation were
multiplied by a correction factor ($\sim 1.75$).  A similar correction
was obtained by Malumuth et al.  (\ct{mal92}; 1.6), Bardelli et
al. (\ct{bar94}; 1.87), and Quintana et al. (\ct{qui00}; 1.57).
External errors, which are however not relevant in the study of
internal dynamics, cannot be estimated since only two previous
redshifts are available for ABCG\,209.  

The catalogue of the spectroscopic sample is presented in Table
~\ref{catalogue}, which includes: identification number of each
galaxy, ID (Col.~1); right ascension and declination (J2000),
$\alpha$ and $\delta$ (Col.~2 and ~3); V magnitude (Col.~4); B--R
colour (Col.~5); heliocentric corrected redshift $\mathrm{z}$ (Col.~6) with
the uncertainty $\mathrm{\Delta z}$ (Col.~7);

Our spectroscopic sample is 60\% complete for V$<$20 mag and drops
steeply to 30\% completeness at V$<$21 mag; these levels are reached both
for the external and the internal cluster regions. The spatial
distribution of galaxies with measured redshifts does not show any
obvious global bias. Only around the brightest cluster
member (cD galaxy; c.f. La Barbera et al. \ct{lab02})
the spatial coverage is less complete because of geometrical restrictions.
This does not affect our dynamical
analysis because we investigate structures at scales larger than
$\sim$ 0.1 \h.

\begin{table}
        \caption[]{Spectroscopic data.}
         \label{catalogue}
{\footnotesize
              $$ 
           \begin{array}{c c c c c c c}
            \hline
            \noalign{\smallskip}
            \hline
            \noalign{\smallskip}

\mathrm{ID} & {\mathrm{\alpha}} & \mathrm{\delta}  & \mathrm{V} & B-R  & z & 
\mathrm{\Delta} z\\
            \hline
            \noalign{\smallskip}   
  1 & 01\ 31\ 33.81 & -13\ 32\ 22.9 & 18.32 & .... & 0.2191 & 0.0003 \\
  2 & 01\ 31\ 33.82 & -13\ 38\ 28.5 & 19.27 & 2.43 & 0.2075 & 0.0002 \\
  3 & 01\ 31\ 33.86 & -13\ 32\ 43.4 & 19.46 & .... & 0.1998 & 0.0003 \\
  4 & 01\ 31\ 34.11 & -13\ 32\ 26.8 & 19.90 & .... & 0.2000 & 0.0004 \\
  5 & 01\ 31\ 34.26 & -13\ 38\ 13.1 & 20.17 & 2.24 & 0.2145 & 0.0005 \\
  6 & 01\ 31\ 35.37 & -13\ 31\ 18.9 & 17.36 & .... & 0.2068 & 0.0004 \\
  7 & 01\ 31\ 35.37 & -13\ 37\ 17.4 & 18.58 & 2.38 & 0.2087 & 0.0002 \\
  8 & 01\ 31\ 35.42 & -13\ 34\ 52.0 & 19.10 & 2.37 & 0.2073 & 0.0002 \\
  9 & 01\ 31\ 35.61 & -13\ 32\ 51.3 & 20.17 & .... & 0.2090 & 0.0002 \\
 10 & 01\ 31\ 36.51 & -13\ 33\ 45.2 & 19.83 & 2.36 & 0.2072 & 0.0004 \\
 11^{\mathrm{a}} & 01\ 31\ 36.54 & -13\ 37\ 56.1 & 19.89 & 1.37 & 0.2620 & 0.0005 \\
 12 & 01\ 31\ 36.76 & -13\ 33\ 26.1 & 21.10 & 2.21 & 0.2064 & 0.0004 \\ 
 13 & 01\ 31\ 36.87 & -13\ 39\ 19.3 & 19.73 & 2.26 & 0.2039 & 0.0002 \\
 14^{\mathrm{a}} & 01\ 31\ 37.19 & -13\ 30\ 04.3 & 20.65 & .... & 0.3650 & 0.0003 \\
 15 & 01\ 31\ 37.23 & -13\ 32\ 09.9 & 20.03 & .... & 0.2084 & 0.0003 \\
 16 & 01\ 31\ 37.38 & -13\ 34\ 46.3 & 18.71 & 2.35 & 0.2134 & 0.0003 \\
 17^{\mathrm{a}} & 01\ 31\ 37.99 & -13\ 36\ 01.6 & 21.38 & 2.67 & 0.3997 & 0.0004 \\ 
 18 & 01\ 31\ 38.06 & -13\ 33\ 35.9 & 19.87 & 2.49 & 0.2073 & 0.0002 \\
 19 & 01\ 31\ 38.69 & -13\ 35\ 54.7 & 20.11 & 2.17 & 0.2084 & 0.0003 \\
 20 & 01\ 31\ 39.06 & -13\ 33\ 39.9 & 19.64 & 2.33 & 0.2120 & 0.0002 \\
 21 & 01\ 31\ 39.89 & -13\ 35\ 45.1 & 18.85 & 1.68 & 0.2073 & 0.0002 \\
 22 & 01\ 31\ 40.05 & -13\ 32\ 08.9 & 19.51 & .... & 0.2096 & 0.0003 \\
 23 & 01\ 31\ 40.17 & -13\ 36\ 06.2 & 19.18 & 2.18 & 0.1995 & 0.0002 \\
 24 & 01\ 31\ 40.58 & -13\ 36\ 32.9 & 19.30 & 2.18 & 0.2052 & 0.0003 \\
 25 & 01\ 31\ 40.76 & -13\ 34\ 17.0 & 18.58 & 2.33 & 0.2051 & 0.0002 \\   
 26 & 01\ 31\ 40.94 & -13\ 37\ 36.5 & 19.10 & 1.66 & 0.2181 & 0.0008 \\
 27 & 01\ 31\ 41.63 & -13\ 37\ 32.3 & 19.39 & 2.28 & 0.2004 & 0.0002 \\
 28 & 01\ 31\ 41.64 & -13\ 38\ 50.2 & 19.22 & 1.55 & 0.2063 & 0.0003 \\
 29 & 01\ 31\ 42.35 & -13\ 39\ 26.0 & 19.07 & 2.19 & 0.2051 & 0.0004 \\
 30 & 01\ 31\ 42.76 & -13\ 38\ 31.6 & 19.76 & 2.31 & 0.2034 & 0.0002 \\
 31 & 01\ 31\ 42.77 & -13\ 35\ 26.6 & ..... & .... & 0.2062 & 0.0007 \\
 32 & 01\ 31\ 43.69 & -13\ 35\ 57.8 & 20.73 & 2.30 & 0.2076 & 0.0003 \\
 33 & 01\ 31\ 43.81 & -13\ 39\ 27.7 & 20.40 & 2.14 & 0.2064 & 0.0002 \\
              \noalign{\smallskip}			    
            \hline					    
            \noalign{\smallskip}			    
            \hline					    
         \end{array}
     $$ 
}
         \end{table}
\addtocounter{table}{-1}
\begin{table}
          \caption[ ]{Continued.}
{\footnotesize
     $$ 
           \begin{array}{c c c c c c c}
            \hline
            \noalign{\smallskip}
            \hline
            \noalign{\smallskip}

\mathrm{ID} & {\mathrm{\alpha}} & \mathrm{\delta}  & \mathrm{V} & B-R  & z & 
\mathrm{\Delta} z\\
            \hline
            \noalign{\smallskip}
 34 & 01\ 31\ 44.47 & -13\ 37\ 03.2 & 19.96 & 1.99 & 0.2027 & 0.0003 \\
 35 & 01\ 31\ 45.24 & -13\ 37\ 39.7 & 18.39 & 2.23 & 0.2060 & 0.0002 \\
 36 & 01\ 31\ 45.61 & -13\ 39\ 02.0 & 19.88 & 2.29 & 0.2087 & 0.0004 \\
 37 & 01\ 31\ 45.87 & -13\ 36\ 38.0 & 18.72 & 2.31 & 0.2066 & 0.0002 \\
 38 & 01\ 31\ 46.15 & -13\ 34\ 56.6 & 18.56 & 2.46 & 0.2084 & 0.0002 \\
 39 & 01\ 31\ 46.33 & -13\ 37\ 24.5 & 20.29 & 2.22 & 0.2105 & 0.0003 \\
 40 & 01\ 31\ 46.58 & -13\ 38\ 40.9 & 19.02 & 2.32 & 0.2167 & 0.0002 \\
 41 & 01\ 31\ 47.26 & -13\ 33\ 10.3 & 19.72 & .... & 0.2038 & 0.0004 \\
 42 & 01\ 31\ 47.69 & -13\ 37\ 50.4 & 18.18 & 2.44 & 0.2125 & 0.0003 \\
 43 & 01\ 31\ 47.91 & -13\ 39\ 07.9 & 18.66 & 2.41 & 0.2097 & 0.0003 \\
 44 & 01\ 31\ 48.01 & -13\ 38\ 26.7 & 19.32 & 2.37 & 0.2161 & 0.0002 \\
 45 & 01\ 31\ 48.20 & -13\ 38\ 12.3 & 20.82 & 2.08 & 0.2188 & 0.0001 \\
 46 & 01\ 31\ 48.45 & -13\ 37\ 43.2 & 20.34 & 2.16 & 0.1979 & 0.0005 \\
 47 & 01\ 31\ 49.21 & -13\ 37\ 35.0 & 19.66 & 2.20 & 0.2039 & 0.0003 \\
 48 & 01\ 31\ 49.38 & -13\ 36\ 06.9 & 20.53 & 2.32 & 0.2133 & 0.0004 \\
 49 & 01\ 31\ 49.47 & -13\ 37\ 26.5 & 18.02 & 1.62 & 0.2140 & 0.0002 \\
 50 & 01\ 31\ 49.64 & -13\ 35\ 21.6 & 19.74 & 2.43 & 0.2061 & 0.0002 \\
 51 & 01\ 31\ 49.84 & -13\ 36\ 11.7 & 20.15 & 2.12 & 0.2123 & 0.0003 \\
 52 & 01\ 31\ 50.43 & -13\ 38\ 35.9 & 19.89 & 2.26 & 0.2111 & 0.0003 \\
 53 & 01\ 31\ 50.64 & -13\ 33\ 36.4 & 18.13 & 2.45 & 0.2064 & 0.0004 \\
 54 & 01\ 31\ 50.89 & -13\ 36\ 04.2 & 18.77 & 2.54 & 0.2078 & 0.0002 \\
 55 & 01\ 31\ 50.98 & -13\ 36\ 49.5 & 19.64 & 2.54 & 0.2042 & 0.0003 \\
 56 & 01\ 31\ 51.15 & -13\ 38\ 12.8 & 18.27 & 2.38 & 0.2183 & 0.0002 \\
 57 & 01\ 31\ 51.34 & -13\ 36\ 56.6 & 18.10 & 2.58 & 0.2068 & 0.0002 \\
 58 & 01\ 31\ 51.58 & -13\ 35\ 07.0 & 18.24 & 2.13 & 0.2001 & 0.0002 \\
 59 & 01\ 31\ 51.73 & -13\ 38\ 40.2 & 19.26 & 2.37 & 0.2184 & 0.0002 \\
 60 & 01\ 31\ 51.82 & -13\ 38\ 30.2 & 20.23 & 2.33 & 0.2028 & 0.0005 \\
 61 & 01\ 31\ 52.31 & -13\ 36\ 57.9 & 17.37 & 3.04 & 0.2024 & 0.0002 \\
 62 & 01\ 31\ 52.54 & -13\ 36\ 40.4 & 17.00 & 2.63 & 0.2097 & 0.0002 \\
 63 & 01\ 31\ 53.34 & -13\ 36\ 31.3 & 18.52 & 2.92 & 0.2094 & 0.0002 \\
 64 & 01\ 31\ 53.86 & -13\ 38\ 21.2 & 19.66 & 1.84 & 0.2170 & 0.0003 \\
 65 & 01\ 31\ 53.87 & -13\ 36\ 13.4 & 18.90 & 2.48 & 0.2085 & 0.0002 \\
 66^{\mathrm{a}} & 01\ 31\ 54.09 & -13\ 39\ 39.0 & 21.03 & 2.73 & 0.4538 & 0.0001 \\
              \noalign{\smallskip}			    
            \hline					    
            \noalign{\smallskip}			    
            \hline					    
         \end{array}
     $$ 
}
         \end{table}
\addtocounter{table}{-1}
\begin{table}
          \caption[ ]{Continued.}
{\footnotesize
     $$ 
           \begin{array}{c c c c c c c}
            \hline
            \noalign{\smallskip}
            \hline
            \noalign{\smallskip}

\mathrm{ID} & {\mathrm{\alpha}} & \mathrm{\delta}  & \mathrm{V} & B-R  & z & 
\mathrm{\Delta} z\\
            \hline
            \noalign{\smallskip}
 67 & 01\ 31\ 54.33 & -13\ 38\ 54.9 & 20.04 & 2.26 & 0.2083 & 0.0004 \\
 68 & 01\ 31\ 55.12 & -13\ 37\ 04.4 & 18.93 & 2.49 & 0.2118 & 0.0003 \\
 69 & 01\ 31\ 55.18 & -13\ 36\ 57.6 & 18.94 & 2.43 & 0.2150 & 0.0002 \\
 70 & 01\ 31\ 55.34 & -13\ 36\ 15.9 & 19.14 & 1.05 & 0.2169 & 0.0003 \\
 71 & 01\ 31\ 55.47 & -13\ 38\ 28.9 & 19.18 & 2.39 & 0.2144 & 0.0002 \\
 72 & 01\ 31\ 55.61 & -13\ 39\ 58.9 & 20.65 & 1.26 & 0.2097 & 0.0002 \\
 73 & 01\ 31\ 55.95 & -13\ 36\ 40.4 & 18.00 & 1.35 & 0.1999 & 0.0005 \\
 74 & 01\ 31\ 56.22 & -13\ 36\ 46.7 & 18.43 & 1.78 & 0.2098 & 0.0003 \\
 75 & 01\ 31\ 56.58 & -13\ 38\ 34.9 & 20.03 & 2.23 & 0.2117 & 0.0002 \\
 76 & 01\ 31\ 56.78 & -13\ 40\ 04.8 & 20.16 & 2.26 & 0.2098 & 0.0004 \\
 77 & 01\ 31\ 56.91 & -13\ 38\ 30.2 & 18.26 & 2.49 & 0.2102 & 0.0002 \\
 78 & 01\ 31\ 57.38 & -13\ 38\ 08.2 & 19.58 & 2.47 & 0.2107 & 0.0003 \\
 79 & 01\ 31\ 57.99 & -13\ 38\ 59.7 & 19.14 & 2.28 & 0.1973 & 0.0003 \\
 80 & 01\ 31\ 58.28 & -13\ 39\ 36.2 & 19.14 & 2.38 & 0.2056 & 0.0002 \\
 81 & 01\ 31\ 58.68 & -13\ 38\ 04.0 & 20.32 & 2.27 & 0.2093 & 0.0004 \\
 82 & 01\ 31\ 59.10 & -13\ 39\ 26.4 & 18.97 & 2.42 & 0.2104 & 0.0002 \\
 83 & 01\ 32\ 00.34 & -13\ 37\ 56.5 & 20.02 & 2.32 & 0.2056 & 0.0002 \\
 84 & 01\ 32\ 01.18 & -13\ 35\ 17.5 & 19.65 & 2.29 & 0.2172 & 0.0003 \\   
 85^{\mathrm{a}} & 01\ 32\ 01.56 & -13\ 32\ 21.1 & 19.95 & .... & 0.2617 & 0.0003 \\ 
 86 & 01\ 32\ 01.66 & -13\ 35\ 33.8 & 18.36 & 2.49 & 0.2069 & 0.0002 \\
 87 & 01\ 32\ 01.82 & -13\ 33\ 34.0 & 18.60 & 2.52 & 0.2083 & 0.0003 \\ 
 88 & 01\ 32\ 01.84 & -13\ 36\ 15.7 & ....  & 2.33 & 0.2131 & 0.0004 \\
 89 & 01\ 32\ 01.91 & -13\ 35\ 31.2 & 18.65 & .... & 0.1984 & 0.0002 \\
 90 & 01\ 32\ 02.18 & -13\ 35\ 51.0 & 19.06 & 2.10 & 0.2091 & 0.0003 \\
 91 & 01\ 32\ 02.47 & -13\ 32\ 13.0 & 19.90 & .... & 0.1982 & 0.0004 \\
 92 & 01\ 32\ 02.94 & -13\ 39\ 39.4 & 19.48 & 2.48 & 0.2110 & 0.0002 \\
 93 & 01\ 32\ 03.07 & -13\ 36\ 44.0 & 19.95 & 2.33 & 0.2102 & 0.0005 \\
 94 & 01\ 32\ 03.50 & -13\ 31\ 59.1 & 19.65 & .... & 0.1971 & 0.0003 \\
 95 & 01\ 32\ 03.96 & -13\ 35\ 54.4 & 19.92 & 2.33 & 0.2137 & 0.0003 \\
 96 & 01\ 32\ 03.97 & -13\ 38\ 01.2 & 20.31 & 2.16 & 0.2116 & 0.0003 \\
 97 & 01\ 32\ 04.29 & -13\ 39\ 53.9 & 17.82 & 2.55 & 0.2125 & 0.0002 \\
 98 & 01\ 32\ 04.35 & -13\ 37\ 26.3 & 18.36 & 2.50 & 0.2061 & 0.0002 \\
 99 & 01\ 32\ 05.00 & -13\ 40\ 35.1 & 19.77 & 2.34 & 0.2132 & 0.0002 \\
              \noalign{\smallskip}			    
            \hline					    
            \noalign{\smallskip}			    
            \hline					    
         \end{array}
     $$ 
}
         \end{table}
\addtocounter{table}{-1}
\begin{table}
          \caption[ ]{Continued.}
{\footnotesize
     $$ 
           \begin{array}{c c c c c c c}
            \hline
            \noalign{\smallskip}
            \hline
            \noalign{\smallskip}

\mathrm{ID} & {\mathrm{\alpha}} & \mathrm{\delta}  & \mathrm{V} & B-R  & z & 
\mathrm{\Delta} z\\
            \hline
            \noalign{\smallskip}
100 & 01\ 32\ 05.02 & -13\ 33\ 42.0 & 20.34 & 2.30 & 0.2087 & 0.0003 \\ 
101 & 01\ 32\ 05.05 & -13\ 37\ 33.3 & 19.28 & 2.35 & 0.2074 & 0.0002 \\
102^{\mathrm{b}} & 01\ 32\ 07.29 & -13\ 37\ 30.3 & 19.06 & 1.63 & 0.1744 & 0.0006 \\
103 & 01\ 32\ 07.98 & -13\ 41\ 31.4 & 19.86 & 2.23 & 0.1997 & 0.0002 \\
104 & 01\ 32\ 10.37 & -13\ 37\ 24.3 & 19.54 & .... & 0.2161 & 0.0002 \\
105 & 01\ 32\ 12.34 & -13\ 34\ 21.1 & 19.74 & .... & 0.2134 & 0.0004 \\
106 & 01\ 32\ 12.51 & -13\ 41\ 18.1 & ..... & .... & 0.2145 & 0.0003 \\
107 & 01\ 32\ 13.19 & -13\ 39\ 46.0 & ..... & .... & 0.2087 & 0.0003 \\
108^{\mathrm{a}} & 01\ 32\ 13.22 & -13\ 31\ 09.1 & 22.34 & .... & 0.2597 & 0.0005 \\
109 & 01\ 32\ 14.04 & -13\ 38\ 08.5 & 17.90 & .... & 0.2175 & 0.0002 \\
110 & 01\ 32\ 14.52 & -13\ 32\ 29.1 & 19.39 & .... & 0.2145 & 0.0002 \\
111 & 01\ 32\ 15.00 & -13\ 41\ 13.9 & ..... & .... & 0.2158 & 0.0003 \\
112 & 01\ 32\ 15.56 & -13\ 37\ 49.2 & 18.94 & .... & 0.2149 & 0.0002 \\
113 & 01\ 32\ 15.84 & -13\ 35\ 41.0 & 20.71 & .... & 0.2124 & 0.0004 \\
114 & 01\ 32\ 16.00 & -13\ 38\ 34.6 & 19.57 & .... & 0.2203 & 0.0003 \\
115 & 01\ 32\ 16.20 & -13\ 32\ 38.8 & 18.97 & .... & 0.2034 & 0.0003 \\
116 & 01\ 32\ 16.76 & -13\ 35\ 03.7 & 20.81 & .... & 0.2138 & 0.0004 \\
117 & 01\ 32\ 16.98 & -13\ 33\ 01.1 & 20.64 & .... & 0.2117 & 0.0004 \\
118 & 01\ 32\ 17.26 & -13\ 38\ 06.1 & 19.74 & .... & 0.2105 & 0.0003 \\
119 & 01\ 32\ 17.48 & -13\ 38\ 42.3 & 19.84 & .... & 0.1963 & 0.0006 \\
            \noalign{\smallskip}			    
            \hline					    
            \noalign{\smallskip}			    
            \hline					    
         \end{array}
     $$ 
}
\begin{list}{}{}  
\item[$^{\mathrm{a}}$] Background galaxy.
\item[$^{\mathrm{b}}$] Foreground galaxy.
\end{list}
   \end{table}

\section{Dynamical analysis}
\label{sec:24}

\subsection{Member selection and global properties}
\label{sec:241}

ABCG\,209 appears as a well isolated peak in the redshift space. The
analysis of the velocity distribution based on the one--dimensional
adaptive kernel technique (Pisani \ct{pis93}, as implemented by
Fadda et al. \ct{fad96} and Girardi et al. \ct{gir96}) confirms
the existence of a single peak at $\mathrm{z\sim 0.209}$. 

\begin{figure*}
\centering
\includegraphics[width=16cm]{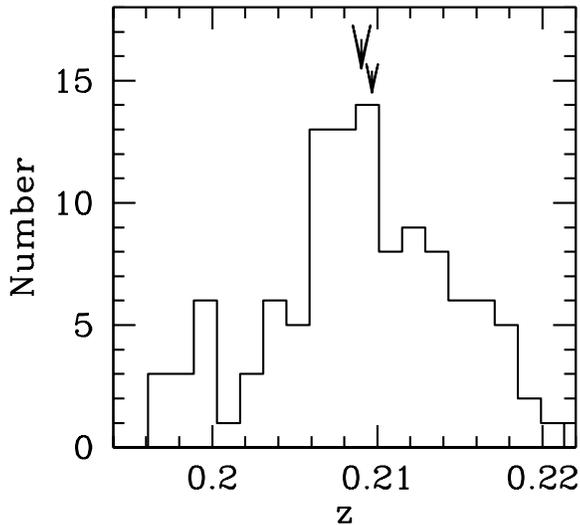}
\vspace{-8cm}
\caption
{Distribution of redshifts of the
cluster members.  The big and small arrows indicate the mean cluster
redshift and the redshift of the cD galaxy, respectively.
}
\label{fighisto}
\end{figure*}

Fig.~\ref{fighisto} shows the redshift distribution of the 112 cluster
members. The mean redshift in the present sample is $\mathrm{<z>=0.2090\pm 0.0004}$,
as derived by the biweight estimator (Beers et al. \ct{bee90}).

In order to determine the cluster center, we applied the
two--dimensional adaptive kernel technique to galaxy positions. The
center of the most dense peak ($\alpha$= 01 31 52.70, $\delta$= -13 36
41.9) is close to the position of cD galaxy ($\alpha$= 01 31 52.54,
$\delta$= -13 36 40.4). We estimated the line--of--sight (LOS)
velocity dispersion, $\mathrm{\sigma_v}$, by using the biweight
estimator (ROSTAT package; Beers et al. \ct{bee90}). By applying the
relativistic correction and the usual correction for velocity errors
(Danese et al. \ct{dan80}), we obtained
$\mathrm{\sigma_v=1394^{+88}_{-99}}$ \kss, where errors were estimated
with the bootstrap method.

\begin{figure*}
\hspace{1cm}
\vspace{-2cm}
\includegraphics[width=16cm]{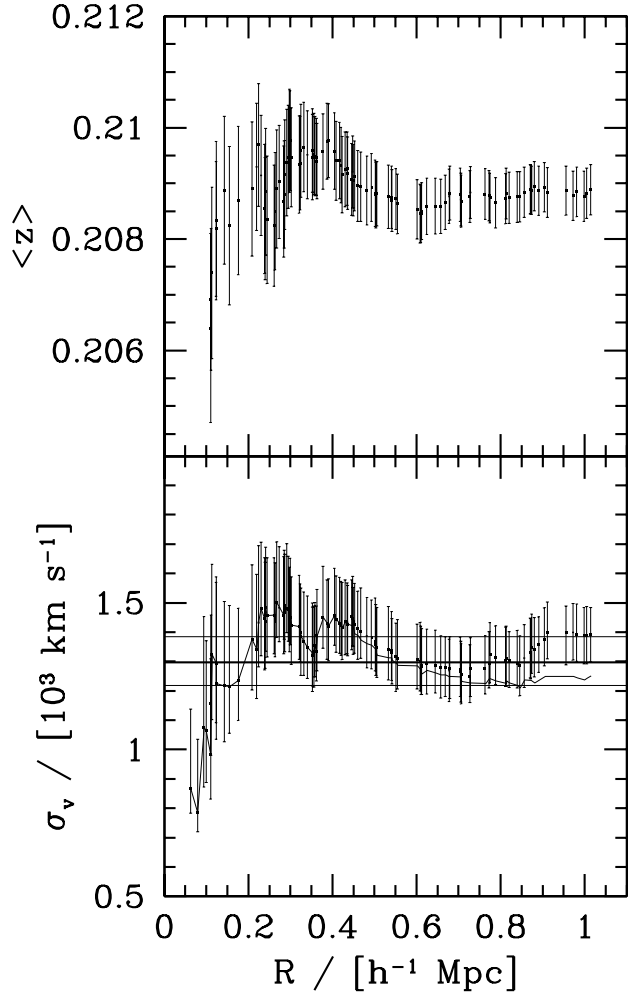}
\caption
{Integrated mean redshift and LOS velocity dispersion profiles (upper
and lower panel, respectively), where the mean and dispersion at a
given (projected) radius from the cluster center is estimated by
considering all galaxies within that radius. The error bars at the
$68\%$ c.l. are shown. In the lower panel, the faint line gives the
profile after the rejection of seven possible interlopers according to
the ``shifting gapper'' method (see text in Sect.~\ref{sec:242}), and
the horizontal lines represent X--ray temperature with the respective
errors (cf. Sect.~\ref{sec:253}) transformed in $\mathrm{\sigma_v}$
imposing $\mathrm{\beta_{spec}=1}$, where
$\mathrm{\beta_{spec}=\sigma_v^2/(kT/\mu m_p)}$, with $\mathrm{\mu}$
the mean molecular weight and $\mathrm{m_p}$ the proton mass. }
\label{figprof}
\end{figure*}

In order to check for possible variation of $\mathrm{<z>}$ and $\mathrm{\sigma_v}$ with
increasing radius we plot the integrated  mean velocity and 
velocity dispersion profiles in Fig.~\ref{figprof}. The measure of
mean redshift and velocity dispersion sharply vary in the internal
cluster region although the large associated errors do not allow to
claim for a statistically significant variation. On the other hand, 
in the external cluster regions, where the number of galaxies is larger,
the estimates of $\mathrm{<z>}$ and $\mathrm{\sigma_v}$ are quite robust.

Assuming that the system is in dynamical equilibrium, the value of
$\mathrm{\sigma_v}$ leads to a value of the radius of the collapsed,
quasi--virialized region $\mathrm{R_{vir}\sim 1.78}$ \h (cf. Eq.~(1)
of Girardi \& Mezzetti 2001). Therefore our spectroscopic data sample
covers about half of the virial region ($\sim$ 1 \hh).  Under the same
assumption, we estimated the mass of the system by applying the virial
method. In particular, for the surface term correction to the standard
virial mass we assumed a value of $20\%$ (cf. Girardi \& Mezzetti
\ct{gir01}), obtaining
$\mathrm{M(<R_{vir})=2.25^{+0.63}_{-0.65}\times10^{15}}$ \msun.

\subsection{Possible contamination effects}
\label{sec:242}

We further explored the reliability of $\mathrm{\sigma_v}$ related to the
possibility of contamination by interlopers.

\begin{figure*}[h]
\hspace{1cm}
\vspace{-8cm}
\includegraphics[width=16cm]{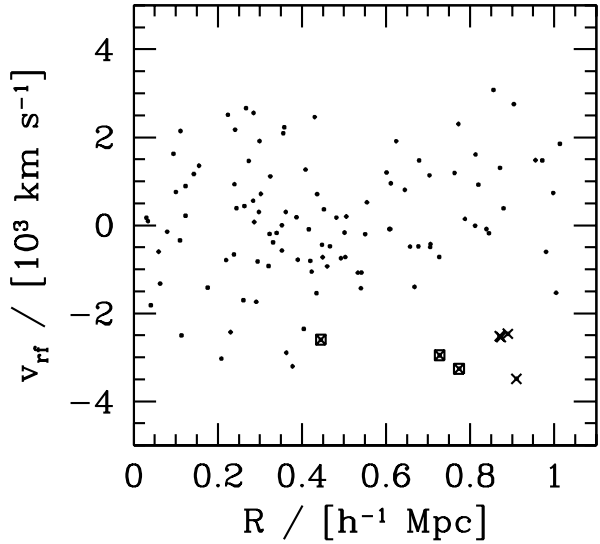}
\caption
{Velocity in the cluster rest frame vs. (projected) clustercentric
distance for the 112 selected members.  The standard application of
the ``shifting gapper'' method would reject galaxies as cluster
members indicated by squares, and a more conservative application also
galaxies indicated by crosses.
}
\label{figvd}
\end{figure*}

The cluster rest--frame velocity vs. projected clustercentric distance
is shown in Fig.~\ref{figvd}.
Although no obvious case of outliers is present, we applied the
procedure of the ``shifting gapper'' by Fadda et al. (\ct{fad96}).
This procedure rejects galaxies that are too far in velocity from the
main body of galaxies of the whole cluster within a fixed bin,
shifting along the distance from the cluster center.  According
to the prescriptions in Fadda et al. (\ct{fad96}), we used a gap of
$1000$ \ks and a bin of 0.4 \hh, or large enough to include at least 15
galaxies.

In this way we rejected three galaxies (indicated by squares in
Fig.~\ref{figvd}). However, for ABCG\,209 the results are too much
sensitive to small changes of the adopted parameters. For instance,
with a bin of $0.5$ \h no galaxy was rejected, while seven galaxies
were rejected with a bin of $0.3$ \h (cf. crosses in
Fig.~\ref{figvd}). In the last case, we obtained a value of
$\mathrm{\sigma_v=1250^{+84}_{-98}}$ \kss, which is slightly smaller
than-- (although fully consistent with--) the value computed in
Sect.~\ref{sec:241} (cf. also the velocity dispersion profile in
Fig.~\ref{figprof}).  With this value of the velocity dispersion, the
computation for the mass within $\mathrm{R_{vir}=1.59}$ \h leads to a
value of total mass
$\mathrm{M(<R_{vir})=1.62^{+0.48}_{-0.46}\times10^{15}}$ \msun.

\begin{figure*}[h]
\hspace{1cm}
\vspace{-8cm}
\includegraphics[width=16cm]{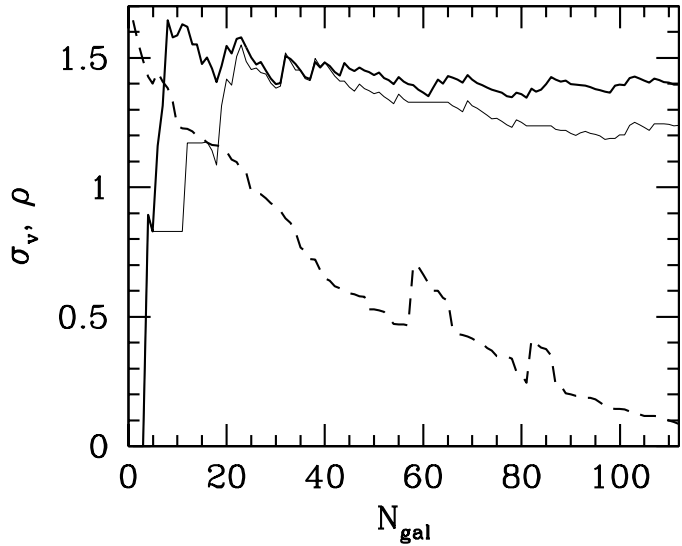}
\caption
{Integrated LOS velocity dispersion profile $\mathrm{\sigma_v}$ (solid line, in
units of $10^3$ \kss) and galaxy number density $\rho$ (dashed line, in
arbitrary units) along the sequences beginning with the center of the
cluster.  The thinner line gives the $\mathrm{\sigma_v}$--profile after the
rejection of
galaxies belonging to the two small density peaks.}
\label{figkittler}
\end{figure*}

We determined the galaxy density and the integrated LOS velocity
dispersion along the sequence of galaxies with decreasing density
beginning with the cluster center (Kittler \ct{kit76}; Pisani
\ct{pis96}).  As shown in Fig.~\ref{figkittler} the density profile
has only two minor peaks, possibly due to non complete sampling, and
their rejection leads to a small variation in the estimate of
$\mathrm{\sigma_v}$ and in the behaviour of velocity dispersion
profile (cf. with Figure~2 of Girardi et al. \ct{gir96} where the
large effect caused by a close system is shown).

We conclude that the contamination by interlopers cannot explain the
high value of velocity dispersion, which is therefore connected to the
peculiarity of the internal dynamics of the cluster itself. In fact, a
high value of $\mathrm{\sigma_v}$ is already present in the internal cluster
region, namely within $\sim$ 0.2--0.3 \h (cf. Fig.~\ref{figprof}), where the
contamination is expected to be negligible.

\subsection{Velocity distribution}
\label{sec:243}

In order to better characterise the velocity distribution, we
considered its kurtosis $K$, skewness $S$, scaled
tail index $STI$, and the probability associated to the
W--test $P(W)$, (cf. Shapiro \& Wilk \ct{sha65}).

For the kurtosis and the skewness we found $K=2.61\pm0.45$,
$S=-0.20\pm0.23$, respectively, i.e. values that are consistent with a
Gaussian distribution (reference value $K$=3, $S$=0).  The $STI$
indicates the amount of the elongation in a sample relative to the
standard Gaussian. This is an alternative to the classical kurtosis
estimator, based on the data set percentiles as determined from the
order statistics (e.g., Rosenberger \& Gasko \ct{ros83}).  The
Gaussian, which is by definition neutrally elongated, has
$STI=1.0$. Heavier--tailed distributions have $STI \sim$ 1.25 (see,
e.g., Beers et al. \ct{bee91}).  For our data $STI=1.254$,
indicating a heavier--tailed distribution, with 95\%-99\% c.l.
(cf. Table~2 of Bird \& Beers \ct{bir93}).  On the other hand, the
W--test rejected the null hypothesis of a Gaussian parent
distribution, with only a marginal significance at $92.3\%$ c.l.

\begin{figure*}[h]
\hspace{1cm}
\vspace{-8cm}
\includegraphics[width=16cm]{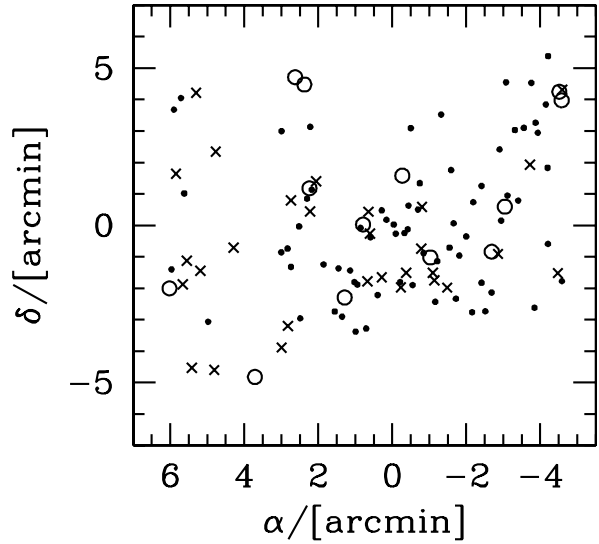}
\caption
{Distribution of member galaxies separated into three clumps according
the one--dimensional KMM test.  The plot is centered on the cluster
center. Open circles, dots and crosses indicate low--,
intermediate--, and high--velocity clumps respectively.}
\label{figkmm}
\end{figure*}

In order to detect possible subclumps in the velocity distribution, we
applied the KMM algorithm (cf. Ashman et al \ct{ash94} and
refs. therein).  By taking the face value of maximum likelihood
statistics, we found a marginal evidence that a mixture of three
Gaussians (of $\mathrm{n_1=13}$, $\mathrm{n_2=70}$, and
$\mathrm{n_3=29}$ members at mean redshift $\mathrm{z_1 = 0.1988}$,
$\mathrm{z_2 = 0.2078}$, and $\mathrm{z_3 = 0.2154}$) is a better
description of velocity distribution (at $91.2\%$ c.l.).  For the
clumps we estimated a velocity dispersion of $\mathrm{\sigma_{v1} =
337}$ \ks, $\mathrm{\sigma_{v2} = 668}$ \ks, and $\mathrm{\sigma_{v3} =
545}$ \ks.  Fig.~\ref{figkmm} shows the spatial distributions of the
three clumps. The second and the third clumps are clearly spatially
segregated.

\subsection{Velocity field}
\label{sec:244}

The above result suggested to investigate the velocity field in more
detail.  To this aim, we divided galaxies into a low-- and a
high--velocity sample (with respect to the mean redshift).  As in
Fig.~\ref{figkmm}, low-- and high--velocity galaxies are clearly
segregated in the E--W direction, the two distributions being
different at the $99.88\%$ c.l., according to the two--dimensional
Kolmogorov--Smirnov test (hereafter 2DKS--test; cf. Fasano \&
Franceschini \ct{fas87}, as implemented by Press et al. \ct{pre92}).

\begin{figure*}[h]
\hspace{1cm}
\vspace{-8cm}
\includegraphics[width=16cm]{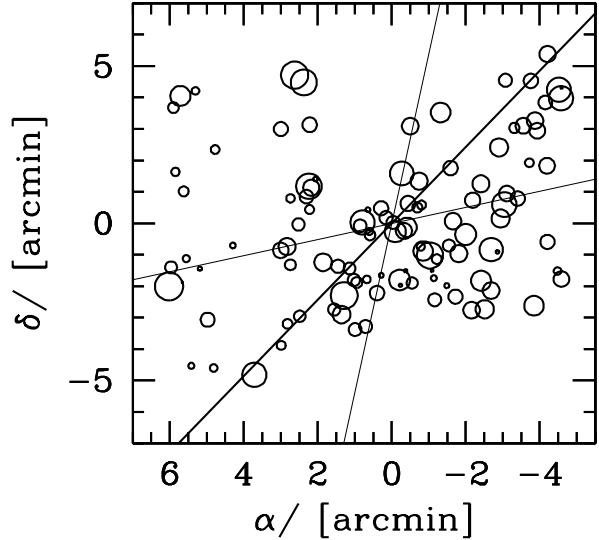}
\caption
{Spatial distribution of member galaxies: the larger the circle, the
smaller is the radial velocity.  The figure is centered on the cluster
center. The solid and the thin lines indicate the position angle of
the cluster velocity gradient and relative errors, respectively.}
\label{figgrad}
\end{figure*}

We then looked for a possible velocity gradient by means of a
multi--linear fit (e.g.,  implemented by NAG Fortran Workstation
Handbook, \ct{nag86}) to the observed velocity field (cf. also den
Hartog \& Katgert \ct{den96}; Girardi et al. \ct{gir96}).  We
found marginal evidence (c.l. $95.2\%$) for the presence of a velocity
gradient in the direction SE--NW, at position
angle PA$=141^{+29}_{-37}$ degrees (cf. Fig.~\ref{figgrad}).

In order to assess the significance of the velocity gradient, we
performed 1000 Monte Carlo simulations by randomly shuffling the
galaxy velocities and determined the coefficient of multiple
determination ($R^2$) for each of them. We then defined the
significance of the velocity gradient as the fraction of times in
which the $R^2$ of the simulation was smaller than the observed $R^2$.

\subsection{3D substructure analysis}
\label{sec:245}

In order to check for the presence of substructure, we combined
velocity and position information by computing the
$\Delta$--statistics \footnote{For each galaxy, the deviation $\delta$
is defined as $\mathrm{\delta^2 = (11/\sigma^2)[(\overline v_l - \overline
v)^2+(\sigma_l - \sigma)^2]}$, where subscript l denotes the average
over the 10 neighbours of the galaxy.  $\Delta$ is the sum of the
$\delta$ of the individual galaxies.} devised by Dressler \& Schectman
(\ct{dre88}) .  We found a value of 162 for the $\Delta$ parameter,
which gives the cumulative deviation of the local kinematical
parameters (velocity mean and velocity dispersion) from the global
cluster parameters.  The significance of substructure was checked by
running 1000 Monte Carlo simulations, randomly shuffling the galaxy
velocities, obtaining a significance level of $98.7\%$.

This indicates that the cluster has a complex structure.

\begin{figure*}
\hspace{1cm}
\vspace{-8cm}
\includegraphics[width=16cm]{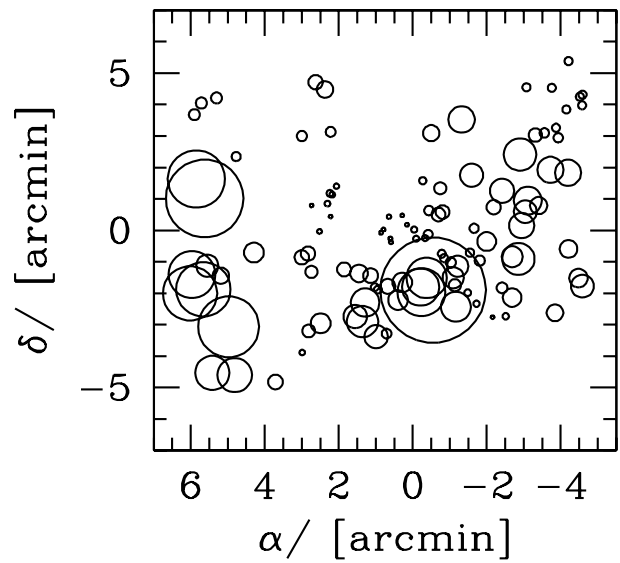}
\caption
{Spatial distribution on the sky of the 112 cluster members, each marked
by a circle: the larger the circle, the larger is the deviation
$\delta$ of the local parameters from the global cluster
parameters. The figure shows evidence for substructure according to
the Dressler \& Schectman test.
The plot is centered on the cluster center. 
}
\label{figds}
\end{figure*}

The technique by Dressler \& Schectman does not allow a direct
identification of galaxies belonging to the detected substructure;
however it can roughly identify the positions of substructures.
To this aim, in Fig.~\ref{figds} the galaxies are marked by circles
whose diameter is proportional to the deviation $\delta$ of the
individual parameters (position and velocity) from the mean cluster
parameters.

A group of galaxies with high velocity is the likely cause of large
values of $\delta$ in the external East cluster region.  The other
possible substructure lies in the well sampled cluster region, closer
than $1$ arcmin SW to the cluster center .

\section{Galaxy populations and the hot gas} 
\label{sec:25}

\subsection{Luminosity and colour segregation}
\label{sec:251}

In order to unravel a possible luminosity segregation, we divided the
sample in a low and a high--luminosity subsamples by using the median
V--magnitude$=19.46$ (53 and 54 galaxies, respectively), and applied them
the standard means test and F--test (Press et al. \ct{pre92}).
We found no significant difference between the two samples.

\begin{figure*}
\hspace{1cm}
\vspace{-8cm}
\includegraphics[width=16cm]{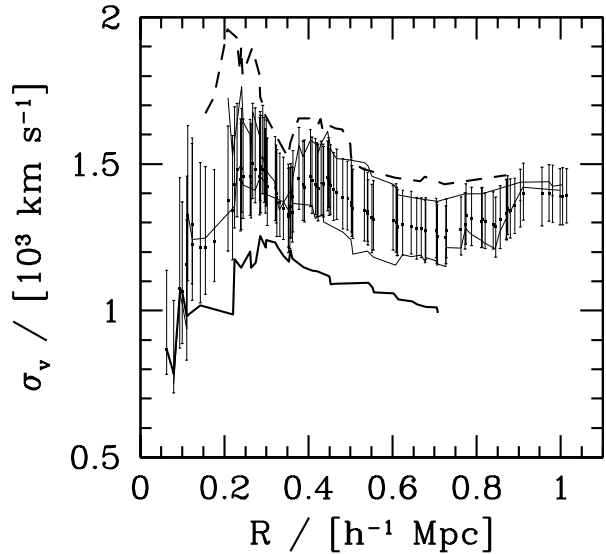}
\caption
{Integrated LOS velocity dispersion profile as in Fig.~\ref{figprof}.
The solid and dashed lines give the profiles of redder and bluer
galaxies, respectively.  The faint lines give the profiles of bright
and faint galaxies (upper and lower lines, respectively).  }
\label{figprofcol}
\end{figure*}

\begin{figure*}
\hspace{1cm}
\vspace{-8cm}
\includegraphics[width=16cm]{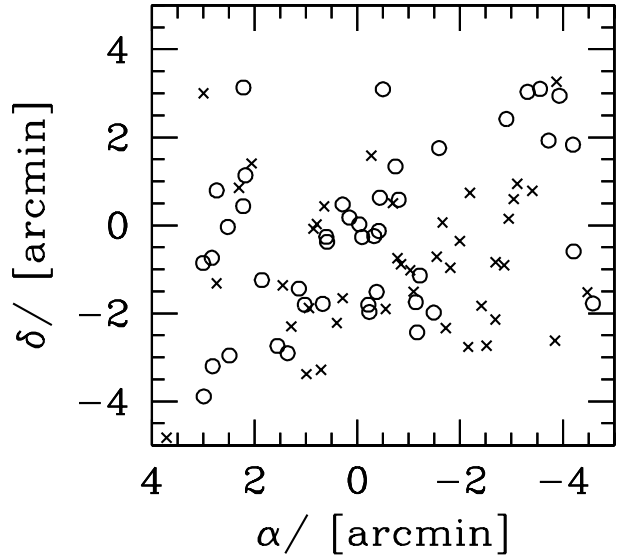}
\caption
{Spatial distribution of the 85 member galaxies having B--R colour
information: circles and crosses give redder and bluer galaxies,
respectively. The plot is centered on the cluster center.
}
\label{figcol}
\end{figure*}

We also looked for possible colour segregation, by dividing the sample
in a blue and a red subsamples relative to the median colour B--R$=2.32$
of the spectroscopic sample (41 and 44 galaxies, respectively).  There
is a slight differences in the peaks of the velocity distributions,
$\mathrm{z_{blue}=0.2076}$ and $\mathrm{z_{red}=0.2095}$, which gives a marginal
probability of difference according to the means test ($93.5\%$).
The velocity dispersions are different being
$\mathrm{\sigma_{v,blue}=1462^{+158}_{-145}}$ \ks larger than
$\mathrm{\sigma_{v,red}=993^{+126}_{-88}}$ \ks (at the $98.6\%$ c.l., according
to the F--test), see also Fig.~\ref{figprofcol}.
Moreover, the two subsamples differ in the distribution of galaxy
positions ($97.4\%$ according to the 2DKS test), 
with blue galaxies lying mainly in the SW--region (cf. Fig.~\ref{figcol}).

\begin{figure*}
\hspace{1cm}
\vspace{-8cm}
\includegraphics[width=16cm]{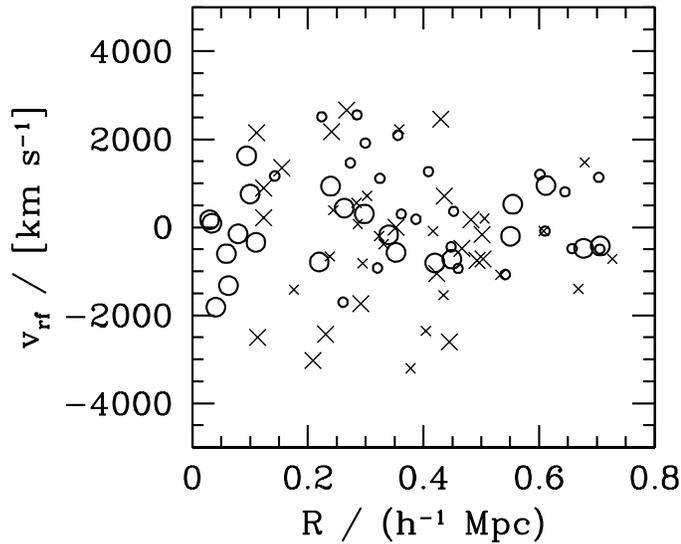}
\caption
{Velocity in the cluster rest--frame vs. (projected) clustercentric distance
for the 85 members having B--R colour information. 
Large circles, small circles, small crosses, and large crosses
indicate galaxies belonging to the four quartiles of the colour distribution,
from reddest to bluest galaxies. 
}
\label{figvdcol}
\end{figure*}

The reddest galaxies (B--R$>2.43$) are characterised by a still
smaller velocity dispersion (732 \kss) and the bluest (B--R$<2.19$)
by a larger velocity dispersion (1689 \kss).  Fig.~\ref{figvdcol}
plots velocities vs. clustercentric distance for the four quartiles of
the colour distribution: the reddest galaxies appear well concentrated
around the mean cluster velocity. From a more quantitative point of
view, there is a significant negative correlation between the B--R
colour and the absolute value of the velocity in the cluster rest
frame $\mathrm{|v_{rf}|}$ at the $99.3\%$ c.l. (for a Kendall correlation
coefficient of $-0.18$).

The velocity of the cD galaxy (z$=0.2095$) shows no evidence of
peculiarity according to the Indicator test by Gebhardt \& Beers
(\ct{geb91}). Moreover, the cD galaxy shows an elongation in the
NE--SW direction.

\begin{figure*}
\centering
\includegraphics[width=0.8\textwidth]{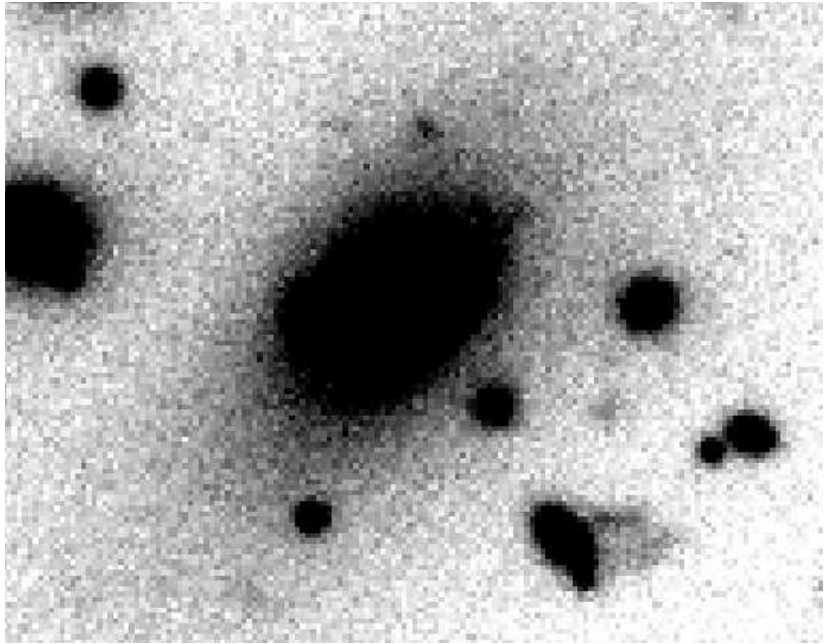}
\caption
{V--band image of cD galaxy ($0.7^\prime \times 0.5^\prime$), with North at top and East to left.}
\label{cD}
\end{figure*}

\subsection{Substructures with photometric data}
\label{sec:252}

We performed a two--dimensional analysis to detect subclumps using the
photometric sample of galaxies for which we have B--R colours. This
allows us to take advantage of the larger size of the photometric
sample compared to the spectroscopic one.
Galaxies were selected within $\pm0.5$ mag of the B--R vs. R
colour--magnitude relation and within the completeness limit magnitude
R$=22$ mag.
The colour--magnitude relation was determined on the N=85 spectroscopically 
confirmed cluster members, for which we have B--R colours.
The above selection leads to 3.68 $<$ (B--R) + 0.097 R $<$ 4.68, with
a sample size of N=392 galaxies.

\begin{figure*}
\centering
\includegraphics[width=0.8\textwidth]{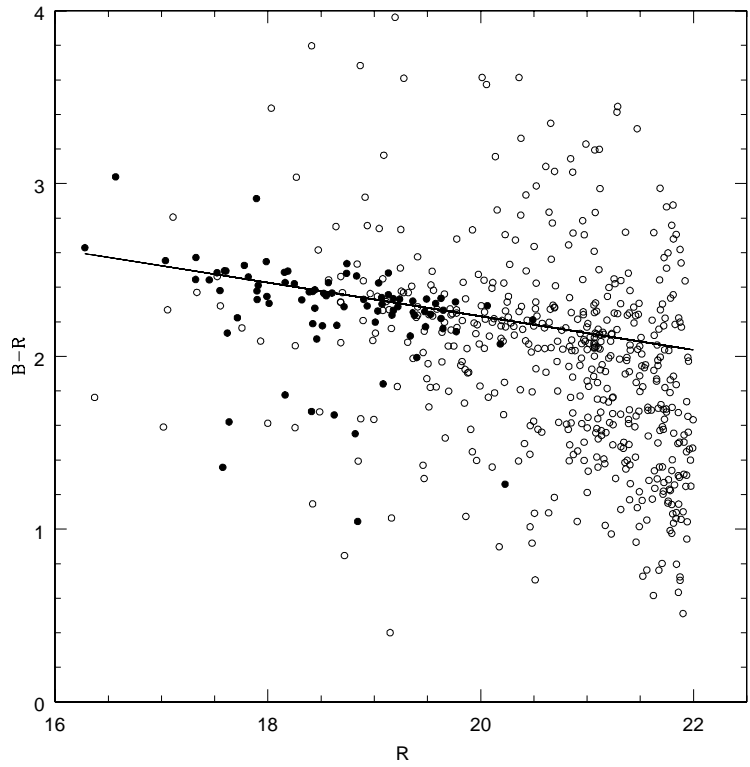}
\caption
{B--R vs. R colour--magnitude diagram for galaxies in the central field of 
ABCG\,209.
Filled circles: spectroscopically confirmed cluster members; open
circles: all galaxies within the completeness limit.}
\label{CM}
\end{figure*}

Fig.~\ref{CM} shows the B--R vs. R colour--magnitude diagram for the
596 galaxies detected in the R--band in the ABCG\,209 central
field. The sample includes also the 85 member galaxies with known
redshift. The colour--magnitude relation was derived by using a linear
regression based on the biweight estimator (see Beers et al. 1990).

\begin{figure*}
\centering
\includegraphics[width=7cm]{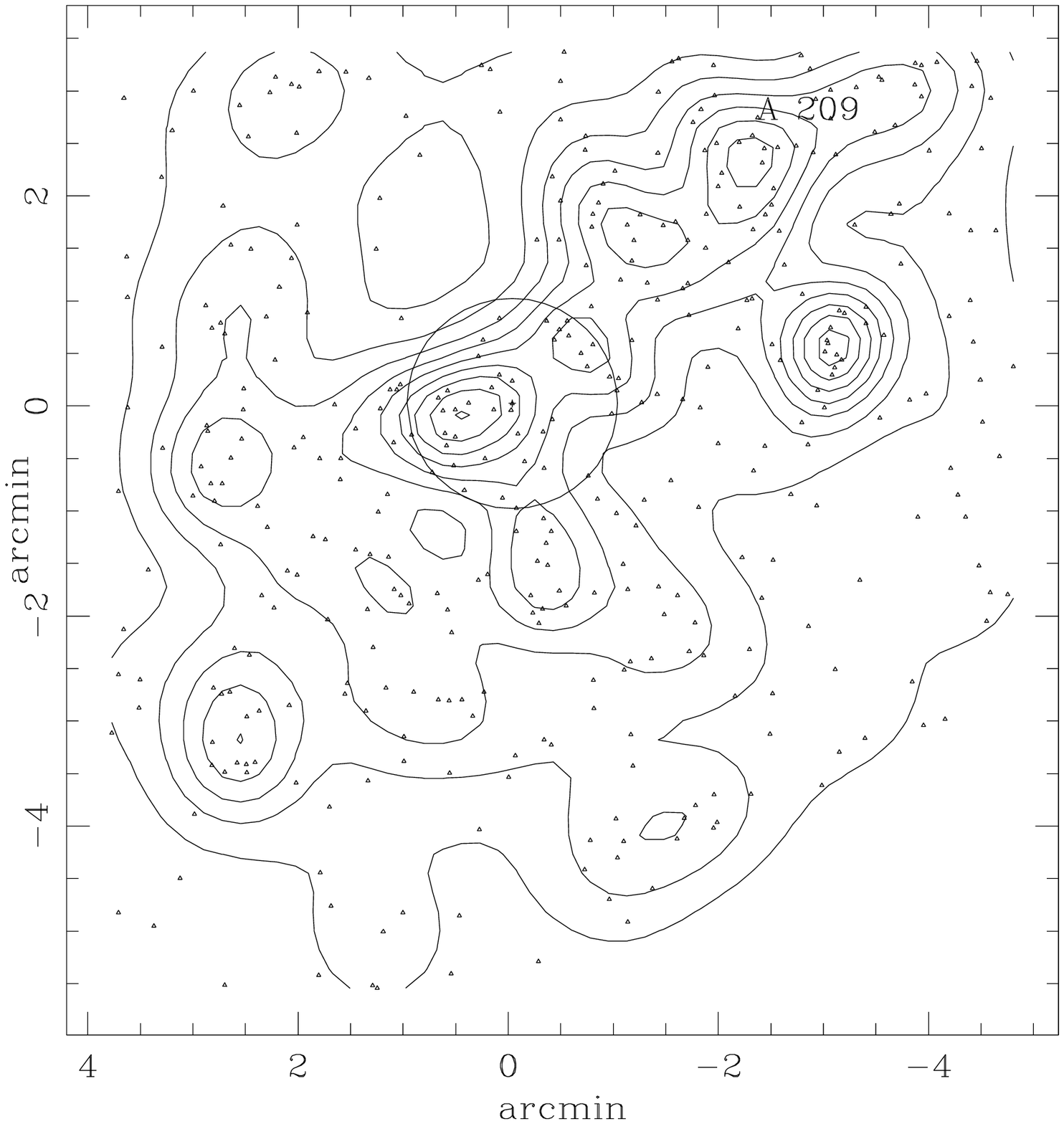}
\vbox{
\hbox{
\includegraphics[width=7cm]{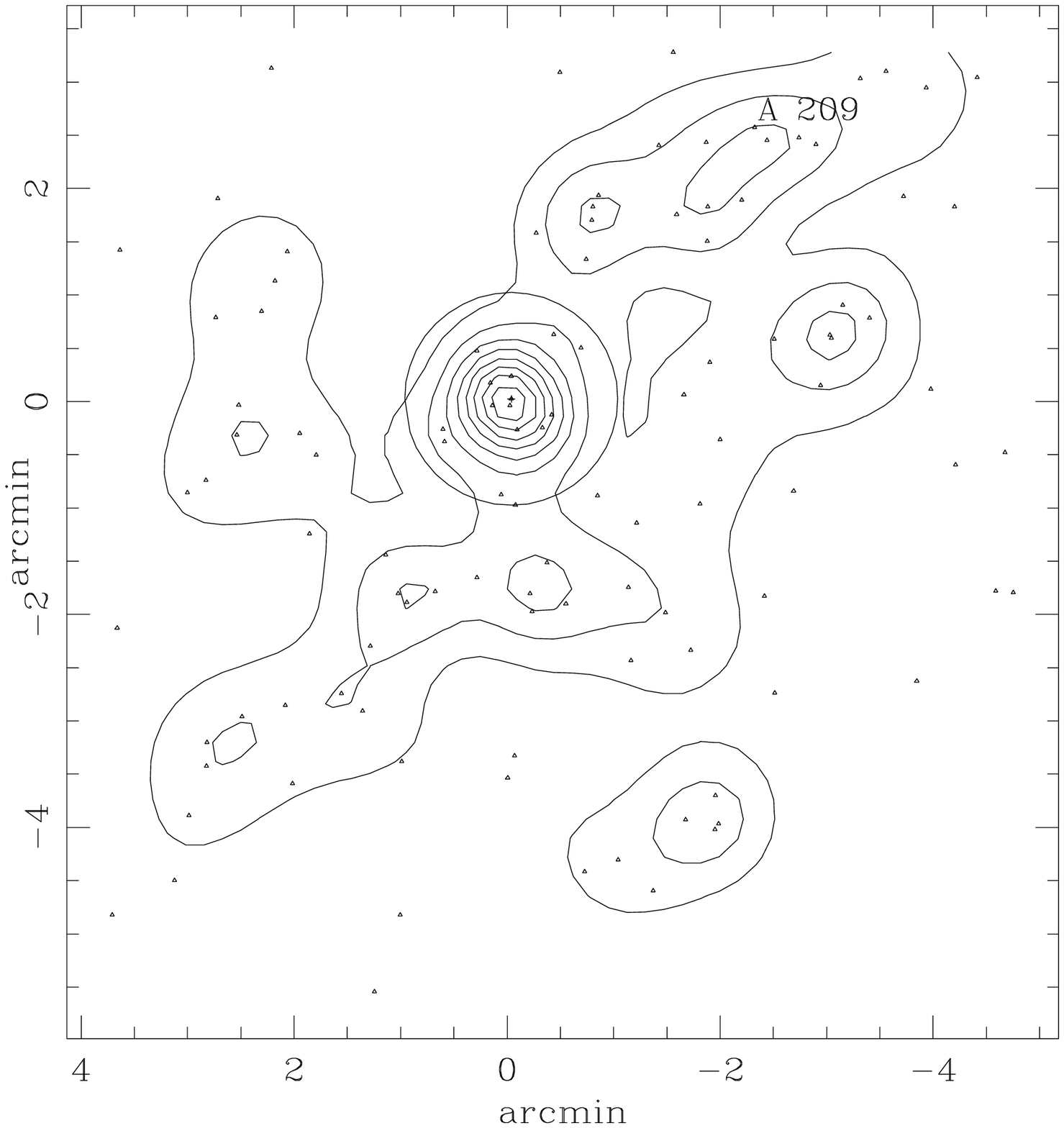}
\includegraphics[width=7cm]{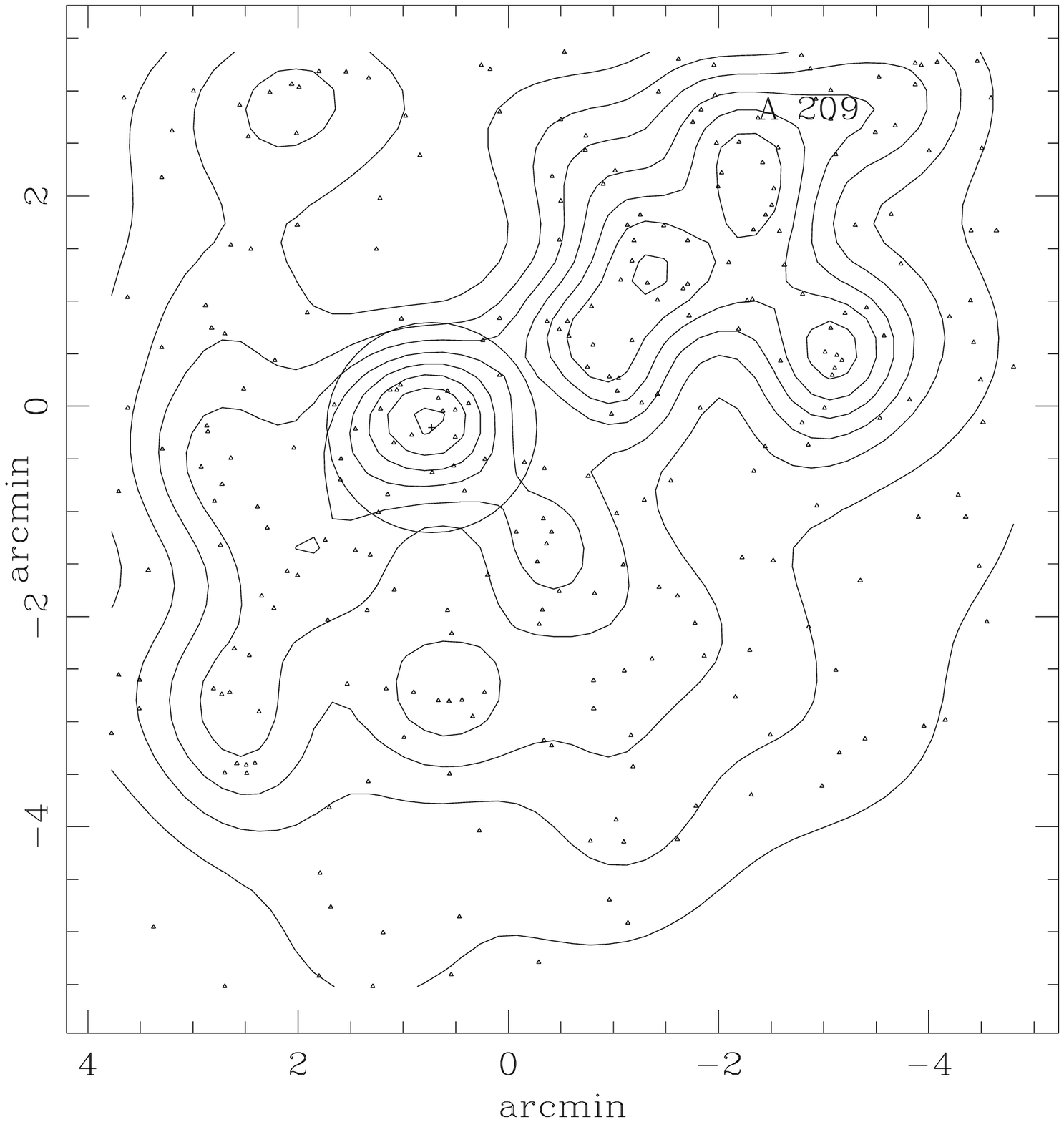}
}}
\caption
{Spatial distribution on the sky and relative isodensity contour map
of likely cluster members selected on the basis of the B--R
colour--magnitude relation, recovered by adaptive--kernel method
(cf. Pisani \ct{pis93},\ct{pis96}).  The plots are centered on the
cluster center.  Top panel: the 392 selected galaxies (see text).  The
1 arcmin circle is centered on the cD galaxy.  Bottom left panel: bright
(R$<19.5$) galaxies. The 1 arcmin circle is again centered on the cD
galaxy.  Bottom right panel: fainter (R$>19.5$) galaxies. The 1 arcmin
circle is centered on the secondary X--ray peak
(cf. Sect.~\ref{sec:253}). In the figures the isodensity contours are
spaced by 1 $\cdot$ 10$^{-4}$ galaxies/arcsec$^2$ in the top panel and
by 5 $\cdot$ 10$^{-5}$ galaxies/arcsec$^2$ in the central and bottom
panels.}
\label{fig2d}
\end{figure*}

Fig.~\ref{fig2d} -- upper panel -- shows the projected galaxy
distribution is clustered and elongated in the SE--NW
direction. Notice that there is no clear clump of galaxies perfectly
centered on the cD galaxy, whose position is coincident with the
center determined in Sect.~\ref{sec:241} from redshift data only.  In
order to investigate this question, we noted that galaxies with
redshift data are brighter than 22 R--mag, being the $\sim 90\%$ of
them brighter than 19.5 mag.  We give the distributions of 110
galaxies and 282 galaxies brighter and fainter than 19.5 mag,
respectively, in left and right lower panels (Fig.~\ref{fig2d}).
Brighter galaxies are centered around the cD, while fainter galaxies
show some clumps aligned in the SE--NW direction; in particular the
main clump, Eastern with respect the cD galaxies, coincides with the
secondary peak found in our analysis of Chandra X--ray data (see
below).

\subsection{Analysis of X--ray data} 
\label{sec:253}

\begin{figure*}
\centering
\includegraphics[width=10cm]{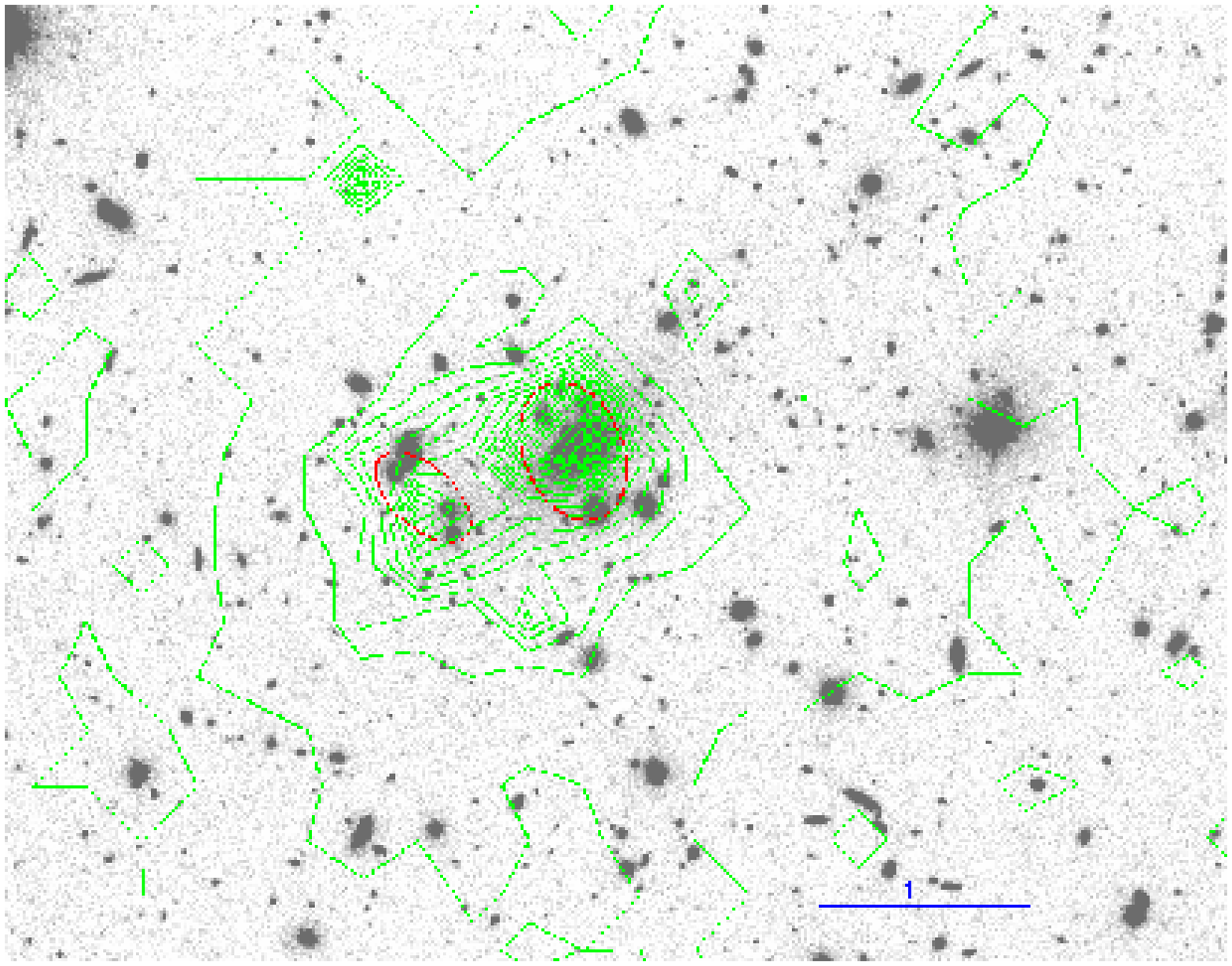}
\caption
{Optical image of ABCG\,209 with, superimposed, the X--ray contour
levels of the Chandra image, with North at top and East to left. The
ellipses represent the two substructures detected by Wavdetect. The
primary X--ray peak coincides with the position of the cD galaxy and
thus with the peak of the bright galaxy distribution
(cf. Fig.~\ref{fig2d} left lower panel).  The secondary X--ray peak
coincides with the main peak of the faint galaxy distribution
(cf. Fig.~\ref{fig2d} right lower panel).}

\label{isofote1}
\end{figure*}

The X--ray analysis of ABCG\,209 was performed by using the Chandra
ACIS--I observation 800030 (exposure ID\#522) stored in the Chandra
archive\footnote{http://asc.harvard.edu/cda/}. 
The pointing has an exposure time of 10 ksec.

Data reduction was performed by using the package CIAO\footnote{CIAO is
freely available at http://asc.harvard.edu/ciao/}(Chandra Interactive
Analysis of Observations). First, we removed events from the level 2
event list with a status not equal to zero and with grades one, five
and seven. Then, we selected all events with energy between 0.3 and 10
keV. In addition, we cleaned bad offsets and examined the data on a
chip by chip basis, filtering out bad columns and removing times when
the count rate exceeded three standard deviations from the mean count
rate per 3.3 second interval. We then cleaned each chip for flickering
pixels, i.e. times where a pixel had events in two sequential 3.3
second intervals. The resulting exposure time for the reduced data is
9.86 ks.

The temperature of the ICM was computed extracting the spectrum of the
cluster within a circular aperture of 3 arcminutes radius around the
cluster center. Freezing the absorbing galactic hydrogen column
density at 1.64 10$^{20}$ cm$^{-2}$, computed from the HI maps by
Dickey \& Lockman (\ct{dic90}), we fitted a Raymond--Smith spectrum using the
CIAO package Sherpa with a $\chi^{2}$ statistics.  We found a best
fitting temperature of $\mathrm{T_X=10.2^{+1.4}_{-1.2}}$ keV.

In order to detect possible substructures in ABCG\,209 we ran the task
CIAO/Wavdetect on a subimage of 1600 by 1600 pixels containing the
cluster. The task was run on different scales in order to search for
substructure with different sizes. The significance threshold was set
at $10^{-6}$. 

The results are shown in Fig.~\ref{isofote1}. Two ellipses are plotted
representing two significant substructures found by Wavdetect. 
The principal one, located at $\alpha$= 01 31 52.7 and $\delta$= -13
36 41, is centered on the cD galaxy, the left one is a secondary
structure located at $\alpha$= 01 31 55.7 and $\delta$= -13 36 54,
about 50 arcseconds ($\sim$120 $\mathrm{h^{-1}}$ kpc) East of the cD (cf. also
Fig.~\ref{fig2d}).
The secondary clump is well coincident with the Eastern clump detected
by Rizza et al.  (\ct{riz98}) by using ROSAT HRI X--ray data, while
we do not find any significant substructure corresponding to their
Western excess.

\section{Formation and evolution of ABCG\,209}
\label{sec:26}

The value we obtained for the LOS velocity dispersion
$\mathrm{\sigma_v\sim 1400}$ \ks (Sect.~\ref{sec:241}) is high compared to the
typical, although similar values are sometimes found in clusters at
intermediate redshifts (cf. Fadda et al. \ct{fad96}; Mazure et
al. \ct{maz96}; Girardi \& Mezzetti \ct{gir01}).  In Sect.~\ref{sec:242} we
show that this high value of $\mathrm{\sigma_v}$ is not due to obvious
interlopers: also a very restrictive application of the ``shifting
gapper'' or the rejection of minor peaks in galaxy density lead to
only a slightly smaller value of $\mathrm{\sigma_v\sim1250}$ \kss.
The high global value of $\mathrm{\sigma_v=1250}$--$1400$ \ks is
consistent with the high value of $\mathrm{T_X \sim 10}$ keV coming
from the X--ray analysis (which, assuming $\mathrm{\beta_{spec}=1}$,
would correspond to $\mathrm{\sigma_v\sim1300}$ \ks,
cf. Fig.~\ref{figprof}) and with the high value of
$\mathrm{L_X(0.1-2.4\ keV)\sim14\;h_{50}^{-2}\;10^{44}}$ erg
$\mathrm{s^{-1}}$ (Ebeling et al. \ct{ebe96}; cf. with
$\mathrm{L_X}$--$\mathrm{\sigma_v}$ relation by e.g., Wu et
al. \ct{wu99}; Girardi \& Mezzetti \ct{gir01}).

Therefore, on the basis of the global properties only, one could
assume that ABCG\,209 is not far from dynamical equilibrium and rely on
large virial mass estimate here computed
$\mathrm{M(<1.78}$ \hh$ )=2.25^{+0.63}_{-0.65}\times10^{15}$ \msun or
$\mathrm{M(<1.59}$ \hh$ )=1.62^{+0.48}_{-0.46}\times10^{15}$ \msun.

On the other hand, the analysis of the integral velocity dispersion
profile shows that a high value of $\mathrm{\sigma_v}$ is already reached
within the central cluster region of 0.2--0.3 \hh.
This suggests the possibility that a mix of clumps at different mean
velocity causes the high value of the velocity dispersion.

A deeper analysis shows that ABCG\,209 is currently undergoing a
dynamical evolution.
We find evidence for a preferential SE--NW direction as indicated
by: a) the presence of a velocity gradient; b) the elongation in the
spatial distribution of the colour--selected likely cluster members; c)
the elongation of the X--ray contour levels in the Chandra image; d) the
elongation of the cD galaxy.

In particular, velocity gradients are rarely found in clusters (e.g., 
den Hartog \& Katgert \ct{den96}) and could be produced by rotation,
by presence of internal substructures,
and by presence of other structures on larger
scales such as nearby clusters, surrounding superclusters, or filaments
(e.g., West \ct{wes94}; Praton \& Schneider \ct{pra94}). 

The elongation of the cD galaxies, aligned along
the major axis of the cluster and of the surrounding LSS (e.g.,
Binggeli \ct{bin82}; Dantas et al. \ct{dan97}; Durret et
al. \ct{dur98}), can be explained if BCMs form by
coalescence of the central brightest galaxies of the merging
subclusters (Johnstone et al. \ct{joh91}).

Other evidence that this cluster is far from dynamical equilibrium
comes out from deviation of velocity distribution from Gaussian, 
spatial and kinematical segregation of members with different B--R colour, and
evidence of substructure as given by the Dressler--Schectman test
(\ct{dre88}).  In particular, although a difference in $\mathrm{\sigma_v}$
between blue and red 
members is common in all clusters (e.g., Carlberg
et al. \ct{car97}), a displacement in mean velocity or in position
center is more probably associated with a situation of non equilibrium
(e.g., Bruzendorf \& Meusinger \ct{bru99}).

Possible subclumps, if any, cannot be
easily separated by using 112 cluster members only. The separation in
three clumps by using one--dimensional KMM test and the visual
inspection of Dressler--Schectman map of Fig.~\ref{figds} represent
only two tentative approaches (cf. Sects.~\ref{sec:243} and \ref{sec:245}).

A further step in the detection of possible subclumps is achieved by
using the two--dimensional complete sample of all $\sim400$
colour--selected cluster members. In fact, several clumps appear well
aligned in the SE--NW direction, and, in particular, the main clump 
revealed by faint galaxies (R$>19.5$ mag) coincides with
the secondary X--ray peak as found in our analysis of Chandra data
(cf. also Rizza et al. \ct{riz98}). This result suggests that one or
several minor clumps are merging along the SE--NW direction, with a
main clump hosting the cD galaxy.  In particular, the presence of a
significant velocity gradient suggests that the merging plane is not
parallel to the plane of the sky.

The presence of a secondary clump as a well distinct unit in
X--ray data suggests that this clump and maybe the whole cluster might be
in a pre--merger phase.

The strong luminosity segregation found for colour--selected galaxies
could however suggests a possible alternative dynamical scenario.  In
fact, very appealingly, galaxies of different luminosity could trace
the dynamics of cluster mergers in a different way.  A first evidence
was given by Biviano et al.  (\ct{biv96}): they found that the two
central dominant galaxies of the Coma cluster are surrounded by
luminous galaxies, accompanied by the two main X--ray peaks, while the
distribution of faint galaxies tend to form a structure not centered
with one of the two dominant galaxies, but rather coincident with a
secondary peak detected in X--ray.  The observational scenario of
ABCG\,209 has some similarities with the situation of Coma: bright
galaxies (R$<19.5$ mag) are concentrated around the cD galaxy, which
coincides with the primary X--ray peak, while faint galaxies
(R$>19.5$ mag) show several peaks, the main of which, Eastern with
respect the cD galaxy, is confirmed by the position of the secondary
X--ray peak.
Therefore, following Biviano et al., we might speculate that the
merging is in an advanced phase, where faint galaxies trace the
forming structure of the cluster, while more luminous galaxies still
trace the remnant of the core--halo structure of a pre--merging clump,
which could be so dense to survive for a long time after the merging
(as suggested by numerical simulations Gonz\'alez--Casado et
al. \ct{gon94}).  
An extended Radio--emission would support an advanced merging phase,
but its presence is still uncertain due to the existence of strong
discrete sources (Giovannini et al. \ct{gio99}). The comparison of
present Radio--image with our galaxy distribution shows that the
diffuse source is located around the central cD galaxy with an
extension toward North.

Unfortunately, redshift data are available only for luminous galaxies
and we cannot investigate the nature of subclumps inferred from the
two--dimensional distribution of faint galaxies.
At the same time, available X--ray data are not deep enough to look
for possible variations of temperature in the region of interest.

\section{Summary}
\label{sec:27}

In order to study the internal dynamics of the rich galaxy cluster
ABCG\,209, we obtained spectra for 159 objects in the cluster region
based on MOS observations carried out at the ESO New Technology
Telescope.  Out of these spectra, we analysed 119 galaxies: 
112 turn out to be cluster members, 1 is foreground and 6 are
background galaxies.

ABCG\,209 appears as a well isolated peak in the redshift distribution
centered at $z=0.209$, characterized by a very high value of the LOS
velocity dispersion: $\mathrm{\sigma_v=1250}$--$1400$ \kss, that
results in a virial mass of $\mathrm{M=1.6}$--$2.2\times 10^{15}$\msun
within R$\mathrm{_{vir}}$.  The analysis of the velocity dispersion
profile show that such high value of $\mathrm{\sigma_v}$ is already
reached in the central cluster region ($<0.2$--0.3 \hh).

The main results of the present study may be summarised as follows.

\begin{itemize}

\item ABCG\,209 is characterised by a preferential SE--NW direction as
indicated by: a) the presence of a velocity gradient in the velocity
field; b) the elongation in the spatial distribution of colour--selected
likely cluster members; c) the elongation of the X--ray contour levels in 
the Chandra image; d) the elongation of the cD galaxy.

\item We find significant deviations of velocity distribution from Gaussian.

\item Red and blue members are spatially and kinematically segregated.

\item There is significant evidence of substructure, as shown by the
Dressler \& Schectman test.

\item The two--dimensional distribution of the colour--selected likely
members shows a strong luminosity segregation: bright galaxies
$\mathrm{R<19.5}$ are centered around the cD galaxy, while faint galaxies
$\mathrm{R>19.5}$ show some clumps. The main one, Eastern with respect to the
cD galaxy, is well coincident with the secondary X--ray peak.

\end{itemize}

This observational scenario suggests that ABCG\,209 is presently
undergoing strong dynamical evolution. Present results suggest the
merging of two or more subclumps along the SE--NW direction in a plane
which is not parallel to the plane of sky, but cannot discriminate
between two alternative pictures.  The merging might be in a very
early dynamical status, where clumps are still in the pre--merging
phase, or in a more advanced status, where luminous galaxies trace the
remnant of the core--halo structure of a pre--merging clump hosting
the cD galaxy.

\large
\chapter{\Large Luminosity function}
\footnotetext[1]{\footnotesize The content of this
chapter is published in Mercurio, A., Massarotti, M., Merluzzi, P.,
Girardi, M., Busarello, G., \& La Barbera, F. 2003, A\&A, 408, 57}
\setcounter{footnote}{1}
\label{cap:3}
\markboth{Chapter 3}{Chapter 3}
\normalsize

In this chapter we derive the luminosity functions in three bands
(BVR) for ABCG\,209. The data cover an area of $\sim$ 78 arcmin$^2$ in
the B and R bands, while a mosaic of three pointings was obtained in
the V band, covering an area of approximately 160 arcmin$^2$. The
galaxy sample is complete to B = 22.8 ($\mathrm{N_{gal}}$ = 339), V =
22.5 ($\mathrm{N_{gal}}$ = 1078) and R = 22.0 ($\mathrm{N_{gal}}$ =
679). Although the fit of a single Schechter function cannot be
rejected in any band, the luminosity functions are better described by
a sum of two Schechter functions for bright and faint galaxies,
respectively. There is an indication for a presence of a dip in the
luminosity functions in the range V = 20.5--21.5 and R = 20.0--21.0.
We find a marked luminosity segregation, in the sense that the number
ratio of bright--to--faint galaxies decreases by a factor 4 from the
center to outer regions. Our analysis supports the idea that ABCG\,209
is an evolved cluster, resulting from the merger of two or more
sub--clusters.

\section{Introduction}
\label{intro}

Galaxy luminosity function (LF) is a powerful tool to constrain galaxy
formation and evolution, since it is directly related to the galaxy
mass function and hence to the spectrum of initial perturbations.
Hierarchical clustering models predict a mass distribution
characterised by a cut--off above a given mass M$^*$ and well
described by a power law at low masses (Press \& Schechter
\ct{pre74}). Starting from these results Schechter (\ct{sch76})
analysed the luminosity distribution of 14 galaxy clusters observed by
Oemler (\ct{oem74}), by introducing an analytical description in the
form:

\begin{equation}
\mathrm{
\phi(L)dL =
\phi^*\left(\frac{L}{L^*}\right)^{\alpha}
e^{-\left(\frac{L}{L^*}\right)} d\left(\frac{L}{L^*}\right) \ .
}
\label{eqsch}
\end{equation}

\noindent
In the Schechter function $\phi^*$ is a normalization density and the
shape of the LF is described by L$^*$, a characteristic cut--off
luminosity, and $\alpha$, the faint--end slope of the distribution. He
used this function to describe the global LF of all galaxy types and
suggested the value $\alpha = -5/4$.

Although investigated in several works, the universality of the LF
faint--end slope is still controversial. The value of the faint--end
slope turns out to be $\alpha\sim-1$ for field galaxies (e.g.,
Efstathiou et al.~\ct{efs88}; Loveday et al.~\ct{lov92}), while
clusters and groups seem to have steeper slopes, -1.8 $< \alpha<$ -1.3
(e.g., De Propris et al. \ct{dep95}; Lumsden et al. \ct{lum97};
Valotto et al. \ct{val97}), suggesting the presence of a larger
number of dwarf galaxies (but see also e.g., Lugger \ct{lug86};
Colless \ct{col89}; Trentham \ct{tre98a}). Changes in the slope of
the faint--end of the LF in clusters can be related to environmental
effects. An increase of the steepening of the LF faint--end in the
cluster outer regions was actually observed (Andreon \ct{and01}) and
explained taking into account that the various dynamical processes
which can destroy dwarf galaxies act preferentially in the
higher--density cores.

Lopez--Cruz et al. (\ct{lop97}) showed that clusters with a flat LF
($\alpha\sim-1$) are a homogeneous class of rich clusters with a
single dominant galaxy, symmetrical single peaked X--ray emission and
high gas masses. Irregular clusters have a steeper faint--end, in
particular, the LFs of ABCG\,1569 and Coma which present
substructures, can be suitably fitted with the sum of two Schechter
functions with $\alpha=-1$ and $\alpha \geq -1.4$ (Lopez--Cruz et al.
\ct{lop97}). Trentham (\ct{tre97}) also suggested that the
faint--end slope of the LF flattens as clusters evolve because of the
destruction of dwarf galaxies by merging with giants galaxies.

The density of the environment seems to affect the distribution of
galaxy luminosity in the sense that Schechter fits are poor for
galaxies in dense environment, where there is an indication of a dip
(e.g., Driver et al. \ct{dri94}; Biviano et al. \ct{biv95}; Wilson
et al. \ct{wil97}; Molinari et al. \ct{mol98}; Garilli et
al. \ct{gar99}; N$\mathrm{\ddot{a}}$slund et al. \ct{nas00}; Yagi
et al. \ct{yag02}). Trentham \& Hodgking (\ct{tre02}) identified
two types of galaxy LF, one for dynamically evolved regions
(i.e. region with a high elliptical galaxy fraction, a high galaxy
density, and a short crossing time), such as Virgo cluster and Coma
cluster, and one for unevolved regions, such as the Ursa Major cluster
and the Local Group. A dip is present in the LF of Virgo and Coma and
is absent in LFs of Ursa Major and Local Group.

The differences in shape of the LFs from cluster to cluster could be
explained assuming that the total LF is the sum of type specific
luminosity functions (hereafter TSLFs), each with its universal shape
for a specific type of galaxies (Binggeli et al. \ct{bin88}). The
total LF then assumes a final shape which can be different from
cluster to cluster according to the mixture of different galaxy
types. 

Therefore, different mixtures of galaxies, induced by
cluster--related processes, may be at the origin of the presence and
of the different shape of dips seen in cluster LFs and may be
responsible for the differences seen in the total LFs among field,
groups and galaxy clusters.

Indeed, dips are found in several clusters, occurring roughly at the
same absolute magnitude (M$\mathrm{_{R,dip}} \sim$ -19.4 or $\sim$
M$^*$ + 2.5), within a range of about one magnitude, suggesting that
clusters have comparable galaxy population. However the dips may have
different shapes, and also depend on the cluster region. This could be
related to the relative abundances of galaxy types, which depend on
the global properties of each cluster and on the local density (Durret
et al. \ct{dur99}).

In order to further investigate the cluster dynamical state, and to
discriminate between the pictures, presented in the previous chapter,
we derived the LFs by using new photometric data for ABCG\,209 based
on ESO--NTT imaging in the B, V and R wavebands. The new photometric
data are presented in Sect. \ref{sec:2}. In Sect. \ref{sec:3} we
describe the data reduction, and the photometric calibrations. The
aperture photometry is presented in Sect.\ref{sec:4}, whereas
Sect. \ref{sec:5} deals with the LFs and Sect. \ref{sec:6} with the
spatial distribution of galaxies of different
luminosity. Sect. \ref{sec:7} is dedicated to the summary and the
discussion of the results. In this work we assume H$_0$ = 70 \ks Mpc,
$\Omega_m$ = 0.3, $\Omega_{\Lambda}$ = 0.7. According to this
cosmology, 1 arcmin corresponds to 0.205 Mpc at z = 0.209.

\section{Observations}
\label{sec:2}


   	\begin{figure*}
	 \centering
   	\includegraphics[width=0.7\textwidth]{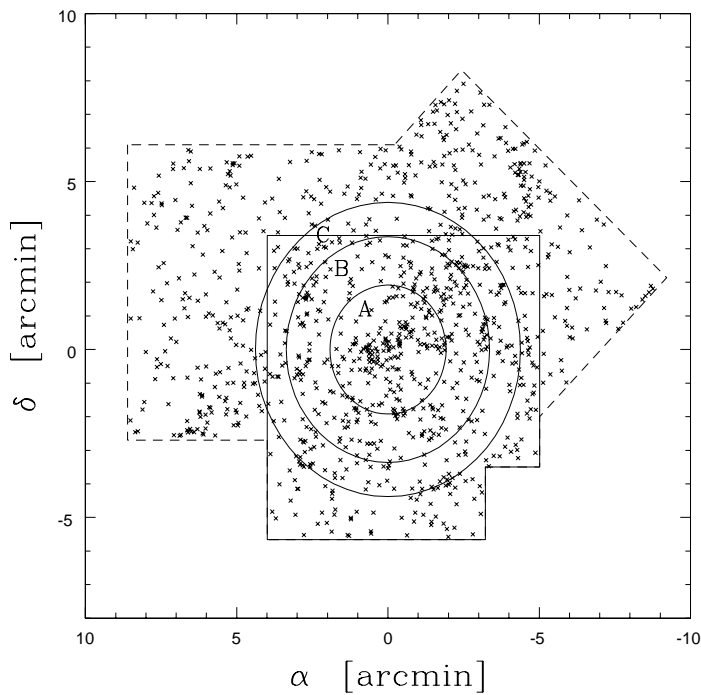}
	\caption{Area covered by the photometry in the region of
   ABCG\,209. All the galaxies brighter than V = 21.5 are marked with
   crosses. The central field (solid contour) was observed in B, V and
   R bands, while the two adjacent fields (dashed contours) were
   observed only in V band. Circles mark the regions analysed in
   Sect. \ref{sec:61}. North is up and East is on the left. The
   origin of the coordinates coincides with the cluster center
   ($\alpha_{2000}$ = 01 31 52.7, $\delta_{2000}$ = -13 36 41.9). A
   region around a bright star was not considered in the analysis
   (right bottom corner of central field).}
	\label{cluster}
	\end{figure*}

A field of $9.2^\prime \times 8.6^\prime$ (1.9 $\times$ 1.8
$\mathrm{h^{-2}_{70}}$ Mpc$^2$), was observed in the B, V and R bands around
the cluster center. In order to sample the cluster at large distance
from the center, we observed other two fields in V band. The total
observed area in V band, accounting for overlapping, is $\sim160$
arcmin$^2$ and is shown in Fig. \ref{cluster}. The relevant
information on the photometry are summarized in Table
\ref{inf}. Standard stars from Landolt (\ct{lan92}) were also
observed before and after the scientific exposures, and were used for
the photometric calibration.

   \begin{table}[h]
      \caption[]{Information on the photometric observations.}
         \label{inf}
{\footnotesize 
    $$
           \begin{array}{c c c c}
            \hline
            \noalign{\smallskip}
            \mathrm{Band} & \mathrm{T_{exp}} & \mathrm{Seeing} & \mathrm{Scale} \\
                 & \mathrm{ks} & \mathrm{arcsec} &\mathrm{arcsec/pxl}\\
            \noalign{\smallskip}
            \hline
            \noalign{\smallskip}
            \mathrm{B} &1.8 & 0.80  & 0.267  \\
            \mathrm{V} &1.26 \times 3 & 0.80 & 0.267   \\
            \mathrm{R} &0.9& 0.90 & 0.267\\
            \noalign{\smallskip}
            \hline
         \end{array}
     $$
}
   \end{table}

\section{Data reduction and photometric calibration}
\label{sec:3}

Standard procedures were employed for bias subtraction, flat--field
correction, and cosmic ray rejection using the IRAF~\footnote{IRAF is
distributed by the National Optical Astronomy Observatories, which are
operated by the Association of Universities for Research in Astronomy,
Inc., under cooperative agreement with the National Science
Foundation.} package. For each waveband the flat--field was obtained
by combining twilight sky exposures. After bias subtraction and
flat--field correction the images were combined using the IRAF task
IMCOMBINE with the CRREJECT algorithm. Residual cosmic rays and hot
pixels were interpolated applying the IRAF task COSMICRAYS. The
resulting images show a uniform background with typical r.m.s. of
2.3\%, 1.4\%, and 1.1\% for the B, V, and R bands, respectively.

   \begin{table}
      \caption[]{Results of the photometric calibration of BVR
      data.}
         \label{PC}
{\footnotesize 
    $$
   \begin{array}{c c c c c c}
   \hline
   \noalign{\smallskip}
   \mathrm{Band} & \mathrm{C} & \gamma & \mathrm{A} & \mathrm{ZP} & \sigma  \\
   \noalign{\smallskip}
   \hline
   \noalign{\smallskip}
   \mathrm{B} & \mathrm{B - V} & -0.037 \pm 0.008 &  0.322 \pm 0.010 & 24.885 \pm 0.015 & 0.016\\
   \mathrm{V} & \mathrm{B - V} &  0.035 \pm 0.009 &  0.204 \pm 0.013 & 25.277 \pm 0.021 & 0.020\\
   \mathrm{R} & \mathrm{V - R} &  0.010 \pm 0.017 &  0.145 \pm 0.014 & 25.346 \pm 0.023 & 0.026\\
            \noalign{\smallskip}
            \hline
         \end{array}
     $$
}
   \end{table}

The photometric calibration was performed into the
Johnson--Kron--Cousins photometric system by using the Landolt standard
fields. The instrumental magnitudes of the stars were measured in a
fixed aperture by using the IRAF packages APPHOT and DAOPHOT. The
aperture size was chosen in order to i) enclose the total flux, ii)
obtain the maximum signal--to--noise ratio. By comparing the
magnitudes of the stars in different apertures, we found that a
reasonable compromise is achieved with an aperture of radius 10 pixels
(cf. Howell \ct{how89}). For each band, we adopted the following
calibration relation:

\begin{equation}
\mathrm{M} = \mathrm{M^{'}} + \gamma \cdot \mathrm{C} - \mathrm{A}
\cdot \mathrm{X} + \mathrm{ZP}  \ , 
\label{eq1}
\end{equation}

\noindent
where M and C are the magnitudes and colours of the standard stars,
M$^{'}$ is the instrumental magnitude, $\gamma$ is the coefficient of
the colour term, A is the extinction coefficient, X is the airmass and
ZP is the zero--point. The quantities $\gamma$, A and ZP were derived
by a least square procedure with the IRAF task FITPARAMS. The results
of the photometric calibrations are reported in Table \ref{PC}.
Unless otherwise stated, errors on estimated quantities are given at
68\% confidence level (hereafter c.l.).

As a test we compare the (B--R, V--R) diagram of the Landolt
standard stars with that of the stars in the cluster field. We
measured the magnitude of these stars by using the software SExtractor
(Bertin \& Arnouts \ct{ber96}). We verified that the observed
distribution of stars in our images matches that of the Landolt stars
in the (B--R, B--V) plane, proving the accuracy of the photometric
calibration.

\section{Aperture photometry}
\label{sec:4}

For each image, a photometric catalog was derived by using the
software SExtractor (Bertin \& Arnouts \ct{ber96}). We measured
magnitudes within a fixed aperture of 5.0$^{''}$, corresponding to
$\sim$ 17 kpc at z = 0.209, and Kron magnitudes (Kron \ct{kro80}), for
which we used an adaptive aperture with diameter $a \cdot r_K$, where
$r_K$ is the Kron radius and $a$ is a constant. We chose $a$ = 2.5,
yielding $\sim$ 94\% of the total source flux within the adaptive
aperture (Bertin \& Arnouts \ct{ber96}). The aperture magnitudes were
used in order to derive galaxy colours. The different seeing of the
R--band image introduces a systematic shift, indipendent from the
total magnitude, of $\sim$ 0.03 mag, that is smaller than the
photometric errors on colours. The measured magnitudes were corrected
for galactic extinction following Schlegel et al. (\ct{sch98}). The
uncertainties on the magnitudes were obtained by adding in quadrature
both the uncertainties estimated by SExtractor and the uncertainties
on the photometric calibrations.

   \begin{figure*}[h]
   \centering
   \includegraphics[width=0.7\textwidth]{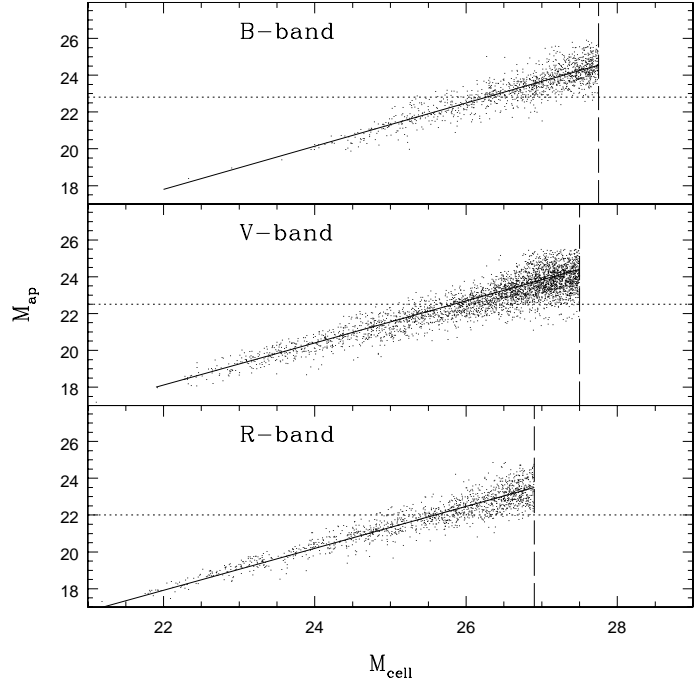}
    \caption{Completeness magnitudes of the B--, V--, and R--band
   images are estimated by comparing magnitudes in the fixed aperture
   (M$_{\mathrm{ap}}$) and in the detection cell
   (M$_{\mathrm{cell}}$). The dotted lines represent the completeness
   limits, the dashed lines mark the limits in the detection cell, and
   solid lines are the linear relations between M$_{\mathrm{ap}}$ 
   and M$_{\mathrm{cell}}$.}
      \label{comple}
    \end{figure*}

The completeness magnitudes were derived following the method of
Garilli et al. (\ct{gar99}), as shown in Fig. \ref{comple}. We
estimated the completeness magnitude as the magnitude where galaxies
start to be lost because they are fainter than the brightness
threshold in the detection cell. The completeness magnitudes are
B$_\mathrm{C}$ = 22.8 (N$\mathrm{_{gal}}$ = 339), V$_\mathrm{C}$ =
22.5 (N$\mathrm{_{gal}}$ = 1078), and R$_\mathrm{C}$ = 22.0
(N$\mathrm{_{gal}}$ = 679).

In order to derive the LF we considered the objects brighter than the
completeness limit and adopted the Kron magnitude, since this is the
best estimate of the total magnitude. The star/galaxy classification
was based on the SExtractor stellar index (SG), defining as stars the
sources with SG $\geq$ 0.98.

By examination of the distribution of sources classified as stars and
galaxies in the magnitude-FWHM plane, and the number-magnitude
distribution of sources classified as stars, we believe the
classification to be efficient up to the completeness magnitude of the
catalogues, where stars constitutes $\sim 12.5$\% of all sources.

   \begin{figure*}[h] 
   \centering
   \includegraphics[width=0.5\textwidth,angle=-90]{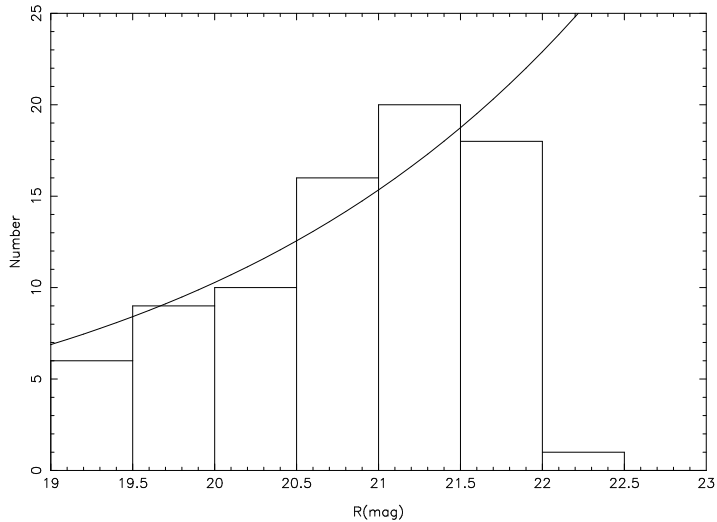}
   \caption{Number--magnitude distribution of sources classified by
   SExtractor as stars. The continuous line represents the
   best--fitting power--law function to the counts for 19.0$<$ R $<$
   21.5.} 
   \label{stars_Sex} 
   \end{figure*}

Figure \ref{stars_Sex} shows the number-magnitude distribution of
sources classified as stars by SExtractor in the R--band. For 19.0$<$
R $<$ 21.5 the distribution behaves as a power--law as would be
expected (see Groenewegen et al. \ct{gro02}). By extrapolating the
power--law to magnitude R = 22.0, we estimate that the level of
contamination by stars in the range 21.5--22.0 is minimal ($\sim$ 4
stars in the whole observed area).

The catalogs for the central region of the cluster were
cross--correlated by using the IRAF task XYXYMATCH. For each object,
colours were derived within the fixed aperture.

\section{Luminosity Functions}
\label{sec:5}

In order to measure the cluster LF in each band we used all the galaxy
photometric data up to the completeness magnitude, and removed the
interlopers by statistically subtracting the background
contamination. We used galaxy counts in B, V and R bands from the
ESO--Sculptor Survey (Arnouts et al. \ct{arn97}; de Lapparent et
al. \ct{del03}) kindly provided to us by V. de Lapparent. These data
cover an area of $\sim$ 729 arcmin$^2$ observed with EMMI
instrument. The data reduction was performed by using the procedures
similar to those adopted in the present work. In particular, the
photometric catalog was obtained by SExtractor and the total
magnitudes were estimated from the Kron magnitudes defined by adopting
the same aperture.

We assumed Poisson statistics for the background and cluster field
galaxy counts. The errors on the cluster LFs were computed by adding
in quadrature Poissonian fluctuations. Fluctuation of background
galaxy counts are no simply Poissonian, besause of the galaxy
correlation function. The galaxy counts variance due to the angular
correlation function, for each bin of magnitude, can be estimated by
using Eq. (5) in Huang et al. \ct{hua97}. By using this formula, we
verify that the background fluctuations do not influence our estimate
of the best-fit parameters of the luminosity function and the errors
on this parameters (there is only a change of 0.1\% in the confidence
levels). The errors on the cluster LFs are dominated by Poissonian
errors on observed counts, since the field of view of our observations
corresponds to an area with equivalent radius of $\sim$ 1 Mpc
$h_{70}^{-1}$, that is 10 times smaller than the area covered by the
background counts and 7 times smaller than the scale of the angular
correlation function.

We derived the LFs for the central field (see Fig. \ref{cluster}) in
the B, V and R bands by fitting the galaxies counts with a single
Schechter function. In V and R bands the fits were also computed with
the sum of two Schechter functions in order to describe the bright and
the faint populations (Sect. \ref{sec:351}). We compared the LFs with
counts obtained selecting red sequence galaxies
(Sect. \ref{sec:352}). All the fit parameters and the $\chi^2$
statistics, are listed in Table \ref{fitsLF}.

\subsection{Multiband analysis}
\label{sec:351}

\begin{figure*} 
   \vspace{-2cm}
   \centering
   \includegraphics[width=1.0\textwidth]{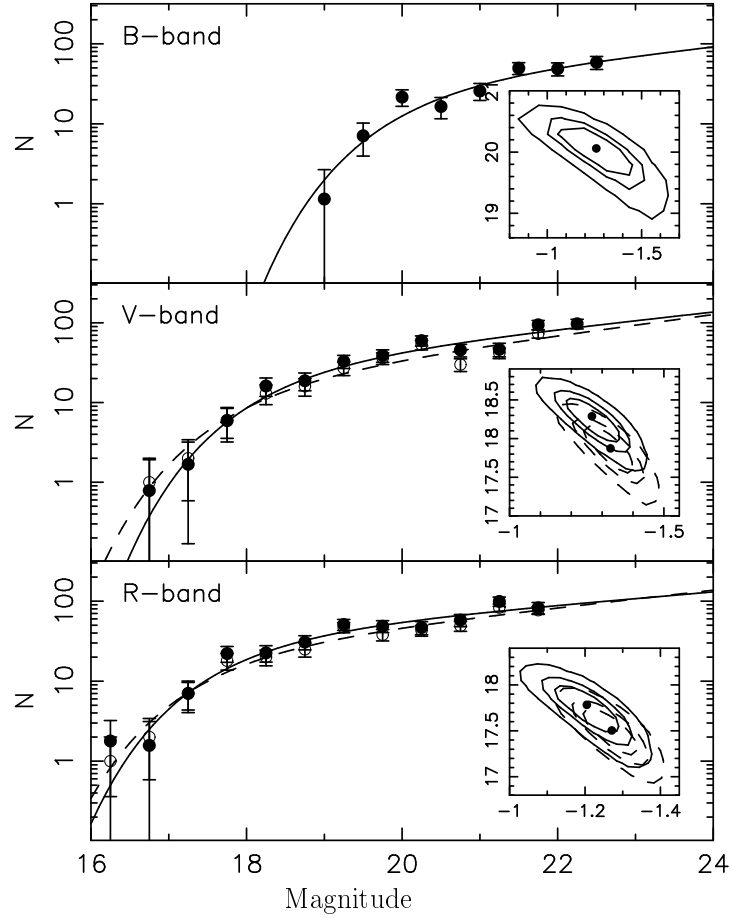}
   \vspace{-4cm} 
   \caption{Luminosity function in the B, V and R bands in the central
   field of 9.2$^\prime \times$ 8.6$^\prime$. In V and R bands filled
   circles represent counts obtained from the photometric catalog with
   a statistical background subtraction, open circles are the counts
   of galaxies belonging to the red sequence of the CM relation (see
   Sect. \ref{sec:352}). Solid and dashed lines are the Schechter
   fits to the filled and open circles respectively. In the small
   panels the 1, 2 and 3$\sigma$ c.l.  of the best--fit parameters
   $\alpha$ and M$^*$ are shown.}
   \label{centLF} 
   \end{figure*}

Figure \ref{centLF} (solid lines) shows the LFs in the B, V and R
bands for the central cluster region, obtained by a weighted
parametric fit of the Schechter function to the statistically
background--subtracted galaxy counts (filled circles). The parameters
of the fit are: B$^*$ = 20.06, $\mathrm{\alpha_B}$ = -1.26, V$^*$ =
18.29, $\mathrm{\alpha_V}$ = -1.27, R$^*$ = 17.78, $\mathrm{\alpha_R}$
= -1.20. We evaluated the quality of the fits by means of the $\chi^2$
statistics (see Table \ref{fitsLF}). The single Schechter function
gives a fair representation of the global distribution of the data,
that is the single Schechter fit cannot be rejected in any band even
at the $10\%$ c.l. .

   \begin{table} 
{\footnotesize 
     \caption[]{Fits to the Luminosity Functions. Errors on the
      $\mathrm{M^*}$ and $\alpha$ parameters can be obtained by the
      confidence contours shown in Fig. \ref{centLF}.}
\label{fitsLF}
     $$
\hspace{-2.0cm}
           \begin{array}{c| c c c c| c c c c }
            \hline
            \noalign{\smallskip}
            \mathrm{Band} & \multicolumn{4}{c|}{\mathrm{~~~~~Single~Schechter~function}~~~~~} & \multicolumn{4}{c}{\mathrm{~~~~~Two~Schechter~functions^{(a)}~}}\\
		\noalign{\smallskip}
		\hline
		\noalign{\smallskip}

 		 &~~~~~~\mathrm{M^*}~~~~~~&~~~~\alpha~~~~&~~~~~\mathrm{\chi^{2(b)}_{\nu}}~~~~~&\mathrm{P(\chi^2>\chi^2_{\nu})}~&~~~\mathrm{M}^*_{faint}~~~&~~\alpha_{faint}~~&~~~\mathrm{\chi^2_{\nu}}~~~&\mathrm{P(\chi^2>\chi^2_{\nu})}\\

            \noalign{\smallskip}
            \hline
            \noalign{\smallskip}
            \mathrm{B} & -21.03 & -1.26 & 0.96 & 46\% &        &       &      &      \\
            \mathrm{V^{c}} & -22.03 & -1.25 & 1.04 & 40\% &        &       &      &      \\
            \mathrm{V} & -22.18 & -1.27 & 1.46 & 15\% & -18.72 & -2.00 & 1.09 & 37\% \\
            \mathrm{R} & -22.48 & -1.20 & 1.24 & 27\% & -19.14 & -1.24 & 1.06 & 39\% \\
            \noalign{\smallskip}
            \hline
            \noalign{\smallskip}
            \multicolumn{1}{c} {\mathrm{    }} & \multicolumn{4}{c}{\mathrm{~~~~~Galaxies~on~the~red~sequence}~~~~~} & \multicolumn{4}{c}{}\\
            \noalign{\smallskip}
            \hline
            \mathrm{V} & -22.60 & -1.33 & 2.10 &  2\% &   -18.68   &  -2.19  & 1.68      &  9 \%   \\
            \mathrm{R} & -22.76 & -1.27 & 1.30 & 23\% &   -19.15     &  -1.39     &  1.09    &  37 \% \\
            \noalign{\smallskip}
            \hline
         \end{array}
    $$
\hspace{-2.0cm}
\begin{list}{}{}  
\item[$\mathrm{^{a}}$] The bright--end LF is fixed (see text).
\item[$\mathrm{^{b}}$] The reduced $\mathrm{\chi^2}$.
\item[$\mathrm{^{c}}$] Total observed field (see Sect. \ref{sec:6}).
\end{list}
}
   \end{table}

On the other hand, there is indication of a dip in the
distribution at V $\sim$ 21.0 and R $\sim$ 20.5.  According to the
fitted single Schechter function, there should be 121 and 125 galaxies
in the range V = 20.5--21.5 and R = 20.0--21.0 respectively, whereas
in our counts we find $\sim 92 \pm 10$ and $\sim 105 \pm 10$
galaxies. If we define the dip amplitude as:

\begin{equation}
\mathrm{
A=\frac{N_e-N_o}{N_e} \ ,
}
\label{dipa}
\end{equation}

\noindent
where $\mathrm{N_e}$ and $\mathrm{N_o}$ are the expected and observed
number of galaxies in the dip magnitude range, we obtain in V (R) band
A$=24 \pm 8 \%$ ($16 \pm 8 \%$).  The position of dips are: M$\mathrm{_V}$
$\sim$ -19.5 and M$\mathrm{_R}$ $\sim$ -19.8. These values were obtained by
converting apparent into absolute magnitudes using k--corrections for
early--type galaxies from Poggianti (\ct{pog97}). The B--band data
are not deep enough to sample the dip position.
 
The presence of a dip was discussed in the literature by considering
the properties of the TSLFs (e.g., Binggeli et al. \ct{bin88}) in
nearby clusters. In a study of the Virgo cluster, Sandage et
al. (\ct{san85}) showed the presence of two distinct classes of
galaxies (normal and dwarfs), with different dynamical and luminosity
evolution. Since the bright--end of the LF is well studied, we can use
{\it a priori} the information about the LF shape for bright galaxies
to fit our data with two different Schechter functions, representing
bright and faint galaxies. We assumed a Schechter model with
R$^*\mathrm{_{bright}}$ = 18.0 and $\alpha\mathrm{_{R,bright}}$ = -1.0
(from N$\mathrm{\ddot{a}}$slund et al. \ct{nas00} results, scaled
according to the adopted cosmology). Using the V--R colour term as
derived from Eq.~(\ref{EQCM}), we also fix V$^*\mathrm{_{bright}}$ =
18.7 and $\alpha\mathrm{_{V,bright}}$ = -1.0.

\begin{figure*} 
   \vspace{-2cm}
   \centering
   \includegraphics[width=1.0\textwidth]{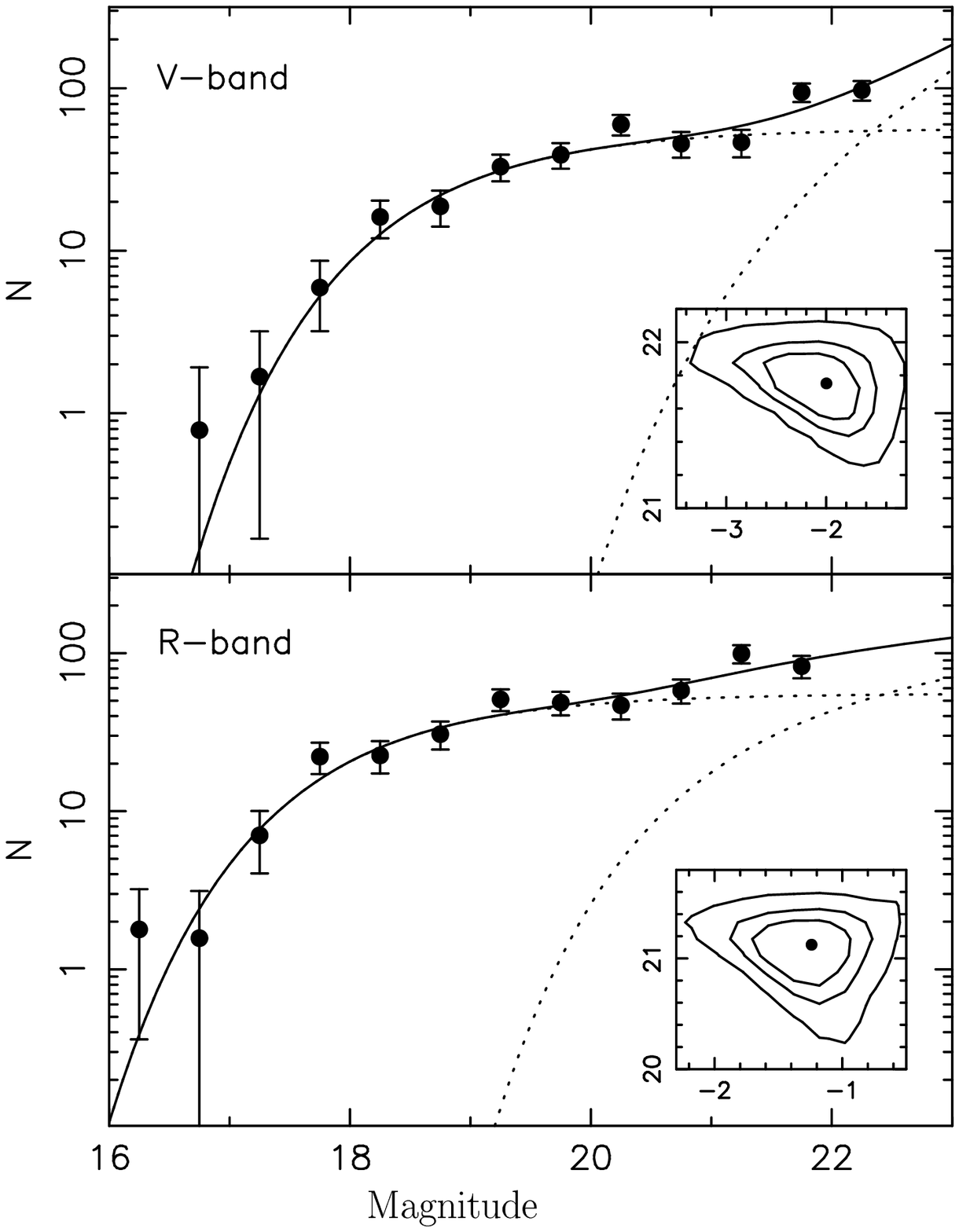}
   \vspace{-4.0cm}
   \caption{Luminosity function in the V and R bands in the central
            field of 9.2$^\prime \times$ 8.6$^\prime$. The dashed
            lines represent the Schechter functions for bright and
            faint galaxies separately (see text). The continuous line
            represents the sum of the two functions. In the small
            panels the 1, 2 and 3$\sigma$ c.l. of the best--fit
            parameters $\alpha$ and M$^*$ are shown.}
            \label{2LF} \end{figure*}

Figure \ref{2LF} shows the cluster LFs in V and R bands modelled with
the two Schechter functions. The fit procedures yields
V$^*\mathrm{_{faint}}$ = 21.75, R$^*\mathrm{_{faint}}$ = 21.12. For
the slope of the faint galaxies function the V-- and R--band data
provide poor constraints, so that we have only an indication of a
steep faint--end. According to the $\chi^2$ statistics (see Table
\ref{fitsLF}), combining two Schechter functions for bright and faint
galaxies the quality of the fit increases both in V and R bands.

\subsection{Galaxies on the red sequence}
\label{sec:352}

\begin{figure*}
\centering
\includegraphics[width=0.8\textwidth]{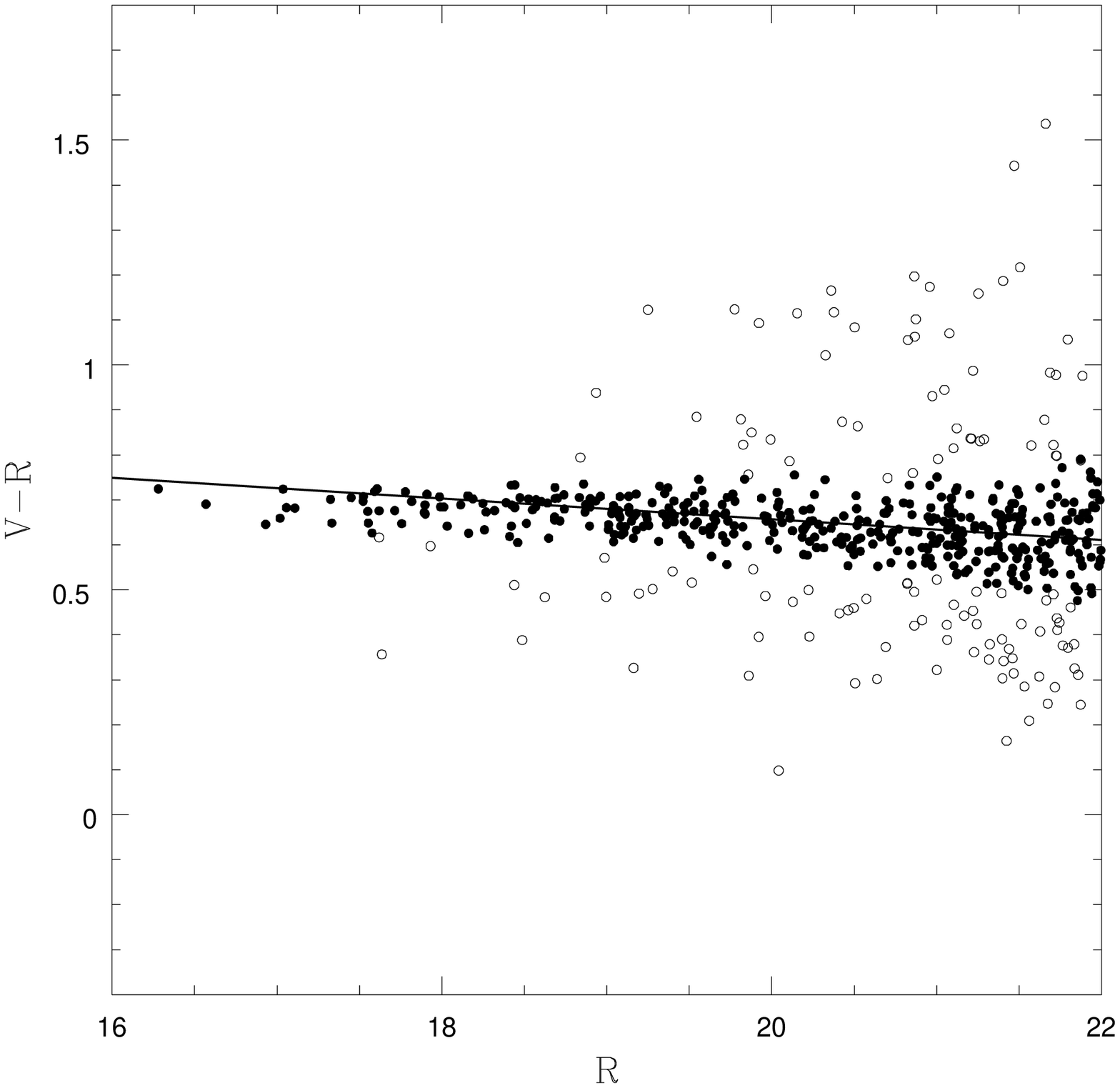}
\caption{V--R vs. R CM diagram for all the galaxies within the
completeness limit in the central field of ABCG\,209. Galaxies of the
red sequence (see solid line) are plotted as filled circles. The
solid line defines the CM sequence for spectroscopically confirmed
cluster members.}
\label{CM_phot}
\end{figure*}

We obtained the Colour--Magnitude (CM) relation by fitting the
photometric data of the spectroscopically confirmed cluster members
(see Chapter~\ref{cap:2}) with a biweight algorithm (Beers et al. \ct{bee90}):

\begin{equation}
\mathrm{
(V-R)_{CM} = - 0.023 \cdot R + 1.117 \ .
}
\label{EQCM}
\end{equation}
\noindent
By using Eq.~(\ref{EQCM}), we defined as sequence galaxies the sources
lying in the region inside the curves:

\begin{equation}
\mathrm{
(V-R)_{seq} = (V-R)_{CM} \pm (\sqrt{\sigma_V^2 +\sigma_R^2} + 0.05) \ ,
}
\label{EQCMS}
\end{equation}
\noindent
where we took into account the photometric uncertainty at 1$\sigma$
both on the V ($\mathrm{\sigma_V}$) and on the R magnitude
($\mathrm{\sigma_R}$) as well as the intrinsic dispersion of the CM relation
(Moretti et al. \ct{mor99}). In Fig. \ref{CM_phot} the red sequence
galaxies are marked with filled circles.

By selecting galaxies on the CMR (within the observed scatter), the
background contamination is expected to be negligible.  So we can
directly compare galaxy counts (open circles in Fig. \ref{centLF})
with those derived in Sect. \ref{sec:351} (filled circles in
Fig. \ref{centLF}). In both V and R bands the counts obtained with the
two different approaches are very similar.

Figure \ref{centLF} (dashed lines) shows the LFs in the V and R
bands for the central cluster region, obtained by a weighted
parametric fit of the Schechter function to the red galaxy counts
(open circles). The parameters of the fit are: V$^*$ = 17.87,
$\mathrm{\alpha_V}$ = -1.33, R$^*$ = 17.50, $\mathrm{\alpha_R}$ =
-1.27. According to the $\chi^2$ statistics (see Table \ref{fitsLF})
we can reject the fit with a single Schechter function in the V band
at 98\% c.l. At the same time, the dip amplitude in the V band, A$=26
\pm 9 \%$, is unchanged respect to the case of the global LF
(Sect. \ref{sec:351}). It turns out that, independently from the fitted
Schechter functions, the ratio ($\sim 60 \%$) of observed counts in
the bins inside the dip and in the bins adjacent to the dip regions is
the same for both galaxy samples, while the overall distribution of
red galaxies cannot be described with a single Schechter
function. According to the $\chi^2$ statistics (see Table
\ref{fitsLF}), combining two Schechter functions the quality of the
fit increases both in V and R bands, and becomes acceptable also in
the V band.

\section{Luminosity segregation}
\label{sec:6}

The data in the V band, covering an area of $\sim$ 160 arcmin$^2$,
corresponding to a circular region with equivalent radius 0.6
R$_{\mathrm{vir}}$ (R$_{\mathrm{vir}}$ = 2.5 h$^{-1}_{70}$ Mpc; see
Sect.~\ref{sec:241}), allow to study the environmental dependence of
LF and the spatial distribution of galaxies as a function of the
clustercentric distance.

Figure \ref{LFsegr_fit} (upper panel) shows the V--band LF in the
whole observed area, modelled by using a weighted parametric fit to a
single Schechter function, with best fit values V$^*$ = 18.45
(M$\mathrm{^*_V}$ = -22.03) and $\mathrm{\alpha_V}$ = -1.25. The LF shape is
very similar to that obtained in the central field (Fig. \ref{centLF})
and also in this case the Schechter function overestimates the
observed counts in the range V = 20.5--21.5. The dip amplitude is
A$=14 \pm 7 \%$.

\subsection{The LF in different cluster regions}
\label{sec:61}

   \begin{table}
     \caption[]{Best fit value of the LFs measured in three regions
      around the center of the cluster.} 
    \label{LFsegr}
{\footnotesize 
     $$
           \begin{array}{c c c c c c }
            \hline
            \noalign{\smallskip}
    \mathrm{Region} & \mathrm{Area (h^{-2}_{70} Mpc^2)} & \mathrm{V}^* & \mathrm{Luminosity \ ratio}& \mathrm{\chi^2_{\nu}} &\mathrm{P(\chi^2>\chi^2_{\nu})} \\
            \noalign{\smallskip}
            \hline
            \noalign{\smallskip}
    \mathrm{A} & 0.50   & 18.09 \pm 0.73 & 12.7 \pm 3.0 & 1.12 & 35\%\\
    \mathrm{B} & 1.00   & 18.49 \pm 0.44 & 6.9  \pm 1.4 & 1.08 & 37\% \\
    \mathrm{C} & 1.00   & 18.88 \pm 0.43 & 6.8  \pm 2.3 & 0.85 & 56\% \\
            \noalign{\smallskip}
            \hline
         \end{array}
     $$
}
   \end{table}

\begin{figure*}
   \vspace{-2cm} \centering
   \includegraphics[width=1.0\textwidth]{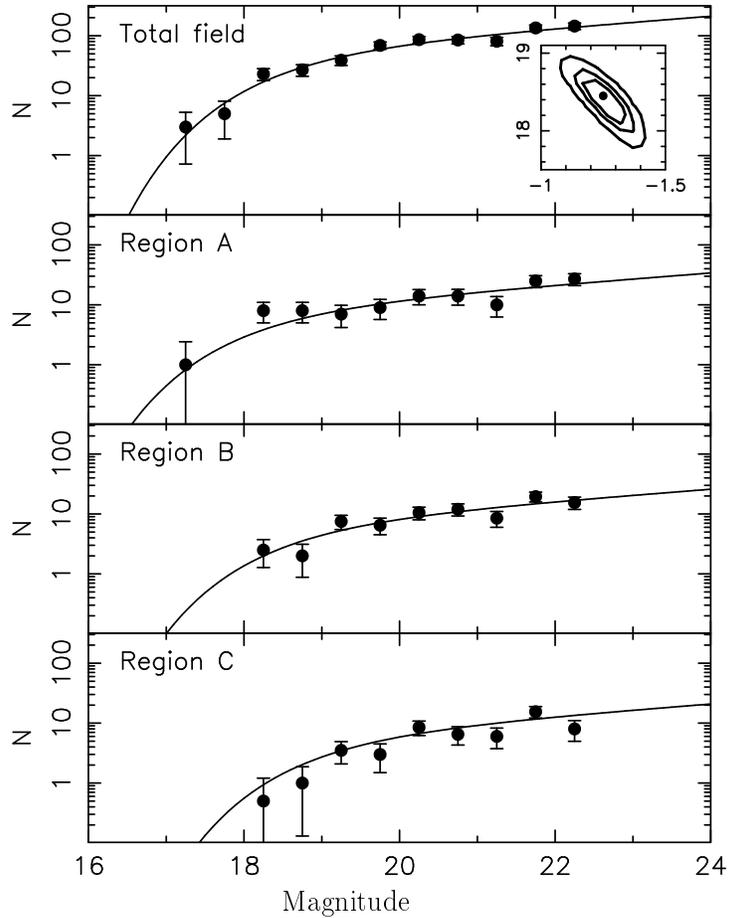}
   \vspace{-4.0cm}
    \caption{Luminosity function in the V band. In the upper panel are
    shown the LF in the whole observed area (160 arcmin$^2$) and the
    1, 2 and 3$\sigma$ c.l. of the best--fit parameters $\alpha$ and
    M$^*$ (small panel). The continuous line represents the fitted
    Schechter functions. In the other panels the LFs in the three
    cluster regions (see text) are shown. Continuous lines are the fit
    of the Schechter functions obtained by fixing the faint--end slope
    at the best fit value of the LF fitted on the whole observed
    area.}
     \label{LFsegr_fit}
    \end{figure*}

In order to compare the abundances of galaxies of various luminosities
in different regions we computed the LFs in three areas at different
distances from the center~\footnote{The center of the cluster was
derived by a two--dimensional adaptive kernel technique from the
spectroscopically confirmed cluster members (for details see
Sect.~\ref{sec:241}) and coincides with the cD position.}. First we
considered the cluster counts in an area of 0.5 h$^{-2}_{70}$ Mpc$^2$
around the center of the cluster (region A), and then in two
concentric circular rings around the first central region, each in an
area of 1.0 h$^{-2}_{70}$ Mpc$^2$, respectively at $\sim$ 2$^\prime$
and $\sim$ 4.7$^\prime$ from the center (region B and C). We fitted
the counts using a single Schechter function with $\alpha$ fixed at
the best fit value obtained from the LF computed over the whole
observed field. Table \ref{LFsegr} reports the relative fit values in
the different regions and Fig. \ref{LFsegr_fit} shows the fitted
functions. We also measured the luminosity weighted ratio of the
number of objects brighter and fainter than the V = 21.0 (Table
\ref{LFsegr}), that is the magnitude where the dip occurs. This
luminosity ratio increases from region B to the center by a factor
$1.8 \pm 0.6$, indicating a significant luminosity segregation,
whereas the ratio does not vary from the region B to the region C.

\subsection{Spatial distribution of galaxies}
\label{sec:62}

\begin{figure*}
   \includegraphics[width=0.5\textwidth,angle=-90]{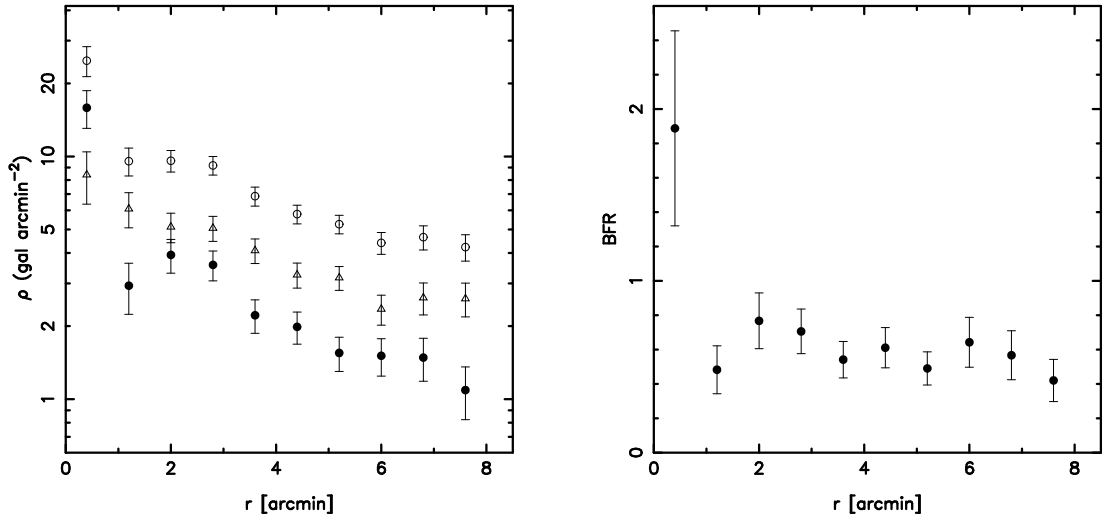}
    \caption{The background--subtracted radial profile of ABCG\,209 is
    shown in the left panel. Open circles represent the distribution
    of all galaxies in the cluster, filled circles are bright galaxies
    (M$\mathrm{^*_V \le }$ -19.5) and open triangles are faint
    galaxies (M$\mathrm{^*_V > }$ -19.5). The number of bright
    galaxies in the first and in the second bin is $\sim$ 32 and
    $\sim$ 17 respectively, while the number of faint galaxies is
    $\sim$ 15 and $\sim$ 30. The bright--to--faint ratio as a
    function of the clustercentric distance is shown in the right
    panel.}
      \label{pdens}
    \end{figure*}

In Fig. \ref{pdens} (left panel) the number density radial profile is
shown, i.e. the number of galaxies measured in concentric rings around
the center. It has been computed for the whole galaxy population (open
circles) and for galaxies brighter (filled circles) and fainter
(triangles) than V = 21.0, the magnitude where the luminosity
distribution shows the distinctive upturn. The number of faint
galaxies is determined up to the completeness limit. The densities
are statistically background--subtracted by using background galaxy
density measured in the ESO--Sculptor Survey (Arnouts et
al. \ct{arn97}). Errors are assumed to be Poissonian.

The radial profile of all galaxies is centrally peaked, showing the
highest gradient for bright galaxies (Fig. \ref{pdens}, left
panel). Indeed, the bright galaxy density decreases by a factor $\sim
5.4$ between the first and the second bin, to be compared with a
decrease of a factor $\sim 1.4$ for faint galaxies and of a factor
$\sim 2.6$ when all galaxies are considered. For faint galaxies (V $>$
21.0), the distribution decreases slowly with clustercentric distance.

Fig. \ref{pdens} (right panel) shows the bright--to--faint ratio (BFR)
as a function of the clustercentric distance. The BFR shows a maximum
at the center, where the number of bright galaxies is two times the
number of faint galaxies, and decreases rapidly between the first and
the second bin. Outside the central region of the cluster (r $>$ 1
arcmin) the BFR distribution is flat and there are roughly two faint
galaxies for each bright galaxy.

\section{Summary and discussion}
\label{sec:7}

The analysis performed on ABCG\,209 can be summarized by the following
results:

\begin{description}

\item[1)] the fit with a single Schechter function cannot be
rejected in any band; at the same time, the luminosity distributions
are better described by a sum of two Schechter functions for bright
and for faint galaxies respectively;

\item[2)] there is an indication of a dip in V-- and R--band LFs,
and this features is also present when considering only galaxies that
lie on the red sequence;

\item[3)] the faint--end slope seems to be steeper going from redder
to bluer wavelengths;

\item[4)] there is a luminosity segregation, in the sense that the
V$^*$ value increases from the center to the outer regions and the
luminosity ratio of bright--to--faint galaxies decreases by a factor
$\sim 2$ from the center to outer regions, whereas the number ratio
decreases by a factor $\sim 4$.

\end{description}

The presence of a dip was found in the LF of many clusters. In Table
\ref{DIP} we reported the positions of the dips for some of these
clusters with 0.0$<$ z $<$ 0.3 in R--band absolute magnitudes
(according to the adopted cosmology). The literature data were
transformed to R--band absolute magnitudes by using k-- and
evolutionary correction, and average colours for early--type galaxies
of Poggianti (\ct{pog97}).

   \begin{table}

      \caption[]{Compilation of dip positions in the restframe R band
      in cluster LFs from literature. Data were transformed to R--band
      absolute magnitudes by using k-- and evolutionary correction,
      and average colours for early--type galaxies of Poggianti
      (\ct{pog97}).}
         \label{DIP}
\smallskip

\smallskip

\smallskip

{\footnotesize
     $$
           \begin{array}{c c c  c }
            \hline
            \noalign{\smallskip}
             \mathrm{Cluster \ name} & \mathrm{Redshift}  &  \mathrm{~~~~Dip \ position} & \mathrm{Reference} \\
            \noalign{\smallskip}
            \hline
            \noalign{\smallskip}

            \mathrm{Virgo}        & 0.0040 & -18.9 & \mathrm{Sandage \ et \ al. (1985)}\\
            \mathrm{ABCG\,194}    & 0.018  & -19.8 & \mathrm{Yagi \ et \ al. (2002)}\\
            \mathrm{Coma}         & 0.0232 & -19.8 & \mathrm{Biviano \ et \ al. (1995)} \\
            \mathrm{ABCG\,496}    & 0.0331 & -19.8 & \mathrm{Durret \ et \ al. (2000)} \\
            \mathrm{ABCG\,2063}   & 0.035  & -19.3 & \mathrm{Yagi \ et \ al. (2002)} \\
            \mathrm{ABCG\,576}    & 0.038  & -18.6 & \mathrm{Mohr \ et \ al. (1996)} \\
            \mathrm{Shapley 8}    & 0.0482 & -19.5 & \mathrm{Metcalfe \ et \ al. (1994)} \\
            \mathrm{ABCG\,754}    & 0.053  & -19.3 & \mathrm{Yagi \ et \ al. (2002)} \\
            \mathrm{ABCG\,85(z)}^{\mathrm{a}}   & 0.0555 & -19.9 & \mathrm{Durret \ et \ al. (1999)} \\
            \mathrm{ABCG\,85} & 0.0555 & -19.3 & \mathrm{Durret \ et \ al. (1999)} \\
            \mathrm{ABCG\,2670}   & 0.076  & -19.3 & \mathrm{Yagi \ et \ al. (2002)} \\
            \mathrm{ABCG\,1689}   & 0.181  & -18.7 & \mathrm{Wilson \ et \ al. (1997)} \\
            \mathrm{ABCG\,665}& 0.182  & -18.7 & \mathrm{Wilson \ et \ al. (1997)} \\
            \mathrm{ABCG\,963}& 0.206  & -19.0 & \mathrm{Driver \ et \ al. (1994)} \\
            \mathrm{ABCG\,209}& 0.209  & -19.6 & \mathrm{This \, work}           \\
           \mathrm{MS\,2255.7+2039} & 0.288 & -19.6 & \mathrm{N\ddot{a}slund \ et \ al. \ (2000)}\\

            \noalign{\smallskip}
            \hline
         \end{array}
     $$
\smallskip
\begin{list}{}{}  
\item[$\mathrm{^{a}}$] LF fitted only on the spectroscopically
confirmed cluster galaxies.
\end{list}
}
   \end{table}

Dips are found at comparable absolute magnitudes, in different
clusters, suggesting that these clusters may have similar galaxy
population. However, the dips have not necessarily the same shape:
the dips found in the the LF of Shapley 8 and Coma are broader
than those found in Virgo, ABCG\,963, and ABCG\,85. As already
mentioned in the introduction, this can be explained with the
hypothesis of an universal TSLF (Binggeli et al. \ct{bin88}),
whereas the different relative abundances of galaxy types, induced by
cluster--related processes, are at the origin of the different dip
shapes (see also Andreon \ct{and98}).

Moreover, as shown in Table \ref{DIP}, there is no correlation between
the dip position and the redshift, suggesting little or no
evolutionary effect in the dip, in agreement with
N$\mathrm{\ddot{a}}$slund et al. (\ct{nas00}), which directly
compare the LFs of Coma, ABCG\,963 and MS2255. They detected no
qualitative difference between nearby and distant clusters.

Studying the R--band LFs for a photometric sample of 10 clusters at
different redshifts and with different richness classes, Yagi et
al. (\ct{yag02}) demonstrated that the dips seen in the LFs are
almost entirely due to r$^{1/4}$--like galaxies. This evidence is in
agreement with those shown in Fig. \ref{centLF} (open circles).  In
fact by comparing the counts of red sequence galaxies with those
obtained with a statistical background subtraction, the dip in V band
appears more pronounced.

Environmental effects can be at the origin of the different positions
of the dips seen in different clusters, since the transition from the
bright--dominated to the faint--dominated parts of the LF can occur at
different magnitudes in different environment. The cluster
morphological mix and the morphology--density relation (Dressler
\ct{dre80}) should give rise to LFs with different shape when
subdividing a sample galaxies respect to the cluster environments.

The effect of the environment can be seen in ABCG\,209 by comparing
the shape of the LF in V band in different regions around the cluster
center (see Fig. \ref{LFsegr_fit} and Sect. \ref{sec:61}). The fitted
M$^*$ value is shifted toward fainter magnitudes going from the inner
(A) to the outer (C) region and the luminosity ratio of
bright--to--faint galaxies decreases by a factor $\sim 2$, indicating
a luminosity segregation.

The bright galaxies are markedly segregated in the inner 0.2
h$^{-1}_{70}$, around the cD galaxy. This suggest that bright galaxies
could trace the remnant of the core--halo structure of a pre--merging
clump.

Although ABCG\,209 is a cD--like cluster, with cD galaxy located in
the center of a main X--ray peak, it shows an elongation and asymmetry
in the galaxy distribution (Chapter~\ref{cap:2}). Moreover, the
faint--end slope turns out to be $\alpha < -1$ at more than 3$\sigma$
c.l. in both V and R bands, thus reconciling the asymmetric properties
of X--ray emission with the non flat--LF shape of irregular systems as
found by Lopez--Cruz et al. (\ct{lop97}).

These results allow to discriminate between the two possible formation
scenarios suggested by the dynamical analysis
(Chapter~\ref{cap:2}). We conclude that ABCG\,209 is an evolved
cluster, resulting from the merger of two or more sub--clusters, while
the elongation and asymmetry of the galaxy distribution (of the X--ray
emission) and the shape of the LFs show that ABCG\,209 is not yet a
fully relaxed system.

Our analysis is in agreement with the existence of i) an universal LF
for bright galaxies, which is well described by a flat Schechter
function, and ii) a steep Schechter function for faint galaxies. A
definitive conclusion regarding the faint--end slope of the LF needs
deeper photometry able to sample the luminosity distribution of dwarf
galaxies.

\large
\chapter{\Large Galaxy evolution in different environments}
\footnotetext[1]{\footnotesize
The content of this chapter is accepted for publication in A\&A. The
authors are: C. P. Haines, A. Mercurio, P. Merluzzi, F. La Barbera,
M. Massarotti, G. Busarello, \& M. Girardi.}
\setcounter{footnote}{1}
\label{cap:4}
\markboth{Chapter 4}{Chapter 4}
\normalsize

In this chapter we examine the environmental effects on the
photometric properties of galaxies for the rich galaxy cluster ABCG
209 at $z=0.209$. We use archive CFHT optical imaging of a
\mbox{$42\times28\,{\rm arcmin}^{2}$} field centred on the cluster to
produce a galaxy sample complete to $B=25.0$ and $R=24.5$. Both the
composite and red sequence galaxy luminosity functions are found to be
dependent on the local galaxy surface density, their faint-end slopes
becoming shallower with increasing density. We explain this as a
combination of the morphology-density relation, and dwarf galaxies
being cannibalised and/or disrupted by the cD galaxy and the ICM in
the cluster core. The $B-R$ colour of the red sequence itself appears
0.02\,mag redder for the highest-density regions, indicative of their
stellar populations being marginally ($<5$\%) older or ($<20$\%) more
metal-rich. This may be due to the galaxies themselves forming
earliest in the rarest overdensities marked by rich clusters, or their
star formation being suppressed earliest by the ICM.

\section{Introduction}
\label{cap:41}

The effect of local environment on the formation and evolution of
galaxies is one of the most pressing issues in cosmology today, with
evidence that environmental processes affect the mass distribution of
galaxies, as well as their star formation and morphological
characteristics. Rich clusters provide a unique opportunity to study
these environmental effects, providing large numbers of galaxies at
the same redshift which have been exposed to a wide variety of
environments. Numerous authors have studied the environmental effects
of rich clusters on the luminosity functions, star formation rates and
morphologies of the constituent galaxies (e.g. Lopez--Cruz et al.
\ct{lop97}; Balogh, Navarro \& Morris \ct{bal00}; Hogg et al.
\ct{hogg}; Treu et al. \ct{treu}).

The galaxy luminosity function (LF), which describes the number of
galaxies per unit volume as a function of luminosity, is a powerful
tool for examining galaxy formation and evolution, since it can be
directly related to the galaxy mass function. The Press-Schechter
prescription for the hierarchical assembly of galaxies predicts a
simple analytical formula for the mass distribution of the form
\mbox{${\rm n(M)dM}\propto {\rm M}^{\alpha}\exp(-{\rm M/M}^{*})$}
(Press \& Schechter \ct{pre74}) which reproduces well the results of
N-body simulations of the growth of dark matter halos. This relation
was confirmed by the observations of Schechter (\ct{sch76}), that also
suggested that the shape of the galaxy LF is universal, and only the
multiplicative constant $\psi^{*}$ differs between clusters.


Numerous studies have since been made to determine the galaxy LF: some
confirming the validity of the ``{\em universal luminosity function
hypothesis}'' (e.g. Lugger \ct{lug86}; Colless \ct{col89}; Trentham
\ct{tre98a}); whereas others indicate that the LF is instead dependent
on the environment of the galaxies, resulting in significant
differences between the LFs from cluster to cluster and between
cluster and field (e.g. Lopez--Cruz et al. \ct{lop97}; Valotto et
al. \ct{val97}). More recent studies, based on large surveys,
empirically suggest that the galaxy LF is indeed dependent on
environment, with dynamically-evolved, rich clusters and clusters with
central dominant galaxies having brighter characteristic luminosities
and shallower faint-end slopes than poorer clusters (Lopez--Cruz et
al. \ct{lop97}; De Propris et al.
\ct{depropris}). In particular these differences can be explained
by considering the composite LF as the sum of type-specific luminosity
functions (TSLFs), each with its universal shape for a specific type
of galaxies (e.g. ellipticals, spirals, dwarfs). The shape of the
composite LF can then vary from cluster to cluster, and from clusters
to field, according to the mixture of different galaxy types in each
environment resulting from the morphology-density relation (Dressler
et al. \ct{dressler87}; Binggeli et al. \ct{bin88}).

The galaxies most synonymous with the cluster environment are the
bulge-dominated, passively-evolving galaxies which make up the cluster
red sequence. Studies of low-redshift clusters (e.g. Bower, Lucey \&
Ellis \ct{bower}) indicate that, irrespective of the richness or
morphology of the cluster, all clusters have red sequences, whose
k-corrected slopes, scatters and colours are indistinguishable. Other
studies find similar regularities between other physical properties of
the early-type galaxies (surface-brightness, mass-to-light ratio,
radius, velocity distribution, metallicity) resulting in the
Faber-Jackson (Faber \& Jackson \ct{faber}) and Kormendy
(\ct{kormendy}) relations and the fundamental plane (Djorgovski \&
Davis \ct{djorgovski}; Dressler et al.
\ct{dressler87}). These all indicate that the early-type galaxies
which make up the red sequences form a homogeneous population, not
only within each cluster, but from cluster to cluster, and also that
red sequences are universal and homogeneous features of galaxy
clusters, at least at \mbox{$z\lesssim0.2$}.

The photometric evolution of the cluster red sequence with redshift
has been studied by a number of authors (e.g. Arag\'{o}n-Salamanca
et al. \ct{aragon}; Ellis et al. \ct{ellis}; Stanford et al.
\ct{stanford98}; Kodama et al. \ct{kodama98}) for clusters out to
\mbox{$z\simeq1.2$} indicating i) that the red sequence remains a universal
feature of clusters; ii) that the stellar populations of its
constituent early-type galaxies are formed in a single, short burst at
an early epoch \mbox{$(z_{f}\gtrsim2)$}; and iii) that the galaxy
colours have evolved passively ever since. This is confirmed by the
spectra of red-sequence galaxies which, in nearby clusters, are best
fit by simple stellar populations of age \mbox{9--12\,Gyr}, resulting
in those galaxies having characteristically strong spectral breaks at
4000\,\AA, and correspondingly red \mbox{$U-V$} colours.

Cosmological models of structure formation indicate that the densest
regions of the universe, corresponding to the rarest overdensities in
the primordial density field, have collapsed earliest, and contain the
most massive objects, i.e. rich galaxy clusters. Detailed cosmological
simulations which follow the formation and evolution of galaxies
(e.g. Kauffmann \ct{kauffmann}; Blanton et al.
\ct{blanton00}) predict that galaxies in these high-density regions
are older and more luminous than those in typical density (i.e. field)
regions. Having initially (\mbox{$z\sim5$}) been the most likely place
for the formation of stars and galaxies, the high-density environment
of rich cluster cores is later (\mbox{$z\sim$1--2}) filled with
shock-heated virialised gas that does not easily cool and collapse
(Blanton et al. \ct{blanton99}), inhibiting both the formation of
stars and galaxies (Blanton et al. \ct{blanton00}), which instead
occurs most efficiently in increasingly less massive dark matter
halos. These models are able to successfully predict the observed
cluster-centric star formation and colour gradients, and the
morphology-density relation of $z<0.5$ clusters (Balogh et al.
\ct{bal00}; Diaferio et al. \ct{dia01}; Springel et al.
\ct{spr01}; Okamoto \& Nagashima \ct{oka03}), although it is
not clear whether they predict trends with density of colour or
star formation for galaxies of a fixed morphology and luminosity. They
show that these gradients in galaxy properties naturally arise in
hierarchical models, because mixing is incomplete during cluster
assembly, and the positions of galaxies within the clusters are
correlated with the epoch at which they were accreted.

To examine the effect of environment on galaxy properties, in
particular the galaxy luminosity function, and the cluster red
sequence, we have performed a photometric study of the galaxy cluster
ABCG\,209 using archive wide-field $B$ and $R$-band imaging, which
allows the photometric properties of the cluster galaxies to be
followed out to radii of
\mbox{3--4$h_{70}^{-1}$\,Mpc}. 

The cluster was initially chosen for its richness, allowing its
internal velocity field and dynamical properties to be studied in
great detail, and also for its known substructure, allowing the effect
of cluster dynamics and evolution on the properties of its member
galaxies to be examined. All the results presented in previous
chapters indicate that ABCG\,209 is undergoing strong dynamic
evolution with the merging of two or more sub-clumps along the SE-NW
direction. This dynamical situation is confirmed by the weak lensing
analysis of Dahle et al. (\ct{dah02}) which shows the cluster to be
highly elongated in the N-S direction, and two significant peaks, the
largest being coincident with the cluster centre and cD galaxy, and
the second separated by 6\,arcmin in the north direction.

The photometric data are presented in Section~\ref{sec:42}. As this
cluster has been found to be significantly elongated along the SE-NW
direction, we examine the effect of the cluster environment by
measuring galaxy properties as a function of local surface density
rather than cluster-centric radius. We describe the method for
determining the cluster environment in Sect.~\ref{sec:43}. The ability
to successfully subtract the background field population is vital to
study the global properties of cluster galaxies from photometric data
alone, and we describe the approach to this problem in
Sect.~\ref{sec:44}. The composite galaxy LF is presented in
Sect.~\ref{sec:45}, and the properties of the red sequence galaxies
described in Sect.~\ref{sec:46}. The effects of environment on the
blue galaxy fraction and mean galaxy colours are presented in
Sects.~\ref{sec:47} and \ref{sec:48}. Sect.~\ref{sec:49} is dedicated
to the summary and discussion of the results. In this work we assume
\mbox{H$_{0}$=70\,km\,s$^{-1}$\,Mpc$^{-1}$}, \mbox{$\Omega_{m}=0.3$},
and \mbox{$\Omega_{\Lambda}=0.7$}. In this cosmology, the lookback
time to the cluster ABCG\,209 at \mbox{$z=0.209$} is 2.5\,Gyr.

\section{The data}
\label{sec:42}

The data were obtained from the Canada-France-Hawaii telescope (CFHT)
science archive (PI. J.-P. Kneib), comprising wide-field $B$- and
$R$-band imaging centred on the cluster ABCG\,209. The observations were
made on 14--16 November 1999, using the CFHT12K mosaic camera, an
instrument made up of 12 \mbox{$4096\times2048$} CCDs, set at the
prime focus of the 3.6-m CFHT. The CCDs have a pixel scale of
$0.206^{\prime\prime}$, resulting in a total field of view of
\mbox{$42\times28\,{\rm arcmin}^{2}$}, corresponding to
\mbox{$8.6\times5.7\,h^{-2}_{70}{\rm Mpc}^{2}$} at the cluster
redshift. The total exposure times for both $B$- and $R$-band images
are 7\,200s, made up of eight 900s $B$-band and twelve 600s $R$-band
exposures, jittered to cover the gaps between the CCDs.

Standard procedures were used to bias-subtract the images, using bias
exposures and the overscan regions of each CCD. Saturated pixels and
bleed trails from bright stars are identified and interpolated
across. The images were then flat-fielded using a superflat made up of
science images taken using the same camera/filter setup of cluster
ABCG\,209, as well as those of the clusters A\,68, A\,383, A\,963,
CL\,0818 and CL\,0819 which were also observed as part of the same
observing program. For each image, the sky is subtracted by fitting a
quadratic surface to each CCD. The sky is then modelled using a
\mbox{$256\times256$} pixel median filter, rejecting all pixels
\mbox{$>3\sigma$} from the median (i.e. galaxies).

Cosmic rays are removed by comparison of each image with a registered
reference image (taken to be the previous exposure). After masking off
those pixels \mbox{$3\sigma$} above the median in the reference image
(taken to be sources) each pixel which has a value \mbox{$3\sigma$}
higher than its value in the reference image, is flagged as a cosmic
ray and interpolated across. The individual images are corrected for
airmass using the prescribed values from the CFHT website of
\mbox{$\alpha(B)=-0.17$} and \mbox{$\alpha(R)=-0.06$}. The photometry
of non-saturated guide stars is found to agree between exposures to an
rms level of 0.003\,mag, indicating the observations were taken in
photometric conditions.

The images are registered and coadded in a two-step process. Each of
the 12 CCDs are initially registered and coadded, using the positions
of sources from the 2nd Guide Star Catalogue (GSC2) to determine
integer-pixel offsets between the jittered exposures. As the exposures
are jittered to cover the gaps between the CCDs, there is some overlap
between the coadded images of adjacent CCDs. This is used to create
linear transforms allowing 12 CCDs to be {\em ``stitched''} together
to form an interim coadded image. This interim image is then used as a
reference to which each individual exposure is registered. The
registered exposures are then coadded after masking out the gaps
between CCDs and bad columns, using $3\sigma$ clipping. The distortion
produced by the camera optics is modelled and removed as a
quartic-polynomial fit to the positions of GSC2 sources in a reference
tangential plane astrometry.

The photometric calibration was performed into the
Johnson-Kron-Cousins photometric system using observations of
$\sim300$ secondary standard stars \mbox{($14<R<17$)} in fields 6 and
7 of Galad\'{\i}-Enr\'{\i}quez, Trullols \& Jordi (\ct{galadi}). The
fields also contain equatorial standard stars of Landolt
(\ct{lan92}) tying the photometric calibration to the Landolt
standards, while allowing the uncertainty of the zeropoint and
colour-terms to be reduced and measured in a statistical manner. The
median photometric uncertainty for each standard star was
\mbox{$\Delta(B)=0.033$}, and \mbox{$\Delta(R)=0.021$}. The
zero-points and colour terms were fitted using a weighted
least-squares procedure, and are shown in Table~\ref{photometry} along
with the observed FWHMs of each image.
\begin{table}
\begin{tabular}{cccc} \hline
    Band FWHM & Colour & Zeropoint & Colour Term \\ \hline
$B$\hspace{0.5cm}1.02 & $B-R$  & $25.720\pm0.008$ & $-0.0180\pm0.005$ \\
$R$\hspace{0.5cm}0.73 & $B-R$  & $25.983\pm0.005$ & $\;\;\,0.0002\pm0.004$ \\ \hline
\end{tabular}
\caption{Photometric parameters of optical data.}
\label{photometry}
\end{table}
Object detection was performed using SExtractor in two-image mode
(Bertin \& Arnouts \ct{ber96}) for sources with 4 contiguous pixels
1$\sigma$ over the background level in the $R$-band image. The total
$B$ and $R$ magnitudes were taken to be the Kron magnitude, for which
we used an adaptive elliptical aperture with equivalent diameter
$a.r_{K}$, where $r_{K}$ is the Kron radius, and $a$ is fixed at a
constant value of 2.5. $B-R$ colours were determined using fixed
apertures of 5\,arcsec diameter (corresponding to \mbox{$\sim17$}\,kpc
at \mbox{$z=0.209$}), after correcting for the differing seeing of the
$R$- and $B$-band images. The measured magnitudes were corrected for
galactic extinction following Schlegel, Finkbeiner \& Davis
(\ct{sch98}) measured as \mbox{${\rm E}(B-V)=0.019$}, giving
\mbox{${\rm A}(B)=0.083$} and \mbox{${\rm A}(R)=0.051$}. The
uncertainties in the magnitudes were obtained by adding in quadrature
both the uncertainties estimated by SExtractor and the uncertainties
of the photometric calibrations.

\begin{figure}[t]
\centerline{{\resizebox{\hsize}{!}{\includegraphics{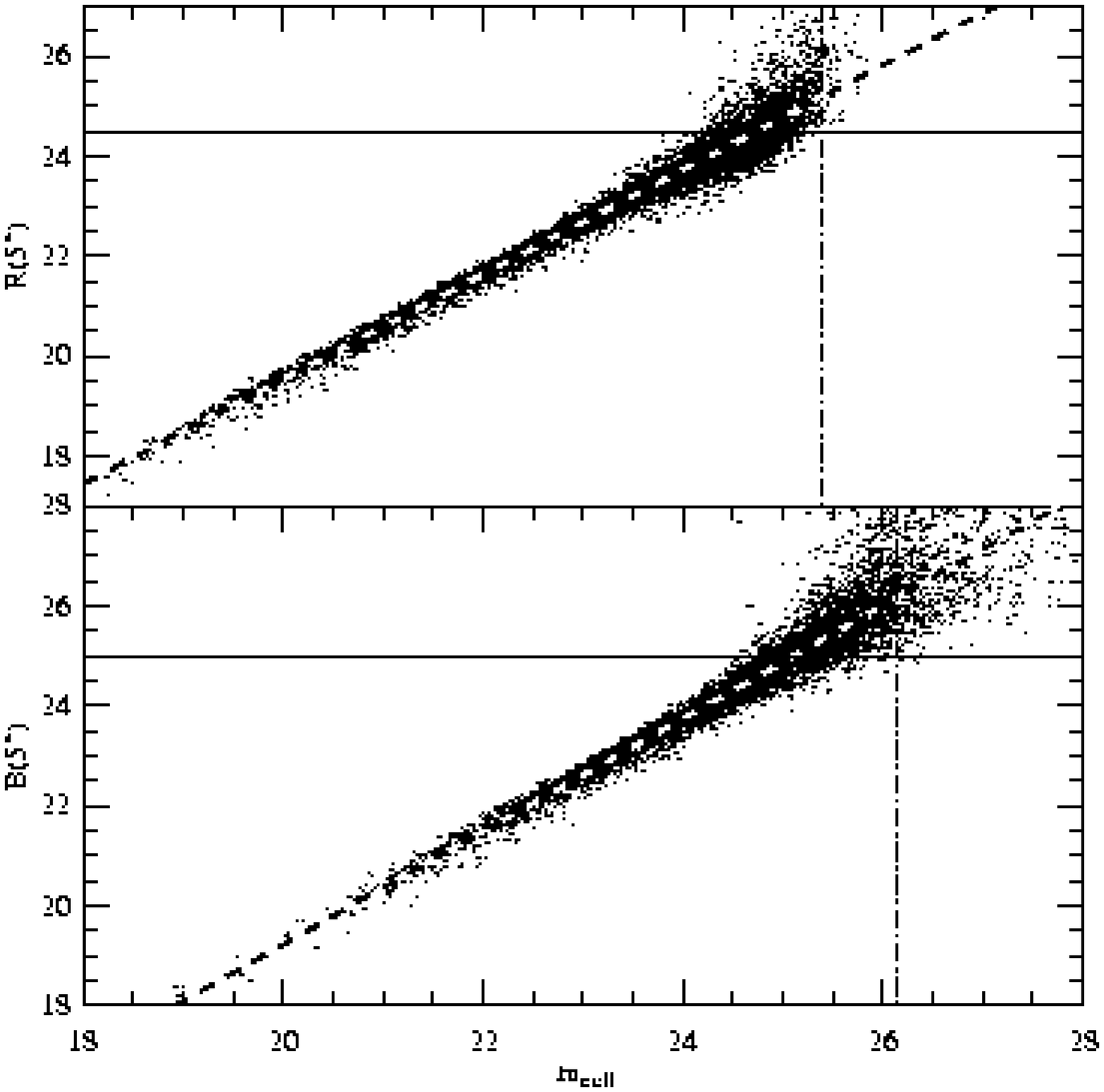}}}}
\caption{Completeness magnitudes of the $B$- and $R$-band images, as estimated by comparing the magnitudes in the fixed aperture (m$_{ap}$) and in the detection cell (m$_{cell}$). The horizontal solid lines represent the completeness limits, the vertical dot-dashed lines mark the limits in the detection cell, and the dashed lines are the linear relations between m$_{ap}$ and m$_{cell}$.}
\label{maglimit}
\end{figure}

The completeness magnitudes were derived following the method of
Garilli, Maccagni \& Andreon (\ct{gar99}), as shown in
Fig.~\ref{maglimit}, which compares the magnitudes in the fixed
aperture (m$_{ap}$) and in the detection cell (m$_{cell}$) which in
our case is an aperture of area 4\,pixels. The magnitude limits at
which galaxies are lost due to being fainter than the threshold in the
detection cell are determined, and are indicated by the vertical
dot-dashed lines.  There is a correspondence between $m_{cell}$ and
m$_{ap}$ as shown by the dashed line, which has a certain scatter that
depends essentially on the galaxy profile and the photometric
errors. To minimise biases due to low-surface brightness galaxies, the
completeness magnitude limits are chosen to consider this dispersion,
and are indicated by the horizontal solid lines at \mbox{$B=25.0$} and
\mbox{$R=24.5$}. For the analyses of the cluster red sequence, we
consider a magnitude limit of \mbox{$R=23.0$}, in order that we remain
complete for galaxies with \mbox{$B-R\sim2.0$}.

\subsection{Star - Galaxy Separation}
\label{sec:421}

\begin{figure}[t]
\centerline{{\resizebox{6cm}{!}{\includegraphics{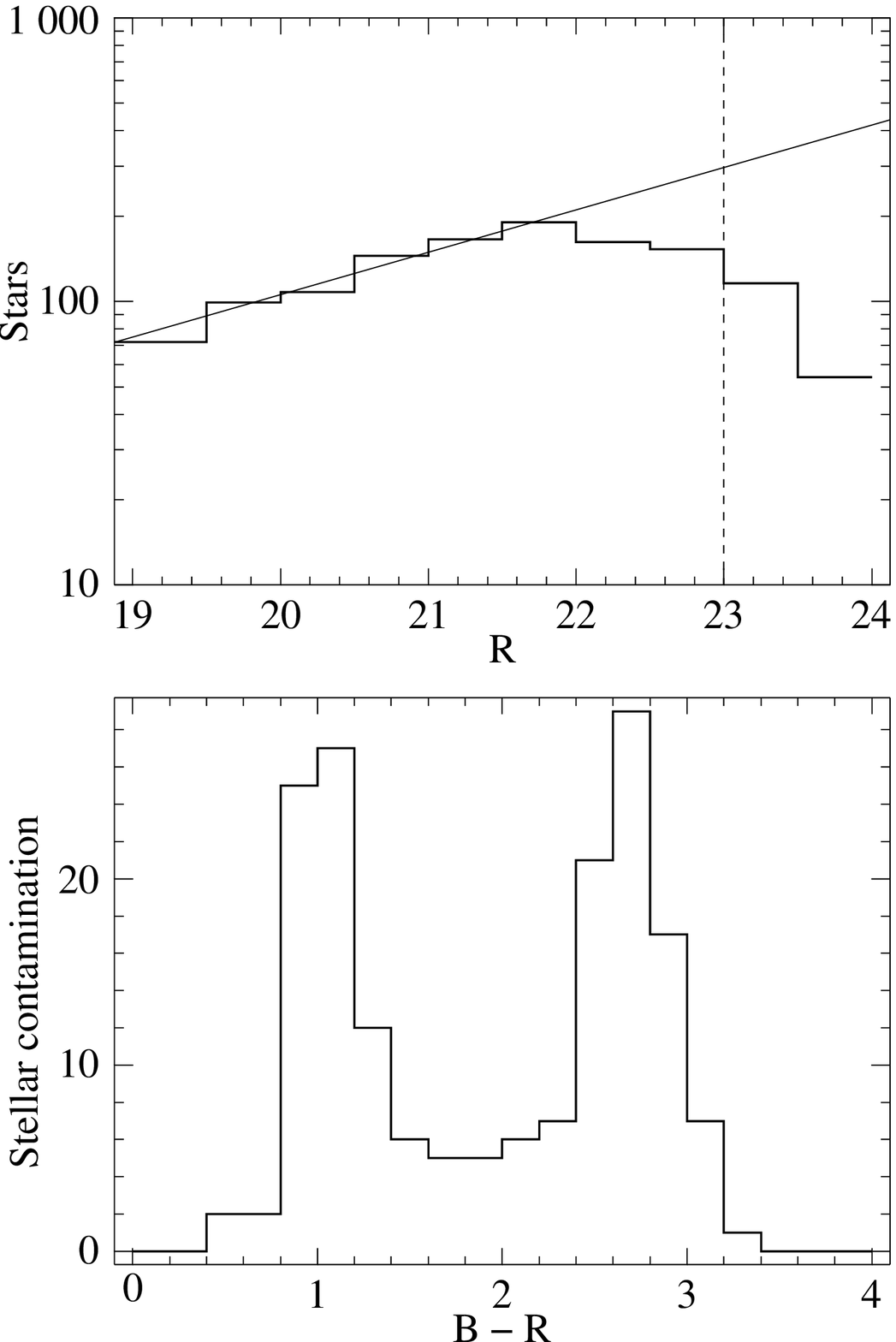}}}}
\caption{Photometric properties of sources classified by SExtractor as being stars. {\bf (top panel)} Number-magnitude distribution of stars. The best-fitting power-law function to the counts for \mbox{$19<R<22$} is indicated, as is the $R=23$ magnitude limit used for analysis of the red sequence galaxy population. {\bf (bottom panel)} $B-R$ colour distribution of \mbox{$19<R<22$} stars, normalised to match the expected level of stellar contamination for \mbox{$22<R<23$}.} 
\label{stars}
\end{figure}

Star-galaxy separation is performed using the SExtractor stellarity
index, with stars defined as sources with stellarities
\mbox{$\geq0.98$}. By examination of the distributions of sources
classified as stars and galaxies in the magnitude-FWHM plane, and the
number-magnitude distribution of sources classified as stars, we
believe the classification to be efficient to \mbox{$R\sim22$} and
\mbox{$B\sim23$}, where stars constitute 12\% of all sources.

\begin{figure*}[t]
\centerline{{\resizebox{\hsize}{!}{\includegraphics[angle=-90]{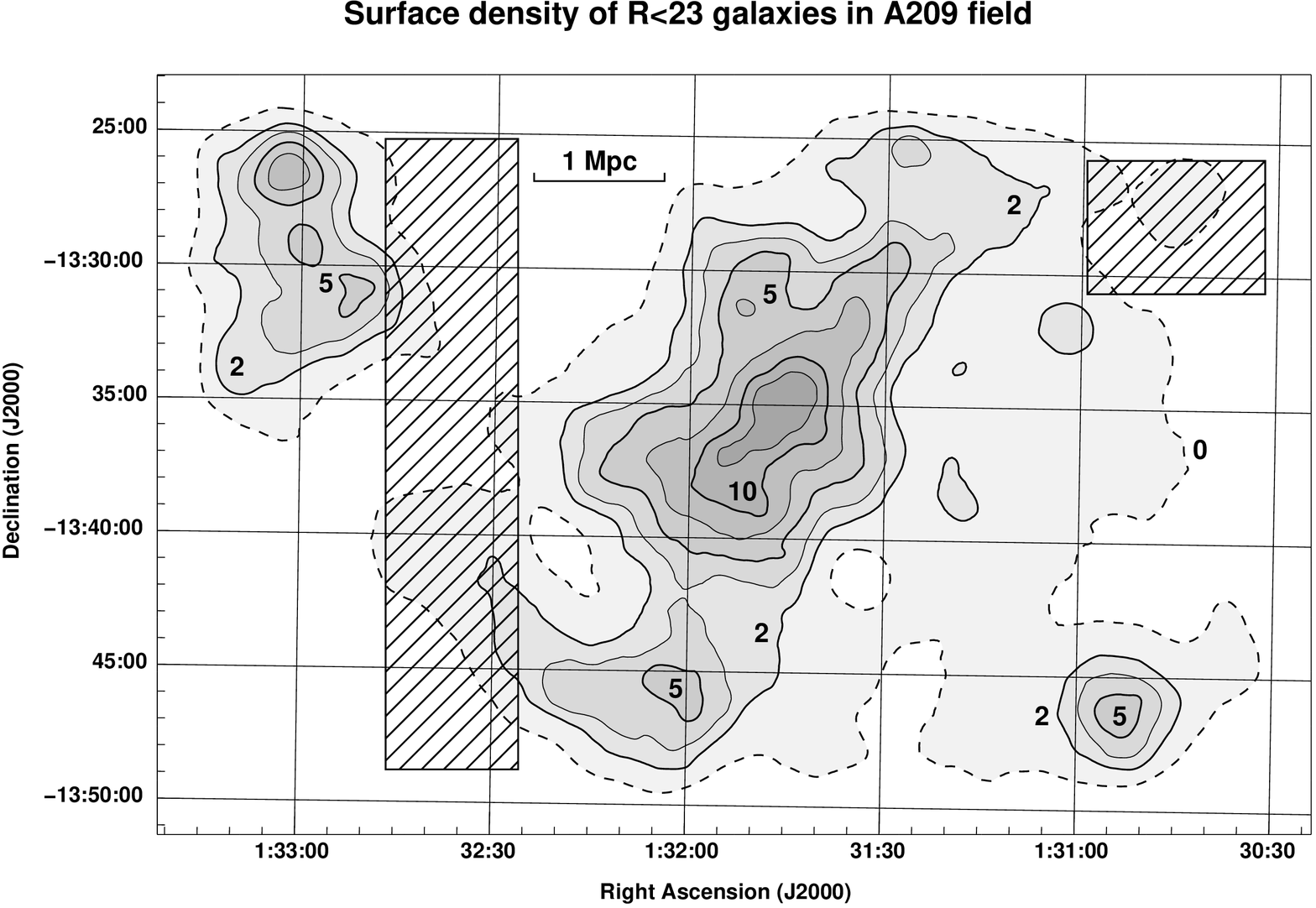}}}}
\caption{The surface density of $R<23.0$ galaxies in the field of ABCG\,209. The dashed contour corresponds to the field galaxy number density, while the solid contours correspond respectively to 2, 3.3, 5, 7.5, 10, and 12.5 cluster galaxies\,arcmin$^{-2}$. The shaded boxes indicate the regions used to represent the field galaxy population.}
\label{density}
\end{figure*}

Figure~\ref{stars}(top) shows the number-magnitude distribution of
sources classified as stars by SExtractor. For \mbox{$19<R<22$} the
distribution behaves as a power-law as would be expected (see
Groenewegen et al. \ct{gro02}). At brighter magnitudes, stars
become saturated, while at fainter magnitudes, the classifier begins
to misclassify stellar sources as galaxies. By extrapolating the
power-law to fainter magnitudes, as indicated by the solid diagonal
line, we estimate the level of contamination by stars in the range
\mbox{$22<R<23$} to be \mbox{$\approx200$} over the whole field
(0.18\,stars\,arcmin$^{-2}$), or $\sim5\%$ of all sources. However,
since the contamination from faint stars should affect those regions
used to define the field galaxy population at the same level as for
the cluster regions, when we correct for field galaxy contamination,
the contamination due to faint stars should automatically cancel out
also.

Figure~\ref{stars}(bottom) shows the $B-R$ colour distribution of
stars in the range \mbox{$19<R<22$}. The y-axis is normalised to
represent the expected level of contamination by stars in the range
\mbox{$22<R<23$} over the whole field, i.e. 200 in total, assuming the
colour distribution remains constant. The distribution is clearly
bimodal with two sharp peaks at \mbox{$B-R\sim1.0$} and
\mbox{$B-R\sim2.7$} indicating the dominant populations of white and
red dwarf stars (Groenewegen et al. \ct{gro02}). Fortunately,
as regards the study of the red sequence galaxy population of ABCG\,209,
at these magnitudes the red sequence corresponds approximately to
\mbox{$2.0<B-R<2.2$}, midway between the two peaks in the colour
distribution of stars. Hence, the expected level of contamination by
stars to the red sequence galaxy population is minimal, corresponding
to \mbox{$\sim6$} stars over the whole mosaic image with
\mbox{$22<R<23$}.

\section{Defining the Cluster Environment}
\label{sec:43}

To study the effect of the cluster environment on galaxies in the
vicinity of ABCG\,209, the local surface density of galaxies, $\Sigma$,
is determined across the CFHT images. This is achieved using an
adaptive kernel estimator (Pisani \ct{pis93}; \ct{pis96}),
in which each galaxy is represented by an exponential kernel,
\mbox{$K(r)\propto\exp(-r/r_{0})$}, whose width, $r_{0}$ is
proportional to $\Sigma^{-1/2}$, thus ensuring greater resolution
where it is needed in the high-density regions, and more smoothing in
the low-density regions where the signal-to-noise levels are much
lower. For this study, the surface number density of \mbox{$R<23.0$}
galaxies is considered, the magnitude limit to which cluster red
sequence galaxies can be detected in both passbands. The local density
for each galaxy is initially determined using an exponential kernel
whose width is fixed to 60\,arcsec, and then iteratively recalculated
using adaptive kernels, before corrected for field contamination. The
resultant surface number density map of the ABCG\,209 field is shown in
Fig.~\ref{density}. The dashed contour corresponds to the density of
field galaxies, while the thick contours correspond to 2, 5 and 10\,
cluster galaxies arcmin$^{-2}$, the densities used to separate the
three cluster environments described below.

The spatial distribution shows the complex structure of this cluster,
characterised by the clear elongation in the SE-NW direction (see
Chapter \ref{cap:2}), indicative of the cluster having gone
through a recent merger event. There are also present three other
subclumps, the first one about 10\,arcmin south along the cluster
elongation, the second at about 18\,arcmin to the north-east, and the
last one, the smallest, at about 18\,arcmin to the south-west. The
colour information on galaxies in these peaks confirm that these could
be groups/clusters at the same redshift as ABCG\,209, each peak
containing galaxies belonging to the cluster red sequence of
ABCG\,209. Given the complexity of the structure of ABCG\,209, it is
important that to study the effect of environment on galaxy
properties, we measure them as a function of local galaxy number
surface density rather than cluster-centric radius as is usual in
these type of studies.

For the following analyses on the effect of the cluster environment on
its constituent galaxy population we define three regions selected
according to the local surface number density. Firstly we consider a
high-density region with \mbox{$\Sigma>10\,$gals arcmin$^{-2}$}, which
corresponds to the cluster core out to a radius of
\mbox{$\sim$500\,kpc}, and covers \mbox{16.0\,arcmin$^{2}$}. Next we
consider intermediate- \mbox{($5<\Sigma<10\,$gals arcmin$^{-2}$)} and
low-density \mbox{($2<\Sigma<5\,$gals arcmin$^{-2}$)} regions which
probe the cluster periphery with median galaxy cluster-centric radii
of 1.3 and 2.2\,Mpc, and cover 68.0 and \mbox{189.6\,arcmin$^{2}$}
respectively. It should be stated that in each case we are studying
the cluster galaxy population and not that of the field, and the
low-density environment represents an overdense region.

\section{Statistical Field Galaxy Subtraction}
\label{sec:44}

\begin{figure}[t]
\centerline{{\resizebox{\hsize}{!}{\includegraphics{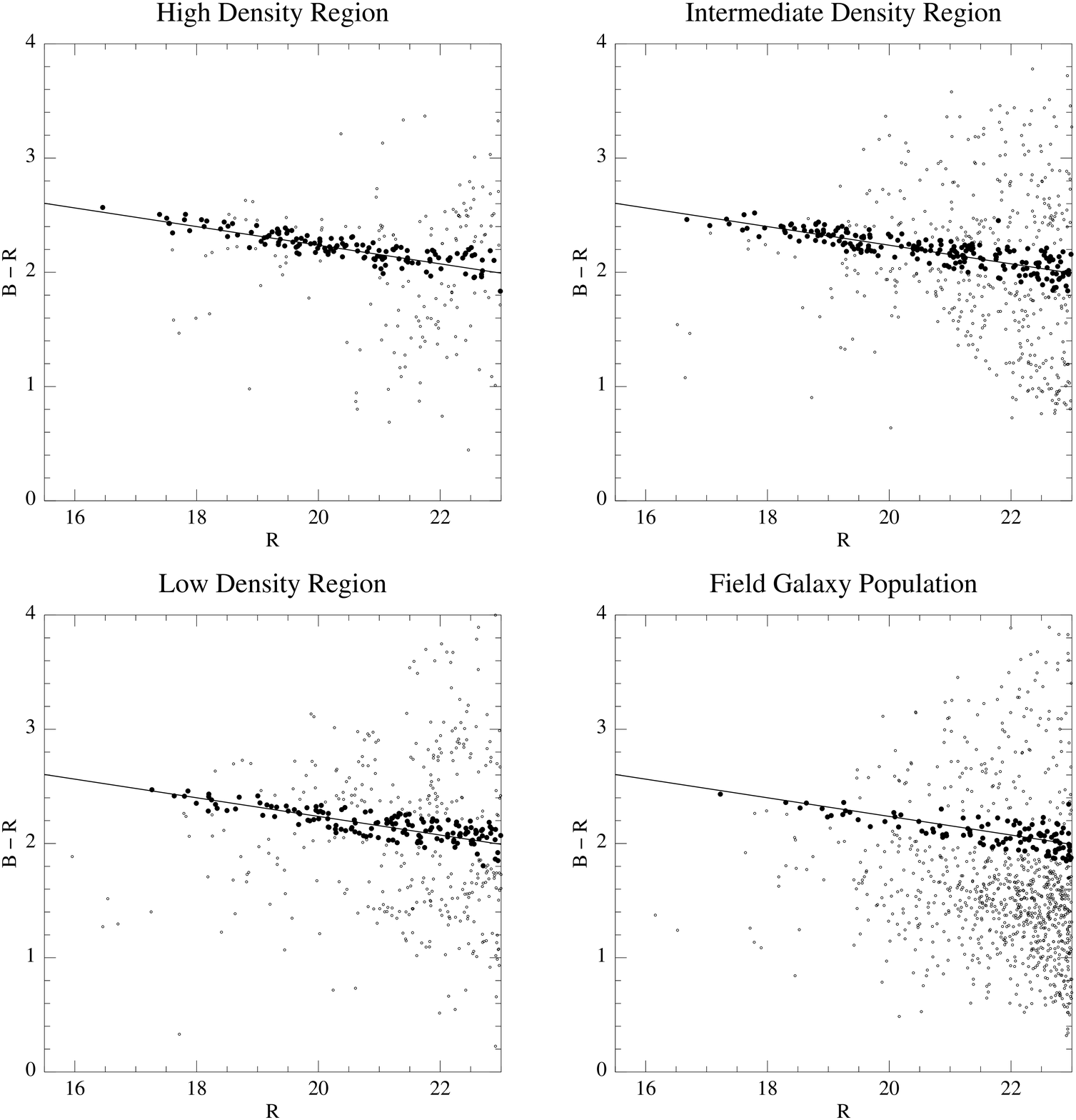}}}}
\caption{The \mbox{$B-R/R$} colour-magnitude diagrams of the cluster galaxy population in the three cluster regions corresponding to high-, intermediate-, and low-density environments. Each of the three diagrams is the result of a Monte-Carlo realisation in which field galaxies are subtracted statistically. For comparison the C-M diagram of field galaxies is shown also. In each plot the solid line indicates the best-fitting C-M relation of Eq.~\ref{CMrelation}, and galaxies identified as belonging to the red sequence through Eq.~\ref{caustic} are indicated by solid circles.}
\label{CMs}
\end{figure}

To accurately measure the properties of the cluster population as a
function of environment requires the foreground / background
contamination to be estimated efficiently and corrected. Given the
lack of spectroscopic information, we estimate the field galaxy
contamination as a function of magnitude and colour (see also Kodama
\& Bower \ct{kodama}).

The sample of field galaxies is taken from two regions within the
CFHT12K field indicated in Fig.~\ref{density} by the shaded
boxes. These areas were identified as regions where the local galaxy
surface density reached a stable minimum level over a wide area
indicative of the field.  They also have galaxy number counts within
$1\sigma$ of the ESO-Sculptor survey of Arnouts et
al. (\ct{arn97}) which covers an area of
\mbox{$0.24\times1.52\,$deg$^{2}$}, and is complete to \mbox{$B=24.5$}
and \mbox{$R=23.5$}. The two {\em ``field''} regions in the ABCG\,209
field cover in total \mbox{160.9\,arcmin$^{2}$}, and contain 965
galaxies to \mbox{$R=23.0$}, resulting in a number surface density of
\mbox{6.0\,gals.arcmin$^{-2}$}, as indicated in Fig.~\ref{density} by
the dashed contour.

A two-dimensional distribution histogram of these field galaxies is
built with bins of width 0.25\,mag in \mbox{$B-R$} colour and 0.5\,mag
in $R$ magnitude. Similar histograms are the constructed for each of
the three cluster regions, which also obviously include a field galaxy
contamination. The field histogram is normalised to match the area
within each cluster region, and for each bin \mbox{$(i,j)$} the number
of field galaxies \mbox{$N_{i,j}^{field}$} and the cluster plus field
\mbox{$N_{i,j}^{cluster+field}$} are counted. The former can be larger
than the latter for low-density bins due to low number statistics, in
which case the excess number of field galaxies are redistributed to
neighbouring bins with equal weight until
\mbox{$N_{i,j}^{field}<N_{i,j}^{cluster+field}$} is satisfied for all
bins. For each galaxy in the cluster region the probability that it
belongs to the field is then defined as:
\begin{equation}
P(field)=\frac{N_{i,j}^{field}}{N_{i,j}^{cluster+field}},
\end{equation}
where the numbers are taken from the bin that the galaxy belongs
to. For each cluster region, 100 Monte-Carlo simulations of the
cluster population are realised and averaged.

The \mbox{$B-R/R$} C-M diagrams for Monte-Carlo realisations of the
cluster galaxy population for each of the three cluster regions
corresponding to high-, intermediate-, and low-density environments
are shown in Fig.~\ref{CMs}, along with that for the field region used
for the statistical field subtraction. The different distributions of
the cluster and field populations are obvious, with the red sequence
prominent in both high- and intermediate-density regions.
 
\section{The Galaxy Luminosity Function}
\label{sec:45}

\begin{figure}[t]
\centerline{{\resizebox{\hsize}{!}{\includegraphics{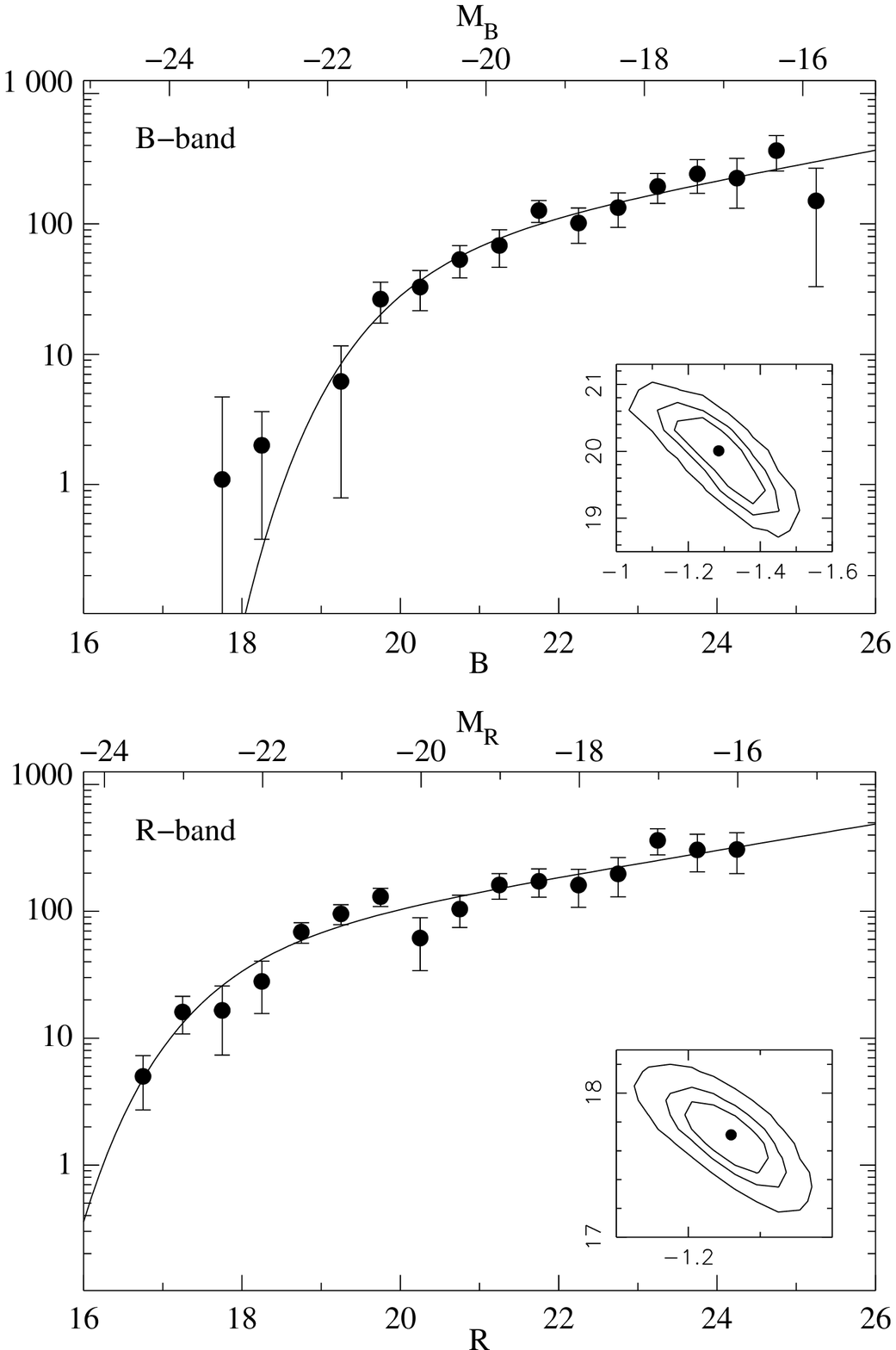}}}}
\caption{The $B$- and $R$-band LFs of galaxies in the virial region. The solid curve indicates the best-fitting Schechter function whose parameters are indicated in Table~\ref{schechter}. In the small panels, the 1, 2 and $3\sigma$ confidence levels of the best-fitting parameters $\alpha$ and $M^{*}$ are shown.}   
\label{lumfuncs}
\end{figure}

\begin{figure}[t]
\centerline{{\resizebox{\hsize}{!}{\includegraphics[angle=-90]{LF_dens_B.ps}}}}
\caption{The $B$-band LFs of galaxies in the three cluster regions corresponding to high-, intermediate- and low-density environments.}
\label{Blumfuncs}
\end{figure}

\begin{figure}
\centerline{{\resizebox{7cm}{!}{\includegraphics{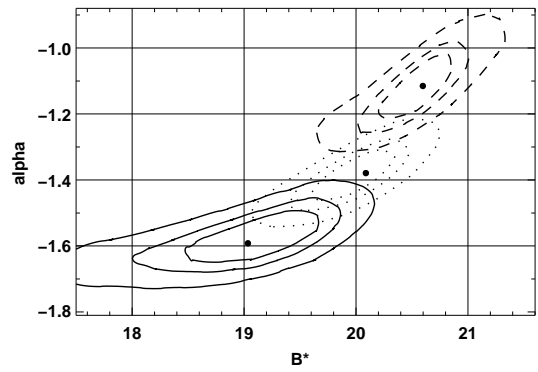}}}}
\caption{The 1, 2 and $3\sigma$ confidence limits for the best-fitting Schechter parameters $\alpha$ and $B^{*}$ for the three cluster regions corresponding to low- (solid contours), intermediate- (dotted) and high-density (dashed) environments.}
\label{contB}
\end{figure}

\begin{figure}[t]
\centerline{{\resizebox{\hsize}{!}{\includegraphics[angle=-90]{LF_dens_R.ps}}}}
\caption{The $R$-band LFs of galaxies in the three cluster regions corresponding to high-, intermediate- and low-density environments.}
\label{Rlumfuncs}
\end{figure}

\begin{figure}
\centerline{{\resizebox{7cm}{!}{\includegraphics{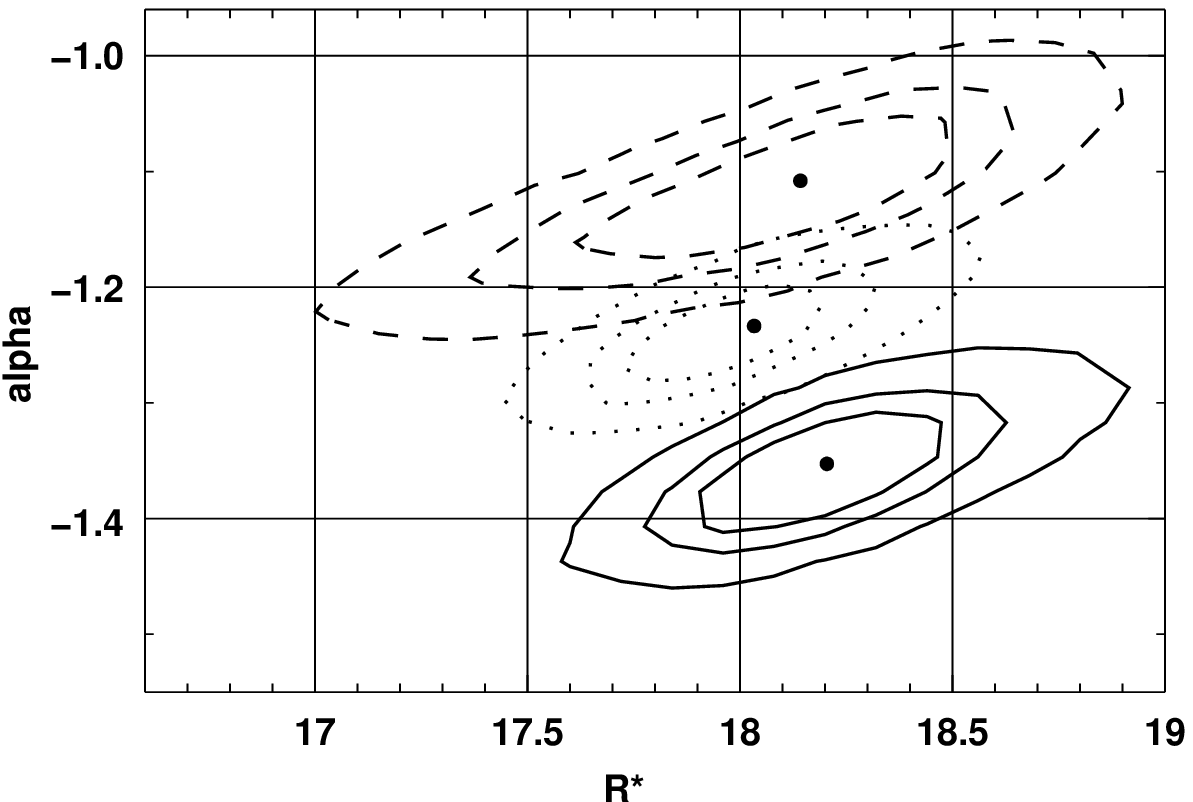}}}}
\caption{Confidence limits for the best-fitting Schechter parameters $\alpha$ and $R^{*}$ for the three cluster regions. Contours as for Fig.~\ref{contB}.}
\label{contR}
\end{figure}

In order to measure the cluster LF in each band we used all the galaxy
photometric data up to the completeness magnitudes and removed the
interlopers by statistically subtracting the background contamination,
as determined from the two field regions. The errors on the cluster
LFs are computed by adding in quadrature the Poisson fluctuations for
the galaxy counts in both the cluster and field regions, and
fluctuations due to photometric errors on Kron magnitudes.

We compute the cluster LFs for both $B$ and $R$ bands for galaxies
within one virial radius, and examine the effect of the cluster
environment by determining the LFs for each of the three cluster
regions corresponding to \mbox{high-,} intermediate- and low-density
environments. For each cluster LF, we fit the observed galaxy counts
with a single Schechter function. Absolute magnitudes are determined
using the k-corrections for early-type galaxies from Poggianti
(\ct{pog97}). All the fit parameters and associated $\chi^{2}$
statistics are listed in Table~\ref{schechter}.

\begin{table}
\begin{center}
\begin{tabular}{cccccc} \hline
 & Region & $B^{*}$ & M$^{*}$ & $\alpha$ & $\chi^{2}_{\nu}$ \\ \hline 
$\!\!B\!\!$ & $r<{\rm R}_{vir}$         & 20.01 & -21.07 & -1.28 & 0.59 \\ \hline
$\!\!B\!\!$ & $\Sigma>10$               & 20.60 & -20.48 & -1.11 & 0.61 \\
$\!\!B\!\!$ & $\!5\!<\!\Sigma\!<\!10\!$ & 20.09 & -20.99 & -1.38 & 0.92 \\
$\!\!B\!\!$ & $\!2\!<\!\Sigma\!<\!5\!$  & 19.03 & -22.05 & -1.50 & 0.65 \\ \hline

 & Region & $R^{*}$ & & &  \\ \hline 
$\!\!R\!\!$ & $r<{\rm R}_{vir}$         & 17.71 & -22.55 & -1.26 & 1.07 \\ \hline
$\!\!R\!\!$ & $\Sigma>10$               & 18.14 & -22.12 & -1.11 & 0.85 \\
$\!\!R\!\!$ & $\!5\!<\!\Sigma\!<\!10\!$ & 18.03 & -22.23 & -1.23 & 0.88 \\
$\!\!R\!\!$ & $\!2\!<\!\Sigma\!<\!5\!$  & 18.20 & -22.06 & -1.35 & 0.66 \\ \hline

\end{tabular}
\end{center}
\caption{Fits to the LFs for cluster galaxies. Errors on the $M^{*}$ and $\alpha$ parameters are indicated by the confidence contours shown in Figures~\ref{lumfuncs},~\ref{contB} and~\ref{contR}.}
\label{schechter}
\end{table}

\subsection{Galaxies within the Virial Radius}
\label{sec:451}

Figure~\ref{lumfuncs} shows the LFs in the $B$ and $R$ bands for
galaxies within one virial radius (\mbox{$2.5h^{-1}_{70}$\,Mpc}) of
the cluster centre. The solid curves show the best-fitting single
Schechter functions, obtained by weighted parametric fits to the
statistically background-subtracted galaxy counts. According to the
$\chi^{2}$ statistic, the global distributions of the data are well
described by single Schechter functions, although there is an
indication of a dip at \mbox{$R\sim20.5$}, as already noted in the
previous study of the LF in the central region of the cluster
(Chapter \ref{cap:3}).

In that study, where a smaller area was considered, the amplitude of
this dip was calculated by comparing expected and observed counts in
the range \mbox{$R=2$0--21}, and found to be \mbox{${\rm
A}=16\pm8$\%}. For galaxies within the virialised region we obtain a
value for the dip amplitude of \mbox{${\rm A}=32\pm20$\%}, which is
consistent with that previously obtained.

In Chapter \ref{cap:3} we derived the LFs for the central
field of \mbox{$9.2^\prime \times 8.6^\prime$}
(\mbox{$1.9\times1.8h^{-2}_{70}\,{\rm Mpc}^{2}$}) by using EMMI--NTT
images, complete to \mbox{$B=22.8$} and \mbox{$R=22.0$}. In this
previous study we subtracted statistically the field contamination
using background counts in $B$- and $R$-bands from the ESO-Sculptor
Survey (Arnouts et al. \ct{arn97}; de Lapparent et al.
\ct{del03}). The parameters of the best-fitting Schechter
functions were: \mbox{$B^{*}=20.06$}, \mbox{$\alpha_{B}=-1.26$}, and
\mbox{$R^{*}=17.78$}, \mbox{$\alpha_{R}=-1.20$}, which are fully
consistent with those derived here.

\subsection{The effect of environment}
\label{sec:452}

To examine the effect of environment on the galaxy LF, we have
determined the $B$- and $R$-band LFs for galaxies in three regions
selected according to their local density. Figures~\ref{Blumfuncs}
and~\ref{Rlumfuncs} show respectively the $B$- and $R$-band LFs in the
three different cluster regions, corresponding to high-,
intermediate-, and low-density environments. Each LF is modelled
through use of a weighted parametric fit to a single Schechter
function, the results of which are presented in
Table~\ref{schechter}. 
measured by fixing $\alpha$ to the value determined for the region
within one virial radius, and fitting a Schechter function to the
bright end of the galaxy LF ($B<22$ and $R<20$) for each region.

As for the region within one virial radius, the single Schechter
function gives a good representation of the global distribution of the
data for each cluster environment in both $B$- and $R$-bands. In both
high- and low-density regions, no dip is apparent at
\mbox{$R=2$0--21}, although the observed counts are marginally below
that predicted from the Schechter function. Only in the
intermediate-density region is a significant dip apparent, where a
deficit of \mbox{$R=20.$0--20.5} galaxies significant at the $2\sigma$
level is observed with respect to the fitted Schechter function.

Figures~\ref{contB} and~\ref{contR} show the confidence contours for
the best fitting Schechter functions for $\alpha$ and $B^{*}$, and
$\alpha$ and $R^{*}$ respectively, for each of the three cluster
regions, allowing the trends with density to be followed. In both $B$
and $R$ bands, the faint-end slope becomes significantly steeper from
high- to low-density environments, the values of $\alpha$ determined
for the high- and low-density regions being inconsistent at more than
the $3\sigma$ confidence level.

\section{Properties of the Red Sequence Galaxies}
\label{sec:46}

We determined the colour-magnitude (CM) relation of galaxies in the
cluster ABCG\,209 by averaging over 100 Monte-Carlo realisations of the
cluster population, and fitting the resultant photometric data of the
$\sim480$ $R<21$ galaxies within one virial radius with the biweight
algorithm of Beers et al. (\ct{bee90}) obtaining
\begin{equation}
(B-R)_{CM} = 3.867\pm0.006 - 0.0815\pm0.0098\times R.
\label{CMrelation}
\end{equation}
Here the errors quoted consider the uncertainty due to background
subtraction, photometric errors, and uncertainty from the fitting
itself as measured using the scale value of the biweight
algorithm. The $B-R$ colour dispersion around the red sequence is
measured as a function of magnitude, $\sigma(R)$, for galaxies in each
magnitude bin over the range $18<R<23$. It is found to be consistent
with being due to an intrinsic colour dispersion, $\sigma_{int}$ and
the photometric error, $\sigma_{B-R}^{2}(R)$ added in quadrature,
\begin{equation}
\sigma(R)^{2}=\sigma_{int}^{2}+\sigma_{B-R}^{2}(R).
\end{equation}
The intrinsic dispersion around the red sequence is found to be
$\sigma_{int}=0.10$\,mag, which is much higher than the dispersion
level of 0.05\,mag found for the Coma and Virgo clusters (Bower, Lucey
\& Ellis \ct{bower}), and which is often quoted as being typical for
low redshift rich clusters. However, in a similar study of the $B-R/R$
CM-relation for 11 $0.07<z<0.16$ clusters Pimbblet et al.
(\ct{pimbblet}) obtain typical dispersion levels of 0.06--0.08, and
observe three clusters with dispersion levels higher than ours.

Using this equation, we then defined those galaxies lying within the region between the curves:
\begin{equation}
(B-R)_{\pm} = (B-R)_{CM} \pm \sqrt{\sigma^{2}_{int} + \sigma^{2}_{B-R}}
\label{caustic}
\end{equation}
as being red sequence galaxies. 

\begin{figure}[t]
\centerline{{\resizebox{\hsize}{!}{\includegraphics{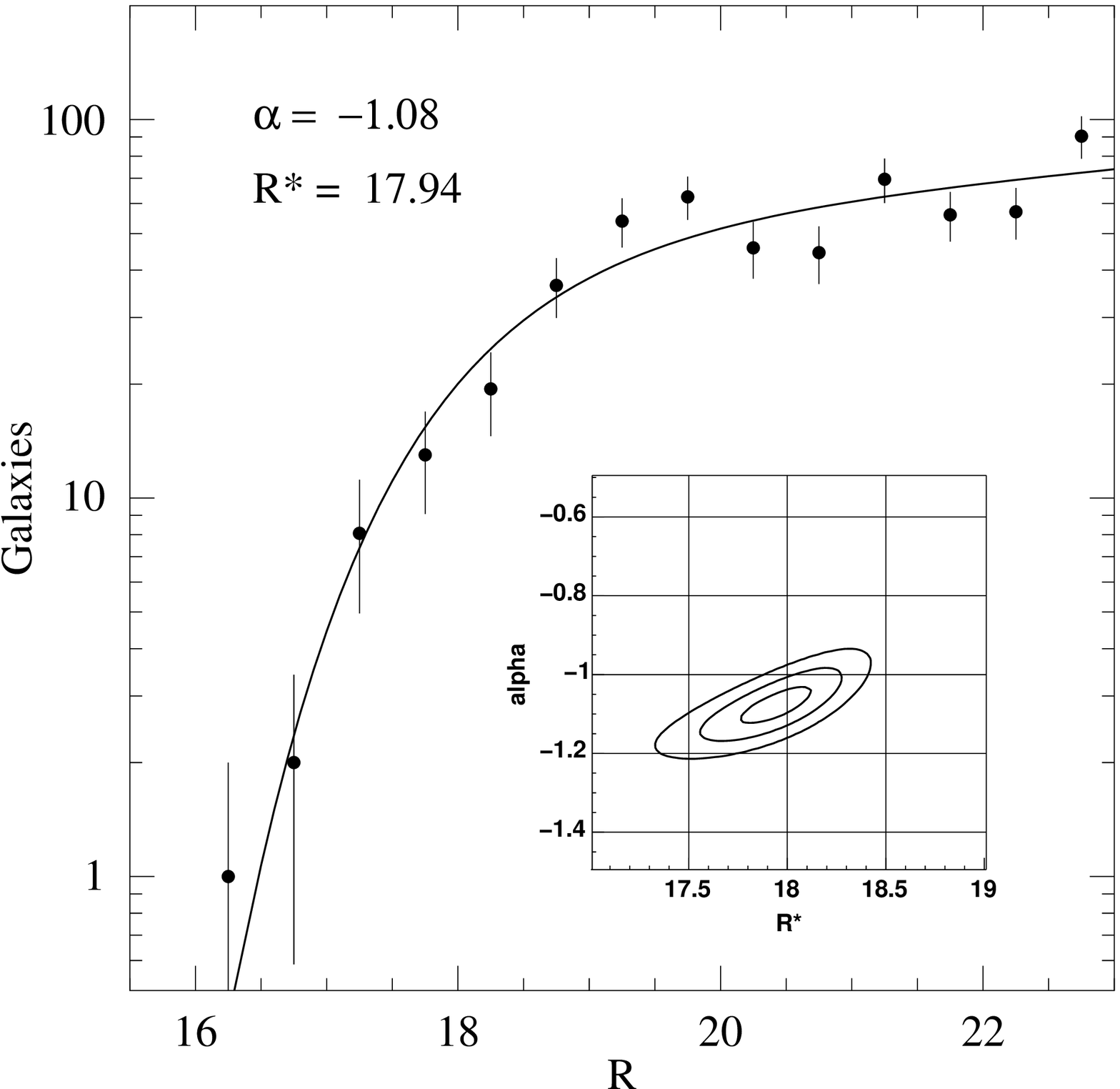}}}}
\caption{The $R$-band LF of red sequence galaxies in the virial region. The solid curve indicates the best-fitting Schechter function whose parameters are indicated in the top-left of each plot. In the small panels, the 1, 2 and $3\sigma$ confidence levels of the best-fitting parameters $\alpha$ and $R^{*}$ are shown.}   
\label{CMlum}
\end{figure}

As for the composite luminosity function, we determine the $R$-band
luminosity function of red sequence galaxies within one virial radius
of the cluster centre, as shown in Figure~\ref{CMlum}.

\begin{figure*}[t]
\centerline{{\resizebox{\hsize}{!}{\includegraphics{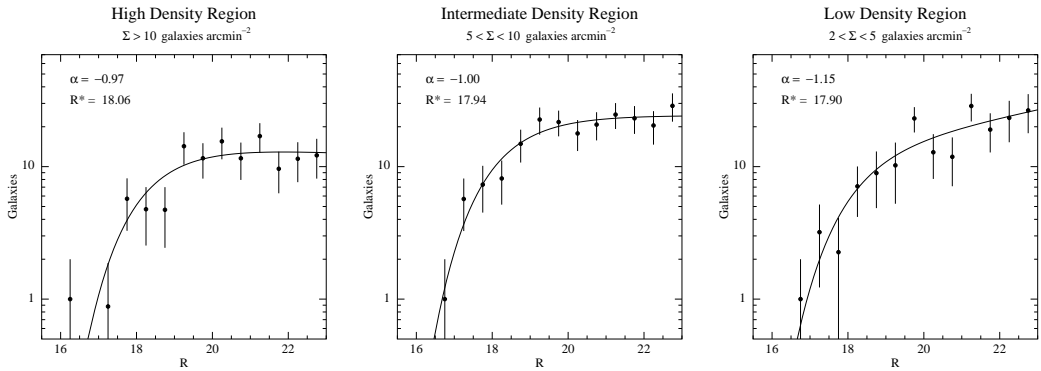}}}}
\caption{The $R$-band LFs of red sequence galaxies in the three cluster regions corresponding to high-, intermediate- and low-density environments. The solid curve indicates the best-fitting Schechter function whose parameters are indicated in the top-left of each plot.}
\label{CMlums}
\end{figure*}

\begin{figure}[t]
\centerline{{\resizebox{6cm}{!}{\includegraphics{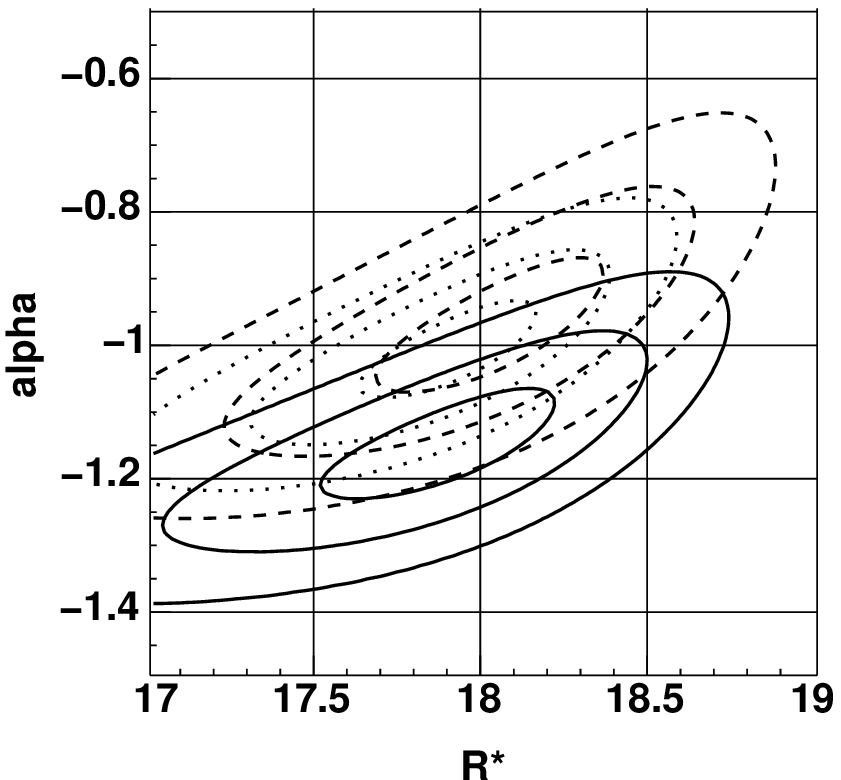}}}}
\caption{Confidence limits for the best-fitting Schechter parameters $\alpha$ and $R^{*}$ for the three cluster regions. Contours as for Fig.~\ref{contB}.}
\label{max}
\end{figure}

\subsection{The effect of environment on the C-M relation}
\label{sec:461}

We also determine the C-M relation in the high- and
intermediate-density regions separately, to see if there is any
environmental effect on the relation. By fixing the slope at -0.0815,
we find the red sequence to be $0.022\pm0.014$\,mag redder in the
high-density region than for the intermediate-density region. The
effect of environment on the red sequence slope was also examined by
leaving the slope as a free parameter, but no significant change was
observed. It was not possible to extend this analysis to the
low-density region, as the red sequence was poorly constrained due to
a much larger contamination by outliers.

The $R$-band luminosity functions of red sequence galaxies (as
selected according to Eq.~\ref{caustic}) in each of the three cluster
environments is shown in Figure~\ref{CMlums}, and the best fitting
parameters shown in Table~\ref{CMlumparams}. The solid curve shows the
best-fitting Schechter function found through a maximum likelihood
analysis whose parameters $\alpha$ and $R^{*}$ are indicated in the
top-left of each plot. The errors for each bin are determined as the
uncertainty from the field galaxy subtraction and the Poisson noise
due to galaxy counts in the cluster and field regions, all added in
quadrature.

\begin{table}
\begin{center}
\begin{tabular}{ccccc}\hline
Region & $R^{*}$ & $M^{*}$ & $\alpha$ & $\chi_{\nu}^{2}$\\ \hline
$r<{\rm R}_{vir}$ & 17.94 & -22.22 & -1.08 & 1.61\\ \hline
$\Sigma>10$   & 18.04 & -22.30 & -0.97 & 0.95\\
$5<\Sigma<10$ & 17.94 & -22.20 & -1.00 & 0.45\\
$2<\Sigma<5$  & 17.90 & -22.24 & -1.15 & 0.90\\ \hline
\end{tabular}
\end{center}
\caption{Fits to the red sequence galaxy LFs. Errors on the $R^{*}$ and $\alpha$ parameters are indicated by the confidence contours shown in Figs.~\ref{CMlum} and~\ref{max}.}
\label{CMlumparams}
\end{table}

For the cluster red sequence we consider a magnitude limit of
$R=23.0$. At this magnitude, red sequence galaxies have $B-R\sim2.0$,
and so are at the completeness limit of the $B$-band image, and have
typical uncertainties of $\Delta(R)\sim0.06$ and
$\Delta(B-R)\sim0.11$. As discussed earlier, the expected
contamination of stars is minimal due to their colour-distribution,
and for the three cluster regions corresponding to high-,
intermediate-, and low-density environments, should be at the level of
0.1, 0.3 and 0.8 stars respectively.

Figure~\ref{max} shows the confidence contours for the best-fitting
Schechter parameters $\alpha$ and $R^{*}$ for each of the three
cluster regions, allowing the trends with density to be followed. As
for the overall galaxy luminosity function, the $R$-band luminosity
function of red sequence galaxies has a steeper faint-end slope,
$\alpha$, in the low-density regions than for their high-density
counterparts, although the effect here is smaller, and significant
only at the $1.7\sigma$ level. Also in each of the three cluster
environments, the faint-end slope for the overall galaxy luminosity
function is steeper than that for the luminosity function of red
sequence galaxies only. This is due to the red sequence galaxies
dominating at bright magnitudes, while at faint magnitudes they
constitute only $\sim25\%$ of the cluster population.

\section{Blue Galaxy Fraction}
\label{sec:47}

\begin{figure}[t]
\centerline{{\resizebox{\hsize}{!}{\includegraphics{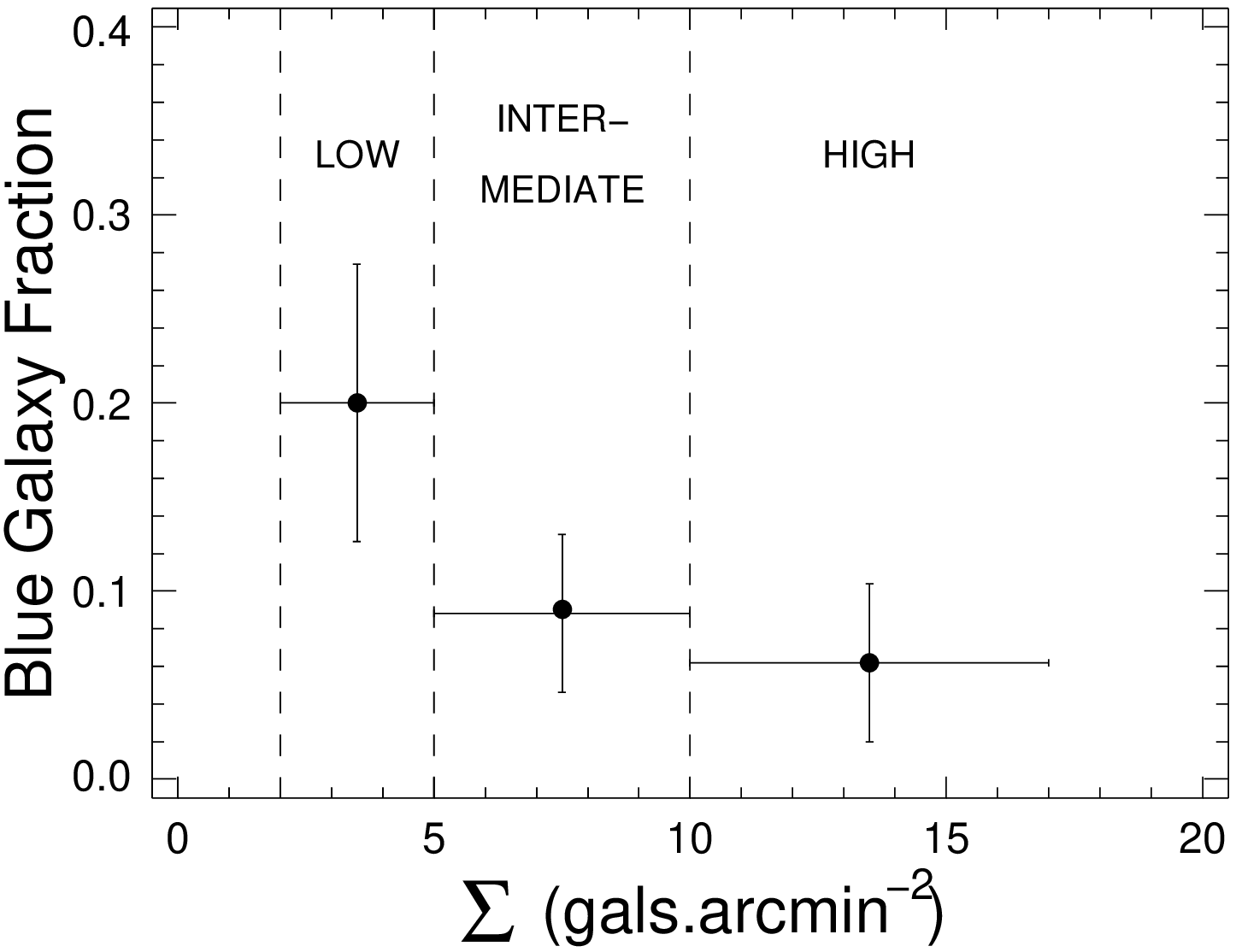}}}}
\caption{The blue galaxy fraction as a function of local number density.}
\label{blue}
\end{figure}

One of the classic observations of the study of galaxy evolution is
the Butcher-Oemler (\ct{but84}) effect in which the fraction of blue
galaxies in clusters is observed to increase with redshift. Here
instead we consider the effect of environment on the fraction of blue
galaxies. We define the blue galaxy fraction as the fraction of $R<20$
($\sim$M$^{*}\!+\!2$) galaxies with rest-frame $B-V$ colours at least
0.2\,mag bluer than that of the C-M relation, as for the original
studies of Butcher \& Oemler (\ct{but84}). We estimate the
corresponding change in $B-R$ colour at the cluster redshift by
firstly considering two model galaxies at $z=0$ (Bruzual \& Charlot
\ct{bruzual}), one chosen to be a typical red-sequence galaxy
(10\,Gyr old, $\tau=0.1$\,Gyr, $Z=0.02$), and the second reduced in
age until its $B-V$ colour becomes 0.2\,magnitude bluer. After
redshifting both galaxies to the cluster redshift, the difference in
their observed $B-R$ colour is found to be 0.447\,mag. Hence we
consider the blue galaxy fraction to be the fraction of $R<20$
galaxies having $B-R < 3.420 - 0.0815\times R$. Figure~\ref{blue}
shows the resulting blue galaxy fractions in the high-, intermediate-,
and low-density environments after averaging over 100 Monte-Carlo
realisations of the cluster population. We find the blue galaxy
fractions are low in both the high- ($0.062\pm0.042$) and
intermediate-density ($0.088\pm0.042$) regions, but increases
dramatically to $0.200\pm0.072$ in the low-density region. Kodama \&
Bower (\ct{kodama}) examine the dependence of the blue galaxy
fraction on radius for 7 clusters at $0.2<z<0.43$, and observe a
general trend for $f_{B}$ to increase with radius, the effect becoming
more prominent for the higher redshift clusters. For each cluster, the
$f_{B}$ in the cluster core ($r\lesssim0.5$\,Mpc), remains low at
$\sim0.10$, whereas at cluster-centric radii of 1--2\,Mpc, the $f_{B}$
increases dramatically to 0.2--0.6.

\section{Galaxy colours}
\label{sec:48}

\begin{figure}[t]
\centerline{{\resizebox{\hsize}{!}{\includegraphics{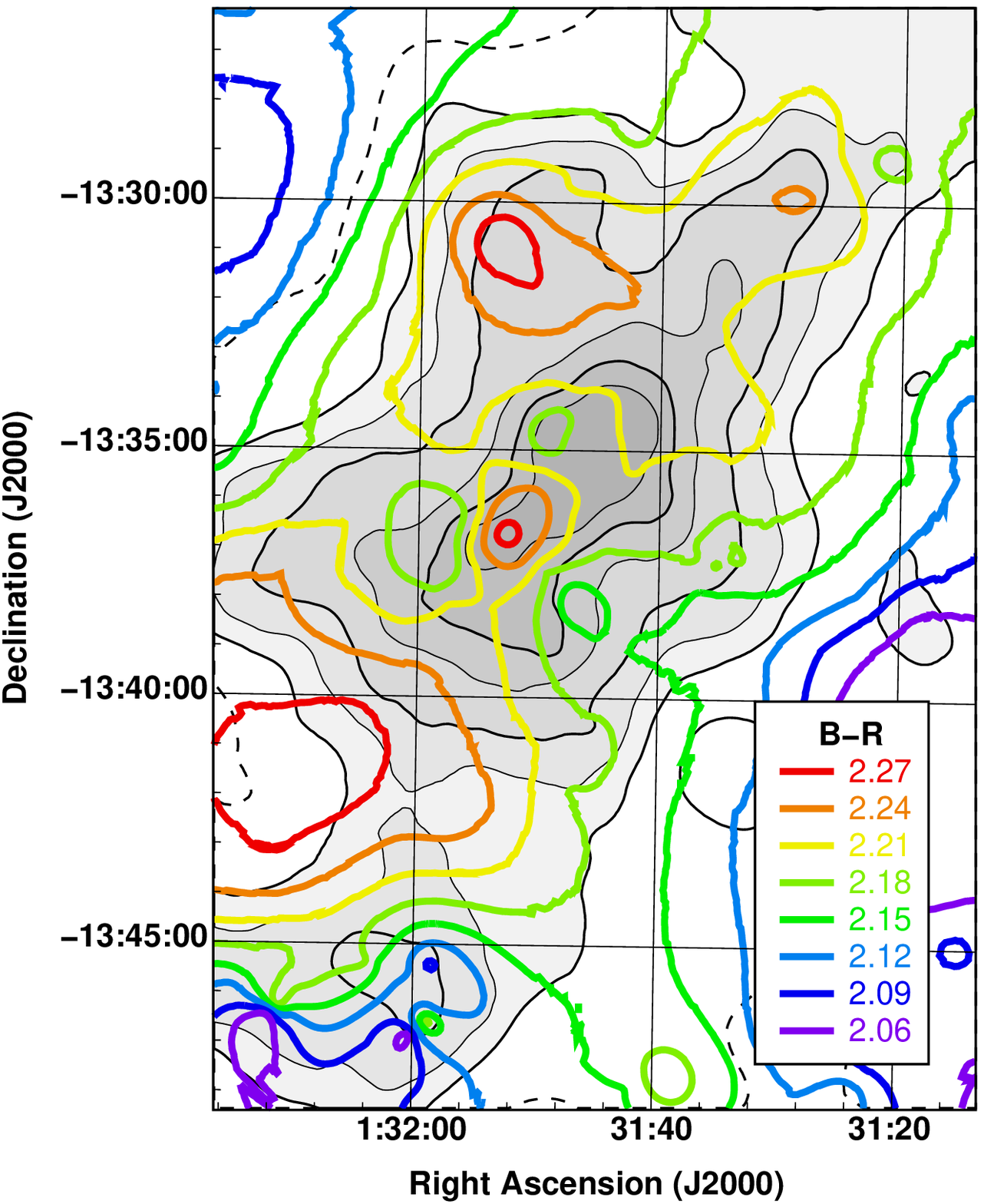}}}}
\caption{The mean galaxy colour of the cluster galaxy population as a function of spatial position (see text).}
\label{colour}
\end{figure}

To gain some further insight into the effect of cluster environment
and also the particular dynamical state of ABCG\,209, we plot in
Figure~\ref{colour} the mean $B-R$ colour of $R<21$ cluster galaxies
as a function of spatial position. This is determined as for the
surface density map of Fig.~\ref{density} through the adaptive kernel
approach (Pisani \ct{pis93}; \ct{pis96}), in which each
galaxy is weighted according to the probability that it is a cluster
member, $1-P(f)$, so that for a particular point, $\mathbf{x}$, the
local mean galaxy colour is given by:
\begin{equation}
\overline{(B\!-\!R)}_{\mathbf{x}} = \frac{\sum_{i=1}^{N} (B\!-\!R)^{CM}_{i}.(1\!-\!P(f)_{i}).K(\mathbf{x}\!-\!\mathbf{x}_{i};r_{i})}{ \sum_{i=1}^{N} (1\!-\!P(f)_{i}).K(\mathbf{x}\!-\!\mathbf{x}_{i};r_{i})}
\end{equation}
where $\mathbf{x}_{i}$, $P(f)_{i}$ and $r_{i}$ are the position, probability of being a field galaxy, and kernel width respectively of galaxy $i$ out of $N$. And $(B-R)_{i}^{CM}$ is the $B-R$ colour for each galaxy after accounting for the luminosity-dependent effect of the slope of the CM-relation through
\begin{equation}
(B-R)_{i}^{CM}=(B-R)_{i}+0.0815 \times (R_{i}-19).
\end{equation}
The resultant mean galaxy colour map $[B\!-\!R](\mathbf{x})$ is shown
by the coloured contours, with the red/orange contours indicating
regions which have redder cluster galaxies on average, and blue/purple
contours indicating regions which have more blue cluster galaxies. To
see how the colour of these cluster galaxies are related to their
environment, the contours are overlaid upon the same contour map of
the surface number density of $R<23.0$ galaxies as shown in
Fig.~\ref{density}.

What is most important is that only the cluster galaxy population is
being examined, and so the fact that field galaxies are bluer on
average than cluster galaxies should not result in a density-dependent
colour gradient, as would be the case if all galaxies were
considered. Instead, any surface density-dependent colour gradient
should be purely the result of galaxies in the low-density
environments on the cluster periphery having differing colours on
average to those in high-density environments. Secondly, as it is the
mean $B-R$ colour of $R<21$ galaxies that is being measured, it is the
effect of the cluster environment on only the luminous cluster
galaxies ($L\gtrsim0.1L^{*}$) that is being examined.

A clear overall density-dependence on galaxy colour is apparent, with
galaxies in the high-density regions having
$\mu(B\!-\!R)\sim2$.21--2.30, while galaxies in the low-density
regions have $\mu(B\!-\!R)\sim2$.14--2.18. This we take to be the
manifestation of the same effect observed with the blue galaxy
fraction, with higher blue galaxy fractions on the cluster
peripheries, thus reducing the mean $B-R$ colour in those regions. 
second contributory effect will be the combination of the slope in the
C-M relation, and the concentration of the most luminous red sequence
galaxies towards the cluster core, which are redder than their fainter
counterparts.

Substructure in the mean $B-R$ galaxy colours is apparent, and appears
related to the dynamical state of the system, being aligned with the
direction of elongation in the galaxy number surface density. The
reddest mean $B-R$ galaxy colours are observed at the very centre of
the cluster, coincident with the cD galaxy and a concentration of the
brightest $R<19$ galaxies, as would be expected. There also appear
regions of red galaxies on either side of the cluster core aligned
with the overall elongation of the cluster. As the overall galaxy
surface density is greater in the NW direction from the cluster core,
we would suggest that this is where the merging clump is found, some
4--5\,arcmin from the centre of ABCG\,209, corresponding to $\sim1$\,Mpc
at $z=0.209$.

In contrast, perpendicular to the axis of elongation appear regions
close to the cluster centre containing bluer than average
galaxies. The concentration to the east appears centred on a bright
face-on spiral (\mbox{$R=17.71$}, \mbox{$B-R=1.47$}), which has been
spectroscopically confirmed as a cluster member (Chapter \ref{cap:2}).
  
\section{Discussion}
\label{sec:49}

We have examined the effect of cluster environment, as measured in
terms of the local surface density of \mbox{$R<23.0$} galaxies, on the
global properties of the cluster galaxies, through their luminosity
functions, colour-magnitude relations, and average colours. For this
study we have considered three cluster environments, a high-density
region sampling the cluster core, and intermediate- and low-density
regions which sample the cluster periphery.

\subsection{The Galaxy Luminosity Function}

The LFs for galaxies within the virialised region are found to be well
described by single Schechter functions to M$_{R}$, M$_{B}\sim-16$,
although there is an indication of a dip at $R=2$0--20.5
(M$_{R}=-20$). Dips in the LF at such absolute magnitudes appear
common for low-redshift clusters (see Chapter \ref{cap:3}),
and appear stronger for richer, early-type dominated clusters (Yagi
et al. \ct{yag02}). 

The faint-end slope, $\alpha$, shows a strong dependence on
environment, becoming steeper at $>3\sigma$ significance level from
high- to low-density environments. We explain this trend as the
combination of two related effects: a manifestation of the
morphology-density relation whereby the fraction of early-type
galaxies which have shallow faint-end slopes increases with density,
at the expense of late-type galaxies which have steep faint-end
slopes; and a luminosity-segregation due to dwarf galaxies being
cannibalised and disrupted by the cD galaxy and the ICM in the cluster
core (Lopez--Cruz \ct{lop97}).

To separate the two effects, we consider the TSLF of galaxies
belonging to the cluster red sequence, as these are predominately
early-type galaxies, and so any trends with environment should be
independent of the morphology-density relation. Some
luminosity-segregation is observed, with a reduction in the fraction
of dwarf galaxies in the high-density regions, as manifested by their
shallower TSLF faint-end slopes. This effect is smaller than that for
the overall LF, being significant only at the $1.7\sigma$ level,
indicating that both luminosity-segregation and the morphology-density
relation drive the observed trends in the overall
LF. Luminosity-segregation is predicted by simulations for early-type
galaxies in clusters, with bright early-type galaxies much more
concentrated than their faint counterparts which follow the
distribution of the dark matter mass profile (Springel et al.
\ct{spr01}).

In a study of 45 low-redshift \mbox{$(0.04<z<0.18)$ }clusters,
Lopez--Cruz et al. (\ct{lop97}) find that for the seven that
are rich, dynamically-evolved clusters, characterised by the presence
of a cD galaxy, their composite LFs are all well described by single
Schechter functions with shallow faint-ends
\mbox{($\alpha\approx$-1.0)}. In contrast other clusters, often
poorer, but in particular those not containing a cD galaxy, require
two Schechter functions to fit their LFs, including a steep faint-end
slope to model the dwarf population (see also Parolin, Molinari \&
Chincarini \ct{parolin}). We thus indicate that the shallow
faint-end slope observed in the high-density region of ABCG\,209 is
related to the presence of the central dominant galaxy. cD galaxies
are regarded as physically different to elliptical galaxies, with a
different formation history, and are the product of dynamic processes
which take place during the formation of their host cluster. They are
thought to be built up through galactic cannibalism, or the
accumulation of tidal debris. Lopez--Cruz et al. (\ct{lop97})
propose that the flatness of the faint-end slope in clusters
containing cD galaxies results from the disruption of a large fraction
of dwarf galaxies during the early stages of cluster evolution, the
stars and gas of which are cannibalised by the cD galaxy, and the
remainder redistributed into the intracluster medium.
 
\subsection{Environmental Effects on the Red Sequence}

As well as having an affect on the TSLF of red sequence galaxies, the
cluster environment could have an affect on the colour or slope of the
red sequence itself, through the mean ages or metallicities of the
galaxies. To examine this possibility the red sequence was fitted for
the high- and intermediate-density regions independently. The red
sequence was found to be \mbox{$0.022\pm0.014$\,mag} redder in the
high-density region than for the intermediate-density region by fixing
the slope. In contrast no correlation between the slope of the red
sequence and environment was observed. A similar effect is observed
for bulge-dominated galaxies taken from the Sloan Digital Sky Survey
(SDSS) (Hogg et al. 2003), in which the modal $^{0.1}[g-r]$ residual
colour to the best-fitting C-M relation is \mbox{0.01--0.02\,mag}
redder in the highest-density environments (corresponding to cluster
cores) than in their low-density counterparts. In a study of 11 X-ray
luminous clusters at $0.07<z<0.16$ Pimbblet et al. (\ct{pimbblet})
examine the red sequence as a function of cluster-centric radius and
local galaxy density, and observe that the relation becomes
progressively bluer as cluster-centric radius is increased out to
3\,Mpc, and as the local surface density is decreased at a rate of
$d(B-R)/d\log_{10}(\Sigma)=-0.08\pm0.01$.
In a study of the cluster A\,2390 at $z=0.23$ Abraham et al.
(\ct{abraham}) also observe the normalised colour of red sequence
galaxies, $(g-r)_{r=19}$, to become increasingly blue with increasing
cluster-centric radius, $r_{p}$, as \mbox{$(g-r)_{r=19}=1.05-0.079\log
r_{p}$}.

These results indicate that the environment can affect the colour of a
red sequence galaxy, so that red sequence galaxies are older and/or
have higher metallicities in denser environments. To quantify this
effect we consider a model red sequence galaxy as a 7.5\,Gyr old (at
\mbox{$z=0.209$}, corresponding to 10\,Gyr at the present epoch),
\mbox{$\tau=0.01$\,Gyr}, solar-metallicity stellar population (Bruzual
\& Charlot \ct{bruzual}). By varying its age and metallicity
independently, we find that to reproduce the observed reddening in the
C-M relation, galaxies in the high-density region must be on average
500\,Myr older or 20\% more metal-rich than their intermediate-density
counterparts. By studying the spectra of 22\,000 luminous, red,
bulge-dominated galaxies from the SDSS, Eisenstein et al.
(\ct{eisenstein}) indicate that red sequence galaxies in
high-density regions are marginally older and more metal-rich than
their counterparts in low-density environments, with both effects
coming in at the same level.

This result is understandable in terms of cosmological models of
structure formation, in which galaxies form earliest in the
highest-density regions corresponding to the cores of rich
clusters. Not only do the galaxies form earliest here, but the bulk of
their star formation is complete by $z\sim1$, by which point the
cluster core is filled by shock-heated virialised gas which does not
easily cool or collapse, suppressing the further formation of stars
and galaxies (Blanton et al. \ct{blanton99};
\ct{blanton00}). Diaferio et al. (\ct{dia01}) shows that as
mixing of the galaxy population is incomplete during cluster assembly,
the positions of galaxies within the cluster are correlated with the
epoch at which they were accreted. Hence galaxies in the cluster
periphery are accreted later, and so have their star formation
suppressed later, resulting in younger mean stellar populations. It
should be considered however that this is a small-scale affect, and
that this result in fact confirms that the red sequence galaxy
population is remarkably homogeneous across all environments.

\subsection{Galaxy Colours}

As well as considering the effect of the cluster environment on galaxy
morphologies and LFs, its effect on star formation has been examined
by numerous authors (e.g. Balogh et al. \ct{bal00}; Ellingson et al.
\ct{ell01}; Lewis et al. \ct{lewis}). They find that
star formation is consistently suppressed relative to field levels for
cluster galaxies as far as twice the virial radius from the cluster
centre, and that for the majority of galaxies in the cluster core
star formation is virtually zero. For photometric studies it is
possible to qualitatively measure the effect of the cluster
environment on star formation through measurement of the blue galaxy
fraction --- the fraction of luminous \mbox{($M_{R}^{*}<-20$)}
galaxies whose colours indicate they are undergoing star formation
typical of late-type galaxies. We find that the blue galaxy fraction
decreases monotonically with density, in agreement with other studies
(e.g. Abraham et al. \ct{abraham}; Kodama \& Bower \ct{kodama}).

The observed trends of steepening of the faint-end slope, faintening
of the characteristic luminosity, and increasing blue galaxy fraction,
from high- to low-density environments, are all manifestations of the
well known morphology-density relation (Dressler \ct{dre80};
Dressler et al. \ct{dressler97}), where the fraction of early-type
galaxies decreases smoothly and monotonically from the cluster core to
the periphery, while the fraction of late-type galaxies increases in
the same manner. The observed trends in the composite LF simply
reflect this morphology-density relation: the galaxy population in the
cluster core is dominated by early-type galaxies and so the composite
LF resembles that of this type of galaxy, with a shallow faint-end
slope and a bright characteristic luminosity; whereas in lower density
regions the fraction of late-type galaxies increases, and so the
composite LF increasingly resembles that of the late-type TSLF, with a
steep faint-end slope and a fainter characteristic magnitude (Binggeli
et al. \ct{bin88}).

Finally, we examined the effect of the cluster environment on galaxies
through measuring the mean colour of luminous (\mbox{$R<21$}) cluster
galaxies as a function of their spatial position, as shown in
Fig.~\ref{colour}. This shows more clearly than any other result, the
complex effects of the cluster environment and dynamics on their
constituent galaxies. ABCG\,209 appears a dynamically young cluster,
with a significant elongation in the SE-NW direction, the result of a
recent merger with smaller clumps. To measure the effect of the
cluster dynamics on the galaxy population, their properties should be
measured against the parameter most easily related back to the
cluster dynamics, their spatial position. As the cluster is
significantly elongated, it should be possible to distinguish between
whether the local density or cluster-centric distance is more
important in defining the properties of galaxies, and indeed it is
clear that the mean galaxy colour correlates most with the local
density rather than the distance of the galaxy from the cluster
core. It should be considered through that as the system appears
elongated due to being the merger of two or more clumps, the notion
of a cluster-centric radius loses much of its validity, as where does
the centre of the system lie during a cluster merger? The location of
the main cluster and the secondary merging clump appear confirmed by
Fig.~\ref{colour}, with the reddest galaxies concentrated around the
cD galaxy (main cluster) and a more diffuse region 5\,arcmin to the
north coincident with the structure predicted from weak lensing
analysis (Dahle et al. \ct{dah02}).  The effect of the preferential
SE-NW direction for ABCG\,209 is apparent in the presence of bright blue
galaxies near the cD galaxy perpendicular to the axis and hence
unaffected by the cluster merger, and an extension of red galaxies to
the SE which may indicate the infall of galaxies into the cluster
along a filament. This preferential SE-NW direction appears related
to the large-scale structure in which A209 is embedded, with two rich
(Abell class R=3) clusters A\,222 at \mbox{$z=0.211$} and A\,223 at
\mbox{$z=0.2070$} are located \mbox{$1.5^{\circ}$} (15\,Mpc) to the
NW along this preferential axis.

Cluster dynamics and large-scale structure clearly have a strong
influence on galaxy evolution, and it would be interesting to search
for direct evidence of their effect on the star formation histories of
galaxies through spectroscopic observations of galaxies in the
secondary clump, in the form of post-starburst signatures.

\large
\chapter{\Large Transformations of
galaxies in ABCG\,209}
\footnotetext[1]{\footnotesize
The content of this chapter is submitted for publication in A\&A. The
authors are: A. Mercurio, G. Busarello, P. Merluzzi, F. La Barbera,
M. Girardi, C. P. Haines.}
\setcounter{footnote}{1}
\label{cap:5}
\markboth{Chapter 5}{Chapter 5}
\normalsize

In this chapter we analyse the properties of galaxy populations in the rich
Abell cluster ABCG\,209 at redshift z$\sim$0.21, on the basis of
spectral classification of 102 member galaxies. We take advantage of
available structural parameters to study separately the properties of
bulge-dominated and disk-dominated galaxies.  The star formation
histories of the cluster galaxy populations are investigated by using
line strengths and the 4000 \AA \ \ break, through a comparison to
stellar population synthesis models. The dynamical properties of
different spectral classes are examined in order to infer the past
merging history of ABCG\,209. The cluster is characterized by the
presence of two components: an old galaxy population, formed very
early (z$_f \gtrsim $ 3.5), and a younger (z$_f \gtrsim $ 1.2)
population of infalling galaxies. We find evidence of a merger with an
infalling group of galaxies occurred 3.5-4.5 Gyr ago. The correlation
between the position of the young H$_\delta$-strong galaxies and the
X-ray flux shows that the hot intracluster medium triggered a
starburst in this galaxy population $\sim$ 3 Gyr ago.

\section{Introduction}

The influence of environment on the formation and the evolution of
galaxies remains one of the most pressing issues in cosmology. The
study of galaxy populations in rich clusters offers a unique
opportunity to observe directly galaxy evolution, and environmental
effects on star formation, by providing large numbers of galaxies at
the same redshift which have been exposed to a wide variety of
environments. In particular, galaxy populations in rich clusters are
different from those in poorer environments, suggesting that some
mechanism working on the cluster population is absent in the low
density or field environments.

The first evidence for significant evolution in the dense environments
of rich clusters over the past $\sim$ 5 Gyr was the discovery of the
Butcher-Oemler effect (Butcher \& Oemler \ct{but78}, \ct{but84}),
in which higher redshift clusters tend to have a larger fraction of
blue galaxies than those at the present epoch (but see also La Barbera
et al. \ct{lab03}). This purely photometric result was subsequently
confirmed through spectroscopic observations (e.g., Dressler \& Gunn
\ct{dre82}, \ct{dre83}, \ct{dre92}; Lavery \& Henry
\ct{lav86}; Couch \& Sharples \ct{cou87}), which have provided key
information concerning the nature of blue galaxies, showing that
several of them have strong emission line spectra, generally due to
the presence of active star formation. They have also shown the
existence of a class of galaxies, first identified by Dressler \& Gunn
(\ct{dre83}), which exhibit strong Balmer absorption lines without
detectable emissions. Dressler \& Gunn inferred the presence of a
substantial population of A--type stars in these galaxies and
concluded that a fraction of distant blue populations are
``post--starburst'' galaxies with abruptly truncated star
formation. In this scenario, however, the most surprising result is
that many red galaxies exhibit post-starburst spectra, indicative of a
recent enhancement of star formation (e.g. Couch \& Sharples
\ct{cou87}; Dressler \& Gunn \ct{dre92}).

Numerous studies have since been made to understand the evolution of
these galaxies and the connection with the galaxy populations observed
in clusters today (e.g. Couch \& Sharples \ct{cou87}; Barger et
al. \ct{bar96}; Couch et al. \ct{cou94}; Abraham et
al. \ct{abraham}; Morris et al. \ct{mor98}; Poggianti et
al. \ct{pog99}; Balogh et al. \ct{bal99}; Ellingson et
al. \ct{ell01}). The emerging picture is that of a galaxy population
formed early (z $\gtrsim$ 2) in the cluster's history, having
characteristically strong 4000 \AA \ \ breaks and red colours. Then
the clusters grow by infall of field galaxies. This accretion process
causes the truncation of star formation, possibly with an associated
starburst. As this transformation proceeds, these galaxies might be
identified with normal looking spirals, then as galaxies with strong
Balmer absorption spectra, and finally as S0 galaxies, which have
retained some of their disk structure but have ceased active star
formation (Dressler et al. \ct{dressler97}).

On the other hand, there is observational evidence for a strong
connection between galaxy properties and the presence of
substructures, which indicates that cluster merger phenomena are still
ongoing. Caldwell et al. (\ct{cal93}) and Caldwell \& Rose
(\ct{cal97}) found that post-starburst galaxies are located
preferentially near a secondary peak in the X-ray emission of the Coma
cluster and suggested that merging between Coma and a group of
galaxies plays a vital role in triggering a secondary starburst in
galaxies (but see also Nichol, Miller, \& Goto \ct{nic03}).

Numerical simulations have shown that cluster mergers may trigger
starbursts, and can affect greatly star formation histories. Tomita et
al. (\ct{tom96}) argued that in a merging cluster some galaxies may
experience a rapid increase of external pressure, passing through
regions which are overdense in intra-cluster gas, leading to a
starburst. An excess of star--forming galaxies is expected in the
regions between the colliding sub-clusters. As shown by Bekki
(\ct{bek99}), mergers induce a time--dependent tidal gravitational
field that stimulates non--axisymmetric perturbations in disk
galaxies, driving efficient transfer of gas to the central region, and
finally triggering a secondary starburst in the central part of these
galaxies. Roettiger et al. (\ct{roe96}) showed that during
cluster--cluster mergers a bow shock forms on the edges of infalling
subclusters, that protects the gas--rich subcluster galaxies from ram
pressure stripping. This protection fails at core crossing, and
galaxies initiate a burst of star formation.

The above results show that episodes of star formation in cluster
galaxies can be driven both by the accretion of field galaxies into
the cluster and by cluster-cluster merging. In order to disentangle
the two effects it is crucial to relate the star formation history of
cluster galaxies to global properties of clusters, such as mass and
dynamical state.

To examine the effect of environment and dynamics on galaxy
properties, in particular on star formation, we have performed a
spectroscopic investigation of luminous galaxies in the galaxy cluster
ABCG\,209 at z=0.21 (Mercurio et al.  \ct{mer03a} and references
therein) using EMMI-NTT spectra. ABCG\,209 is a rich (richness class
R=3, Abell et al. \ct{abe89}) , X-ray luminous
(L$_{\mathrm{X}}$=(0.1--2.4\,keV)$\sim2.7\times10^{45}\,h^{-2}_{70}\,$erg
s$^{-1}$, Ebeling et al. \ct{ebe96}; T$_{\mathrm{X}}\sim10$ keV,
Rizza et al. \ct{riz98}), and massive cluster
($\mathrm{M(<R_{vir})=2.25^{+0.63}_{-0.65}\times10^{15}}$ \msun,
Mercurio et al.  \ct{mer03a}). It is characterized by the presence
of substructure, allowing the effect of cluster dynamics and evolution
on the properties of its member galaxies to be examined. Evidence in
favour of the cluster undergoing strong dynamical evolution is found
in the form of a velocity gradient along a SE-NW axis, which is the
same preferential direction found from the elongation in the spatial
distribution of galaxies and of the X-ray flux as well as that of the
cD galaxy, and in the presence of substructures. These substructures
are manifest both in the elongation and asymmetry of the X-ray
emission, with two main clumps (Rizza et al. \ct{riz98}) and from
the analysis of the velocity distribution (Mercurio et
al. \ct{mer03a}). Moreover, the young dynamical state of the cluster
is indicated by the possible presence of a radio halo (Giovannini,
Tordi \& Feretti \ct{gio99}), which has been suggested to be the
result of a recent cluster merger, through the acceleration of
relativistic particles by the merger shocks (Feretti \ct{fer02}).

The data are presented in Sect.~\ref{sec:52}, and the spectroscopic
measurement in Sect.~\ref{sec:53}. In Sect.~\ref{sec:54} we discuss the
spectral classification comparing data with models with different
metallicities and star formation histories. We study separately the
properties of disks and spheroids, on the basis of the Sersic index,
in Sect.~\ref{sec:55}. We derive the age distribution of different
spectral classes in Sect.~\ref{sec:56}. The analysis of the velocity
distribution and of possible segregation is presented in
Sect.~\ref{sec:57}. Finally we summarize and discuss the results in
Sect.~\ref{sec:58}.

Throughout the chapter, we use the convention that equivalent widths are
positive for absorption lines and negative for emission lines.

We adopt a flat cosmology with $\mathrm{\Omega_M=0.3}$,
$\mathrm{\Omega_{\Lambda}=0.7}$, and H$_0$ = 70 $h_{70}$ km s$^{-1}$
Mpc$^{-1}$. With this cosmological model, the age of the universe is
13.5 Gyr, and the look-back time at z=0.21 is 2.5 Gyr.

\section{The data}
\label{sec:52}

The observations of the galaxy cluster ABCG\,209 consist of
spectroscopic and photometric data, collected at the ESO New
Technology Telescope (NTT) with the ESO Multi Mode Instrument (EMMI)
in October 2001, and archive Canada-France-Hawaii Telescope (CFHT)
images.

Spectroscopic data have been obtained in the multi--object
spectroscopy (MOS) mode of EMMI. Targets were randomly selected by
using preliminary R--band images (T$\mathrm{_{exp}}$=180 s) to
construct the multislit plates. A total of 112 cluster members were
observed in four fields (field of view $5^\prime \times 8.6^\prime$),
with different position angles on the sky.  We exposed the masks with
integration times from 0.75 to 3 hr with the EMMI--Grism\#2, yielding
a dispersion of $\sim2.8$ \AA/pix and a resolution of $\sim 8$ \AA \ \
FWHM, in the spectral range 385 -- 900 nm.  Reduction procedures used
for the spectroscopic data are described in Mercurio et
al. (\ct{mer03a}). In order to perform the flux calibration, the
spectrophotometric standard star LTT 7987 from Hamuy et
al. (\ct{ham94}) was also observed.

The photometric data were collected on November 1999 (PI. J.-P. Kneib)
using the CFH12K mosaic camera, consisting of B-- and R--band wide
field images, centred on the cluster ABCG\,209 and covering a total
field of view of $42^\prime \times 28^\prime$ (8.6 $\times$ 5.7
$\mathrm{h^{-2}_{70}}$ Mpc$^2$ at the cluster redshift). The CCDs have
a pixel scale of 0.206 arcsec. The total exposure times for both B and
R images were 7200s and the seeing was 1.02$^{\prime\prime}$ in B and
0.73$^{\prime\prime}$ in R.  Reduction procedures, photometric
calibration and catalogue extraction are described in Haines et
al. (\ct{hai03}).

We classified each member galaxy on the basis of the measured
equivalent widths of the atomic features [OII], H$_{\delta A}$, [OIII]
, H$_{\alpha}$ and of the strength of the 4000 \AA \ \ break (see
Sect. \ref{sec:3} and Sect. \ref{sec:4}). Out of 112 cluster members,
we were not able to obtain the spectral classification for 10
galaxies, because the spectra covered a too short wavelength range on
the CCD.  We also studied the properties of bulge-dominated and
disk-dominated galaxies separately, based on the value of the Sersic
index $n$ in the R--band. The Sersic index $n$ was derived in La
Barbera et al. (\ct{lab02}, \ct{lab03}). Table \ref{phot_cat}
presents the photometric properties of the member galaxies.

\begin{table}
        \caption[]{\footnotesize Photometric data. Running number for
         galaxies in the present sample, ID (Col.~1); R-band Kron
         magnitude (Col.~2); B-R colour (Col.~3); heliocentric
         corrected redshift $\mathrm{z}$ (Col.~4); Sersic index $n$
         (Col.~5). Errors in magnitudes and colours are rounded to
         the second decimal place.}
         \label{phot_cat}
{\scriptsize
         $$ 
\hspace{-0.5cm}
           \begin{array}{c c c c c|c c c c c}
            \hline
            \noalign{\smallskip}
            \hline
            \noalign{\smallskip}
\mathrm{ID} & \mathrm{R} & 
\mathrm{B-R} & \mathrm{z} & n&\mathrm{ID} & \mathrm{R} & 
\mathrm{B-R} & \mathrm{z} & n\\
            \hline
            \noalign{\smallskip}   
  1& 17.77\pm0.01& 1.69\pm0.01& 0.2191\pm0.0003& 2.8& 39& 18.54\pm0.01& 2.39\pm0.01& 0.2161\pm0.0002& 1.9\\
  2& 18.49\pm0.01& 2.40\pm0.01& 0.2075\pm0.0002& 2.3& 40& 19.64\pm0.01& 2.21\pm0.02& 0.1979\pm0.0005& 2.3\\ 
  3& 18.54\pm0.01& 2.37\pm0.01& 0.1998\pm0.0003& 2.8& 41& 18.88\pm0.01& 2.21\pm0.01& 0.2039\pm0.0003& 3.3\\ 
  4& 18.99\pm0.01& 1.82\pm0.01& 0.2000\pm0.0004& 3.6& 42& 19.76\pm0.01& 2.29\pm0.02& 0.2133\pm0.0004& 2.7\\ 
  5& 19.42\pm0.01& 2.30\pm0.02& 0.2145\pm0.0005& 1.6& 43& 17.57\pm0.01& 1.55\pm0.01& 0.2140\pm0.0002& 1.4\\ 
  6& 16.62\pm0.01& 2.43\pm0.01& 0.2068\pm0.0004& 3.2& 44& 18.94\pm0.01& 2.42\pm0.01& 0.2061\pm0.0002& 4.0\\ 
  7& 17.74\pm0.01& 2.49\pm0.01& 0.2087\pm0.0002& 3.1& 45& 19.41\pm0.01& 2.05\pm0.02& 0.2123\pm0.0003& 3.0\\ 
  8& 18.21\pm0.01& 2.37\pm0.01& 0.2073\pm0.0002& 8.4& 46& 17.28\pm0.01& 2.43\pm0.01& 0.2064\pm0.0004& 5.6\\ 
  9& 19.28\pm0.01& 2.41\pm0.01& 0.2090\pm0.0002& 4.5& 47& 18.03\pm0.01& 2.43\pm0.01& 0.2078\pm0.0002& 3.4\\ 
 10& 19.07\pm0.01& 2.33\pm0.01& 0.2072\pm0.0004& 2.4& 48& 19.06\pm0.01& 2.27\pm0.02& 0.2042\pm0.0003& 3.4\\ 
 11& 18.95\pm0.01& 2.38\pm0.01& 0.2039\pm0.0002& 4.6& 49& 17.55\pm0.01& 2.31\pm0.01& 0.2183\pm0.0002& 6.3\\ 
 12& 19.17\pm0.01& 2.37\pm0.01& 0.2084\pm0.0003& 4.2& 50& 17.49\pm0.01& 2.40\pm0.01& 0.2068\pm0.0002& 3.7\\ 
 13& 17.90\pm0.01& 2.36\pm0.01& 0.2134\pm0.0003& 4.2& 51& 17.55\pm0.01& 2.16\pm0.01& 0.2001\pm0.0002& 4.0\\ 
 14& 19.06\pm0.01& 2.48\pm0.01& 0.2073\pm0.0002& 1.8& 52& 18.46\pm0.01& 2.37\pm0.01& 0.2184\pm0.0002& 5.4\\ 
 15& 19.40\pm0.01& 2.20\pm0.02& 0.2084\pm0.0003& 3.1& 53& 19.44\pm0.01& 2.33\pm0.02& 0.2028\pm0.0005& 3.8\\ 
 16& 18.85\pm0.01& 2.32\pm0.01& 0.2120\pm0.0002& 3.9& 54& 17.19\pm0.01& 2.35\pm0.01& 0.2024\pm0.0002& 5.3\\ 
 17& 18.35\pm0.01& 1.66\pm0.01& 0.2073\pm0.0002& 1.5& 55& 16.41\pm0.01& 2.54\pm0.01& 0.2097\pm0.0002& 6.6\\ 
 18& 18.67\pm0.01& 2.33\pm0.01& 0.2096\pm0.0003& 4.6& 56& 18.46\pm0.01& 2.28\pm0.01& 0.2094\pm0.0002& 2.9\\ 
 19& 18.46\pm0.01& 2.18\pm0.01& 0.1995\pm0.0002& 5.2& 57& 19.00\pm0.01& 1.85\pm0.01& 0.2170\pm0.0003& 1.4\\ 
 20& 18.61\pm0.01& 2.19\pm0.01& 0.2052\pm0.0003& 1.6& 58& 18.11\pm0.01& 2.42\pm0.01& 0.2085\pm0.0002& 2.5\\ 
 21& 17.83\pm0.01& 2.33\pm0.01& 0.2051\pm0.0002& 3.0& 59& 19.29\pm0.01& 2.23\pm0.01& 0.2083\pm0.0004& 4.8\\ 
 22& 18.53\pm0.01& 1.66\pm0.01& 0.2181\pm0.0008& 3.2& 60& 18.58\pm0.01& 2.44\pm0.01& 0.2118\pm0.0003& 7.3\\ 
 23& 18.60\pm0.01& 2.28\pm0.01& 0.2004\pm0.0002& 5.6& 61& 18.40\pm0.01& 2.41\pm0.01& 0.2150\pm0.0002& 8.4\\ 
 24& 18.81\pm0.01& 1.58\pm0.01& 0.2063\pm0.0003& 2.3& 62& 18.82\pm0.01& 0.95\pm0.01& 0.2188\pm0.0001& 1.2\\ 
 25& 18.28\pm0.01& 2.28\pm0.01& 0.2051\pm0.0004& 7.7& 63& 18.34\pm0.01& 2.35\pm0.01& 0.2144\pm0.0002& 4.8\\ 
 26& 18.95\pm0.01& 2.30\pm0.01& 0.2034\pm0.0002& 5.0& 64& 17.66\pm0.01& 1.43\pm0.01& 0.1999\pm0.0005& 0.8\\ 
 27& 20.00\pm0.01& 2.34\pm0.02& 0.2076\pm0.0003& 7.9& 65& 18.10\pm0.01& 2.11\pm0.01& 0.2098\pm0.0003& 1.4\\ 
 28& 19.66\pm0.01& 2.16\pm0.02& 0.2064\pm0.0002& 4.0& 66& 19.28\pm0.01& 2.35\pm0.02& 0.2117\pm0.0002& 4.8\\ 
 29& 19.32\pm0.01& 2.02\pm0.01& 0.2027\pm0.0003& 0.8& 67& 19.40\pm0.01& 2.29\pm0.02& 0.2098\pm0.0004& 4.0\\ 
 30& 17.64\pm0.01& 2.25\pm0.01& 0.2060\pm0.0002& 3.6& 68& 17.46\pm0.01& 2.44\pm0.01& 0.2102\pm0.0002& 4.9\\ 
 31& 19.11\pm0.01& 2.36\pm0.01& 0.2087\pm0.0004& 3.1& 69& 18.76\pm0.01& 2.43\pm0.01& 0.2107\pm0.0003& 5.1\\ 
 32& 18.07\pm0.01& 2.37\pm0.01& 0.2066\pm0.0002& 2.8& 70& 18.33\pm0.01& 2.38\pm0.01& 0.1973\pm0.0003& 2.8\\ 
 33& 17.75\pm0.01& 2.43\pm0.01& 0.2084\pm0.0002& 2.3& 71& 18.34\pm0.01& 2.39\pm0.01& 0.2056\pm0.0002& 2.9\\ 
 34& 19.53\pm0.01& 2.23\pm0.02& 0.2105\pm0.0003& 4.1& 72& 19.61\pm0.01& 2.20\pm0.02& 0.2093\pm0.0004& 7.7\\ 
 35& 18.24\pm0.01& 2.32\pm0.01& 0.2167\pm0.0002& 4.1& 73& 18.17\pm0.01& 2.37\pm0.01& 0.2104\pm0.0002& 8.4\\ 
 36& 18.88\pm0.01& 2.24\pm0.01& 0.2038\pm0.0004& 3.9&74& 19.30\pm0.01& 2.30\pm0.01& 0.2056\pm0.0002& 2.3 \\ 
 37& 17.34\pm0.01& 2.47\pm0.01& 0.2125\pm0.0003& 6.5&75& 18.87\pm0.01& 2.32\pm0.01& 0.2172\pm0.0003& 5.5\\  
 38& 17.76\pm0.01& 2.47\pm0.01& 0.2097\pm0.0003& 3.9&76& 17.47\pm0.01& 2.45\pm0.01& 0.2069\pm0.0002& 3.0\\ 
        \noalign{\smallskip}	     		    
            \hline			    		    
         \end{array}
     $$ 
}
         \end{table}

\addtocounter{table}{-1}
\begin{table}
          \caption[ ]{\footnotesize Continued.}
{\scriptsize
         $$ 
\hspace{-0.5cm}
           \begin{array}{c c c c c|c c c c c}
            \hline
            \noalign{\smallskip}
            \hline
            \noalign{\smallskip}
\mathrm{ID} & \mathrm{R} & 
\mathrm{B-R} & \mathrm{z} & n&\mathrm{ID} & \mathrm{R} & 
\mathrm{B-R} & \mathrm{z} & n\\

77& 17.75\pm0.01& 2.38\pm0.01& 0.2083\pm0.0003& 5.0&90& 19.16\pm0.01& 2.23\pm0.01& 0.1997\pm0.0002& 2.2\\
78& 19.49\pm0.01& 2.16\pm0.02& 0.2131\pm0.0004& 2.2&91& 18.74\pm0.01& 2.39\pm0.01& 0.2161\pm0.0002& 2.9\\   
79& 18.04\pm0.01& 1.91\pm0.01& 0.1984\pm0.0002& 1.9&92& 19.10\pm0.01& 1.72\pm0.01& 0.2134\pm0.0004& 0.7\\   
80& 18.40\pm0.01& 2.10\pm0.01& 0.2091\pm0.0003& 2.6&93& 17.21\pm0.01& 2.44\pm0.01& 0.2175\pm0.0002& 4.1\\   
81& 19.31\pm0.01& 1.76\pm0.01& 0.1982\pm0.0004& 0.8&94& 18.70\pm0.01& 2.03\pm0.01& 0.2145\pm0.0002& 1.4\\   
82& 18.71\pm0.01& 2.46\pm0.01& 0.2110\pm0.0002& 3.2&95& 17.96\pm0.01& 2.37\pm0.01& 0.2158\pm0.0003& 2.9\\   
83& 19.19\pm0.01& 2.21\pm0.02& 0.2102\pm0.0005& 5.6&96& 18.15\pm0.01& 2.40\pm0.01& 0.2149\pm0.0002& 2.8\\   
84& 19.04\pm0.01& 1.74\pm0.01& 0.1971\pm0.0003& 1.1&97& 19.98\pm0.01& 2.05\pm0.02& 0.2124\pm0.0004& 2.7\\   
85& 19.13\pm0.01& 2.32\pm0.01& 0.2137\pm0.0003& 2.0&98& 18.82\pm0.01& 2.39\pm0.01& 0.2203\pm0.0003& 2.8\\   
86& 17.20\pm0.01& 2.44\pm0.01& 0.2125\pm0.0002& 9.2&99& 18.19\pm0.01& 2.24\pm0.01& 0.2034\pm0.0003& 3.7\\   
87& 17.56\pm0.01& 2.47\pm0.01& 0.2061\pm0.0002& 5.0&100& 20.11\pm0.01& 1.98\pm0.02& 0.2138\pm0.0004& 1.2\\  
88& 19.61\pm0.01& 2.22\pm0.02& 0.2087\pm0.0003& 2.7&101& 19.88\pm0.01& 2.23\pm0.02& 0.2117\pm0.0004& 3.2\\  
89& 18.56\pm0.01& 2.33\pm0.01& 0.2074\pm0.0002& 7.5&102& 19.46\pm0.01& 1.28\pm0.01& 0.1963\pm0.0006& 1.0\\
        \noalign{\smallskip}	     		    
            \hline			    		    
        \noalign{\smallskip}	     		    
            \hline			    		    
         \end{array}
     $$ 
}
         \end{table}
\section{Derivation of line indices}
\label{sec:53}

In order to measure line indices, galaxy spectra were corrected for
galactic extinction following Schlegel et al. (\ct{sch98}),
flux--calibrated and wavelength--calibrated by using the
IRAF~\footnote{IRAF is distributed by the National Optical Astronomy
Observatories, which are operated by the Association of Universities
for Research in Astronomy, Inc., under cooperative agreement with the
National Science Foundation.} package ONEDSPEC.

The flux calibration was performed using observations of the
spectrophotometric standard LTT 7987, whose spectrum was acquired
before and after each scientific exposure.  The sensitivity function,
derived by using the tasks STANDARD and SENSFUNC (rms $\sim$ 0.03),
was applied to the galaxy spectra by using the task CALIBRATE. Within
this task, the exposures are corrected for the atmospheric extinction,
divided by the exposure time, and finally transformed using the
sensitivity curve. The calibrated spectra were then corrected for the
measured velocity dispersion by using DISPCOR.

In order to asses the quality of the flux calibration, we compared the
V-R colours as derived from photometry (Mercurio et al. \ct{mer03b})
with those measured from the spectra of 41 galaxies for which the
observed wavelength range is greater than 4750--8750 \AA. The
resulting mean difference in colours turns out to be $\Delta$(V-R) =
0.06$\pm$0.08, proving that the derived fluxes are correct.


We measured the equivalent widths of the following atomic and
molecular features: H$_{\delta A}$, H$_{\gamma A}$, Fe4531, H$_\beta$,
Fe5015, Mg$_1$, Mg$_2$, Mg$b$, Fe5270, Fe5335. We adopted the
definition of the extended Lick system (Worthey et al. \ct{wor94},
Worthey \& Ottaviani \ct{wor97}; Trager et al. \ct{tra98}), where
the spectral indices are determined in terms of a central feature
bandpass bracketed by two pseudo-continuum bandpasses at a resolution
of $\sim$9 \AA \ \ FWHM, that is similar to
ours\footnote{\footnotesize The difference of 1 \AA \ \ in resolution
between our data and the extended Lick standard results in a
difference in line indices amounting to about 1\% of the uncertainty
of the measurements.}.  Following the convention, atomic indices are
expressed in Angstroms of equivalent width, while molecular indices
are expressed in magnitudes.

We also measured the equivalent widths for the emission lines [OII],
[OIII] and H$_{\alpha}$ (see Table \ref{indices} for the definition of
wavelength ranges), as well as the strength of the 4000 \AA \ \ break
D$_n$(4000). We adopted the definition of D$_n$(4000) by Balogh et
al. (\ct{bal99}) as the ratio of the average flux in the narrow
bands 3850--3950 and 4000--4100 \AA. The original definition of this
index by Bruzual (\ct{bru83}) uses wider bands
(3750--3950 and 4050--4250 \AA), and hence it is more sensitive to
reddening effects.

\begin{table}
     \caption[]{Wavelength ranges for equivalent widths of emission lines, 
                expressed in Angstrom.} 
    \label{indices}
    $$
           \begin{array}{c c c c}
            \hline
            \noalign{\smallskip}
    \mathrm{Name} & \mathrm{Index \ bandpass} & \mathrm{Blue \ continuum} & \mathrm{Red \ continuum}\\
            \noalign{\smallskip}
            \hline
            \noalign{\smallskip}
\mathrm{[OII]}      & 3713-3741 & 3653-3713 & 3741-3801\\
\mathrm{[OIII]}     & 4997-5017 & 4872-4932 & 5050-5120\\
\mathrm{H_{\alpha}} & 6555-6575 & 6510-6540 & 6590-6620\\
            \noalign{\smallskip}
            \hline
         \end{array}
     $$
   \end{table}

The derivation of equivalent widths and the catalogue of spectroscopic
measurements are presented in Appendix A.

\section{Spectral classification}
\label{sec:54}

\begin{figure*}
  \centering 
\includegraphics[width=0.7\textwidth]{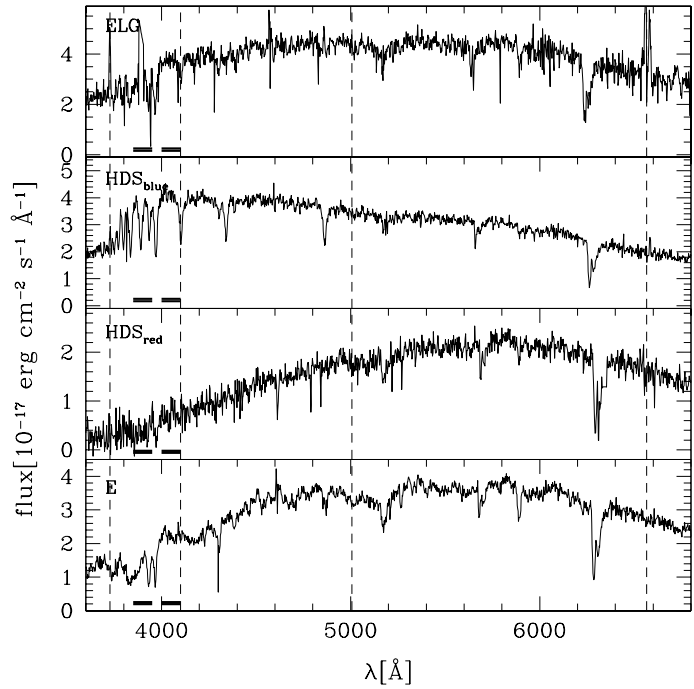} 
\caption{Rest-frame spectra of galaxies belonging to different
spectral classes. Lines OII, H$_{\delta}$, OIII, H$_{\alpha}$ are
highlighted by dashed lines. Wavelength ranges adopted to measure
D$_n$(4000) are also indicated by black lines.}
  \label{spectra}
\end{figure*}

Useful tool to classify galaxies are the diagram
D$_n$(4000)-[H$_\delta$] (or (B-R)-[H$_\delta$]; e.g. Couch \&
Sharples \ct{cou87}, Barger et al. \ct{bar96}) and the emission
line measurements (e.g. [OII]). In this way galaxies can be divided in
four classes: emission line (ELG), blue H$_{\delta}$ strong
(HDS$_{\mathrm{blue}}$), red H$_{\delta}$ strong
(HDS$_{\mathrm{red}}$) and passive (E) galaxies (see Fig.~\ref{spectra}).

We remark that there is not a unique criterion in literature for the
definition of spectral classes of galaxies, mainly because there is no
general agreement on the definition of indices, and because the
spectra vary in terms of S/N and resolution. Moreover, the comparison
of spectral indices among different works is not straightforward since
line indices are sensitive to the definition of the bands and to the
method used to measure their strengths. Therefore, we derived spectral
indices for both data and models with the same procedures and we used
models themselves to define the spectral classes (see below).

\subsection{Galaxy population models}

To investigate the nature of galaxy populations in ABCG\,209, we used
the code GISSEL03 (Bruzual \& Charlot \ct{bru03}), by adopting a
Salpeter (\ct{sal55}) initial mass function (IMF), with a stellar
mass range from 0.1 to 100 $\mathrm{M_{\odot}}$. Results vary only
slightly choosing different IMFs (but maintaining the same mass
range).

The GISSEL03 code allows the spectral evolution of stellar populations
to be computed at a resolution of 3 \AA (FWHM), with a sampling of 1
\AA \ \ across the wavelength range 3200-9500 \AA. Differently from
previous models (Bruzual \& Charlot \ct{bruzual}), this code provides
galaxy spectra with different metallicities, in the range
Z=(0.005--2.5)Z$_\odot$, which are required to break the
age-metallicity degeneracy, through a comparison of age and
metallicity sensitive indices (see Bruzual \& Charlot \ct{bru03} for
details).

Models were first degraded and resampled in order to match data and
then D$_n$(4000) and rest-frame equivalent widths were computed in the
same manner as for observations. The free parameters in our analysis
are the star formation rate and the age of the galaxies, while the
adopted metallicities were Z=0.4Z$_\odot$, Z$_\odot$, and
2.5Z$_\odot$.

\subsection{Emission line galaxies}

We define as emission line those galaxies showing in the spectra the
emission lines [OII], [OIII] and H$_{\alpha}$. We decided to use
simultaneously these three lines and not only [OII], because i) we
lack information on [OII] for more than half of spectra (because of
the available wavelength range), ii) [OII] is heavily obscured by
dust, and iii) [OII] is found at wavelengths where the noise is
higher.

This class contains star-forming (SF) and short starburst (SSB)
galaxies (Balogh et al. \ct{bal99}). We interpret these as systems
undergoing a star formation, that can be described by a model with
exponentially decaying SFR model with decay parameter
$\mu$=0.01\footnote{\footnotesize $\mu$=1-exp(1.0 Gyr/$\tau$)} (Barger
at al. \ct{bar96}). SSB galaxies, on the contrary, have experienced
a large increase of star formation over a short time, and can be
described by models with an initial burst of 100 Myr.

We detect emission lines in $\sim$ 7\% of cluster members.

\subsection{H$_{\delta}$-strong (HDS) galaxies}

\begin{figure*}
  \centering 
\includegraphics[width=0.7\textwidth,bb= 10 330 377 698,clip]{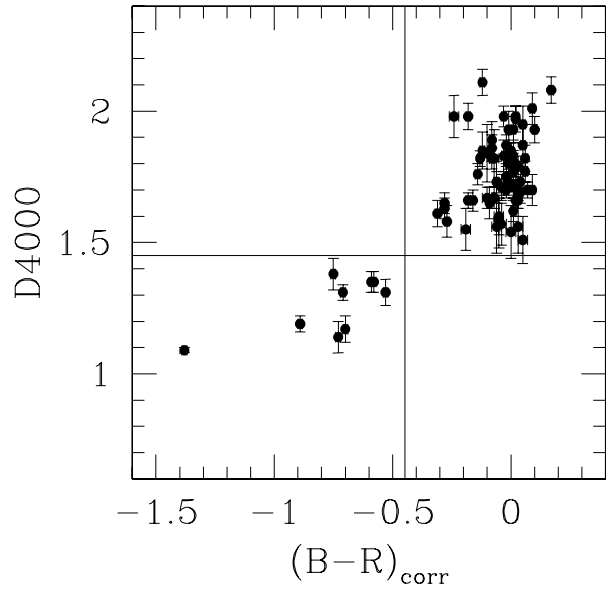} 
\vspace{-1.0cm} 
\caption{Observed distribution of D$_{n}$(4000) versus
(B--R)$_{\mathrm{corr}}$ (see text) for 76 member galaxies for which
we can measure D$_{n}$(4000). The vertical line indicates the
separation between blue and red galaxies as defined in Haines et
al. (\ct{hai03}) and the horizontal line is the cut for the break at
4000 \AA, adopted by Balogh et al. (\ct{bal99}).}
  \label{BR_D4000}
\end{figure*}

In order to investigate the physical properties of HDS galaxies, we
divided the sample into blue and red galaxies, according either to the
break at 4000 \AA \ \ or, when D$_n$(4000) cannot be measured, to B-R
colours, (B-R)$_{\mathrm{corr}}$, corrected for the colour-magnitude
(CM) relation\footnote{\footnotesize (B-R)$_{\mathrm{corr}}$=
(B-R)-(3.867 - 0.0815$\cdot$R)}.

We define as ``blue'' the galaxies with rest-frame B-V colours at
least 0.2\,mag bluer than that of the CM relation, as for the original
studies of Butcher \& Oemler (\ct{but84}). We estimate the
corresponding change in B-R colour at the cluster redshift by
firstly considering two model galaxies at $z=0$ (Bruzual \& Charlot
\ct{bru03}), one chosen to be a typical early-type galaxy 10\,Gyr
old with solar metallicity, and the second reduced in age until its
B-V colour becomes 0.2\,magnitude bluer. After moving both galaxies
to the cluster redshift, the difference in their observed B-R colour
is found to be 0.447\,mag (see Haines et al. \ct{hai03} for
details).  Thus we define as ``blue'' the galaxies with
(B-R)$_{\mathrm{corr}}$ $<$ -0.447.

The relation between D$_{n}$(4000) and (B-R)$_{\mathrm{corr}}$
(Fig.~\ref{BR_D4000}) shows that there is a clear separation between
blue and red galaxies and that we can adopt D$_{n}$(4000) = 1.45 as a
cut for the measurement of 4000 \AA \ \ break. This value corresponds
to that adopted by Balogh et al. (\ct{bal99}), for the
classification of galaxies in the CNOC sample.

   \begin{figure*}
   \centering
   \includegraphics[width=1.0\textwidth,bb= 3 400 577 696,clip]{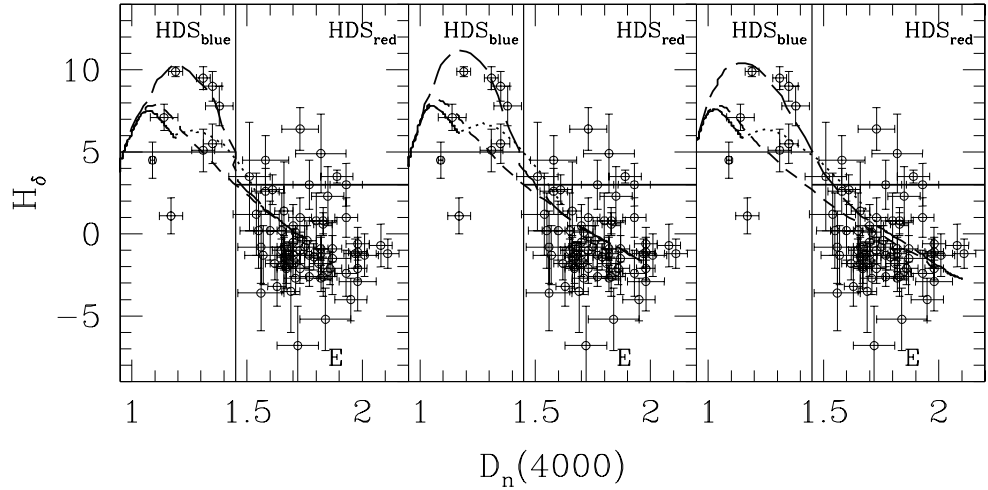}
    \caption{Equivalent width of H$_{\delta}$ versus D$_n$(4000) \AA \
      \ for 76 member galaxies compared to different GISSEL03
      models. Left panel shows models with Z=0.4Z$_\odot$, while the
      central and right panels refer to Z$_\odot$ and 2.5Z$_\odot$,
      respectively. In each panel, the short dashed curves represent a
      model with exponential declining SFR with $\tau$=1.0 Gyr, while
      for the continuous and dotted lines the decay parameter is
      $\mu$=0.01. For the last one, the star formation is truncated at
      t=10 Gyr. The long dashed lines represent a model with an initial
      100 Myr burst.}
   \label{hd_D4000}
   \end{figure*}

In Fig.~\ref{hd_D4000} we plot, in the D$_n$(4000)--H$_\delta$ plane,
the data (open circles) of 76 galaxies, for which we have information
on both D$_n$(4000) and H$_\delta$, with superimposed the output of
four different models: i) a model with an initial burst of star
formation lasting 100 Myr (long dashed line); ii) a model with an
exponentially decaying SFR with decay parameter $\mu$=0.01 (continuous
line) lasting 11 Gyr and iii) truncated at 10 Gyr (dotted line); iv) a
model with an exponentially decaying SFR with time scale $\tau$ =1.0
Gyr (short dashed line).

For blue galaxies, to reach [H$_\delta$] $>$ 5.0 \AA, the galaxy light
must be dominated by A-- and early F--type stars, whereas the presence
of O-- and B--type stars, which have weak intrinsic H$_\delta$
absorption, reduces this equivalent width. Moreover, the equivalent
width of H$_\delta$ is inversely related to the duration of star
formation. For this reason, the blue, H$_\delta$--strong galaxies are
expected to be the result of a short starburst or of a recently
terminated star formation (e.g. Dressler \& Gunn \ct{dre83}; Barger
et al. \ct{bar96}). By using these models when D$_n$(4000) is equal
to 1.45, the H$_\delta$ equivalent width is 5.0 \AA, so we decided to
use [H$_\delta$] $>$ 5.0 \AA \ \ as the selection criterion for blue
HDS.

On the contrary, normal early-type galaxies in the red half of the
plane generally show little or no signs of star formation, and can be
reproduced by a model with an exponentially decaying SFR with time
scale $\tau$ =1.0 Gyr. In this case, the model with D$_n$(4000) $>$
1.45 has [H$_\delta$] $<$ 3.0 \AA. Therefore, for red HDS galaxies we
adopted a threshold of 3 \AA. Note that these selection criteria are
equivalent to those adopted by Balogh et al. (\ct{bal99}), derived
by using PEGASE models. None of the GISSEL03 models used here
describes well red HDS galaxies and require the introduction of other
ingredients, as discussed in Sect.~\ref{sec:6}.

5\% of cluster galaxies are classified as blue HDS galaxies and 7\% as
red HDS galaxies.

\subsection{Passive galaxies}

Passive galaxies are red with no emission lines and [H$\delta$] $<$
3.0 \AA.  These are galaxies without significant star formation, are
mostly ellipticals or S0s and can be modelled by an exponential star
formation rate with $\tau$ = 1.0 Gyr.

81\% of member galaxies are classified as E galaxies.

\section{Comparison with structural properties}
\label{sec:55}

We analysed the spectral properties of disk and spheroidal galaxies,
separately (see La Barbera et al. \ct{lab03}). We define as
spheroids the objects with the Sersic index $n>2$, and the remaining
as disks. This corresponds to distinguishing between galaxies with a
low bulge fraction ($<$20\%) and those with a more prominent bulge
component (van Dokkum et al. \ct{van98}).

   \begin{figure*}
   \centering
   \includegraphics[width=0.7\textwidth,bb= 17 336 371 705,clip]{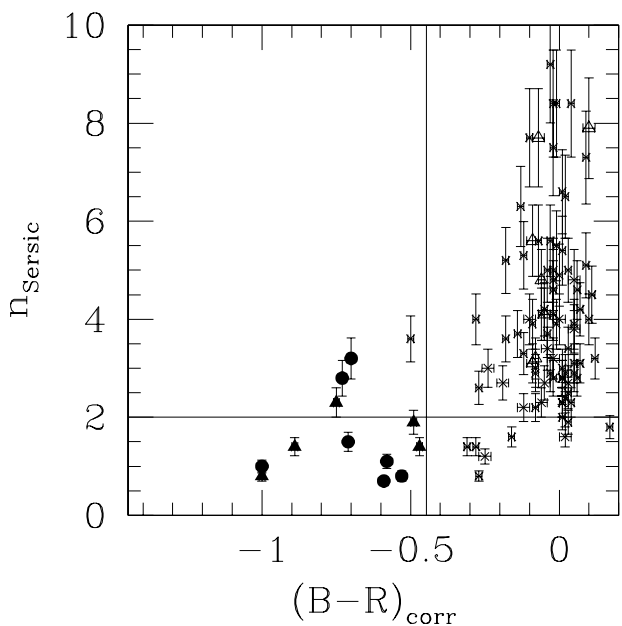}
	\vspace{-1.0cm}
    \caption{Comparison between B--R galaxy colours and Sersic indices
    (n$_{\mathrm{Sersic}}$). The vertical line marks the separation
    between blue and red galaxies (see text), while the horizontal line
    is the separation between disks and spheroids. Filled circles and
    triangles indicates ELG and HDS blue galaxies, respectively, while
    crosses and open triangles are for E and HDS red galaxies.}
   \label{BR_nsersic}
   \end{figure*}

\begin{figure*}
\centering
\includegraphics[width=0.8\textwidth]{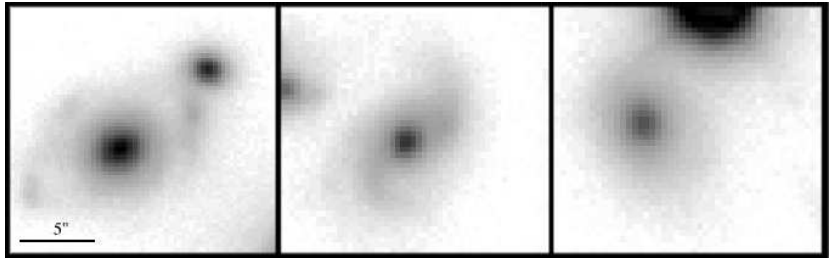}
\caption
{R--band image of the four galaxies discussed in the text, with North
at top and East to left. Scale is indicated in the left panel.}
\label{galaxies}
\end{figure*}

In Fig.~\ref{BR_nsersic} the Sersic index is plotted against the B-R
colour corrected for the effect of the CM relation, and different
symbols are used to distinguish the different galaxy spectral types.
This plot shows that there is a marked correspondence between spectral
and structural properties: blue disks are ELG or HDS$_{\mathrm{blue}}$
galaxies while red spheroids are E or HDS$_{\mathrm{red}}$ galaxies.
The converse is also true with some exceptions. Noticeably, one
HDS$_{\mathrm{blue}}$ and two ELG galaxies have $2<n<4$, and one E
galaxy has blue colour. By a visual inspection of R--band images, the
two ELGs show signs of spiral arms (Fig.~\ref{galaxies} left and
central panels). We note that one of those is also close to the E
galaxy with blue colour (see Fig.~\ref{BR_nsersic} and
Fig.~\ref{galaxies} left panel). Since the HDS$_{\mathrm{blue}}$
galaxy is close to a bright star (Fig.~\ref{galaxies} right panel), it
could be missclassified.

For the eight red E galaxies having Sersic index less than 2 we
suggest that they are a population of disk galaxies. van den Berg,
Pierce \& Tully (\ct{van90}), classifying 182 galaxies in the Virgo
cluster, have shown the existence of ``anemic'' spiral galaxies,
i.e. spiral galaxies deficient in neutral hydrogen. They have shown
that these observations may be accounted for by assuming that gas has
been stripped from the outer parts of these objects. Following van den
Berg \ct{van91} we recognize these galaxies as Ab-spirals and
exclude them from the sample of ellipticals.

HDS$_{\mathrm{red}}$ galaxies have all Sersic indices greater than 3,
so they seem to be early-type galaxies.

\section{Distribution of ages}
\label{sec:56}

In order to follow the evolution of galaxies belonging to different
spectral classes, we compare the strengths of various spectral indices
in the observed galaxy spectra to stellar population models with
different star formation rates and metallicities.

Bruzual \& Charlot (\ct{bru03}) defined those indices that are most
suitable in their models to investigate galaxy populations. They
suggested two sets of spectral indices that should be fitted
simultaneously in order to break the age-metallicity degeneracy: one
of them being primary sensitive to the star formation history
(D$_n$(4000), H$_\beta$, H$_{\gamma A}$ and H$_{\delta A}$) and the
other primary sensitive to metallicity ([MgFe]$^\prime$,[Mg$_1$Fe] and
[Mg$_2$Fe]). The latter indices, in fact, are sensitive to metallicity
but do not depend sensitively on changes in alpha-element to iron
abundance ratios (see Thomas et al. \ct{tho03} and Bruzual \&
Charlot \ct{bru03} for details).

For these reasons we use models with different star formation
histories in order to fit simultaneously the observed strength of
D$_n$(4000), H$_{\delta A}$, H$_{\gamma A}$, H$_\beta$,
[MgFe]$^\prime$, [Mg$_1$Fe], and [Mg$_2$Fe].

   \begin{figure*}
   \centering
   \includegraphics[width=1.0\textwidth,bb= 3 400 577 696,clip]{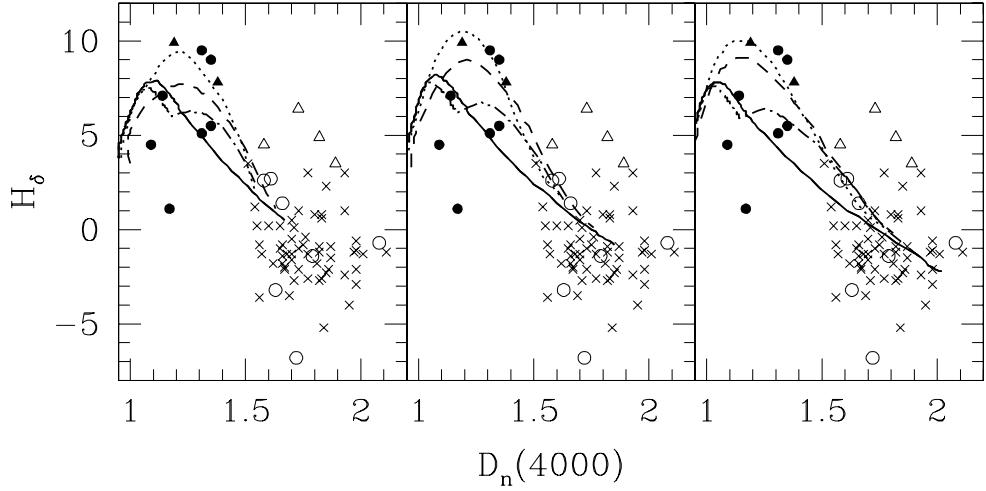}
    \caption{Equivalent width of H$_{\delta}$ versus the break at 4000
      \AA \ \ for galaxies of different spectral types and GISSEL03
      models. Filled and open circles, filled and open triangles, and
      crosses denote ELG, Ab--spiral, HDS$_{\mathrm{blue}}$,
      HDS$_{\mathrm{red}}$ and E galaxies, respectively. The left
      panel shows models with Z=0.4Z$_\odot$, central and right panels
      1Z$_\odot$ and 2.5Z$_\odot$, respectively. In each panel it is
      plotted a model with exponential declining SFR with time scale
      $\tau$=1.0 Gyr lasting 11 Gyr (continuous line), with
      superimposed a burst at t=9 Gyr, involving 10\% (dashed line) or
      40\% (dotted line) of the total mass, and a model with
      continuous star formation truncated at t=10 Gyr (dot-dashed
      line).}
   \label{hd_D4000_bis}
   \end{figure*}

\begin{figure*}
   \centering
   \includegraphics[width=0.5\textwidth,bb= 7 428 304 710,clip]{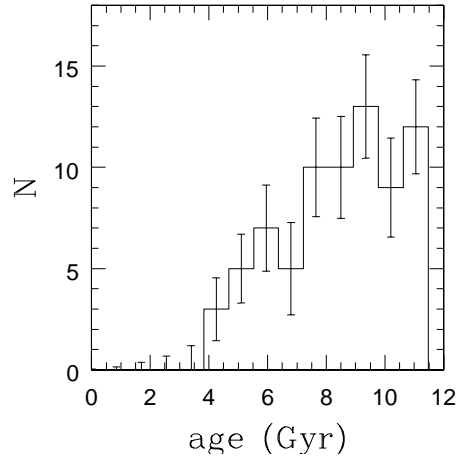}
    \caption{Distribution of best-fitted ages of E galaxies
      (continuous line), by using a model with exponential declining
      SFR with time scale $\tau$=1.0 Gyr. Errors on ages are obtained
      performing 1000 simulations accounting for
      equivalent width measurement errors.}
   \label{eta_tau1}
   \end{figure*}

E-galaxies (crosses in Fig.\ref{hd_D4000_bis}) exhibit moderate
dispersion in D$_n$(4000), caused by a combination of age and
metallicity effects and little star formation. They are well described
by an exponentially declining star formation with time scale
$\tau$=1.0 Gyr (Merluzzi et al. \ct{merl03} and references
therein). Assuming this SFR and metallicities Z=0.4
Z$_\odot$,Z$_\odot$,2.5 Z$_\odot$, the distribution of ages for
early-type galaxies, (fig. \ref{eta_tau1}), shows that the age of the
bulk of ellipticals in ABCG\,209 is greater than $\sim$ 7 Gyr and that
there is a tail toward younger ages up to $\sim$ 4 Gry. This
translates into a formation epoch z$_f \gtrsim$ 1.6 for the majority
of ellipticals and z$_f \gtrsim$ 0.7 for the younger population.  The
KMM algorithm (cf. Ashman et al. \ct{ash94} and refs. therein)
suggests that a mixture of two Gaussians (with $\mathrm{n_1=19}$, and
$\mathrm{n_2=55}$ members) is a better description of this age
distribution (although only at $\sim 90\%$ c.l.).  In this case, the
mean ages of the two distributions are $\mathrm{t_1 = 5.8 \pm 0.8}$
Gyr, and $\mathrm{t_{2} = 9.2 \pm 1.2}$ Gyr respectively. Thus, 74.3\%
of early-type galaxies seem to be coeval and to be formed early during
the initial collapse of the cluster, at z$_f \gtrsim$ 3.5. The
remaining 25.7\% is constituted by a younger galaxy population, that
could be formed later (z$_f \gtrsim$ 1.2) or could have experienced an
enhancement in the star formation rate $\sim$ 8.5 Gyr ago. This is in
remarkable agreement with the trend with redshift of the total stellar
mass in the cluster galaxies, derived from the colour-magnitude
relation by Merluzzi et al. \ct{merl03} (see their Fig. 10).

As shown in Fig.~\ref{hd_D4000_bis} (filled circles), ELG galaxies can
be fitted either by a model with truncated star formation history and
by a model with a burst involving the 40\% of the total mass
(Fig.~\ref{hd_D4000_bis}), whereas the two HDS$_{\mathrm{blue}}$
galaxies, for which we have the measurement of both D$_n$(4000) and
[H$_\delta$], seem to be reproduced only by a model with a short
starburst. The starburst is assumed to begin at t=9 Gyr lasting for
0.1 Gyr. An equivalent width of H$_\delta$ greater than 5 \AA \ \
implies that a starburst occurred in the galaxy before star formation
was quenched, otherwise quiescent star formation activity would
produce a spectrum with weaker Balmer lines (e.g., Couch \& Sharples
\ct{cou87}; Poggianti \& Barbaro \ct{pog96}), as showed also by
the comparison of the two models (dotted and dot- dashed lines) in
Fig. \ref{hd_D4000_bis}. HDS$_{\mathrm{blue}}$ galaxies have EWs
stronger than 5 \AA, and thus must be post-starburst galaxies. The
strength of the lines and the colour of these galaxies indicate an
interruption of the star formation within the last 0.5 Gyr, with a
strong starburst preceding the quenching of star formation (see also
Poggianti et al. \ct{pog99}; Poggianti \& Barbaro \ct{pog96}).

In contrast, the typical time elapsed since the last star formation in
the HDS$_{\mathrm{red}}$ will be in the range 1-2 Gyr, since this
population exhibits weaker H$_{\delta}$ equivalent widths. However,
the observed [H$_{\delta}$] values for HDS$_{\mathrm{red}}$ tend to be
larger than those predicted by both starburst and truncated star
formation models, except from one galaxy that is well reproduced by a
burst model (observed $\sim$ 1 Gyr after the burst). This problem was
pointed out by several authors (e.g. Couch \& Sharples \ct{cou87},
Morris et al. \ct{mor98}, Poggianti \& Barbaro \ct{pog96}, Balogh
et al. \ct{bal99}) and will persist if photometric colours are
considered instead of D$_n$(4000).

Models with IMFs biased toward massive stars are able to reproduce
better these data. However, the stellar population resulting from such
a burst is very short lived ($\sim$ 300 Myr), thus it is unlikely that
many of these red galaxies are undergoing such bursts.  Stronger
[H$_{\delta}$] may also be produced by temporal variations in the
internal reddening in truncated or burst models. Balogh et
al. (\ct{bal99}) have shown that by reddening the model it is
possible to recover the observed data. This supports the idea that
dust obscuration may play an important role in the appearance of these
spectra. Recently, in fact, Shioya, Bekki \& Couch (\ct{shi04})
demonstrates that the reddest HDS galaxies can only be explained by
truncated or starburst models with very heavy dust extinction ($A_V >
$ 0.5 mag).

\section{Dynamical scenario}
\label{sec:57}

The analysis of the velocity distribution for all member galaxies of
ABCG\,209 (Mercurio et al. \ct{mer03a}) shows the existence of a
peak at $<z>=0.2090\pm 0.0004$ with a line--of--sight velocity
dispersion $\sigma_v=1394^{+88}_{-99}$ \kss.

In order to check for possible variations of $<z>$ and $\sigma_v$ for
different spectral types we analysed separately ELG,
HDS$_{\mathrm{blue}}$, Ab-spirals, HDS$_{\mathrm{red}}$, and E
galaxies. All galaxy populations show consistent mean redshift
suggesting that they are really cluster populations and not merely
foreground or background objects. On the other hand, the velocity
dispersions of different classes show significant differences (see
Table~\ref{vel_disp}).

\begin{table}

     \caption[]{Redshifts and velocity dispersion of different
     spectral classes. The biweight estimators (Beers et
     al. \ct{bee90}) were used to derive the values}.
    \label{vel_disp}
    $$
           \begin{array}{c c c c }
            \hline
            \noalign{\smallskip}
    \mathrm{Class} & \mathrm{N_{gal}} &  \mathrm{Redshift} & \mathrm{\sigma_{v}} \\
            \noalign{\smallskip}
            \hline
            \noalign{\smallskip}
\mathrm{ALL}        & 112 & 0.2090\pm0.0004 & 1394_{-99}^{+88}\\
\mathrm{E}          &  74 & 0.2086\pm0.0005 & 1323_{-94}^{+127}\\
\mathrm{HDS_{red}}  &   7 & 0.2087\pm0.0005 & 339_{-46}^{+91}\\
\mathrm{Ab-spirals}  &   8 & 0.2107\pm0.0017 & 1295_{-102}^{+340}\\
\mathrm{HDS_{blue}} &   5 & 0.2071\pm0.0044 & 2257_{-65}^{+1085}\\
\mathrm{ELG}        &   7 & 0.2070\pm0.0038 & 2634_{-53}^{+712}\\
            \noalign{\smallskip}
            \hline
         \end{array}
     $$
   \end{table}

E galaxies define the mean redshift and the velocity dispersion of the
cluster. The very low value of the velocity dispersion of
HDS$_{\mathrm{red}}$ galaxies indicates that they could constitute a
group of galaxies located at the cluster redshift. These
post-starburst galaxies could be the remnant of the core of an
infalling clump, that have suffered significant ram--pressure
stripping in crossing the cluster core, preceded by an instantaneous
burst of star formation. 

The groups of ELG and HDS$_{\mathrm{blue}}$ galaxies exhibit very high
velocity dispersions. Spectral, photometric and structural properties
(cfr. Sect~\ref{sec:5}) of these two classes seem to indicate that
they are strongly related, as it is also indicated by the slight
difference in the velocity dispersion. The high velocity dispersions
of these populations of disk-dominated galaxies indicate that they do
not constitute a single infalling bound galaxy group but are the
results of the growing of the cluster through the accretion of field
galaxies.  These facts are in agreement with a scenario in which
spirals are accreted into the cluster from the field, then star
formation is stopped, the galaxies become gas-deficient, and
eventually undergo a morphological transformation (Couch et
al. \ct{cou98}). This scenario may also account for the existence of
anemic spirals (Ab-spirals in Table~\ref{vel_disp}), whose velocity
dispersion is fully consistent with those of the cluster.

\begin{figure*}
   \centering
   \includegraphics[width=0.7\textwidth]{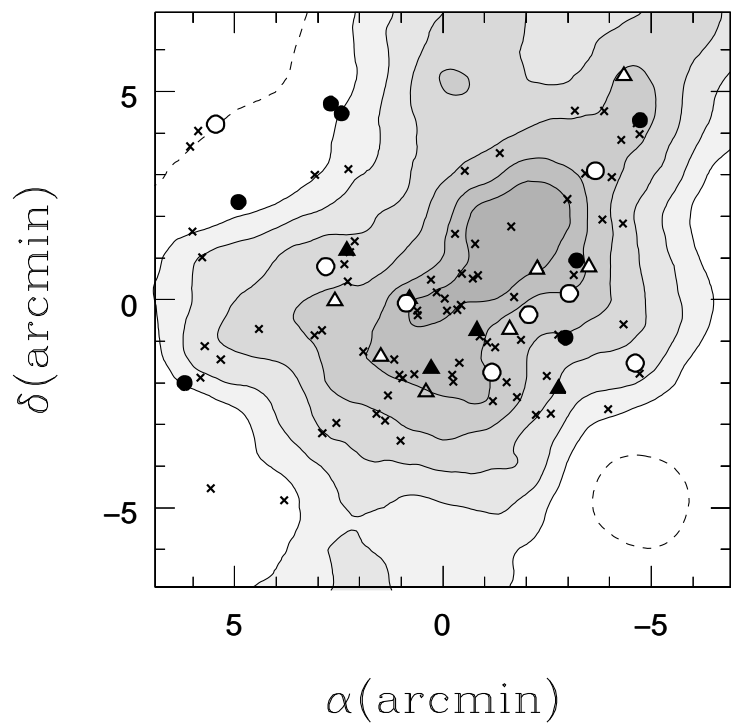}
    \caption{Spatial distribution of different spectral
     classes. Filled and open circles, filled and open triangles, and
     crosses denote ELG, Ab--spiral, HDS$_{\mathrm{blue}}$,
     HDS$_{\mathrm{red}}$ and E galaxies, respectively. The plot is
     centred on the cluster center. The dashed contour corresponds to
     the field galaxy number density, while the solid contours
     correspond respectively to the background subtracted 2, 3.3, 5,
     7.5, 10, and 12.5 galaxies arcmin$^{-2}$}
   \label{spat_distr}
   \end{figure*}

Figure~\ref{spat_distr} shows the spatial distribution of different
spectral classes. Taking into account the SE-NW elongation of the
cluster, ELGs (filled circles) seem to be uniformly distributed in the
outer parts of the cluster, while blue HDS galaxies (filled triangles)
show a preferential location within the cluster. They are concentrated
around the centre of the cluster along a direction perpendicular to
the cluster elongation. Ab--spirals (open circles) and E galaxies
(crosses) are uniformly distributed over the whole cluster plane and
red HDS galaxies (open triangles) lie around the centre along the
cluster elongation.

By using the information about the phase-space distributions, it is
possible to find segregations of the various galaxy populations.
Biviano et al. (\ct{biv02}) searched for segregation by comparing
(R,v)-distributions in the clusters of the ESO Nearby Abell Cluster
Survey, using the combined evidence from projected positions and
relative velocities (they may not be independent). They also stressed
the importance of the estimate of the projected clustercentric
distance R, in order to make the comparison as unbiased as possible.

The spatial distribution of galaxies in ABCG\,209 shows a complex
structure, characterised by an elongation in the SE-NW direction (see
Mercurio et al. \ct{mer03a}). To measure the effect of the cluster
environment on the galaxy population, their spectral properties should
be measured against the parameter most related to the cluster
structure, which is the local galaxy number surface density rather
than cluster-centric radius.

The local surface density (LD) of galaxies was determined across the
CFHT images by Haines et al. (\ct{hai03}). We performed the
comparison of (LD,v)-distributions by means the two--dimensional
Kolmogorov--Smirnov test (hereafter 2DKS--test; Press et
al. \ct{pre92}). The 2DKS test is relatively conservative: if it
indicates that two distributions have a high probability of not being
drawn from the same parent population, other tests (e.g. Rank-Sum
tests or Sign Tests) indicate the same thing, while the contrary is
not always true. Moreover, the 2DKS test turns out to be reliable with
relatively small samples (e.g. Biviano et al. \ct{biv02}).

According to the 2DKS--test, we find significant evidence for
segregation of ELGs with respect to both Es (99.5\%) and
HDS$_{\mathrm{blue}}$ galaxies (97.6\%), where ELGs lie in regions
with lower density than E and HDS$_{\mathrm{blue}}$ galaxies.
Moreover the distributions of HDS$_{\mathrm{blue}}$ and
HDS$_{\mathrm{red}}$ galaxies differ at 90.7\% c.l. and
HDS$_{\mathrm{blue}}$ and Ab-spirals at 92.0\%. These results show
the effects of the cluster environment on the spectral properties of
their constituents and support the scenario in which ELGs are accreted
into the cluster from the field. 

\section{Summary and Discussion}
\label{sec:58}

In order to investigate the nature of the galaxy population in
ABCG\,209, we have performed a detailed study of the photometric and
spectroscopic properties of luminous member galaxies (R $\lesssim$
20.0).  The primary goal of this analysis was to study the relation
between cluster dynamics and evolution of stellar populations, and the
effects of cluster environment on the properties of galaxies.

ABCG\,209 is undergoing strong dynamical evolution with the merging of
at least two clumps along the SE-NW direction, as indicated by i) the
presence of a velocity gradient along a SE-NW axis, ii) the elongation
of X-ray contour levels in Chandra images, and iii) the presence of
substructures manifest both in the detection of two peaks separated by
50 arcsec in the X-ray flux and in the analysis of the velocity
distribution, that showed three sub-clumps (Mercurio et
al. \ct{mer03a}). Moreover, the young dynamical state of the cluster
is indicated by the possible presence of a radio halo (Giovannini,
Tordi \& Feretti \ct{gio99}), which has been suggested to be the
result of a recent cluster merger, through the acceleration of
relativistic particles by the merger shocks (Feretti \ct{fer02}).

The analysis of the spectroscopic and photometric properties of 102
cluster members has shown the presence of five different types of
galaxies: i) passive evolving galaxies (E), which exhibit red colours
and no emission lines, ii) emission line galaxies (ELG), which are
blue and have prominent emission lines, iii-iv) strong H$_\delta$
galaxies, that are characterized by the presence of strong H$_\delta$
absorption lines and can be divided into blue (HDS$_{\mathrm{blue}}$)
and red (HDS$_{\mathrm{red}}$) galaxies, according to the index
D$_n$(4000) and to the B-R colours; v) anemic spirals (Ab-spirals),
that have the same spectral properties of passive evolving galaxies,
but are disk-dominated systems.

E galaxies represent $\sim$ 74\% of the cluster population, and lie
mainly in high density regions. Their age distribution is
characterized by the presence of an ``old population'' formed early
during the initial collapse of the cluster at ages greater than 7 Gyr,
and another one, younger, that could be formed later or could have
experienced an enhancement in the star formation rate.

ELGs lie in low density regions and have high line--of--sight velocity
dispersion. Both the spatial position and the velocity dispersion
suggest that this is a population of galaxies that has recently fallen
into the cluster. This infall has induced a truncation of the star
formation with, possibly, an associated short starburst.

These results are understandable in terms of cosmological models of
structure formation, in which galaxies form earliest in the
highest-density regions corresponding to the cores of rich
clusters. Not only do the galaxies form earliest here, but the bulk of
their star formation is complete by $z\sim1$, at which point the
cluster core is filled by shock-heated virialised gas which does not
easily cool or collapse, suppressing the further formation of stars
and galaxies (Blanton et al. \ct{blanton99};
\ct{blanton00}). Diaferio et al. (\ct{dia01}) shows that as mixing
of the galaxy population is incomplete during cluster assembly, the
positions of galaxies within the cluster are correlated with the epoch
at which they were accreted. Hence galaxies in the cluster periphery
are accreted later, and so have their star formation suppressed later,
resulting in younger mean stellar populations. It should be considered
however that this is a small-scale affect, and that this result in
fact confirms that the red sequence galaxy population is remarkably
homogeneous across all environments.

A similar scenario is suggested also by the velocity dispersion and
the spectral properties of HDS$_{\mathrm{blue}}$ galaxies, which,
however, are found to be aligned in a direction perpendicular to the
cluster elongation, that coincides with the elongation of the X-ray
contour levels in the Chandra images (see Fig. 14 in Mercurio et
al. \ct{mer03a}). As already pointed out by Poggianti et
al. (\ct{pog04}) analysing the position of the strongest k+a galaxies
(having [H$_\delta$] $>$ 5 \AA) in Coma, the correlation between the
location of the post-starburst galaxies and the substructures in the
intracluster medium strongly suggest that the truncation of the star
formation activity in these galaxies, and possibly the previous
starburst, could be the result of an interaction with the hot
intracluster medium. Thus, the origin of HDS$_{\mathrm{blue}}$ could
be a cluster related and, in particular, an ICM-related phenomenon,
closely connected with the dynamical state of the cluster (see
Poggianti et al. \ct{pog04}).

HDS$_{\mathrm{red}}$ galaxies are distributed along the elongation of
the cluster, mainly in intermediate density regions and have a low
velocity dispersion. According to the evolution models, the presence
of strong H$_\delta$ absorption line in their spectra indicates that
these galaxies have experienced a short starburst of star formation in
the past few Gyr. In the starburst model [H$_{\delta}$] and
D$_n$(4000) declines on a timescale of $\sim$ 2 Gyr after the burst
has ceased, regardless of their SFR before the burst. This implies
that, in the galaxies we observe, the burst has occurred no more than
4.5 Gyr ago.

As also indicated by the low value of the velocity dispersion,
HDS$_{\mathrm{red}}$ galaxies could be the remnant of the core of an
infalling clump of galaxies, that have experienced a merger with the
main cluster. This merger may have induced a time--dependent tidal
gravitational field that stimulated non--axisymmetric perturbations in
the galaxies, driving effective transfer of gas to the central
galactic region and, finally, triggering a secondary starburst in the
central part of these galaxies (Bekki \ct{bek99}). The second peak
of the X-ray distribution could be related to the presence of this
small galaxy group.

Considering that the mass is proportional to the cube of the measured
velocity dispersion (see eq. (4) and (11) of Girardi et
al. \ct{gir98}), the ratio between the mass of the cluster as
constituted only by E galaxies and the group of HDS$_{\mathrm{red}}$
galaxies is $\sim$ 0.017. Taffoni et al. (\ct{taf03}) studied the
evolution of dark matter satellites orbiting inside more massive
haloes, by using semi-analytical tools coupled with high-resolution
N-body simulations. They explored the interplay between dynamic
friction and tidal mass loss/evaporation in determining the final fate
of the satellite, showing that, for satellites of intermediate mass
(0.01 M$_\mathrm{h} <$ M$_\mathrm{s,0} < $ 0.1 M$_\mathrm{h}$, where
M$_\mathrm{s,0}$ is the initial mass of the satellite and
M$_\mathrm{h}$ is the mass of the main halo), the dynamical friction is
strong and drives the satellite toward the centre of the main halo,
with significant mass loss. The final fate depends on the
concentration of the satellite, relative to that of the main
halo. Low--concentration satellites are disrupted, while
high--concentration satellites survive, with a final mass that
depends on the decay time.

Although both HDS$_{\mathrm{blue}}$ and HDS$_{\mathrm{red}}$ are
characterized by the presence of strong Balmer lines, their colours
and structural parameters show that they are two different galaxy
populations. Deriving the Sersic index, we found that
HDS$_{\mathrm{blue}}$ galaxies are disk-dominated galaxies while
HDS$_{\mathrm{red}}$ are spheroids. On the contrary, Poggianti et
al. (\ct{pog04}) found that, with the exception of two blue k+a
galaxies compatible with a de Vaucouleurs' profile, all the other blue
and red k+a galaxies in the Coma cluster have exponential or steeper
profiles. This apparent contradiction may be overcome by considering
the time for a HDS$_{\mathrm{blue}}$ galaxy to become red due to the
damping of star formation is shorter than the difference in look-back
time between ABCG\,209 and Coma. It is thus conceivable that the
HDS$_{\mathrm{blue}}$ galaxies we observe at z$\sim$0.21 will turn
into HDS$_{\mathrm{red}}$ galaxies observed Coma.

All the presented results support an evolutionary scenario in which
ABCG\,209 is characterized by the presence of two components: an old
galaxy population, formed very early (z$_f \gtrsim $ 3.5) and a
younger population of infalling galaxies. Moreover, this cluster may
have experienced, 3.5-4.5 Gyr ago, a merging with an infalling
galaxy group.  The merger of the cluster with the infalling group
could have also powered the observed radio halo. In fact, merger
activity and high ICM temperature may be responsible for producing a
radio halo (Liang et al. \ct{lia00}), because merging can provide
enough energy to accelerate the electrons to relativistic energies,
give rising to non-thermal emission. After the shock disappeared,
radio halos may be maintained in situ by electron acceleration in the
residual turbulence.

\section*{Appendix A}
\label{sec:A}

The equivalent width is defined by:

\begin{equation}
\mathrm{EW} = \sum^{N}_{i=1} \frac{\mathrm{F_{Ci}} - \mathrm{F_i}}
{\mathrm{F_{Ci}}} \Delta \lambda \ , 
\label{eq51}
\end{equation}

where F$_i$ is the flux in the pixel $i$, $N$ is the number of pixels
in the integration range, $\Delta \lambda$ is the dispersion in
\AA/pixel and F$_{Ci}$ is the straight line fitted to the blue and the
red pseudo--continuum intervals, evaluated in the pixel $i$ (pixels
with values more than 3$\sigma$ away from the continuum level were
rejected).

An index measured in magnitudes is:

\begin{equation}
\mathrm{Mag} = - 2.5 \mathrm{log} \left[1 - \frac{\mathrm{EW}}
{\mathrm{(\lambda_2-\lambda_1)}}\right] \ , 
\label{eq2}
\end{equation}

where $\lambda_1$ and $\lambda_2$ are the wavelength limits of the
feature bandpass.

The errors of the equivalent width measurement are estimated from
the following relation (Czoske et al. \ct{czo01}):

\begin{equation}
\sigma^2_{EW}=\left(\frac{\mathrm{S}}{\mathrm{N}}\right)^{-2} \left[(N \Delta \lambda - EW) \Delta \lambda + \frac{(N \Delta \lambda - EW)^2}{N_1+N_2} \right] \ ,
\label{eq4}
\end{equation}
where S/N is the signal--to--noise ratio evaluated in the feature bandpass.

The local S/N for each galaxy is obtained by dividing the galaxy
spectrum by its associated noise spectrum, as determined by adding in
quadrature the Poisson noise (obtained through the IRAF task APALL)
and the read-out noise of EMMI. The result is then fitted to remove
residual features, and averaged over the wavelength range of the
spectral feature to obtain the S/N ratio.
 
The catalogue of spectroscopic measurements is presented in Table
~\ref{catalogue2}. 

\begin{landscape}
\begin{table}

        \caption[]{\footnotesize Spectroscopic data. Running number
        for galaxies in the presented sample (see also Table
        \ref{phot_cat}), ID (Col.~1); strength of the 4000 \AA \ \
        break (Col.~2); Cols.~3-15: equivalent widths. A value of
        '0.0' for the equivalent width for an emission-line indicates
        that no emission is observed, whereas '....'  indicates that
        the equivalent width could not be measured as the line does
        not lie within the available wavelength range. The spectral
        classification adopted in the present paper is listed in the
        last column. ELG: emission line galaxies;
        HDS$_{\mathrm{blue}}$: H$_\delta$ strong galaxies with blue
        colour; HDS$_{\mathrm{red}}$: H$_\delta$ strong galaxies with
        red colour; E: passive evolving galaxies; Ab-spirals: passive
        evolving disk-dominated galaxies (n$_{\mathrm{Sersic}} \le$
        2). Galaxy 62 is referred to as AGN because it shows the signature
        of a galactic active nucleus and is excluded from the
        analysis.}

         \label{catalogue2}
{\tiny
              $$
\hspace{-3.5cm}
           \begin{array}{c c c c c c c c c c c c c c c c }
            \hline
            \noalign{\smallskip}
            \hline
            \noalign{\smallskip}
\mathrm{ID} & 
\mathrm{D_n(4000)} & \mathrm{[OII]} & \mathrm{H_{\delta A}} & 
\mathrm{H_{\gamma A}} & \mathrm{Fe4531} & \mathrm{H_{\beta}} & \mathrm{[OIII]}
& \mathrm{Fe5015} & \mathrm{Mg_1} & \mathrm{Mg_2} & \mathrm{Mg}b & 
\mathrm{Fe5270} & \mathrm{Fe5335} & \mathrm{H_{\alpha}} & \mathrm{Spectral}\\
 & \AA & \AA & \AA & \AA & \AA & \AA & \AA & \AA & 
\mathrm{mag} & \mathrm{mag} & \AA & \AA & \AA & \AA & \mathrm{class}\\
            \hline
            \noalign{\smallskip}   
  1&1.14\pm0.06&-10.1\pm 1.2&  7.1\pm 0.8&  0.3\pm 0.8&  4.7\pm 1.0& -0.5\pm 0.6& -0.3\pm 0.4&  5.2\pm 1.2&0.07\pm0.01&0.21\pm0.02&  3.4\pm 0.5&  2.7\pm 0.6&  2.0\pm 0.7& -10.8\pm  0.9& \mathrm{ELG}\\       
  2&1.73\pm0.03&  0.0       & -1.0\pm 0.8& -4.2\pm 0.7&  4.3\pm 0.7&  1.6\pm 0.5&  0.0       &  6.1\pm 0.9&0.16\pm0.01&0.30\pm0.01&  3.6\pm 0.4&  3.6\pm 0.5&  2.7\pm 0.6&   2.5\pm  0.4& \mathrm{E} \\        
  3&1.93\pm0.05&  0.0       &  1.0\pm 1.2& -2.8\pm 1.1&  2.9\pm 1.1&  2.5\pm 0.7&  0.0       &  7.4\pm 1.3&0.15\pm0.02&0.33\pm0.02&  5.6\pm 0.6&  5.4\pm 0.7&  2.2\pm 0.9&   2.7\pm  0.6& \mathrm{E} \\        
  4&  ....     &  ....      & -4.0\pm 2.2&  1.4\pm 1.8&  2.4\pm 2.1&  4.2\pm 1.3&  0.0       & 11.3\pm 2.5&0.12\pm0.03&0.18\pm0.04&  3.9\pm 1.3&  5.5\pm 1.4&  3.4\pm 1.6&   5.5\pm  1.2& \mathrm{E} \\        
  5&1.72\pm0.09&  ....      & -6.8\pm 2.4& -1.8\pm 1.6& -0.4\pm 1.8&  5.3\pm 1.0&  0.0       &  5.4\pm 2.1&0.10\pm0.02&0.27\pm0.03&  1.3\pm 1.0&  5.9\pm 1.0&  2.0\pm 1.3&  -4.2\pm  1.1& \mathrm{Ab-spiral} \\        
  6&1.89\pm0.07& 0.0        &  3.5\pm 0.4& -9.0\pm 0.5&  2.6\pm 0.5&  2.8\pm 0.3&  0.0       & 10.6\pm 0.7&0.19\pm0.01&0.44\pm0.01&  6.9\pm 0.3&  6.4\pm 0.4&  5.8\pm 0.5&   3.6\pm  0.4& \mathrm{HDS_{red}} \\        
  7&1.70\pm0.03& 0.0        & -1.3\pm 0.7& -3.0\pm 0.6&  4.1\pm 0.7&  1.9\pm 0.4&  0.0       &  5.1\pm 0.8&0.16\pm0.01&0.34\pm0.01&  4.3\pm 0.4&  3.1\pm 0.4&  2.2\pm 0.5&   0.3\pm  0.4& \mathrm{E} \\        
  8&1.73\pm0.05&  0.0       &  1.0\pm 1.2& -3.9\pm 1.0&  4.7\pm 1.0&  1.3\pm 0.6&  0.0       &  7.5\pm 1.2&0.19\pm0.02&0.36\pm0.02&  5.2\pm 0.6&  3.9\pm 0.7&  2.3\pm 0.8&   0.4\pm  0.6& \mathrm{E} \\        
  9&  ....     &  ....      & -1.6\pm 1.6& -0.3\pm 1.3&  4.8\pm 1.4&  1.1\pm 0.9&  0.0       &  6.0\pm 1.7&0.17\pm0.02&0.34\pm0.03&  4.1\pm 0.8&  4.4\pm 0.9&  6.2\pm 1.0&   0.4\pm  0.8& \mathrm{E} \\        
 10&1.66\pm0.06&  ....      & -1.8\pm 1.7& -3.5\pm 1.4&  4.2\pm 1.4&  2.0\pm 1.0&  0.0       &  7.0\pm 1.9&0.15\pm0.02&0.31\pm0.03&  4.9\pm 0.8&  3.1\pm 0.9&  4.3\pm 1.1&   4.6\pm  0.7& \mathrm{E} \\        
 11&1.77\pm0.04&  0.0       & -1.3\pm 0.9& -4.6\pm 0.8&  4.4\pm 0.9&  2.8\pm 0.5&  0.0       &  5.4\pm 1.1&0.17\pm0.01&0.32\pm0.02&  5.1\pm 0.5&  3.1\pm 0.6&  1.9\pm 0.6&   1.8\pm  0.5& \mathrm{E} \\        
 12&  ....     &  ....      & -2.7\pm 1.7& -8.4\pm 1.5&  4.4\pm 1.6&  4.0\pm 1.0&  0.0       &  7.5\pm 2.0&0.24\pm0.03&0.31\pm0.03&  4.0\pm 0.9&  8.1\pm 1.0&  2.8\pm 1.2&   1.7\pm  1.0& \mathrm{E} \\        
 13&  ....     &  ....      &  2.8\pm 1.8& -2.8\pm 1.6&  4.4\pm 1.7&  1.0\pm 1.1&  0.0       &  8.3\pm 2.0&0.14\pm0.03&0.27\pm0.03&  2.8\pm 1.0&  7.1\pm 1.0&  4.7\pm 1.2&   1.4\pm  0.9& \mathrm{E} \\ 	  
 14&2.08\pm0.05&  0.0       & -0.7\pm 1.3& -2.8\pm 1.1&  6.2\pm 1.2&  3.2\pm 0.8&  0.0       &  7.1\pm 1.6&0.17\pm0.02&0.33\pm0.02&  4.2\pm 0.7&  3.3\pm 0.8&  1.9\pm 1.0&   4.4\pm  0.6& \mathrm{Ab-spiral} \\        
 15&  ....     &  ....      &  3.9\pm 1.8& -2.1\pm 1.7&  5.7\pm 2.0&  5.8\pm 1.2&  0.0       & 10.2\pm 2.3&0.11\pm0.03&0.25\pm0.04&  3.8\pm 1.2&  3.9\pm 1.3&  6.0\pm 1.5&   4.1\pm  1.2& \mathrm{HDS_{red}}  \\	  
 16&1.80\pm0.05&  ....      &  0.8\pm 1.7& -2.4\pm 1.4&  5.8\pm 1.4&  2.1\pm 0.9&  0.0       &  5.9\pm 1.8&0.14\pm0.02&0.27\pm0.03&  6.3\pm 0.8&  3.3\pm 0.9&  3.8\pm 1.1&   1.7\pm  0.8& \mathrm{E} \\        
 17&1.31\pm0.03&  ....       &  9.5\pm 0.7&  8.0\pm 0.7&  3.0\pm 1.1&  4.8\pm 0.6& -1.0\pm 0.5&  3.6\pm 1.4&0.07\pm0.02&0.15\pm0.02&  3.5\pm 0.7&  4.0\pm 0.7&  3.1\pm 0.9&  -7.5\pm  0.9&\mathrm{ELG}\\   
 18&  ....     &  ....      & -1.1\pm 1.7& -3.8\pm 1.5&  5.8\pm 1.5&  2.5\pm 1.0&  0.0       &  9.0\pm 1.9&0.16\pm0.02&0.34\pm0.03&  6.9\pm 0.9&  4.5\pm 1.0&  1.2\pm 1.3&   2.5\pm  1.0& \mathrm{E} \\        
 19&1.98\pm0.05&  ....      & -0.6\pm 1.0&  1.5\pm 0.8&  6.1\pm 0.8&  2.4\pm 0.5&  0.0       &  4.9\pm 1.0&0.15\pm0.01&0.26\pm0.01&  5.0\pm 0.4&  3.0\pm 0.5&  7.4\pm 0.6&   0.5\pm  0.4& \mathrm{E} \\        
 20&1.66\pm0.04&  0.0       &  1.4\pm 3.0& -8.3\pm 2.7&  4.5\pm 2.9&  6.6\pm 1.7&  0.0       &  7.2\pm 3.4&0.13\pm0.04&0.30\pm0.05&  2.7\pm 1.7&  2.7\pm 1.8&  2.2\pm 2.1&   1.7\pm  1.6& \mathrm{Ab-spiral} \\        
 21&1.82\pm0.04&  0.0       & -1.2\pm 1.1& -3.1\pm 0.9&  4.9\pm 1.1&  2.5\pm 0.7&  0.0       &  7.5\pm 1.2&0.13\pm0.02&0.29\pm0.02&  5.0\pm 0.6&  3.4\pm 0.6&  2.8\pm 0.8&   3.4\pm  0.5& \mathrm{E} \\        
 22&1.17\pm0.05&  ....      &  4.6\pm 2.1&  1.7\pm 1.0&  8.7\pm 1.2& -3.0\pm 0.9& -2.5\pm 0.5&  7.1\pm 1.5&0.04\pm0.02&0.10\pm0.02&  4.7\pm 0.6&  2.5\pm 0.8&  7.2\pm 0.7&  -5.6\pm  0.7& \mathrm{ELG} \\	  
 23&1.67\pm0.04&  0.0       & -2.1\pm 0.9& -4.0\pm 0.8&  4.9\pm 0.9&  2.1\pm 0.6&  0.0       &  6.0\pm 1.1&0.13\pm0.01&0.26\pm0.02&  4.6\pm 0.5&  4.9\pm 0.6&  3.3\pm 0.7&   2.7\pm  0.5& \mathrm{E} \\        
            \noalign{\smallskip}	     		    
            \hline			    		    
         \end{array}
     $$ 
}
         \end{table}
\end{landscape}

\begin{landscape}
\addtocounter{table}{-1}
\begin{table}
          \caption[ ]{\footnotesize Continued.}
{\tiny
     $$ 
\hspace{-3.5cm}
           \begin{array}{c c c c c c c c c c c c c c c c}
            \hline
            \noalign{\smallskip}
            \hline
            \noalign{\smallskip}
\mathrm{ID} &\mathrm{D_n(4000)} &\mathrm{[OII]} &\mathrm{H_{\delta A}} &
\mathrm{H_{\gamma A}} &\mathrm{Fe4531} &\mathrm{H_{\beta}} &\mathrm{[OIII]} 
&\mathrm{Fe5015} &\mathrm{Mg_1} &\mathrm{Mg_2} &\mathrm{Mg}b &
\mathrm{Fe5270} &\mathrm{Fe5335} &\mathrm{H_{\alpha}} & \mathrm{Spectral}\\
 &\AA &\AA &\AA &\AA &\AA &\AA &\AA &\AA &
\mathrm{mag} &\mathrm{mag} &\AA &\AA &\AA &\AA & \mathrm{class}\\
            \hline
            \noalign{\smallskip}   
 24&1.38\pm0.06&  ....      &  7.8\pm 1.2&  7.7\pm 1.1&  4.0\pm 1.5&  7.2\pm 0.8&  0.0       &  7.4\pm 1.8&-.02\pm0.02&0.11\pm0.03&  3.8\pm 0.9&  4.9\pm 1.1& -0.6\pm 1.5&   0.7\pm  1.1& \mathrm{HDS_{blue}} \\  
 25&1.84\pm0.11&  0.0       & -5.2\pm 1.9& -5.5\pm 1.3&  3.7\pm 1.3&  5.3\pm 0.7&  0.0       &  6.7\pm 1.4&0.15\pm0.02&0.30\pm0.02&  4.8\pm 0.6&  3.5\pm 0.7&  3.3\pm 0.9&  ....        & \mathrm{E} \\        
 26&1.70\pm0.03&  ....      & -1.6\pm 1.0& -2.8\pm 0.8&  5.2\pm 1.0&  2.5\pm 0.6&  0.0       &  6.0\pm 1.1&0.11\pm0.01&0.27\pm0.02&  4.6\pm 0.6&  2.5\pm 0.6&  4.2\pm 0.7&   2.9\pm  0.5& \mathrm{E} \\        
 27&  ....     &  ....      &  4.6\pm 2.2& -7.3\pm 2.2& 11.8\pm 1.8& -0.3\pm 1.3&  0.0       &  7.6\pm 2.5&0.18\pm0.03&0.38\pm0.04&  7.6\pm 1.1&  5.9\pm 1.3&  4.9\pm 1.5&   7.1\pm  1.0& \mathrm{HDS_{red}}\\              
 28&1.67\pm0.04&  0.0       & -2.0\pm 1.3& -2.8\pm 1.1&  4.9\pm 1.1&  5.1\pm 0.7&  0.0       &  7.4\pm 1.4&0.13\pm0.02&0.30\pm0.02&  4.8\pm 0.7&  2.8\pm 0.7&  2.8\pm 0.9&   ....       & \mathrm{E} \\         
 29&1.58\pm0.06&  0.0       &  2.6\pm 1.4&  3.9\pm 1.1&  5.6\pm 1.2&  5.8\pm 0.7&  0.0       &  5.4\pm 1.5&0.04\pm0.02&0.16\pm0.02&  3.4\pm 0.7&  4.9\pm 0.8&  3.2\pm 0.9&   3.6\pm  0.7& \mathrm{Ab-spiral} \\   	
 30&1.66\pm0.03&  0.0       & -0.8\pm 0.6& -4.1\pm 0.5&  4.3\pm 0.6&  2.0\pm 0.4&  0.0       &  5.2\pm 0.7&0.12\pm0.01&0.29\pm0.01&  4.9\pm 0.3&  3.1\pm 0.4&  2.1\pm 0.5&  -0.3\pm  0.4& \mathrm{E} \\         
 31&1.87\pm0.08&  0.0       & -1.5\pm 2.2& -1.5\pm 1.6& -1.9\pm 1.8&  4.1\pm 0.9&  0.0       &  6.9\pm 1.7&0.15\pm0.02&0.30\pm0.03&  6.2\pm 0.8&  5.5\pm 0.9&  2.5\pm 1.0&   1.2\pm  0.7& \mathrm{E} \\         
 32&1.83\pm0.04&  0.0       &  0.6\pm 0.7& -4.5\pm 0.7&  4.0\pm 0.7&  3.0\pm 0.5&  0.0       &  7.6\pm 0.9&0.15\pm0.01&0.29\pm0.01&  4.2\pm 0.4&  2.8\pm 0.5&  1.2\pm 0.6&   3.2\pm  0.4& \mathrm{E} \\         
 33&1.83\pm0.06&  0.0       &  0.8\pm 0.8& -3.6\pm 0.8&  3.4\pm 0.8&  2.7\pm 0.5&  0.0       &  5.6\pm 1.0&0.21\pm0.01&0.39\pm0.02&  6.4\pm 0.5&  3.5\pm 0.5&  3.0\pm 0.6&   2.3\pm  0.5& \mathrm{E} \\         
 34&1.58\pm0.10&  0.0       &  4.5\pm 1.5& -3.7\pm 1.3&  8.5\pm 1.3&  4.9\pm 0.7&  0.0       &  7.0\pm 1.5&0.09\pm0.02&0.32\pm0.02&  6.4\pm 0.7&  5.1\pm 0.8&  0.9\pm 1.0&   2.7\pm  0.6& \mathrm{HDS_{red}} \\       
 35&1.73\pm0.04&  0.0       & -2.1\pm 0.9& -3.5\pm 0.8&  5.6\pm 0.8&  3.6\pm 0.5&  0.0       &  8.3\pm 1.0&0.13\pm0.01&0.40\pm0.02&  7.1\pm 0.4&  4.4\pm 0.5&  4.7\pm 0.5&   ....       & \mathrm{E} \\         
 36&1.65\pm0.06&  ....      &  0.2\pm 2.3& -3.5\pm 1.7&  2.7\pm 1.9&  2.5\pm 1.1&  0.0       &  8.8\pm 2.1&0.14\pm0.03&0.32\pm0.03&  4.9\pm 1.0& 12.1\pm 1.0&  3.8\pm 1.3&   4.9\pm  0.9& \mathrm{E} \\         
 37&1.97\pm0.05&  0.0       & -1.2\pm 0.9& -5.1\pm 0.8&  4.8\pm 0.7&  1.4\pm 0.5&  0.0       &  4.0\pm 1.0&0.18\pm0.01&0.43\pm0.02&  6.5\pm 0.4&  2.8\pm 0.5&  2.4\pm 0.6&   2.4\pm  0.5& \mathrm{E} \\         
 38&1.95\pm0.07&  0.0       & -4.0\pm 1.3& -3.3\pm 1.0&  6.7\pm 0.9&  3.2\pm 0.5&  0.0       &  4.5\pm 1.0&0.21\pm0.01&0.39\pm0.02&  6.3\pm 0.5&  4.4\pm 0.5&  2.9\pm 0.7&  -0.4\pm  0.4& \mathrm{E} \\                 
 39&1.79\pm0.04&  0.0       & -1.4\pm 0.8& -4.0\pm 0.7&  4.7\pm 0.8&  3.7\pm 0.5&  0.0       &  7.1\pm 0.9&0.14\pm0.01&0.36\pm0.01&  6.4\pm 0.4&  3.5\pm 0.5&  3.1\pm 0.6&   2.3\pm  0.5& \mathrm{Ab-spiral} \\         
 40&1.56\pm0.10&  ....      & -3.6\pm 2.3&  1.8\pm 1.6&  3.7\pm 1.9&  2.0\pm 1.1&  0.0       &  8.8\pm 2.1&0.04\pm0.03&0.20\pm0.03&  4.1\pm 0.9&  4.0\pm 1.1&  5.3\pm 1.3&   2.1\pm  0.9& \mathrm{E} \\         
 41&1.85\pm0.07&  ....      &  2.3\pm 1.8& -2.2\pm 1.4&  9.0\pm 1.3&  2.2\pm 0.9&  0.0       & 10.4\pm 1.6&0.14\pm0.02&0.29\pm0.03&  4.3\pm 0.7&  2.4\pm 0.9&  3.4\pm 1.0&   4.3\pm  0.6& \mathrm{E} \\         
 42&1.56\pm0.10&  ....      & -0.8\pm 2.2&  3.9\pm 1.6&  0.9\pm 2.0&  3.6\pm 1.3&  0.0       &  1.3\pm 2.6&0.12\pm0.03&0.29\pm0.04&  5.6\pm 1.2&  5.4\pm 1.2&  2.3\pm 1.5&   1.9\pm  1.1& \mathrm{E} \\   
 43&1.19\pm0.03&  0.0       &  9.9\pm 0.3&  6.7\pm 0.3&  2.4\pm 0.5&  7.7\pm 0.3&  0.0       &  4.5\pm 0.6&0.04\pm0.01&0.13\pm0.01&  1.7\pm 0.3&  2.3\pm 0.4&  2.5\pm 0.5&   1.5\pm  0.4& \mathrm{HDS_{blue}} \\   
 44&1.93\pm0.05&  0.0       & -2.4\pm 1.4& -2.2\pm 1.1&  4.9\pm 1.3&  4.2\pm 0.7&  0.0       &  7.1\pm 1.6&0.15\pm0.02&0.38\pm0.02&  5.5\pm 0.7&  3.5\pm 0.8&  3.5\pm 0.9&  -1.9\pm  0.8& \mathrm{E} \\         
 45&1.98\pm0.08&  ....      & -2.9\pm 1.8& -3.7\pm 1.4&  4.7\pm 1.5&  3.6\pm 0.8&  0.0       &  3.6\pm 1.7&0.22\pm0.02&0.28\pm0.02&  7.2\pm 0.7&  6.1\pm 0.8&  4.2\pm 1.0&   2.3\pm  0.8& \mathrm{E} \\         
 46&1.83\pm0.06&  ....      & -2.6\pm 1.2& -4.7\pm 1.0&  6.9\pm 1.1&  3.2\pm 0.6&  0.0       &  9.0\pm 1.2&0.24\pm0.02&0.33\pm0.02&  1.5\pm 0.6&  2.0\pm 0.7&  3.0\pm 0.8&  -1.4\pm  0.6& \mathrm{E} \\        	
 47&1.70\pm0.04&  0.0       & -0.5\pm 0.8& -2.8\pm 0.7&  5.6\pm 0.8&  1.6\pm 0.5&  0.0       &  6.1\pm 1.0&0.17\pm0.01&0.35\pm0.02&  4.5\pm 0.5&  3.9\pm 0.5&  3.6\pm 0.6&   3.4\pm  0.5& \mathrm{E} \\		
 48&1.57\pm0.08&  0.0       & -1.3\pm 1.8& -2.1\pm 1.5&  4.1\pm 1.5&  2.5\pm 0.9&  0.0       &  9.4\pm 1.6&0.19\pm0.02&0.29\pm0.03&  6.1\pm 0.8&  2.1\pm 0.9&  2.9\pm 1.0&   3.4\pm  0.7& \mathrm{E} \\         
 49&1.82\pm0.03&  ....      & -0.9\pm 0.7& -1.6\pm 0.6&  6.0\pm 0.6&  2.6\pm 0.4&  0.0       &  6.0\pm 0.7&0.12\pm0.01&0.30\pm0.01&  4.6\pm 0.3&  3.2\pm 0.4&  3.2\pm 0.5&   ....       & \mathrm{E} \\         
 50&  ....     &  ....      & -1.5\pm 0.7& -5.7\pm 0.6&  4.0\pm 0.6&  2.4\pm 0.4&  0.0       &  5.8\pm 0.7&0.17\pm0.01&0.35\pm0.01&  4.3\pm 0.3&  3.6\pm 0.4&  2.9\pm 0.4&   3.7\pm  0.3& \mathrm{E} \\         
                \noalign{\smallskip}	     		    
            \hline			     		    
         \end{array}
     $$ 
}
         \end{table}
\end{landscape}

\begin{landscape}
\addtocounter{table}{-1}
\begin{table}
          \caption[ ]{\footnotesize Continued.}
{\tiny
     $$ 
\hspace{-3.5cm}
           \begin{array}{c c c c c c c c c c c c c c c c c}
            \hline
            \noalign{\smallskip}
            \hline
            \noalign{\smallskip}
\mathrm{ID} &
\mathrm{D_n(4000)} &\mathrm{[OII]} &\mathrm{H_{\delta A}} &
\mathrm{H_{\gamma A}} &\mathrm{Fe4531} &\mathrm{H_{\beta}} &\mathrm{[OIII]} 
&\mathrm{Fe5015} &\mathrm{Mg_1} &\mathrm{Mg_2} &\mathrm{Mg}b &
\mathrm{Fe5270} &\mathrm{Fe5335} &\mathrm{H_{\alpha}} & \mathrm{Spectral}\\
 &\AA &\AA &\AA &\AA &\AA &\AA &\AA &\AA &
\mathrm{mag} &\mathrm{mag} &\AA &\AA &\AA &\AA & \mathrm{class}\\
            \hline
            \noalign{\smallskip}  
 51&1.65\pm0.04&  0.0       & -1.0\pm 0.7& -1.1\pm 0.6&  3.4\pm 0.7&  3.6\pm 0.4&  0.0       &  6.1\pm 0.8&0.12\pm0.01&0.29\pm0.01&  4.3\pm 0.4&  3.9\pm 0.5&  2.9\pm 0.6&   4.1\pm  0.4& \mathrm{E} \\         
 52&1.81\pm0.04&  0.0       & -1.8\pm 0.9& -3.3\pm 0.8&  3.8\pm 0.8&  2.4\pm 0.5&  0.0       &  5.8\pm 1.0&0.17\pm0.01&0.37\pm0.02&  5.7\pm 0.5&  2.5\pm 0.5&  2.8\pm 0.6&   2.3\pm  0.5& \mathrm{E} \\         
 53&1.51\pm0.09&  0.0       &  2.9\pm 3.3& -5.0\pm 2.3&  4.5\pm 2.2&  1.2\pm 1.2&  0.0       &  5.4\pm 2.3&0.15\pm0.03&0.34\pm0.04&  5.0\pm 1.0&  2.1\pm 1.2&  6.1\pm 1.3&   4.1\pm  0.9& \mathrm{E} \\       
 54&2.11\pm0.05&  0.0       & -1.2\pm 0.9& -2.3\pm 0.8&  4.1\pm 0.8&  3.4\pm 0.5&  0.0       &  7.8\pm 0.9&0.17\pm0.01&0.39\pm0.02&  5.7\pm 0.5&  4.3\pm 0.5&  2.4\pm 0.7&   2.3\pm  0.4& \mathrm{E} \\         
 55&1.77\pm0.04&  0.0       & -2.6\pm 0.6& -6.4\pm 0.6&  4.9\pm 0.6&  3.7\pm 0.4&  0.0       &  5.0\pm 0.7&0.17\pm0.01&0.41\pm0.01&  6.5\pm 0.3&  2.2\pm 0.4&  3.2\pm 0.5&   1.4\pm  0.4& \mathrm{E} \\         
 56&1.86\pm0.05&  0.0       & -2.1\pm 1.0& -4.3\pm 0.8&  2.6\pm 0.8&  3.3\pm 0.5&  0.0       &  4.9\pm 0.9&0.15\pm0.01&0.38\pm0.01&  6.4\pm 0.4&  2.9\pm 0.5&  2.6\pm 0.6&   0.7\pm  0.4& \mathrm{E} \\         
 57&  ....     &  ....      &  4.5\pm 1.3&  0.3\pm 1.2&  4.0\pm 1.4&  1.5\pm 0.8&  0.0       &  4.4\pm 1.6&0.08\pm0.02&0.18\pm0.02&  3.2\pm 0.8&  4.9\pm 0.8&  3.3\pm 1.0&   0.5\pm  0.7& \mathrm{HDS_{blue}} \\   
 58&1.66\pm0.03&  0.0       & -1.4\pm 0.8& -5.6\pm 0.7&  5.0\pm 0.8&  0.6\pm 0.5&  0.0       &  5.5\pm 1.0&0.17\pm0.01&0.39\pm0.02&  4.6\pm 0.5&  3.4\pm 0.5&  3.3\pm 0.7&   2.5\pm  0.5& \mathrm{E} \\         
 59&1.73\pm0.08&  0.0       &  6.4\pm 1.3& -2.9\pm 1.1&  6.4\pm 1.1& -0.4\pm 0.7&  0.0       &  8.9\pm 1.1&0.14\pm0.01&0.29\pm0.02&  5.9\pm 0.6&  3.7\pm 0.6&  2.3\pm 0.8&   1.9\pm  0.5& \mathrm{HDS_{red}} \\       
 60&1.70\pm0.06&  0.0       &  0.5\pm 1.3& -3.3\pm 1.0&  4.8\pm 1.0&  3.0\pm 0.6&  0.0       &  6.1\pm 1.2&0.15\pm0.01&0.39\pm0.02&  7.5\pm 0.5&  4.3\pm 0.6&  4.0\pm 0.7&   1.9\pm  0.5& \mathrm{E} \\         
 61&1.69\pm0.04&  0.0       & -3.5\pm 2.5& -2.5\pm 2.1&  4.8\pm 2.4&  2.3\pm 1.5&  0.0       &  7.7\pm 2.7&0.19\pm0.04&0.34\pm0.05&  5.5\pm 1.3&  2.6\pm 1.6&  4.4\pm 1.7&   2.0\pm  1.2& \mathrm{E} \\         
 62&1.09\pm0.01&-57.6\pm 1.8&  4.5\pm 1.1& -3.4\pm 1.2&  6.5\pm 1.3&-19.1\pm 1.2&-22.9\pm 0.8&-11.4\pm 1.9&0.06\pm0.02&0.05\pm0.02&  0.7\pm 0.9&  1.2\pm 1.0&  0.7\pm 1.1&-123.3\pm  1.6& \mathrm{AGN} \\ 
 63&  ....     &  ....      & -2.2\pm 1.1& -3.2\pm 1.0&  4.4\pm 0.9&  2.5\pm 0.5&  0.0       &  5.1\pm 1.1&0.16\pm0.01&0.37\pm0.02&  5.1\pm 0.5&  4.2\pm 0.6&  2.8\pm 0.7&   1.3\pm  0.5& \mathrm{E} \\         
 64&  ....     &  ....      &  3.2\pm 1.3& -0.1\pm 1.3&  4.4\pm 1.5&  0.6\pm 1.1&  0.0       &  4.8\pm 1.9&0.08\pm0.02&0.20\pm0.03&  2.8\pm 0.9&  4.6\pm 1.0&  2.4\pm 1.3& -10.6\pm  1.1& \mathrm{HDS_{blue}} \\   
 65&1.63\pm0.04&  0.0       & -3.2\pm 1.2& -3.5\pm 1.0&  6.3\pm 1.1&  3.3\pm 0.7&  0.0       &  7.8\pm 1.4&0.15\pm0.02&0.21\pm0.02&  6.9\pm 0.7&  4.5\pm 0.8&  0.8\pm 0.9&  -1.0\pm  0.8& \mathrm{Ab-spiral} \\         
 66&  ....     &  ....      & -1.0\pm 1.6& -2.1\pm 1.3&  4.4\pm 1.2&  1.8\pm 0.8&  0.0       &  7.0\pm 1.5&0.15\pm0.02&0.32\pm0.02&  5.3\pm 0.6&  3.6\pm 0.8&  3.1\pm 0.9&   2.1\pm  0.6& \mathrm{E} \\         
 67&1.54\pm0.10&  ....      &  1.2\pm 2.5&  4.4\pm 1.7& 10.5\pm 1.7&  4.5\pm 1.1&  0.0       & -0.7\pm 2.2&0.17\pm0.03&0.34\pm0.03&  2.9\pm 1.0&  2.7\pm 1.1&  2.1\pm 1.3&   ....       & \mathrm{E} \\         
 68&1.85\pm0.03&  0.0       & -2.3\pm 0.5& -4.3\pm 0.5&  4.6\pm 0.5&  4.2\pm 0.3&  0.0       &  6.5\pm 0.6&0.17\pm0.01&0.38\pm0.01&  5.8\pm 0.3&  3.5\pm 0.3&  3.1\pm 0.4&   1.5\pm  0.3& \mathrm{E} \\         
 69&2.01\pm0.06&  0.0       & -1.3\pm 1.3& -5.8\pm 1.1&  4.3\pm 1.0&  1.4\pm 0.7&  0.0       &  6.5\pm 1.2&0.17\pm0.01&0.40\pm0.02&  7.7\pm 0.5&  4.5\pm 0.6&  2.3\pm 0.7&   2.5\pm  0.5& \mathrm{E} \\         
 70&  ....     &  ....      &  1.2\pm 1.7& -3.1\pm 1.4&  2.9\pm 1.4&  0.0       &  0.0       &  5.1\pm 1.7&0.13\pm0.02&0.23\pm0.02&  3.8\pm 0.7&  2.5\pm 0.8&  3.7\pm 1.0&   1.7\pm  0.5& \mathrm{E} \\         
 71&1.98\pm0.04&  0.0       & -1.1\pm 0.9& -3.0\pm 0.7&  4.6\pm 0.8&  3.3\pm 0.5&  0.0       &  5.2\pm 0.9&0.18\pm0.01&0.36\pm0.02&  5.1\pm 0.4&  3.9\pm 0.5&  2.2\pm 0.6&   ....       & \mathrm{E} \\         
 72&1.82\pm0.11&  ....      &  4.9\pm 2.4&-11.1\pm 2.2&  4.4\pm 2.0&  5.1\pm 1.2&  0.0       &  2.4\pm 2.3&0.20\pm0.03&0.22\pm0.03&  5.5\pm 1.0&  6.0\pm 1.1&  4.3\pm 1.3&   3.6\pm  0.9& \mathrm{HDS_{red}} \\      
 73&1.87\pm0.04&  0.0       & -0.9\pm 0.8& -3.7\pm 0.6&  4.2\pm 0.7&  3.7\pm 0.4&  0.0       &  8.4\pm 0.9&0.16\pm0.01&0.39\pm0.01&  6.1\pm 0.4&  3.9\pm 0.4&  3.3\pm 0.6&   ....       & \mathrm{E} \\         
 74&1.71\pm0.04&  0.0       &  0.1\pm 0.4& -3.6\pm 0.4&  4.3\pm 0.5&  2.6\pm 0.3&  0.0       &  4.8\pm 0.7&0.14\pm0.01&0.22\pm0.01&  2.3\pm 0.3&  4.9\pm 0.4&  3.2\pm 0.5&   4.3\pm  0.3& \mathrm{E} \\         
 75&1.93\pm0.07& 0.0        &  3.0\pm 1.3& -5.6\pm 1.4&  2.0\pm 1.6&  3.3\pm 1.3&  0.0       &  2.6\pm 2.7&0.25\pm0.04&0.79\pm0.05& 16.7\pm 1.0& -1.7\pm 1.6&  0.6\pm 2.0&   ....       & \mathrm{E} \\         
 76&1.62\pm0.04&  0.0       & -1.8\pm 1.0& -4.0\pm 0.9&  2.7\pm 1.1&  6.0\pm 0.7&  0.0       &  3.1\pm 1.5&0.23\pm0.02&0.42\pm0.02&  4.4\pm 0.7&  3.1\pm 0.8&  5.2\pm 0.9&   ....       & \mathrm{E} \\         
 77&1.71\pm0.05&  0.0       & -2.7\pm 1.0& -6.1\pm 0.9&  5.1\pm 1.0&  1.9\pm 0.7&  0.0       &  5.5\pm 1.4&0.20\pm0.02&0.42\pm0.02&  5.6\pm 0.7&  1.4\pm 0.8&  2.5\pm 1.0&   ....       & \mathrm{E} \\         
            \noalign{\smallskip}	     		    
            \hline			    		    
         \end{array}
     $$ 
}
         \end{table}
\end{landscape}

\begin{landscape}
\addtocounter{table}{-1}
\begin{table}
          \caption[ ]{\footnotesize Continued.}
{\tiny
     $$ 
\hspace{-3.5cm}
           \begin{array}{c c c c c c c c c c c c c c c c}
            \hline
            \noalign{\smallskip}
            \hline
            \noalign{\smallskip}
\mathrm{ID} &\mathrm{D_n(4000)} &\mathrm{[OII]} &\mathrm{H_{\delta A}} &
\mathrm{H_{\gamma A}} &\mathrm{Fe4531} &\mathrm{H_{\beta}} &\mathrm{[OIII]} 
&\mathrm{Fe5015} &\mathrm{Mg_1} &\mathrm{Mg_2} &\mathrm{Mg}b &
\mathrm{Fe5270} &\mathrm{Fe5335} &\mathrm{H_{\alpha}} & \mathrm{Spectral}\\
 & &\AA &\AA &\AA &\AA &\AA &\AA &\AA &\AA &
\mathrm{mag} &\mathrm{mag} &\AA &\AA &\AA & \mathrm{class}\\
            \hline
            \noalign{\smallskip}   
 78&  ....     &  ....       &  2.3\pm 2.9&  0.5\pm 3.1&  3.8\pm 3.5&  4.5\pm 2.2& 0.0       & 14.5\pm 5.9&0.02\pm0.08&0.15\pm0.10&  4.8\pm 3.3& 11.0\pm 3.2& -6.9\pm 6.1&   ....       & \mathrm{E} \\         
 79&  ....     &  ....      &  5.1\pm 0.6&  5.0\pm 0.5&  3.1\pm 0.7&  4.5\pm 0.4&  0.0       &  6.6\pm 0.8&0.08\pm0.01&0.19\pm0.01&  3.2\pm 0.4&  3.8\pm 0.5&  3.9\pm 0.6&   2.5\pm  0.5& \mathrm{HDS_{blue}} \\   
 80&  ....     &  ....      &  0.3\pm 1.6& -4.7\pm 1.6&  5.2\pm 1.8&  4.4\pm 1.1&  0.0       & 12.1\pm 2.2&0.10\pm0.03&0.38\pm0.04&  6.1\pm 1.1&  0.6\pm 1.4&  5.2\pm 1.4&  -3.6\pm  1.4& \mathrm{E} \\         
 81&1.31\pm0.05& -6.1\pm 1.7&  5.1\pm 1.3&  3.5\pm 1.3&  4.8\pm 1.6&  1.6\pm 1.1& -0.2\pm 0.8&  0.5\pm 2.3&0.07\pm0.03&0.23\pm0.04&  0.9\pm 1.2&  6.0\pm 1.1&  3.0\pm 1.4&  -4.1\pm  1.3& \mathrm{ELG} \\       
 82&  ....     &  ....      & -2.5\pm 1.3& -4.7\pm 1.1&  4.6\pm 1.0&  2.9\pm 0.6&  0.0       &  7.6\pm 1.2&0.14\pm0.02&0.35\pm0.02&  5.5\pm 0.6&  3.1\pm 0.6&  2.6\pm 0.7&   2.2\pm  0.5& \mathrm{E} \\            
 83&  ....     &  ....      &  3.3\pm 1.6& -3.9\pm 1.7& -1.9\pm 2.3&  2.9\pm 1.5&  0.0       &  5.4\pm 3.6&0.05\pm0.05&0.61\pm0.07& 11.8\pm 1.5&-11.8\pm 2.5&  5.1\pm 2.5&   ....       & \mathrm{HDS_{red}} \\         
 84&1.35\pm0.04& -6.5\pm 1.4&  9.0\pm 0.9&  2.4\pm 0.9&  0.7\pm 1.3&  3.7\pm 0.9& -1.1\pm 0.7&  9.3\pm 1.8&0.02\pm0.02&0.19\pm0.03&  3.9\pm 0.9&  0.0\pm 1.1&  2.1\pm 1.3&  -6.0\pm  1.5& \mathrm{ELG} \\          
 85&1.77\pm0.07&  0.0       &  3.0\pm 1.5&  0.3\pm 1.5&  0.5\pm 1.9&  7.1\pm 1.2&  0.0       & 10.7\pm 2.6&0.18\pm0.04&0.52\pm0.05&  7.2\pm 1.3&  0.8\pm 1.6&  5.5\pm 1.9&  -4.1\pm  2.8& \mathrm{E} \\            
 86&1.98\pm0.04&  ....      & -2.1\pm 0.7& -4.0\pm 0.6&  3.8\pm 0.6&  2.1\pm 0.4&  0.0       &  5.5\pm 0.7&0.16\pm0.01&0.35\pm0.01&  4.5\pm 0.3&  3.2\pm 0.4&  2.4\pm 0.5&   1.3\pm  0.3& \mathrm{E} \\            
 87&1.68\pm0.05&  0.0       & -1.3\pm 0.8& -3.1\pm 0.7&  2.7\pm 0.8&  2.2\pm 0.5&  0.0       & 10.5\pm 0.9&0.20\pm0.01&0.33\pm0.01&  4.7\pm 0.5&  3.5\pm 0.5&  0.6\pm 0.6&   0.0\pm  0.5& \mathrm{E} \\      	     
 88&1.60\pm0.07&  0.0       &  0.2\pm 1.8& -5.4\pm 1.7&  5.0\pm 1.9& -0.4\pm 1.4&  0.0       &  1.6\pm 2.8&0.04\pm0.03&0.17\pm0.04&  2.2\pm 1.5&  6.4\pm 1.5&  4.6\pm 1.9&   6.3\pm  2.0& \mathrm{E} \\            
 89&1.75\pm0.05&  0.0       & -0.8\pm 1.3& -1.6\pm 1.2&  4.8\pm 1.4&  4.3\pm 0.8&  0.0       &  6.5\pm 1.6&0.19\pm0.02&0.31\pm0.03&  4.4\pm 0.8&  4.5\pm 0.9&  1.9\pm 1.1&   0.6\pm  0.9& \mathrm{E} \\            
 90&  ....     &  ....      &  1.7\pm 3.1& -3.5\pm 2.5&  4.8\pm 2.3&  3.7\pm 1.4&  0.0       &  4.7\pm 2.6&0.11\pm0.03&0.25\pm0.04&  5.2\pm 1.1&  2.9\pm 1.3&  3.2\pm 1.5&   2.3\pm  0.8& \mathrm{E} \\            
 91&  ....     &  ....      & -4.1\pm 1.1& -4.7\pm 1.0&  4.1\pm 1.0&  2.9\pm 0.6&  0.0       &  5.1\pm 1.2&0.14\pm0.01&0.37\pm0.02&  6.3\pm 0.5&  4.0\pm 0.6&  3.8\pm 0.8&   1.9\pm  0.6& \mathrm{E} \\  	     
 92&1.35\pm0.04& -7.8\pm 1.9&  5.5\pm 1.2&  4.9\pm 1.2&  3.4\pm 1.5&  0.9\pm 1.0& -0.2\pm 0.7&  3.8\pm 2.1&0.10\pm0.03&0.22\pm0.03&  2.7\pm 1.1&  2.8\pm 1.2&  5.2\pm 1.3& -16.2\pm  1.8& \mathrm{ELG} \\	     
 93&  ....     &  ....      & -1.8\pm 0.7& -6.2\pm 0.6&  5.0\pm 0.7& -0.1\pm 0.4&  0.0       &  4.5\pm 0.8&0.20\pm0.01&0.38\pm0.01&  5.2\pm 0.4&  3.4\pm 0.4&  3.1\pm 0.5&   2.3\pm  0.3& \mathrm{E} \\            
 94&1.61\pm0.05&  ....      &  2.7\pm 0.9& -0.8\pm 0.9&  5.5\pm 1.0&  4.5\pm 0.6&  0.0       &  7.7\pm 1.2&0.11\pm0.02&0.36\pm0.02&  5.2\pm 0.6&  0.9\pm 0.7&  1.6\pm 0.9&  -0.4\pm  0.8& \mathrm{Ab-spiral} \\             
 95&  ....     &  ....      & -1.0\pm 0.9& -3.8\pm 0.8&  4.7\pm 0.8&  3.2\pm 0.5&  0.0       &  9.6\pm 0.9&0.19\pm0.01&0.37\pm0.01&  6.0\pm 0.4&  4.3\pm 0.5&  4.0\pm 0.6&   0.8\pm  0.4& \mathrm{E} \\             
 96&  ....     &  ....      & -1.5\pm 0.9& -4.6\pm 0.8&  3.5\pm 0.9&  4.6\pm 0.6&  0.0       &  3.9\pm 1.2&0.16\pm0.02&0.38\pm0.02&  6.3\pm 0.6&  3.1\pm 0.7&  3.3\pm 0.8&   1.4\pm  0.7& \mathrm{E} \\             
 97&1.55\pm0.08&  ....      &  0.2\pm 2.3& -3.0\pm 2.2&  3.3\pm 2.4&  2.5\pm 1.5&  0.0       & 13.2\pm 2.6&0.18\pm0.04&0.26\pm0.04&  4.5\pm 1.4& 11.7\pm 1.3& 10.4\pm 1.4&   4.9\pm  0.8& \mathrm{E} \\             
 98&1.82\pm0.05&  ....      & -2.7\pm 0.8& -6.8\pm 1.4&  4.4\pm 1.4&  2.2\pm 0.8&  0.0       & 10.2\pm 1.7&0.16\pm0.02&0.31\pm0.03&  1.5\pm 1.0&  5.7\pm 1.1&  3.0\pm 1.3&  -4.0\pm  1.9& \mathrm{E} \\             
 99&1.76\pm0.04&  ....      & -0.4\pm 0.9& -6.6\pm 0.8&  3.8\pm 0.8&  3.7\pm 0.5&  0.0       &  4.4\pm 1.0&0.11\pm0.01&0.30\pm0.02&  4.2\pm 0.5&  3.6\pm 0.5&  3.2\pm 0.6&   2.5\pm  0.5& \mathrm{E} \\             
100&  ....     &  ....      &  3.0\pm 1.6&-10.1\pm 1.8& -0.9\pm 2.2& 10.0\pm 1.4&  0.0       &  9.5\pm 3.8&0.02\pm0.04&0.21\pm0.05&  0.9\pm 1.8&  6.4\pm 1.8&  3.0\pm 2.2&  -3.4\pm  1.8& \mathrm{Ab-spirals} \\	     
101&  ....     &  ....      &  0.3\pm 2.6& -4.4\pm 2.1&  3.4\pm 2.1&  6.7\pm 1.2&  0.0       & 14.6\pm 2.6&0.18\pm0.04&0.34\pm0.05& 10.2\pm 1.2&  6.4\pm 1.4& -3.3\pm 2.2&   1.8\pm  1.6& \mathrm{E} \\	     
102&  ....     &  ....      &  4.5\pm 1.0& -1.0\pm 1.0&  4.0\pm 1.3&  1.8\pm 0.9& -1.6\pm 0.7&  8.8\pm 1.9&0.03\pm0.02&0.11\pm0.03&  3.6\pm 1.0& -0.6\pm 1.2&  2.3\pm 1.3&  -6.9\pm  1.3& \mathrm{ELG}\\                    
            \noalign{\smallskip}	     		   
            \hline
            \noalign{\smallskip}
            \hline
         \end{array}
     $$ 
}
         \end{table}
\end{landscape}

\large
\chapter*{ \huge Final summary}
\markboth{Final summary}{Final summary}
\addcontentsline{toc}{chapter}{\numberline{}Final summary}
\vspace{2.0cm}
\normalsize

This thesis work is focused on the analysis of the galaxy clusters
ABCG\,209, at z$\sim$ 0.2, which is characterized by a strong
dynamical evolution. The data sample used is based mainly on new
optical data (EMMI-NTT: B, V and R band images and MOS spectra),
acquired in October 2001 at the European Southern Observatory in
Chile. Archive optical data (CFHR12k: B and R images), and X-ray
(Chandra) and radio (VLA) observations are also analysed.

The main goal of this analysis is the investigation of the
connection between internal cluster dynamics and star formation
history, aimed at understanding the complex mechanisms of cluster
formation and evolution.

Clusters at intermediate redshifts (z = 0.1-0.3) seem to represent an
optimal compromise for these studies, because they allow to achieve
the accuracy needed to study the connection between the cluster
dynamics and the properties of galaxy populations and, at the same
time, to span look--back times of some Gyr. On the other hand, up to
now complete and detailed multi-band analysis have been performed only
on local clusters, making it difficult to trace a complete scenario of
the cluster evolution.

The combined multi-wavelength analysis performed on a complex
structure such as ABCG\,209, and the comparison with stellar
population evolutionary models allow to precisely characterize
galaxies belonging to different structures and environments, and to
investigate the subclustering properties. This may give further
insight in the understanding of the physics of clusters formation and
evolution.

The cluster was initially chosen for its richness, allowing its
internal velocity field and dynamical properties to be studied in
great detail, and also for its known substructure, allowing the effect
of cluster dynamics and evolution on the properties of its member
galaxies to be examined.  The substructure is manifested by an
elongation and asymmetry in the X-ray emission, characterized by two
main clumps (Rizza et al. \ct{riz98}). Moreover, the young dynamical
state is indicated by the possible presence of a radio halo
(Giovannini, Tordi
\& Feretti \ct{gio99}), which has been suggested to be the result of
a recent cluster merger, through the acceleration of relativistic
particles by the merger shocks (Feretti \ct{fer02}).

The internal dynamics of the cluster has been studied in Chapter 2,
through a spectroscopic survey of 112 cluster members, suggesting the
merging of two or more subclumps along the SE--NW direction in a plane
which is not parallel to the plane of sky. This study alone cannot
discriminate between two alternative pictures: the merging might be
either in a very early dynamical status, where clumps are still in the
pre--merging phase, and in a more advanced status, where luminous
galaxies trace the remnant of the core--halo structure of a
pre--merging clump hosting the cD galaxy.

Recent studies, based on large surveys, empirically suggest that the
galaxy LF is dependent on cluster environment, with
dynamically-evolved, rich clusters and clusters with central dominant
galaxies having brighter characteristic luminosities and shallower
faint-end slopes than less evolved clusters.

The analysis discussed in Chapter \ref{cap:3} has pointed out that,
although ABCG\,209 is a cD--like cluster, with a cD galaxy located in
the center of a main X--ray peak, the faint--end slope of the LF turns
out to be $\alpha < -1$ at more than 3$\sigma$ c.l. in both V and R
bands, thus reconciling the asymmetric properties of X--ray emission
with the non flat LF shape of irregular systems. Moreover, the bright
galaxies are markedly segregated in the inner 0.2 h$^{-1}_{70}$ Mpc,
around the cD galaxy. This suggests that bright galaxies could trace
the remnant of the core--halo structure of a pre--merging clump.

These results allow to discriminate between the two possible formation
scenarios, suggesting that ABCG\,209 is an evolved cluster, resulting
from the merger of two or more sub--clusters, while the elongation and
asymmetry of the galaxy distribution (and of the X--ray emission) and
the shape of the LFs show that ABCG\,209 is not yet a fully relaxed
system.

The effect of cluster environment on the global properties of the
cluster galaxies, as measured in terms of the local surface density of
\mbox{$R<23.0$} galaxies, is examined in Chapter \ref{cap:4} through
the analysis of the luminosity functions, colour-magnitude relations,
and average colours.

The faint-end slope $\alpha$ shows a strong dependence on environment,
becoming steeper at $>3\sigma$ significance level from high- to
low-density regions. Moreover, the cluster environment could have an
effect on the colour or slope of the red sequence itself, through the
mean ages or metallicities of the galaxies. The red sequence was found
to be \mbox{$0.022\pm0.014$\,mag} redder in the high-density region
than for the intermediate-density region (slope is fixed). In
contrast, no correlation between the slope of the red sequence and
environment was observed. By studying the effect of the cluster
environment on star formation, we find that the fraction of blue
galaxies decreases monotonically with the density, in agreement with
other studies (e.g. Abraham et al. \ct{abraham}; Kodama \& Bower
\ct{kodama}).

The observed trends of steepening of the faint-end slope, faintening
of the characteristic luminosity, and increasing blue galaxy fraction,
from high- to low-density environments, are all manifestations of the
morphology-density relation (Dressler \ct{dre80}; Dressler et
al. \ct{dressler97}), where the fraction of early-type galaxies
decreases smoothly and monotonically from the cluster core to the
periphery, while the fraction of late-type galaxies increases in the
same manner.

We also examined the effect of the cluster environment on galaxies by
measuring the mean colour of luminous (\mbox{$R<21$}) cluster galaxies
as a function of their spatial position. This analysis shows clearly
the complex effects of the cluster environment and dynamics on their
constituent galaxies. ABCG\,209 appears a dynamically young cluster,
with a significant elongation in the SE-NW direction, as the result of
a recent merger with smaller clumps. The reddest galaxies are
concentrated around the cD galaxy (main cluster) and a more diffuse
region, 5\,arcmin to the North, is coincident with the structure
predicted from weak lensing analysis (Dahle et al. \ct{dah02}).  The
effect of the preferential SE-NW direction for ABCG\,209 is apparent
from the presence of bright blue galaxies close to the cD galaxy,
displaced perpendicularly to the main cluster axis, and hence
unaffected by the cluster merger. Furthermore an extension of red
galaxies to the SE which may indicate the infall of galaxies into the
cluster, possibly along a filament. This preferential SE-NW direction
appears related to the large-scale structure in which A209 is
embedded, including two rich (Abell class R=3) clusters A\,222 at
\mbox{$z=0.211$} and A\,223 at
\mbox{$z=0.2070$}, located {\mbox{$1.5^{\circ}$}} (15\,Mpc) in the
NW direction along this axis.

Cluster dynamics and large-scale structure clearly have a strong
influence on galaxy evolution, so we have performed a detailed study of
spectroscopic properties of luminous member galaxies in chapter
\ref{cap:5}.

The analysis of the spectroscopic properties of 102 cluster
members supports an evolutionary scenario in wich ABCG\,209 is
characterized by the presence of two components: an old galaxy
population, formed very early (z$_f \gtrsim $ 3.5), and a younger
(z$_f \gtrsim $ 1.2) population of infalling galaxies. We find
evidence of a merger with an infalling group of galaxies occurred
3.5-4.5 Gyr ago, as indicated by the presence of red galaxies with
strong H$_\delta$ equivalent width (HDS$_{\mathrm{red}}$). As also
indicated by the low value of their velocity dispersion,
HDS$_{\mathrm{red}}$ could be the remnant of the core of an infalling
clump that have experimented a merger with the main cluster, which
induced a secondary starburst in the central part of these
galaxies. The merger of the cluster with this infalling group could
have also powered the observed radio halo. In fact, merger activity
and high ICM temperature may be responsible for producing a radio halo
(Liang et al. \ct{lia00}), because merging can provide enough energy
to accelerate the electrons to relativistic energies, giving rise to
non-thermal emission. After the shock disappeared, radio halos may be
maintained in situ by electron acceleration in the residual
turbulence. Moreover, the correlation between the position of the young
H$_\delta$-strong galaxies and the X-ray flux shows that the hot
intracluster medium triggered a starburst in this galaxy population
$\sim$ 3 Gyr ago.

This study shows clearly the importance of multi--band data, for the
characterization of the different cluster components. It is now
crucial to extend this kind of analysis to other clusters at different
(higher) redshifts and with different dynamical properties. To address
the issue of whether clusters are generally young or old, one needs to
have measurements of subclustering properties for a large sample of
clusters and, at the same time, to precisely characterize cluster
components belonging to different structures and environments inside a
single cluster.

Essential ingredients in this study are i) deep wide--field images in
different bands, in order to investigate photometric properties
allowing the effect of environment to be followed, ii) spectra with
high or intermediate resolution for a large number of galaxy members
spanning magnitudes up to M$^*$+3 and covering a large area of the
cluster and iii) space data in order to study the galaxy morphologies
and the lensing effects. These data have to be complemented with X-ray
and radio data, where available, that allow to investigate the status
of the ICM.

\end{document}